\documentclass[aps,prx,floatfix,reprint,showpacs,nofootinbib,superscriptaddress,longbibliography]{revtex4-2}
\usepackage{graphicx} % Required for inserting images
\usepackage{subfigure}  % For subfigures
\usepackage{lipsum}
\usepackage{babel}
\usepackage[normalem]{ulem}
\usepackage{xcolor}
\usepackage{bm}
\usepackage{amsmath}
\usepackage{braket}

\usepackage[unicode=true,
 bookmarks=false,
 breaklinks=false,pdfborder={0 0 1},backref=false,colorlinks=false]{hyperref}
\hypersetup{
 urlcolor=blue,linkcolor=black,citecolor=black
}

\begin{document}
%\title{First-Principles Hydrogen Simulations Challenge Jupiter's Interior  Models}
\title{A Denser Hydrogen Inferred from First-Principles Simulations Challenges Jupiter's Interior Models}

\author{Cesare Cozza}
\email{cesare.cozza@uzh.ch}
\affiliation{Department of Astrophysics, University of Zürich, Winterthurerstrasse 190, 8057 Zürich, Switzerland}

\author{Kousuke Nakano}
\thanks{C.C. and K.N. contributed equally to this work.}
\affiliation{Center for Basic Research on Materials, National Institute for Materials Science (NIMS), 1-2-1 Sengen, Tsukuba, Ibaraki 305-0047, Japan}
\affiliation{Center for Emergent Matter Science (CEMS), RIKEN, 2-1 Hirosawa, Wako, Saitama, 351-0198, Japan.}

\author{Saburo Howard}
\affiliation{Department of Astrophysics, University of Zürich, Winterthurerstrasse 190, 8057 Zürich, Switzerland}

\author{Hao Xie}
\email{qwexiehao@gmail.com}
\affiliation{Department of Astrophysics, University of Zürich, Winterthurerstrasse 190, 8057 Zürich, Switzerland}

\author{Ravit Helled}
\affiliation{Department of Astrophysics, University of Zürich, Winterthurerstrasse 190, 8057 Zürich, Switzerland}

\author{Guglielmo Mazzola}
\email{guglielmo.mazzola@uzh.ch}
\affiliation{Department of Astrophysics, University of Zürich, Winterthurerstrasse 190, 8057 Zürich, Switzerland}

\date{\today}

\begin{abstract}
First-principle modeling of dense hydrogen is crucial in materials and planetary sciences. Despite its apparent simplicity, predicting the ionic and electronic structure of hydrogen is a formidable challenge, and it is connected with the insulator-to-metal transition, a century-old problem in condensed matter.
Accurate simulations of liquid hydrogen are also essential for modeling gas giant planets.
Here we perform an exhaustive study of the equation of state of hydrogen using Density Functional Theory and quantum Monte Carlo simulations.
We find that the pressure predicted by Density Functional Theory may vary qualitatively when using different functionals.
The predictive power of first-principle simulations is restored by validating each functional against higher-level wavefunction theories, represented by computationally intensive variational and diffusion Monte Carlo calculations.
Our simulations provide evidence that hydrogen is denser at planetary conditions, compared to currently used equations of state. For Jupiter, this implies a lower bulk metallicity (i.e., a smaller mass of heavy elements).
Our results further amplify the inconsistency between Jupiter's atmospheric metallicity measured by the Galileo probe and the envelope metallicity inferred from interior models.
\end{abstract}

\maketitle

\section{Introduction}
\label{s:intro}
Compressed hydrogen (H) has been subject of both theoretical and experimental research for nearly a century\cite{mcmahon_properties_2012,helled2020understanding}.
Most experimental studies have focused on exploring the complex phase diagram of hydrogen, particularly the insulator-to-metal transition, which has been referred to as the holy grail of condensed matter physics\cite{wigner,RevModPhys.76.981,dalladay2016evidence, Diaseaal1579,loubeyre2020synchrotron}.
This phase transition, when occurring in the liquid sector\cite{weir1996metallization,Celliers677,knudson2015direct} of the pressure-temperature ($P$-$T$) diagram, has significant implications for planetary science, as it affects the inferred  planetary internal structure and determines the conditions for magnetic field generation in gas giant planets \cite{2018ASSL..448.....L}.
Since the gas giants in the Solar System are predominantly composed of hydrogen, qualitative discoveries and quantitative assessments of the thermodynamics of this simple element are crucial to constraining the compositions and internal structures of these planets\cite[e.g.,][]{helled2020understanding}.
{
Moreover, the equation of state of hydrogen is also important for understanding icy giants,\cite{2020RSPTA.37890474H} and hydrogen-rich exoplanets \cite{2025A&A...693L...7H}}.
Constraining the equation of state (EoS) of hydrogen has become even more important recently as the Juno and Cassini missions provided accurate measurements of the gravitational fields of Jupiter and Saturn \cite{durante2020,iess2019}. The gravitational fields are used to constrain the planetary density profiles\cite{helledhoward2024}. Given the small uncertainties in the gravity field measurements, minor differences in the hydrogen EoS can result in relatively large differences in the inferred internal structure \cite{Miguel2016,howard2023}. Recent developments have opened new questions in planetary science and have affected the way we view giant planets. Interestingly, Jupiter interior models have found that the planet is inhomogeneous in composition \cite{miguel2022}.  In addition, interior models suggest that Jupiter's deep interior is rather metal-poor \cite{Wahl2017,nettelmann2021,miguel2022,militzer2022,howard2023} while atmospheric measurements from the Galileo probe indicate that Jupiter's atmosphere is enriched by a factor of $\sim3$ compared to solar composition. This discrepancy could have several explanations, one of which is that the behavior of hydrogen is not yet understood to within 1\% accuracy \cite{stevenson2020}.
%It is, however, important to understand the sensitivity of the results to the used EoS.
Given the precision of the gravity measurements, it is crucial to ensure that theoretical uncertainties  are compatible with these measurements. As a result, obtaining an accurate H-EoS is of great importance.
\par
{Experiments can only provide partial information. They can track the location of phase transitions, mostly at room temperature, but sometimes providing an inconsistent picture\cite{Celliers677,knudson2015direct,dalladay2016evidence, Diaseaal1579,loubeyre2020synchrotron}. Shock-compression experiments can also be used to benchmark the EoS, but only in a tiny corner of the whole phase diagram.\cite{PhysRevLett.108.091102,2004PhRvB..69n4209K,2009PhRvB..79a4112H} Since performing experiments at
 at planetary conditions is exceptionally hard}, numerical models must be used.
The range of pressures and temperatures encountered from the planetary surface to the center varies by several orders of magnitude, and therefore different theories are combined.
 The choice of method or theory for calculating the EoS depends on the temperature and density conditions.
From ambient to $ \approx$ 0.1 g/cc density, hydrogen is present in a weakly interacting molecular form, and the Saumon Chabrier van Horn (SCvH) EoS \cite{scvh1995} is utilized. Hydrogen becomes a challenging system at higher density, where pressure-induced dissociation occurs,  hence, one needs to treat the system as a genuine many-electron quantum mechanical problem.
In this case, \textit{first-principle} (or \textit{ab initio}) molecular dynamics (MD) simulations are used.

Most methods provide an EoS that relates the  pressure $P$ and density $\rho$ with temperature $T$. For instance, MD simulations fix $\rho$ and $T$ and output the corresponding $P$. While this information could be sufficient to explore  phase diagrams, planetary models, which are often adiabatic,  also require knowledge of the entropy, $S$. This information is computed through post-processing of the MD data.
However, previous studies that used different methods have led to rather different inferred entropy {even for pure hydrogen (i.e. without mixing with other elements), and} considering compatible $P$ vs $\rho$ input data\cite{Nettelmann_2012,militzer2013ab,Miguel2016}.
These differences significantly affected the predicted internal structure of Jupiter where the inferred masses of heavy elements could differ by a factor of $\sim$5. \cite{Miguel2016,Miguel2018erratum}
Recently, in Ref~\cite{xie2025}, we addressed this aspect conclusively, showing that the source of entropy error stems from thermodynamic inconsistencies at the boundaries, where different theories are joined to produce a single EoS. By solving this major methodological issue, we can now address the final fundamental weaknesses of the various hydrogen EoSs, namely, the choice of the electronic structure theory.

There have been several widely used  \textit{ab-initio} hydrogen EoSs for planetary modeling, namely the Chabrier, Mazevet and Soubiran (CMS19)\cite{CMS2019}, Militzer and Hubbard (MH13)\cite{militzer2013ab}, and the Rostock EoS (REOS2\cite{Nettelmann_2012}, REOS3\cite{REOS_2014}).
All these EoSs have a common origin,
 as their  \textit{ab initio} part
 is based on density functional theory (DFT), with the specific choice of the Perdew-Burke-Ernzerhof (PBE) functional as the exchange-correlation (xc) functional.

While in principle DFT is an exact theory (if the exact xc functionals were  known), in practice, the choice of an approximation for the xc functionals  introduces some errors\cite{burke_perspective_2012}.
Although the PBE functional is the most widely used in materials science, due to an excellent balance between providing reasonable results and computational cost, there is no guarantee that it is suitable for a system as peculiar as dense hydrogen.
In fact, it is known that the choice of functional has a significant impact on predicting the phase diagram at low temperatures, such that resorting to higher levels of theory is considered fundamental.\cite{azadi_fate_2013,knudson2015direct,mazzola2018,pierleoni2016liquid}
{Previous works attempted at identify the best performing xc functional against an higher-level theory\cite{PhysRevB.89.184106} or conductivity measurements\cite{PhysRevB.98.174110}.}

In this work, we explore the effect of the functionals on the H-EoS  and show that other  reasonable xc functionals yield pressure outputs that may differ
by $\pm 20\%$, at planetary conditions, compared to PBE, i.e., compared to the most-commonly used EoS tables in planetary science.
This is far exceeding the level of accuracy required for planetary modeling.
In the absence of an experimental benchmark, valid over the whole range of $P$ and $T$ of interest, the only possible way forward to identify the most accurate xc functional is through a comparison with a higher level of theory, quantum Monte Carlo (QMC).

This manuscript is organized as follow. In Sect.~\ref{s:methods} we outline the first-principle methods to compute EoS, In Sect.~\ref{s:resultbench} we compare several DFT functionals and identify the best performing one using QMC calculations.
{We also compute the liquid-liquid transition boundary and the Hugoniot line, and compare with experiments and previous simulations.}
In Sect.~\ref{s:entropy} we use this information to compute a thermodynamically consistent EoS, including entropy, which is then used in the interior models of Sect.~\ref{s:models}.
We discuss the implications for Jupiter modeling in Sect.~\ref{s:conclu}.

\section{Ab-initio electronic structure methods}
\label{s:methods}

The majority of hydrogen in gas giant planets exists in a density regime where a fully quantum treatment of the electrons is essential\cite{mcmahon_properties_2012,Bonitz_2024}.
%The interacting $N$-electron problem is the central theme in materials science and quantum chemistry, and it inevitably requires approximations to be solved \cite{martin_Book2016}.
The key concept with which we want to begin is that being \textit{ab initio}, or \textit{first-principle}, does not mean that a method is exact; rather that it does not require \textit{ad hoc} fits to experimental data. In this Section, we outline the theoretical foundations and the various approximations that must be made to address the electronic structure problem.
The starting point is the (non-relativistic) Hamiltonian of $N$ electrons and $N$ protons in the so-called first quantization formalism:
\begin{align}
    \hat H &= \hat H_e + \hat H_p + \hat H_{ep} ,\label{eq:hamiltonian-h}\\
    \hat H_e &= -\frac{1}{2}\sum_{i=1}^{N} \nabla_i^2 + \sum_{i<j}^{N} { \frac{1}{  |\textbf{r}_i-\textbf{r}_j|}}\,,\label{eq:hamiltonian-e}\\
    \hat H_p &= -\frac{1}{2M}\sum_{i=1}^{N} \nabla_i^2 + \sum_{i<j}^{N} { \frac{1}{|\textbf{R}_i-\textbf{R}_j|}}\,,\label{eq:hamiltonian-p}\\
    \hat H_{ep} &= - \sum_{i, j=1}^N { \frac{1}{|\textbf{r}_i-\textbf{R}_j|}}\,,\label{eq:hamiltonian-ep}
\end{align}
{in  atomic unit}, where $ \textbf{r} = \{\textbf{r}_1,\dots \textbf{r}_N\}$ and ${\bf x} = \{ {\bf R}_1, {\bf R}_2, \cdots, {\bf R}_N\}$, are the electrons and protons coordinates, respectively.
Due to the large mass ratio $M = m_p / m_e$ between the proton and electron, the problem can be greatly simplified by following the ground-state Born-Oppenheimer (BO) approximation, where the electrons are considered to be in their instantaneous ground state, and the protons ${\bf x}$ move according to the corresponding potential energy surface $E_e[{\bf x}]$.
%Notice that the BO holds also in case of a quantum description of the nuclei. In this case, one could adopt path-integral formalism to describe the quantum delocalization of the protons at low temperatures.\cite{marx2009ab,van2021isotope}
In this work, we further treat the protons as classical objects, which is valid for temperatures above 2000 K as relevant to planetary science applications~\cite{van2021isotope}. While our EoS extends also to lower temperatures, where nuclear quantum effects may not be negligible~\cite{zaghoo2018striking} (e.g., at 500 K), we do this to maintain consistency with other EoS tables and to analyze the temperature dependence of the variation between different xc functional  predictions, which can be of independent interest for the high-pressure community.

Even within the Born-Oppenheimer approximation, solving the fixed-nuclei Schrödinger equation
%\begin{equation}
%\label{eq:hamiltonian-onlyelectrons}
%    (\hat H_e + \hat H_{ep}[{\bf x}]) ~\psi({\bf r}) = E_e[{\bf x}]~ %\psi({\bf r}),
%\end{equation}
%where $\psi({\bf r})$ is a $N$-electron non-separable wavefunction, obeying the fermionic anti-symmetry principle,
requires computational resources that grow exponentially with $N$\cite{schuch2009computational}.
%Previous works have shown that one needs to reach a system of about a hundred hydrogen atoms, with appropriate schemes to mitigate electronic finite-size effects, to attain a converged EoS\cite{lorenzen2010}.
Therefore, simulating a system of this size exactly can only be achieved using approximations. Several methods exist, with the key difference being that some methods features uncontrolled approximations, while others can be systematically improved.

\subsection{Density Functional Theory}
Density Functional Theory is the most commonly used \textit{ab initio} simulation method for electronic systems with  $N>100$ electrons.\cite{burke_perspective_2012,dft2015RMP}
Its widespread use relies on a relatively simple theoretical framework and the development of several software packages that allow fast and reproducible calculations~\cite{lejaeghere2016reproducibility,Bosoni_2023}.
The fundamental idea behind DFT is that it is possible to describe the ground-state properties of an $N-$ electron system in terms of the three-dimensional electronic density $\rho_e$ alone, without
any explicit reference to the $3N$-dimensional many-body wave function.\cite{Hohenberg_Kohn1964, Kohn_Sham1965}
%Therefore, one could search for the solution of Eq.~\eqref{eq:hamiltonian-onlyelectrons} in the space of antisimmetrized products of single particle orbitals $\phi({\bf r}_j)$, a much simpler task. Indeed, even if the exact solution of Eq.~\eqref{eq:hamiltonian-onlyelectrons} is a non-separable wavefunction, the wavefunction constructed from the orbitals $\phi({\bf r}_j)$ could yield the same exact electronic density.
%To obtain the single-particle orbitals it is sufficient to work with a one-body Hamiltonian of the form:
%\begin{equation}
%\label{eq:hamiltonian-dft}
%    \hat H_{\textrm{DFT}}[{\bf x}] = -\frac{\hbar^2}{2m_e}\sum_{i=1}^N \nabla_i^2 + \hat H_{ep}[{\bf x}] + \hat V_{\mathrm{xc}}[\rho_e].
%\end{equation}
In practice, in DFT we replace the
%Notice the replacement of the
explicit two-body electron-electron Coulomb term in Eq.~\eqref{eq:hamiltonian-e} with a potential $V_{\mathrm{xc}}$ which should only depends on the electron density.
This approach would be exact if
%the functional form key term in Eq.~\eqref{eq:hamiltonian-dft},
the exchange-correlation functional  $V_{\mathrm{xc}}$ is known.
Unfortunately, this term can only be approximated, such that the DFT solution for the electronic density, and the other properties, is not exact.
For more details, we refer to comprehensive reviews of the DFT method\cite{martin_ElectronicStructure,dft2015RMP}.

There are various theories, each offering different prescriptions for constructing $V_{\mathrm{xc}}$. Each of these is labeled by the name of the xc functional .
PBE functional\cite{PBE1996}, introduced 30 years ago, is the most used in materials science, and the only one used so far to create EoS tables for hydrogen\cite{REOS_2014,CMS2019} and hydrogen-helium\cite{militzer2013ab, CMS2021}
More recent functionals
include the BLYP functional\cite{Becke1988_BLYP, Lee1988_BLYP}, which is most commonly used in chemical problems, functionals with van-der-Waals corrections like vdW-DF\cite{Dion2004_vdw-df,Thonhauser2015_vdw-df} and vdW-DF2,\cite{Lee2010_vdw-df2} and the recently released strongly constrained and appropriately normed (SCAN) functional\cite{Sun2015_SCAN}.
These functionals have been applied to study of dense hydrogen (and hydrogen-helium mixtures)\cite{PhysRevB.89.184106, van2021isotope, PhysRevB.102.174109,PhysRevLett.120.115703,azadi2017role, lu2019towards, hinz2020fully}.

It is crucial to note that (1) the sophistication level or release date of a functional does not necessarily mean it is better than its predecessors, and (2) in general, the accuracy of a functional is system-dependent\cite{burke_perspective_2012}
For most materials science problems, precise experimental data is available, so one can \textit{a posteriori} select which xc functional  performs best for a given problem. This is not the case for hydrogen at order of 100 GPa pressures, where experiments can only provide incomplete -and sometimes contested- results, such as phase boundaries.\cite{Diaseaal1579,Celliers677, knudson2015direct,zaghoo2016evidence,zaghoo2017conductivity, ji2019ultrahigh}

Unfortunately, the choice of the xc functional  has a qualitative impact on the phase diagram of hydrogen and quantitatively alters its EoS. The predicted phase boundaries between insulating solid phases\cite{azadi_fate_2013}, as well as the insulator-to-metal transition\cite{morales_nuclear_2013,knudson2015direct,Geng2019}, can vary by up to 100$\%$ depending on the functional chosen.
Given this body of evidence, there is no rigorous justification for using a PBE-based EoS because (1) PBE is not expected to be the best-performing functional for this system, since it overestimates the molecular bond lenght by $\sim5\%$, and overstabilizes the metallic phase\cite{morales_nuclear_2013,PhysRevB.89.184106,clay2016benchmarking}, and (2) numerical predictions obtained with PBE do not align with experimental results in the solid region\cite{azadi_fate_2013}.

Details of our DFT setup are provided in Appendix~\ref{app:dft}.

\subsection{Quantum Monte Carlo}

While DFT is an incredibly successful technique, it faces challenges describing strong electronic correlations\cite{cohen2011challenges,wagner2012reference,Motta_benchmark}
and non-covalent interactions\cite{dubecky2016noncovalent,doi:10.1063/1.4944633}, due to the approximate functional form to describe the xc effects between the electrons.
For this reason, and especially in high-pressure physics, where experiments do not represent a quantitative benchmark, DFT results should be validated against higher-level theories.

Quantum Monte Carlo is among one of these methods as
it represents the most reliable choice available for medium-sized system (e.g. order of 100-200 electrons) \cite{foulkes_quantum_2001,dubecky2016noncovalent}.
The  main advantages of QMC are: (1)
QMC relies on a many-body theory with a natural and explicit description of electron correlations, and its accuracy is systematically improvable.
(2) QMC gives accurate results exhibiting, at the same time, a comparable scaling of computational cost with system size with {(Kohn-Sham)} DFT (although usually with a much larger prefactor). (3) New supercomputer architectures are becoming more and more suitable for intrinsically parallel techniques such as Monte Carlo rather than for  DFT or quantum chemistry methods.\cite{nakano2020turborvb}
Therefore, the system sizes that can be simulated by QMC are substantially larger compared to the ones of Couple Cluster (CC) theory, which is considered the \textit{gold standard} (in its CCSD(T) formulation\cite{szabo2012modern,doi:10.1021/cr1000173}) for quantum chemistry calculations.
This enables the possibility of performing electronic structure simulations on bulk systems comparable with DFT, and to benchmark xc functionals.

In this work we employ two QMC techiques: Variational Monte Carlo (VMC) and Lattice Regularized Diffusion Monte Carlo (LRDMC)~{\cite{casula2005lrdmc}} implemented in the TurboRVB package~\cite{nakano2020turborvb, nakano2023turbogenius}.
The first is a purely variational method, while the second is a projection method.
The central object of VMC is the trial-wave function $\psi_{\boldsymbol{\theta}} ({\bf r})$ which depends explicitly on the $3N$-dimensional vector of electron positions, ${\bf r}$ (notice that for hydrogen the number of electrons is the same as the number of protons), and parametrically on a set of optimizable parameters $\boldsymbol{\theta}$. The wavefunction also implicitly depends on the ionic coordinates, ${\bf x}$, as it is constructed using an atom-centered basis set.
Appendix \ref{app:vmc-and-lrdmc} presents further details on the functional form of the trial state $\psi_{\boldsymbol{\theta}} ({\bf r})$.

The parameters  $\boldsymbol{\theta}$ are optimized in order to minimize the energy:
\begin{equation}
    E_{\boldsymbol{\theta}}[{\bf x}] = \frac{\langle \psi_{\boldsymbol{\theta}} ({\bf r; {\bf x}}) | H | \psi_{\boldsymbol{\theta}} ({\bf r}; {\bf x}) \rangle}{\langle \psi_{\boldsymbol{\theta}} ({\bf r; {\bf x}}) | \psi_{\boldsymbol{\theta}} ({\bf r}; {\bf x}) \rangle} \ge E_0 [{\bf x}]~,
\end{equation}
where the expectation value needs to be evaluated as $3N$-dimensional integral over the coordinate ${\bf r}$, and $E_0$ is the exact ground state energy.
This high-dimensional integral cannot be computed using quadrature methods but instead with the Monte Carlo method, hence the name of the technique.

The main difference with the DFT framework is that QMC features the exact Hamiltonian (Eq.~\ref{eq:hamiltonian-ep}) while the approximations are transferred to the wavefunction.
Crucially, the trial wave function can be systematically improved both in its functional form and in the number of parameters. For example, one could enhance the description of 2-, 3-, or 4-body electron-ion correlations, or expand the basis set used to define the wave function (see Appendix ~\ref{app:vmc-and-lrdmc}). This mirrors the popular machine learning methods, where accuracy increases with the complexity of the network or model.

Following this parallel with ML,\cite{mazzola_finite-temperature_2012,Hermann_2020,PhysRevB.110.035108,PhysRevResearch.2.033429} VMC also needs efficient and stable methods to optimize the parameters in order to be practical. In our case, the wave function features 13315 tunable parameters (only in the Jastrow factor), which are optimized using the Stochastic Reconfiguration method of Refs.~\cite{sorella1998sr, sorella2007sr}.
The optimized trial wave function obtained from a VMC calculation serves as the starting point for the LRDMC method.

The LRDMC method provides even more accurate results and offers another way to systematically improve the many-body wave function, and has been employed successfully in Refs.~{\cite{Lorenzo.hydrogen.2023, Raghav2023.g2set.benchmark}}. The computational settings employed in the calculations are provided in Appendix ~\ref{app:vmc-and-lrdmc}. Further details of the LRDMC method can be found in Ref.~{\onlinecite{casula2005lrdmc, Becca2017, nakano2020turborvb}}.

Concerning the computational cost, VMC and LRDMC electronic structure calculations are more expensive than DFT calculations by more than an order of magnitude.
Moreover, LRDMC with the lattice discretization parameter of 0.3 bohr is 3 times more expensive than VMC, and therefore has been used mostly to validate the accuracy of VMC, over the full range of density and pressure.
Additionally, direct LRDMC simulations will be used in the range of 0.2–0.5 g/cc and 2000–5000 K for the xc functional benchmarking. This is because the VMC pressure estimate is affected by an error of about 1 GPa compared to LRDMC. While this value is negligible at moderate to high pressure, it cannot be neglected at low pressures. For this reason, we switch to LRDMC in this region of the phase diagram.

%In summary, we computed energies and pressures of 4512 and 832 structures by VMC and LRDMC, respectively.
%These atomic structures are sampled using PBE-driven molecular dynamics (see procedure outlined in Sect.~\ref{ss:bench}). Notice that, since the pressures were computed numerically by fitting the potential energy surfaces computed with 7 volumes for each structure as described in Appendices ~\ref{app:difficutly-in-qmc} and \ref{app:qmc-validation}, we actually computed seven times larger numbers of structures (i.e., 31584 and 5824 structures by VMC and LRDMC calculations, respectively).
For this project, we perfomed 31584 and 5824 single structure calculation by VMC and LRDMC calculations, respectively (see Sect.~\ref{ss:bench}, Appendices ~\ref{app:difficutly-in-qmc} and \ref{app:qmc-validation}).
These calculations were carried out on the Fugaku supercomputer with total CPU costs of approximately 57 million and 25 million CPU hours for the VMC and LRDMC parts, respectively. It is clear that brute-force MD simulations using QMC (while technically possible\cite{mazzola_unexpectedly_2014,zen2015ab,mazzola2018}) would be impractical in the full $\rho$-T range, as they would require hundreds of times more computational resources (note that each DFT-MD simulation at a fixed density and temperature involves around 10,000 electronic structure calculations).

A possible strategy could be to use the QMC dataset as a training set for a machine learning potential\cite{cheng2020evidence, tirelli2022high, PhysRevLett.130.076102,goswami2024hightemperaturemeltingdense}. However, there are outstanding technical challenges\cite{nakano2024efficient} and we hope to address this in future studies.

{
Finally, we note that QMC encompasses a broader range of methods. Even within the same VMC approach, several groups may employ different trial states. For instance, the Slater determinant can be augmented using highly expressive Jastrow operators, as in our work, or by employing backflow transformations, as in the Coupled electron-ions Monte Carlo (CEIMC) method.\cite{pierleoni_coupled_2004,pierleoni2016liquid}
Most VMC calculations require a DFT initialization (see e.g.\cite{nakano2020turborvb,pierleoni2016liquid}).
Neural-network quantum states have also recently been considered as ansatze for the problem of hydrogen.\cite{PhysRevLett.131.126501,linteau2025universal}
}

\begin{figure*}[t]
    \centering
    \includegraphics[width=\textwidth]{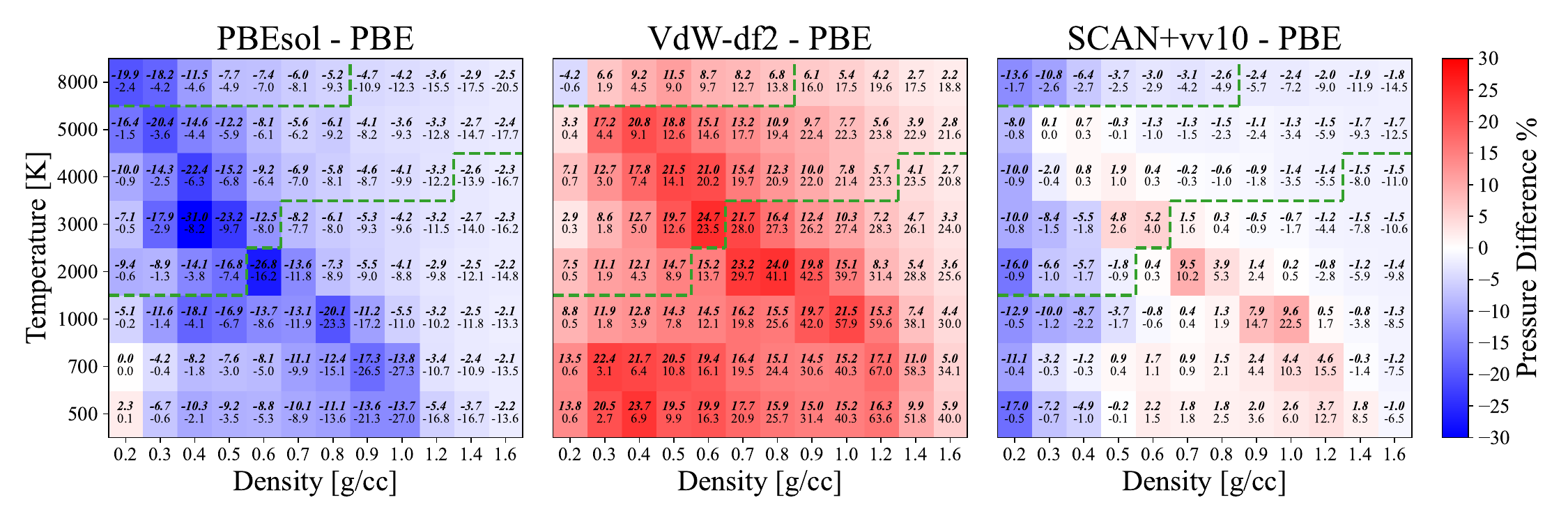}
    \caption{\textbf{Equation of state of hydrogen using selected DFT approximations.}
    We plot the difference between the EoS calculated with three different xc DFT potentials, PBEsol (left), vdW-DF2 (middle), and SCAN+vv10 (right) versus the REOS3 data (which is obtained with the PBE functional).
    In each cell we report the relative (above, in $\%$) and the absolute (below, in GPa) differences.
    The color code refers to the relative values. The green dashed lines highlight the `planetary' region of temperatures and densities which is of particular interest for interior models (see Sect.~\ref{ss:bench}).
    }
    \label{fig:DFTMD}
\end{figure*}
\subsection{Molecular dynamics}

In the previous sections, we introduced two different computational methods to solve the many-body electronic problem at fixed ionic configurations. The output of these methods includes not only the potential energy surface, $E_e[{\bf x}]$, but also the electronic forces and the electronic contribution to the pressure, which is the dominant factor at the densities considered here.

To obtain the EoS, we need to average over the ionic positions.
This task is obtained using molecular dynamics, where the motion of $N$ interacting ions is numerically integrated based on the forces acting between them.\cite{marx2009ab} By employing an appropriate thermostat, the canonical ensemble can be simulated, at a constant temperature $T$.\cite{allen2017computer} For sufficiently large systems and long simulation times, thermodynamic properties such as internal energy and pressure can be determined by averaging their values over the trajectory, and statistical methods can be used to calculate reliable error bars. The forces between ions are computed at each step of the dynamics from the computed electronic structure.
It is important to note that different levels of theory (or  exchange-correlation functionals) produce different forces, which in turn affect the equilibrium distribution generated.

%After an equilibration period, the MD sequentially generates configurations ${\bf x}_i$ distributed according to the Boltzmann weights $ p ({\bf x}_i)\propto e^{-\beta E ({\bf x}_i)}$ ($\beta=1/T$, in atomic units).
%We can calculate the energy $E ({\bf x}_i)$, the ionic forces ${\bf F} ({\bf x}_i)$, and the pressure $P ({\bf x}_i)$ for each configuration ${\bf x}_i$.
Indeed, all the quantities above are understood having also a subscript $\{A,B,C,\cdots\}$ indicating the level of theory we are using to calculate them.
At each thermodynamic point $(\rho, T)$ we will obtain different average values of energy $\langle E \rangle_A$, $\langle E \rangle_B$ etc., and pressure $\langle P \rangle_A$, $\langle P \rangle_B$ etc., not only because the output $E_A({\bf x}_i)$ and $P_A({\bf x}_i)$ is different from $E_B({\bf x}_i)$ and $P_B({\bf x}_i)$ at the same ionic configuration ${\bf x}_i$, but also because the generated configurations will also be different.
{Notice that here we always assume the NVT ensemble, namely the simulations are conducted at fixed volume and temperature, and have the pressure as output.}

%To summarize, we will run different MD using several xc-correlation functionals, and compute different EoS which correspond to these different electronic structure theories.
%Moreover, we will use a dataset of uncorrelated configurations  $\{\bar{\bf x}\}_\textrm{PBE}^{(\rho, T)}$, for each $(\rho, T)$ point, obtained at the DFT-PBE level of theory and will recompute the energies and pressure using our QMC method as well as 15 other different DFT xc functionals.
%This will allows us to select the best performing xc functional against QMC data.

\begin{figure*}[ht]
    \centering
    \includegraphics[width=\textwidth]{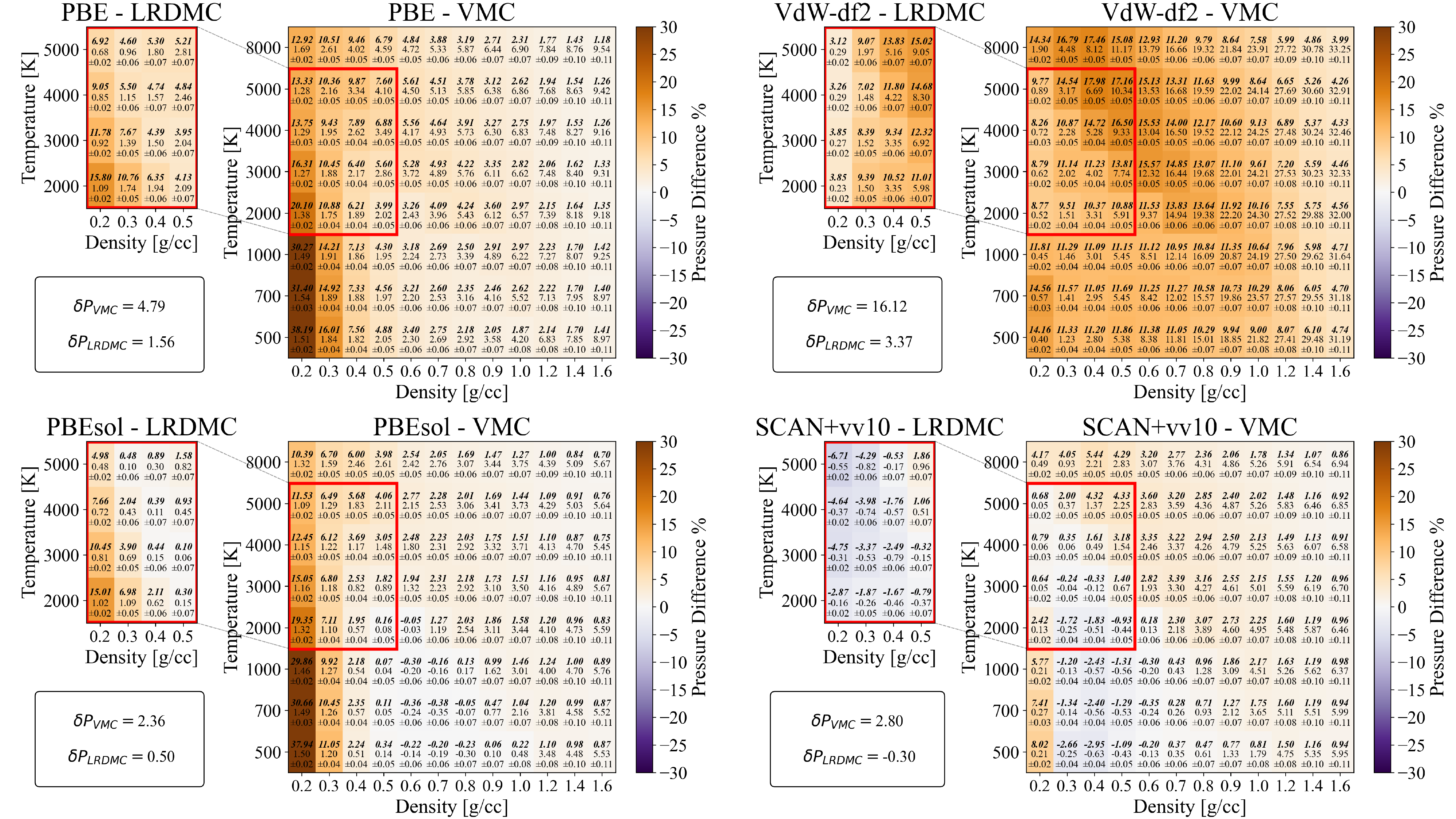}
    \caption{\textbf{Benchmarks of selected DFT xc functionals against VMC and LRDMC pressure data.}
    We plot the pressure error as defined in Eq.~\ref{eq:pressure_metric} of PBE, PBEsol, vdW-DF2, and SCAN+vv10 compared to VMC (full table) and LRDMC (left insets) benchmarks.
     In each cell we report the relative (above, in $\%$) and the absolute (below, in GPa, with associated statistical error) differences.
    The color code refers to the relative values. Lighter colors indicate smaller errors.
   The values in the box are the error averaged only over the `planetary' region, LRDMC where it is present and VMC otherwise. This is the same region highlighted by the green dashed lines in Fig.~\ref{fig:DFTMD}. }
    \label{fig:heatmapQMC}
\end{figure*}

\section{Selecting the most accurate equation of state}
\label{s:resultbench}

\subsection{Direct DFT-MD simulations with various xc functionals}

We perform direct MD simulations using four DFT theories: PBE, PBEsol\cite{PBEsol}, vdW-DF2, and SCAN+vv10\cite{Peng2016_scan+vv10}, across a density range of 0.2–1.6 g/cc and a temperature range of 500–8000 K. We use a 12-point density grid and an 8-point temperature grid, resulting in 96 ($\rho$, T) thermodynamic points.
The first set of simulations, using PBE, is primarily used to validate our setup against REOS3, which also employs the same xc functional  (in this range) but a different DFT code and particle number. We find that our PBE EoS is perfectly consistent with the REOS3 pressure output.

We then run the 96 MD simulations for each of the other three functionals and compare the new EoS against REOS3 in Fig. 1. We observe that the changes in the EoS are significant. The PBEsol functional, a modified version of PBE designed to improve the description of solids\cite{PBEsol}, and that could be better for high-pressure materials\cite{PhysRevB.79.155107}, predicts a denser liquid across the entire range, with peaks of up to 20-30$\%$ denser liquid at 3000-4000 K and 0.4 g/cc (i.e., along Jupiter's adiabatic curve).
For clarity, PBEsol yields a lower pressure than REOS at fixed density, therefore by inverting the $P-\rho$ relation, it predicts a denser liquid at fixed pressure.
In contrast, the vdW-DF2 functional predicts a much lighter fluid across the table. In this case, we find a 20$\%$ lighter fluid at 4000 K and 0.5–0.6 g/cc.
Finally, the SCAN+vv10 functional shows smaller variations compared to REOS3, predicting either a denser or lighter liquid in different regions of the phase diagram.

It is noticeable that the largest deviations from REOS3 occur along a diagonal line in the center of the plot. This is due to the different positions of the metal-to-insulator transition predicted by the various functionals. For example, PBE (which is the basis of REOS3) underestimates the transition pressure compared to vdW-DF2\cite{knudson2015direct}. Between the two predicted phase boundaries, PBE stabilizes an atomic liquid, while vdW-DF2 predicts a molecular one, leading to much differing EoS.

It is also interesting to observe the pressure changes in a subset of ($\rho$, T) values relevant to gas giants. While PBEsol (vdW-DF2) consistently predicts a denser (lighter) fluid, the SCAN+vv10 functional shows a slightly denser fluid across most of the range, except for the density interval between 0.3 and 0.6 g/cc, at about 4000 K, where it predicts a lighter material.

 Note that all four functionals considered so far, PBE, PBEsol, vdW-DF2, and SCAN+vv10 are plausible choices. Indeed, PBE has been the most widely used for describing the solid and liquid phase diagrams in the 00's and the early 10's of this century\cite{bonev2004quantum,morales_evidence_2010,scandolo2003liquid, lorenzen2010, tamblyn2010structure}.
Later, van der Waals corrected functionals, such as the vdW-DF2, has been introduced among the employed for studying this and related systems\cite{morales_nuclear_2013, knudson2015direct, azadi2017role, PhysRevLett.120.115703}.
Finally, the newly introduced family of SCAN functionals, has also been used for hydrogen yielding a liquid-liquid phase boundary which seems to be more in agreement with QMC predictions compared to other functionals.\cite{hinz2020fully,bergermann2024nonmetal}
In this regard, while previous DFT and QMC works have focused on determining the liquid-liquid phase boundary, the calculation of a complete EoS table using functionals different that PBE is a new contribution of the manuscript.
For example, it is known that SCAN improves the description of electronic gaps over PBE. However, this fact may simply improve the prediction for the metallization transition with no guarantees in improving uniformly also the EoS.

\subsection{Benchmarking results}
\label{ss:bench}

\begin{figure}[t]
    \centering
    \includegraphics[width=1.0\columnwidth]{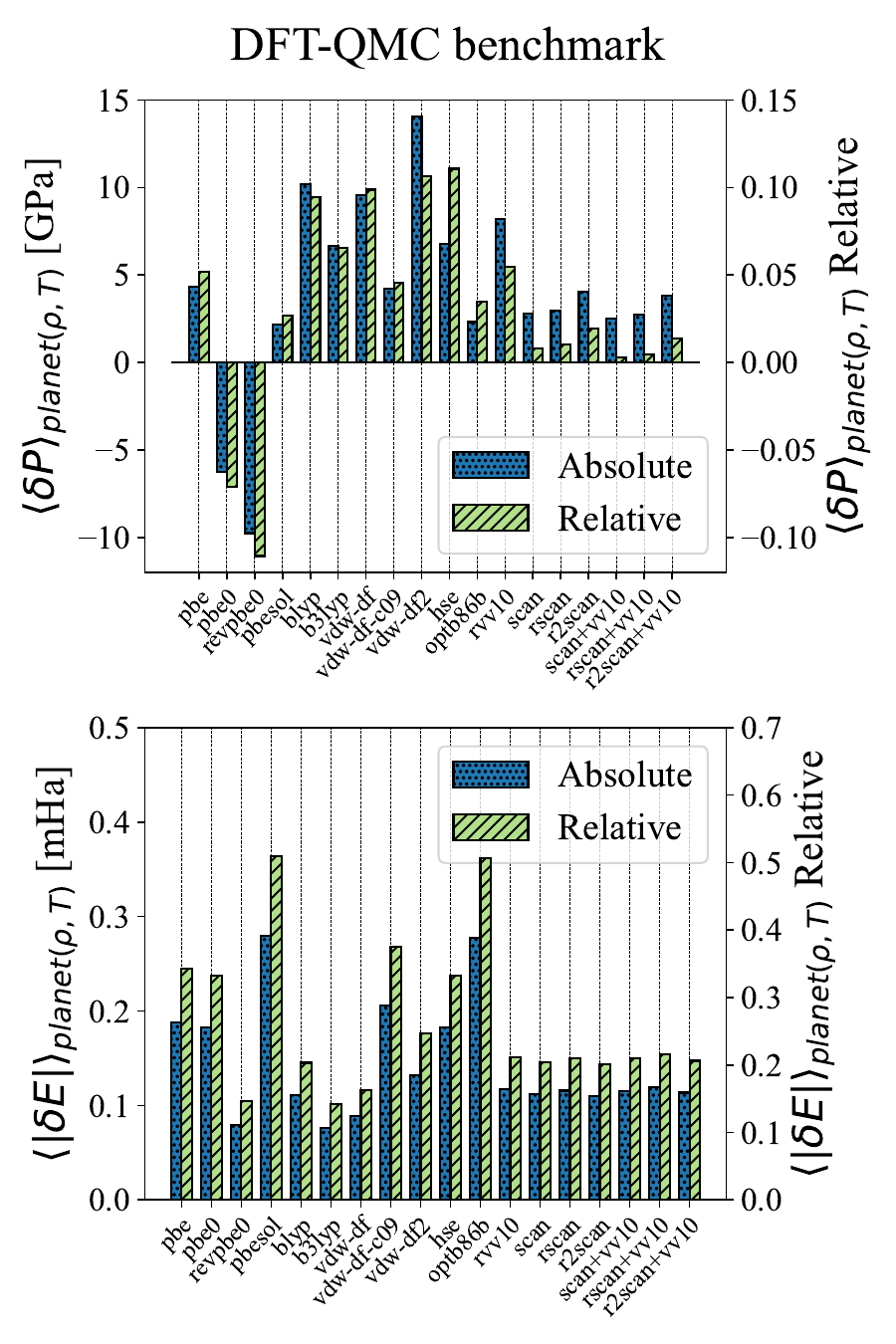}
    \caption{\textbf{Pressure and energy benchmarks.} Results for the benchmark of Pressure and Energy per atom from all the set of DFTxc against QMC results computed as shown in Eq.\ref{eq:scoreP}. The set of $(T,\rho)$  over which the benchmark has been computed is the one highlighted in Fig.\ref{fig:DFTMD} by the green dashed line, with LRDMC results where they are present and VMC elsewhere.
    }
    \label{fig:allbenchs}
\end{figure}

It is not possible to choose a functional in the absence of reliable experimental results or comprehensive data from higher-level theories such as QMC.
In 2014, Clay and collaborators benchmarked pressure, energy, and bond lengths of selected solid structures, as well as representative liquid structures, of several xc functionals against QMC data \cite{PhysRevB.89.184106}.
The liquid structures used in Clay et al.'s work were sampled at T = 1000 K and under three density regimes, corresponding to molecular, atomic, and mixed fluids.
The main difference compared to our work is that (1) we feature a much larger dataset in liquid phase, utilizing 47 uncorrelated 128-atoms structures for each of the 96 thermodynamic conditions in our EoS table. This provides a better understanding of DFT errors across the entire range, especially for planetary science applications.
(2) We test more functionals including those not developed at that time, and this is crucial given that SCAN is the best performing one.
(3) Additionally, another significant difference is in the QMC setup. Clay et al.\cite{PhysRevB.89.184106} employed a different functional form for the trial wavefunction in VMC, a different projective QMC approach, and a smaller system size of 54 protons compared to our current study. Therefore, we expect quantitative differences that may affect the results.

Their main finding of Ref.~\cite{PhysRevB.89.184106} was that, at least for the set of functionals considered at the time, no clear `winner' could be identified, as the performance of most DFT functionals depended on the property being studied and the thermodynamic conditions (e.g., metallic or insulating). Moreover, they concluded that PBE did not perform well for the energetics and properties of molecular bonds.
Finally, other groups have employed QMC at the fixed-node DMC level, focusing on solid structures.\cite{azadi2017nature} This information have also been used to assess the predictive power of various xc functionals. For instance, that dataset suggested that vdW-DF2 perform better than vdW-DF \cite{azadi2017role}.

In this work, for consistency, we do not use previous QMC data but we construct from scratch our validation dataset, following the procedure of Ref.~\cite{PhysRevB.89.184106}, but using a much finer mesh of density and temperatures.
For each of the 96 $\rho$ and $T$ combinations, we extract $M=47$ uncorrelated structures ${\bf x}_i$ from the PBE molecular dynamics and calculate the VMC energy and pressures.
{These structures are not relaxed using any of the exchange-correlation functionals.}
At lower densities, from 0.2 to 0.5 g/cc, and temperatures between 2000 and 5000 K (i.e. for a total of 16 thermodynamic points) we also calculate the LRDMC outputs. This is needed because at low densities the pressure difference between DFT data and VMC falls within the estimated accuracy of VMC, so a more accurate QMC method is needed.
The total number of structured considered is 4512.
We choose not to utilize the VMC forces as benchmark as they can be affected by a residual error~{\cite{Tiihonen2021SCE, nakano2022SCE, nakano2024efficient}}, as explained in Appendix~\ref{app:difficutly-in-qmc}. This error may not be negligible if our task is to provide a quantitative benchmark of DFT, given that we also expect several xc functional  being competitive with each other.

We benchmark {18} functionals, belonging to the set $\mathrm{DFTxc} = \{$ PBE, PBEsol, PBE0\cite{Perdew1996_pbe0}, {revPBE0}\cite{revPBE0}, BLYP, {B3LYP}\cite{B3LYP}, vdW-DF, vdW-DF2, vdw-DF2-c09\cite{Berland2017_vdW-c09}, {optb86b}\cite{optb86b}, rVV10\cite{Vydrov2010_vv10,Sabatini2013_rvv10}, HSE\cite{Heyd2003_HSE}, SCAN, rSCAN\cite{Bartok2019_rscan}, r2SCAN\cite{Furness2020_r2scan}, SCAN+vv10, rSCAN+vv10, r2SCAN+vv10 $\}$.
For each ($\rho,T$) condition, we introduce the following metrics.
The pressure error of a given functional against QMC is defined as
\begin{equation}
\label{eq:pressure_metric}
    \delta P^{(\rho,T)} = \frac{1}{M}  \sum_{{\bf x}_i \in \mathcal{S}^{(\rho,T)}} \left( P^{\mathrm{DFTxc}}({\bf x}_i) - P^{\mathrm{QMC}}({\bf x}_i) \right)
\end{equation}
where $\mathcal{S}^{(\rho,T)}$ is the set of $M$ uncorrelated and physically representative of the thermodynamic conditions ($\rho,T$).
Concering $P^{\mathrm{QMC}}$ will use both pressures calculated with VMC, $P^{\mathrm{VMC}}$ and LRDMC,  $P^{\mathrm{LRDMC}}$, when available.

Assessing the energy error is more cumbersome given that energies are defined up to a offset.
We adopt a strategy similar to Ref.~\cite{PhysRevB.89.184106}, and define the energetic error as
\begin{equation}
\label{eq:energy_metric}
    |\delta E^{(\rho,T)}| = \frac{1}{M}  \sum_{{\bf x}_i \in \mathcal{S}^{(\rho,T)}} \left| E^{\mathrm{DFTxc}}({\bf x}_i) - E^{\mathrm{QMC}}({\bf x}_i) - c_\mathcal{S}\right|,
\end{equation}
where $c_\mathcal{S}$ is a set-dependent constant that is optimized to minimize the total difference $ \sum_{{\bf x}_i \in \mathcal{S}^{(\rho,T)}} \left| E^{\mathrm{DFTxc}}({\bf x}_i) - E^{\mathrm{QMC}}({\bf x}_i) \right|$.
{This offset is therefore different for each xc.}
This metric quantifies how large is the spread of the differences between the DFTxc and the QMC energies on the same dataset. In Fig.~\ref{fig:heatmapQMC} we plot the relative pressure error of the four xc functionals previously considered, i.e, PBE, PBEsol, vdW-DF2, and SCAN+vv10. More are reported in Appendix~\ref{app:more_benchmarks}.

We observe that the PBE pressure errors are almost always greater than 1 GPa and reach about 10 GPa at higher densities. The error is uniformly positive, meaning that PBE consistently predicts slightly higher pressures than QMC. However, considering that absolute pressures are also small at low densities, even a sub-GPa error translates into a significant relative error of about 10$\%$. For this reason, we use LRDMC data in this low-pressure region.

On the other hand, a 10 GPa error at high pressure may be almost negligible. For instance, at T=5000 K, the pressure at the lower end of the density range is about 1 GPa, while at the opposite end (1.6 g/cc), it reaches around 700 GPa.
We believe that plotting the relative error is much more informative, though these considerations need to be taken into account when ranking the DFT xc functionals.

Continuing with our analysis, the vdW-DF2 functional performs comparatively worse, with relative errors of about 20$\%$ under planetary conditions. Nevertheless, this functional performs better than PBE in the low-density regime, where intermolecular distances are relatively large and van der Waals forces become important. However, this xc functional already underperforms compared to PBE at 0.3 g/cc. Overall, this data rule out the possibility of an excessively light EoS, such as the one provided by vdW-DF2.

We observe that PBEsol and SCAN+vv10 perform much better than PBE and vdW-DF2 according to this metric. It is not straightforward to rank these two functionals, as they both show regions where their predicted pressures are reasonably consistent with QMC.

Note that, unlike vdW-DF2, which clearly overestimates QMC pressures, it is not possible to conclude whether the QMC EoS would be lighter or denser than PBEsol or SCAN+vv10. Indeed, what is plotted here are the calculated pressures from datasets $\mathcal{S}^{(\rho,T)}$, which are sampled from a PBE equilibrium distribution. The true, unknown QMC ionic equilibrium distribution is expected to be different under the same conditions, so assigning the QMC thermodynamic average directly as $\langle P^{\textrm{QMC}}(\rho,T) \rangle = 1/M \sum_{{\bf x}_i \in \mathcal{S}^{(\rho,T)}} P^{\mathrm{QMC}}({\bf x}_i)  $ would be inaccurate (see Sect.~\ref{ss:reweight}).
For the same reason, we cannot calculate the QMC correction to the REOS EoS by simply subtracting, for example, the values in Fig. 2 from the REOS.

On the other hand, it is also true that if an xc functional provided uniformly the same pressure, stress, and energy over all the datasets, this functional would also yield the same QMC ionic equilibrium distribution, meaning that its calculated EoS would be equal to the QMC EoS. This validates the entire benchmarking procedure.

Similar plots can be done for the energy, and are featured in Appendix~\ref{app:more_benchmarks}.
To select the best performing DFTxc, for planetary science applications, we need to reduce this large amount of information into a single number.
For the pressure, we define the following average pressure error
\begin{equation}
\label{eq:scoreP}
    \delta P^{QMC} = \frac{1}{\# \mathcal{J}} \sum_{(\rho,T) \in \mathcal{J}} \delta P^{(\rho,T)},
\end{equation}
where $\mathcal{J}$ is the set of $(\rho,T)$ points relevant for planetary models, as defined in Fig.\ref{fig:DFTMD}, and $\# \mathcal{J}$ is the number of elements in the set.
This choice prevents outlier values in the low-temperature solid region from disrupting valuable information for planetary science applications.
The respective metric for the energy follows the definition of Eq.~\ref{eq:scoreP}, exchanging $E \leftrightarrow P$.

Fig.~\ref{fig:allbenchs} shows the final outcome of the pressure and energy benchmarks.
We find that the family of SCAN functional performs better than the rest for the pressure.
While PBEsol performs comparatively similar to SCAN+vv10, it displays a much larger energy error. For this reason, we conclude that SCAN+vv10 is the closest approximation to QMC.
Notice that our results concerning the vdW functionals are consistent with Ref.~\cite{PhysRevB.89.184106}, despite the fact that we focus here on structures sampled at much higher temperatures.
Indeed, we also find that these van der Waals corrected functionals overperforms PBE for energy but are significantly worse as far as the pressure is concerned
In Sect.~\ref{ss:reweight} we provide further evidence that the SCAN+vv10 EoS should replace the PBE-based one for planetary interiors modeling.

\subsection{Benchmarking with the liquid-liquid transition and reweighting of QMC data}
\label{ss:reweight}

We seek another independent validation of the SCAN+vv10 functional using the predicted location of the liquid-liquid phase transition as benchmark\cite{morales_evidence_2010,morales_nuclear_2013,mazzola2018}.
The nature and position of the liquid-liquid transition between a molecular, insulating and a atomic, metallic fluid is still debated experimentally. As mentioned previously,
DFT predictions heavily depend on the functional chosen, while QMC predictions from different groups are nowadays consistent\cite{pierleoni2016liquid,mazzola2018}.
Here we demonstrate that the SCAN+vv10 prediction is closer to the QMC line compared to other functionals.
While liquid-liquid transition using SCAN have already been reported \cite{lu2019towards,hinz2020fully,bergermann2024nonmetal}, they used different simulations set-ups and codes.
Using a finer grid of densities, we trace the position of the liquid-liquid transition at lower temperatures (i.e. 1000 and 1500 K), i.e. where the transition is expected to exhibit a first-order character.
The results, shown in Appendix~\ref{app:llpt},
further validate the choice of our SCAN+vv10 set-up as the best EoS for liquid hydrogen. The predicted SCAN+vv10 transition pressure is only about 10 GPa away from the QMC references\cite{pierleoni2016liquid,mazzola2018}, while PBE differs by about 50 GPa, at 1500 K. Notice that the absolute pressure transitions are $\sim 120$ and $\sim 170$ GPa for PBE and SCAN respectively, such that a difference of 50 GPa is noteworthy.
Finally, as mentioned above, it would be tempting to utilize the calculated QMC pressures to construct a fully-QMC EoS; however this would be inaccurate (i.e targeting a sub-GPa accuracy) as the configurations, are sampled at each $\rho$ and $T$ according to weights $e^{-\beta E_\textrm{PBE} ({\bf x}_i)}$.
In principle, one could adopt a reweighting strategy, where each configuration is weighted with  $e^{-\beta (E_\textrm{VMC} ({\bf x}_i)-E_\textrm{PBE} ({\bf x}_i))}$.
Unfortunately, this estimate is affected by an enhanced statistical error, using only 47 samples for each  ($\rho,T$) parameter.
Despite this statical noise, the underlying signal is compatible with the SCAN+vv10 EoS. This corroborates the choice of the SCAN+vv10 functional for hydrogen.
The results are shown in Appendix~\ref{app:reweighting}.

\subsection{Benchmarking with the Hugoniot}
\label{ss:hugo}
\begin{figure}[t]
    \centering
    \includegraphics[width=0.95\columnwidth]{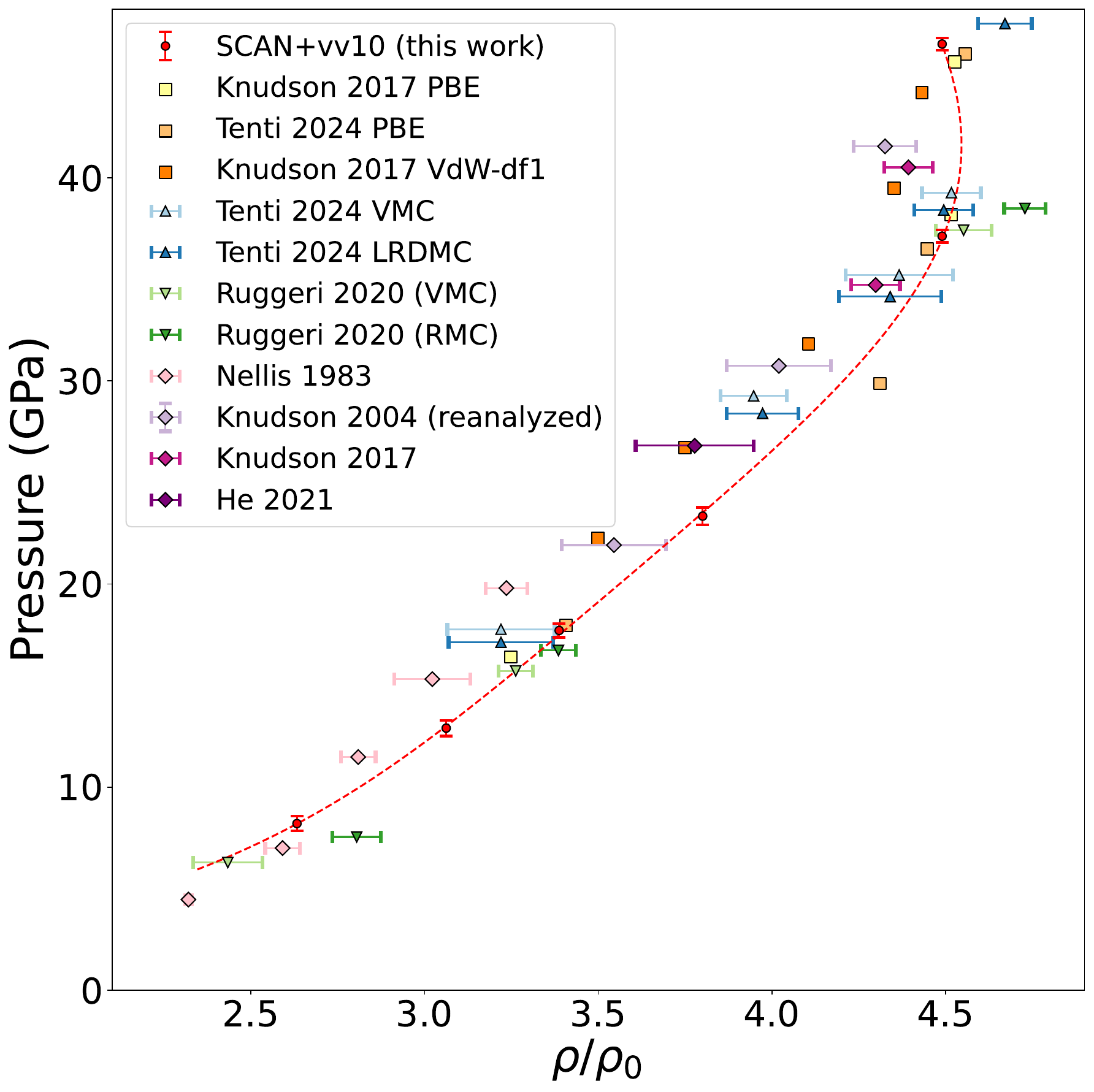}
    \caption{\textbf{Hugoniot.} {Comparison of the Hugoniot equation of state for deuterium. The Hugoniot predicted by our SCAN+vv10 (red dots with line) calculations is presented alongside a range of benchmark data, including experimental measurements showed in tilted squares~\cite{PhysRevLett.118.035501, 2004PhRvB..69n4209K, Knudson_2021, 1983JChPh..79.1480N}, calculations employing alternative DFT functionals showed in squares~\cite{tenti2024hugoniot, PhysRevLett.118.035501}, and QMC results showed in upper and lower triangles~\cite{ruggeri2020quantum, tenti2024hugoniot}, highlighting the accuracy and consistency of our approach.}}
    \label{fig:hugoniot}
\end{figure}

{In principle, the equation of state of hydrogen can be determined from shock experiments.
A system initially at a pressure $P_0$, temperature $T_0$, energy per atom $e_0$, and density $\rho_0$ can be compressed to reach final states with energy $e$, pressure $P$, and density $\rho$, as determined by the Hugoniot-Rankine relation:
\begin{equation}
\label{eq:hugo}
(e - e_0) = \frac{1}{2} (P + P_0) \left( \frac{1}{\rho_0} - \frac{1}{\rho} \right) \quad .
\end{equation}
For the sake of brevity, we refer to Refs.~\cite{PhysRevLett.118.035501, 2015JAP...118s5901B, helled2020understanding} for a detailed explanation of the experiments employed to calculate the necessary quantities. We utilize our SCAN+vv10 data and compute the zeros of the Eq.~\ref{eq:hugo} along several isotherms, ranging from $2000$ to $11000$ K. In Fig.~\ref{fig:hugoniot} compare our results against both experimental data and previous numerical studies. Notice that in this subsection we consider the deuterium isotope, as it is customary in literature.
\\
Overall, our Hugoniot line is consistent with previous DFT-PBE predictions. As observed in Ref.~\cite{PhysRevB.100.075103}, DFT benefits from several error cancellation properties, when computing the zeros of Eq.~\ref{eq:hugo}, making it unsurprising that our SCAN+vv10 results align closely with prior DFT data. Our line is also compatible, within their error bars with VMC predictions from CEIMC,\cite{ruggeri2020quantum} and a machine learning potential besed on VMC data, obtained with the TurboRVB code.\cite{tenti2024hugoniot}
\\
We also analyze the dependence on the variation of the initial condition $e_0$, which can be a source of errors.\cite{ruggeri2020quantum} Here, we chose to maintain consistency by calculating $e_0$ using the same level of electronic structure theory, SCAN+vv10, and including the zero-point energy due to nuclear quantum effects.
The initial density is $\rho_0=0.167$ g/cc (for deuterium), $T_0=22$ K and $p_0\approx 0$ (this value is negligible when inserted in Eq.~\ref{eq:hugo}).
At these conditions we calculate $e_0 = -0.58114(1)$ Ha/atom (see Appendix~\ref{app:e0_hugo}).
In Fig.~\ref{fig:hugoniot} we estimate, and show as the error bars of the SCAN+vv10 Hugoniot line, the impact determined by a shift of $\pm$0.00057 Ha of the $e_0$ value. This shift is the difference in the value of $e_0$ computed with SCAN+vv10 and PBE (see Appendix~\ref{app:e0_hugo}).}

%\section{Implications for Jupiter's interior}
%\label{s:models}

\section{Calculating the full EoS with entropy}
\label{s:entropy}

Planetary modeling also requires calculations of entropy. This introduces an additional issue.
There has been a long-standing entropy discrepancy between the two most widely used EoS models in planetary science \cite{militzer2013ab,REOS_2014}, both based on the same DFT-PBE electronic structure theory.
This conundrum has been recently resolved by some of us in Ref.~\cite{xie2025}. The primary source of error was identified as inconsistencies in matching different theories when constructing the EoS table, resulting in an inconsistent thermodynamic construction.
Interestingly, the error is not caused by thermodynamic integration over a coarse interpolation grid in $(P, T)$ space, nor due to the linear mixing approximation.

\begin{figure}[t]
    \centering
    \includegraphics[width=1.0\columnwidth]{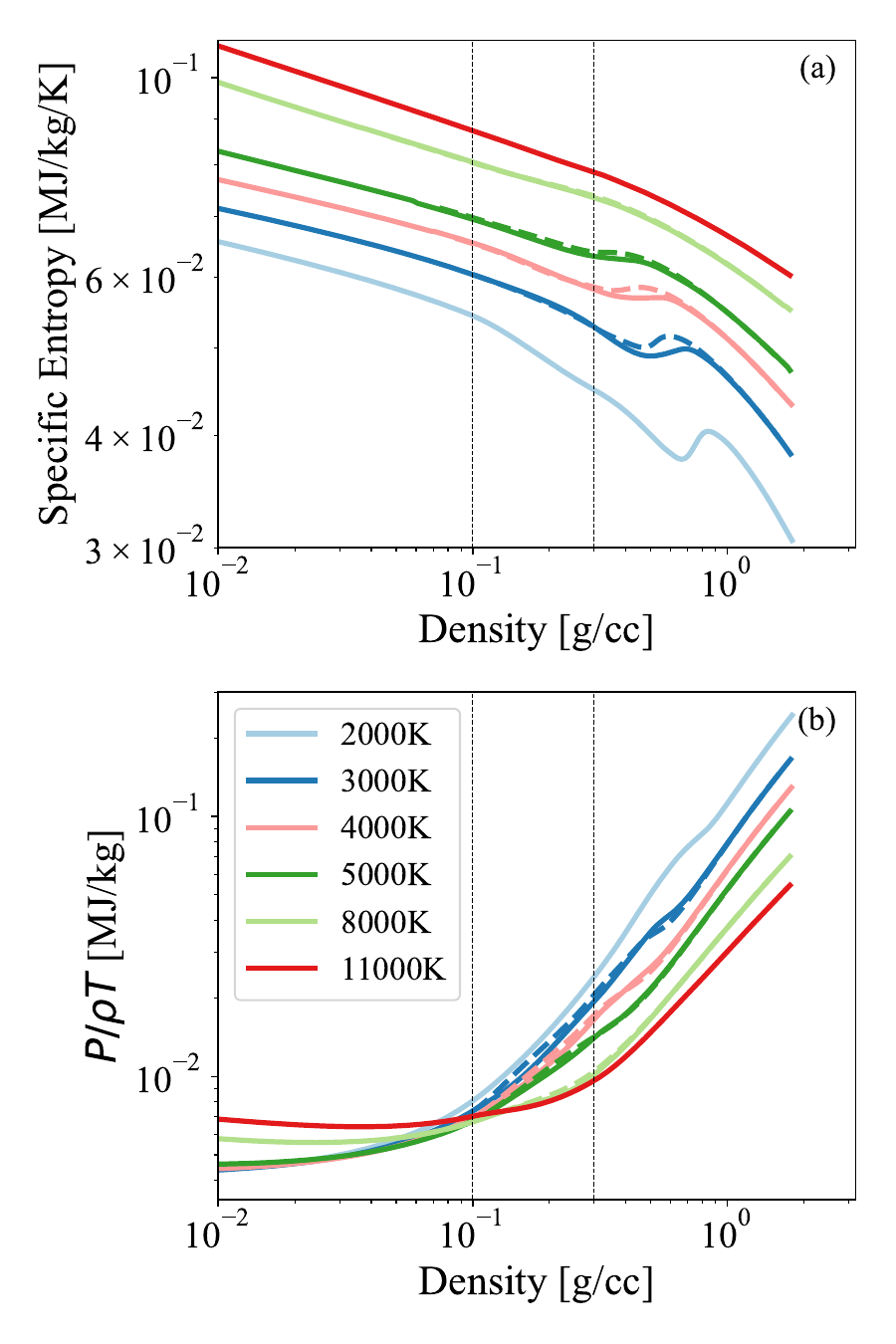}
    \caption{\textbf{SCAN+vv10 EoS. } Several (a) specific entropy, (b) pressure (scaled by $\rho T$) isotherms of our final SCAN+vv10 EoS(solid lines), which consist of the {\it ab initio} data at high densities, the SCvH EoS at low densities, and an interpolation region between them, as indicated by the vertical lines. SCAN+vv10 results are compared with PBE results (below 8000 K) from Xie {\it et al.}~\cite{xie2025} (dashed lines).
        }
    \label{fig:EOS_isoT}
\end{figure}

Here, we apply the same methodology to the SCAN+vv10 data. We interpolate the free energy surface over the \emph{ab initio} range and then perform a smooth connection with the SCvH EoS. The interpolation region is between 0.1 and 0.3 g/cc. As explained in detail in Ref.~\cite{xie2025}, special care must be taken to ensure the \emph{absolute} entropies on both sides are consistently referenced. To do so, we have subtracted a global constant $0.0057 \, \textrm{MJ/kg/K}$ from the originally reported SCvH entropy to make it align with experimental values at ambient temperatures and pressures~\cite{cox1989codata, Lemmon_Thermophysical_Properties}. This constant turns out to be equal to $k_B \ln 2 / m_p$, which accounts for the contribution of proton spin degrees of freedom that are implicitly neglected in the \textit{ab initio} portion of data~\cite{private_comm}.
On the other hand, we calculated the absolute entropy at the reference point $5000$K and $1.4$g/cc in the \textit{ab initio} region to be $0.0494(2) \, \textrm{MJ/kg/K}$, which is nearly equal to the value calculated at the PBE level~\cite{xie2025}. This indicates that PBE and SCAN describe the same kind of atomic liquid at high pressures.

\begin{figure}[t]
    \centering
    \includegraphics[width=1.0\columnwidth]{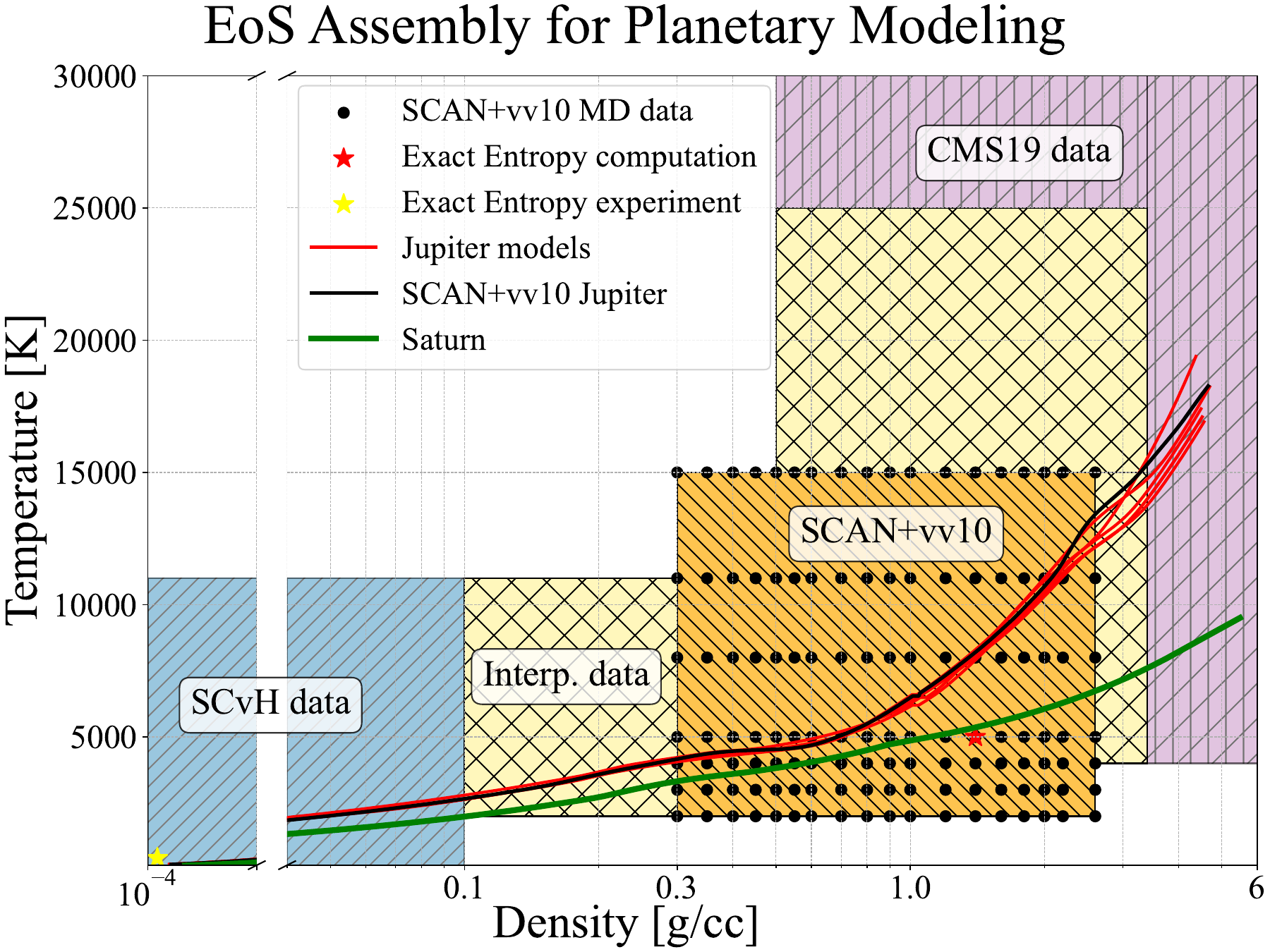}
    \caption{\textbf{Present work EoS table. } {Illustrative figure of the different regions that compose
    our final EoS. At very low $T-\rho$ SCvH data are employed (blue region). In the region between 2000 and 15000 $K$, 0.3 and 2.6 g/cc (orange region) MD simulations employing SCAN+vv10 have been performed in the points highlighted in black. Above 23000 K and 3.4 g/cc CMS19 data have been used (purple region). In the intermediate region between SCAN+vv10 data with the neighbors, SCvH on the left and CMS19 on the up and right side, we have performed accurate interpolation
    to ensure a smooth connection between the different sources of data (yellow region). The Jupiter models showed in the figure are taken from \cite{howard2023} for different EoSs. The presence of a small plateau in Jupiter adiabats is due to interior models assuming a change of composition due to helium rain at pressures between 1.64 and 2.12 Mbar. The Saturn model is taken from \cite{mankovich2021}. We plot in black the Jupiter adiabat obtained using the SCAN+vv10 EoS (see Sect.~\ref{s:models})
      }}
    \label{fig:tableEOS}
\end{figure}

Our final SCAN EoS with entropy is shown in Fig.~\ref{fig:EOS_isoT}, which also includes the thermodynamically consistent EoS recently obtained at the PBE level of theory in Ref.~\cite{xie2025}. Since we use the same procedure to calculate the entropy, this direct comparison allows us to precisely quantify the effect of xc functional on the EoS.
While the difference in Fig.~\ref{fig:EOS_isoT} seems small at this scale, the interior models are very sensitive to changes in thermodynamic properties, such that a few percent difference in the EoS may translate into a much more sizable variation in the model's predictions (see Sec.~\ref{s:models}).

As expected from the previous section, the main differences occur in the intermediate dissociation region, i.e., from $0.3$ to $1$ g/cc. Entropy exhibits the largest deviations, owing to its strong dependence on molecular dissociation.
Notice that we compute the entropy only for $T\ge 2000$ K.

For simplicity, we avoid comparisons with REOS and CMS19 in  Fig.~\ref{fig:EOS_isoT}. Note that both our results and the PBE results of ~\cite{xie2025} predict a non-monotonic behavior of entropy in contrast to the CMS19 results.
This distinctive entropy behavior implies a flatter Jupiter adiabat (see Sec.~\ref{s:models} and Fig.~\ref{fig:tableEOS}).

To produce the final SCAN+vv10 EoS table we have performed MD simulations over a broad range of temperatures and densities  spanning $2000  \le T \le 15000$ K and $0.3  \le \rho \le 2.6 ~\textrm{g/cc}$. The location of the MD points in the EoS can be seen in Fig.~\ref{fig:tableEOS}, depicted as black dots. For temperatures above 11000 K, we employed finite-temperature DFT using 90 bands, as electronic temperature effects become non-negligible in this regime. Additionally, we changed the pseudopotential above 1.8 g/cc, since the core radius of the SG15-ONCV pseudopotential used at lower densities prevents simulations beyond 1.9 g/cc. For the density region $2.0\le\rho\le2.6$, we adopted an ONCV pseudopotential from the PseudoDojo library~\cite{pseudodojo2018}, which features a smaller core radius, allowing us to reach densities up to 2.7 g/cc.

Above 8000 K or about 1 g/cc, PBE and SCAN+vv10 are very consistent across all quantities. As previously mentioned, the absolute entropies in the fully atomic liquid phase are in agreement. For all practical purposes, this means that the SCAN+vv10 EoS can be extended to even higher densities and temperatures by matching it to thermodynamically consistent EoSs calculated with PBE. However, H content in the core is much lower compared to the envelope, which contains the majority of the planet's mass. Residual H-EoS errors in this density regime are therefore expected to have a negligible impact on the models.

The SCAN+vv10 EoS table used to calculate the interior models of Sec.~\ref{s:models},  merge our new SCAN+vv10 results with the CMS19 EoS data~\cite{CMS2019}. An interpolation region is used to smoothly connect the pressure and energy data, while the entropy is recalculated using thermodynamic integration. The initial starting point is selected based on our usual reference condition, at 5000 K and 1.4 g/cc, for which we have a precise and accurate absolute entropy evaluation.

{The full EoS is released in tabular form at \cite{H_SCANvv10_EoS}}.

\begin{figure}[t]
    \centering
    \includegraphics[width=0.95\columnwidth]{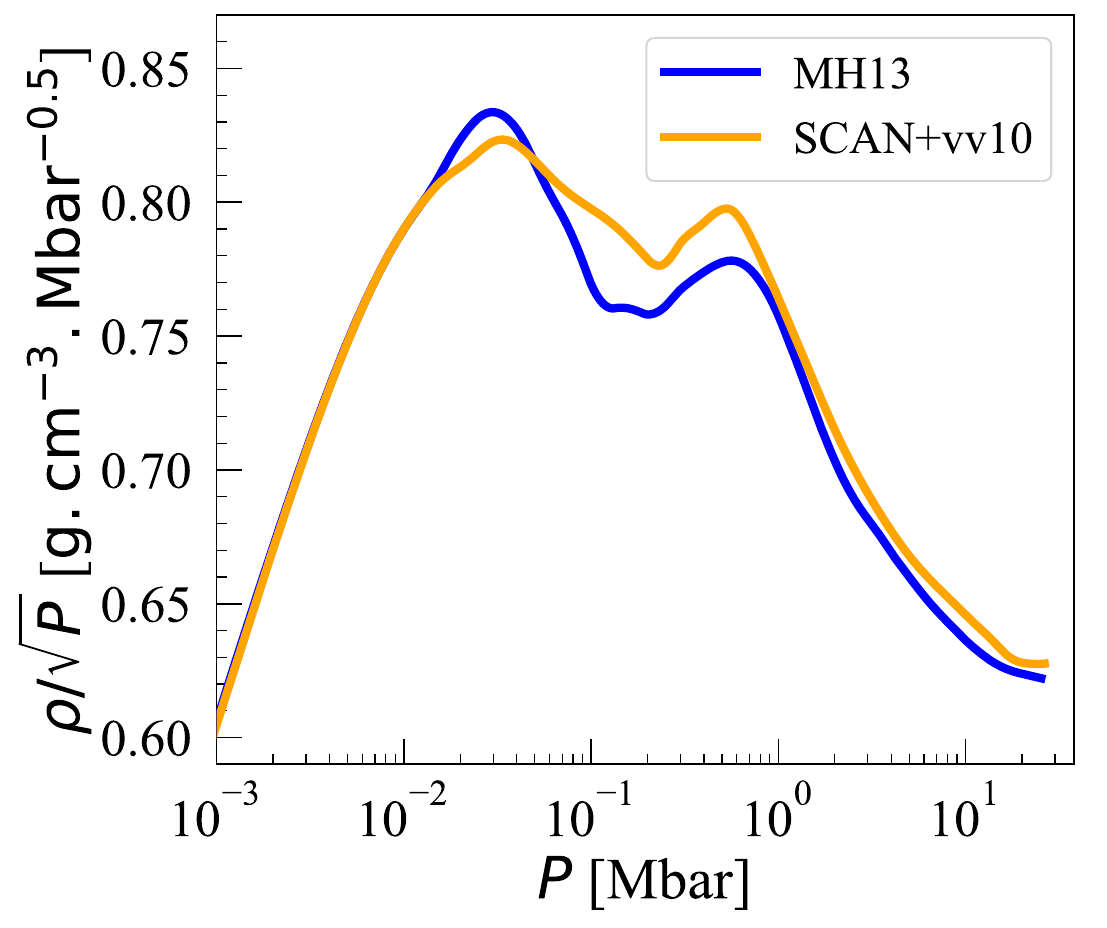}
    \caption{\textbf{Jupiter adiabat.} A comparison of the adiabats obtained with our new SCAN+vv10 EoS and the MH13 EoS \cite{militzer2013ab}. Here we used a homogeneous model with no core and a uniform helium composition, starting at $P=1~$bar and $T=166.1~$K (i.e. Jupiter's conditions).
        }
    \label{fig:saburo_XC_rhoP}
\end{figure}

\section{Implications for Jupiter's interior}
\label{s:models}

\begin{figure*}[t]
    \centering
    \includegraphics[width=1.0\textwidth]{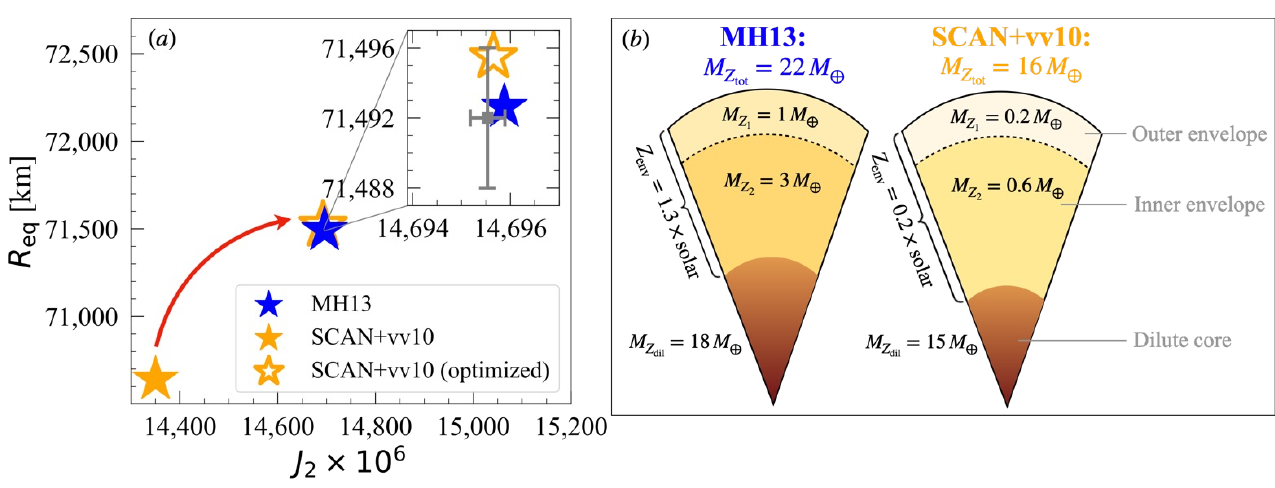}
    \caption{\textbf{Jupiter Internal Structure.}The effect of our new hydrogen EoS on Jupiter's internal structure. Panel (\textit{a}) shows the equatorial radius and $J_2$ for Jupiter interior models using SCAN+vv10 (orange) or MH13 (blue). The orange empty star shows a model which was optimized to match the observed $R_{\rm eq}$ and $J_2$. The grey errorbar shows the observed equatorial radius and the measured $J_2$ value accounting for differential rotation (see \cite{miguel2022,howard2023}).
    Panel (\textit{b}) shows schematics of Jupiter's internal structure using the MH13 (left) and SCAN+vv10 (right) EoSs. We show the inferred total heavy-element mass ($M_{Z_{\rm tot}}$) as well as the heavy-element mass in the different layers of the models ($M_{Z_1}$, $M_{Z_2}$, $M_{Z_{\rm dil}}$) and the envelope metallicity ($Z_{\rm env}$).
        }
    \label{fig:sketchs}
\end{figure*}
In this section, we investigate how the new SCAN+vv10 EoS affects the inferred internal structure and bulk composition of Jupiter. {We performed two distinct comparisons.}

First, we calculated adiabats for pure hydrogen-helium mixtures at Jupiter's conditions ($T_{\rm 1 bar}=166.1~$K). We used a simple homogeneous model, without a core and with a uniform helium composition. We assumed linear mixing with mass fractions of hydrogen and helium set at $X=0.755$ and $Y=0.245$, respectively. We compared the adiabat derived from our new EoS for hydrogen, combined with the SCvH-He EoS \cite{scvh1995} for helium, to the MH13 EoS \cite{militzer2013ab} which is the current standard for H-He mixtures. The results are shown in  Fig.~\ref{fig:saburo_XC_rhoP}. Since $P \propto \rho^2$ in Jupiter's interior \cite{hubbard1975} we compare values of $\rho/ \sqrt{P}$. We find that the differences between the two adiabats are up to 4\%. The SCAN+vv10 EoS yields lower densities only between 0.01 and 0.05~Mbar, and higher densities for pressures above 0.05~Mbar (0.01 Mbar $=$ 1 GPa). This is consistent with Fig.~\ref{fig:DFTMD}.
Fig.~\ref{fig:tableEOS} also shows the Jupiter adiabat (in the $\rho-T$ diagram) inferred with our new EoS. Interestingly, its profile is flatter compared to previous models, suggesting a mildly increasing temperature across a large region in the planetary interior.
\\
The implication of the new EoS on Jupiter's composition and thermal profile should be investigated in future studies.

Overall, our new EoS results in a denser adiabat throughout most of Jupiter's interior. This implies that interior models using the new EoS will yield lower metallicities (less heavy elements). This is because if the density of hydrogen is higher, in order to fit a given density (which is consistent with the gravity data), less heavy elements can be included. {In Appendix~\ref{app:robustness_model}, we also compared adiabats obtained with different helium EoSs and show that the hydrogen EoS is the dominant factor controlling the density profile of Jupiter.}

To determine the differences in the inferred bulk composition of Jupiter with the new EoS, we used a Jupiter model from \cite{howard2023} based on the MH13 EoS (see their Table E.1.). This model assumes the presence of a dilute core, helium differentiation between an outer envelope of molecular hydrogen and an inner envelope of metallic hydrogen, with a high internal entropy (see e.g., \cite{helledhoward2024} for further details). Note that the heavy-element mass fraction (i.e., the envelope metallicity) is assumed to be the same in the inner and outer envelopes.   The left panel of Figure~\ref{fig:sketchs} shows the  inferred equatorial radius and $J_2$ for the Jupiter model with the MH13 EoS (blue star). Using CEPAM \cite{guillot1995}, we calculated a similar Jupiter model with SCAN+vv10 (orange star). Due to the higher densities predicted by our EoS, we obtain a smaller radius and a lower $J_2$ value.
In order to match Jupiter's observed equatorial radius and $J_2$, we adjust the extent of the dilute core and the heavy-element mass fraction in the envelope ($Z_{\rm env}$). The results for Jupiter's internal structure are shown in the right panel of Figure~\ref{fig:sketchs}. We find that: (i) the required modification of the extent of the dilute core decreases the heavy-element mass in this region from $M_{Z_{\rm dil}} = 18~M_{\oplus}$ to $15~M_{\oplus}$, and (ii) the envelope metallicity $Z_{\rm env}$ decreases  from $1.3 \times \rm solar$ to $0.2 \times \rm solar$. The heavy-element mass in the outer envelope is reduced from $M_{Z_1}= 1 \, M_{\oplus}$ to $0.2 \, M_{\oplus}$ and in the inner envelope from $M_{Z_2}= 3 \, M_{\oplus}$ to $0.6 \, M_{\oplus}$.

Overall, using SCAN+vv10 rather than MH13 led to a decrease in the inferred total heavy-element mass in Jupiter, from $M_{Z_{\rm tot}} = 22~M_{\oplus}$ to $16~M_{\oplus}$. Note that although our results correspond to a single optimized model, the trend of lower inferred metallicity for Jupiter using the SCAN+vv10 EoS is robust due to the higher inferred density of hydrogen at Jupiter's conditions.

Our study suggests that the tension between Jupiter's atmospheric metallicity as measured by the Galileo probe and the metallicity inferred by interior models remains, and in fact, with the new EoS the difference between the two values increases. While the Galileo probe measured an atmospheric metallicity of three times solar, our presented Jupiter model predicts an atmospheric metallicity of 0.2 solar, a difference by a factor of 15!
It is therefore clear that the discrepancy between Jupiter's atmosphere and interior metallicities cannot be resolved by improvements in the hydrogen EoS calculations.

\section{Conclusion}
\label{s:conclu}

We calculate the equation of state for hydrogen using a more accurate electronic structure method compared to currently adopted EoSs,\cite{REOS_2014,militzer2013ab,CMS2019,CMS2021} validated against computationally intensive quantum Monte Carlo calculations (these alone utilized 82 million CPU hours). Several density functional theories have been tested, and we identified SCAN+vv10 as the consistently best-performing setup. In addition, we find that other reasonable and previously adopted functionals for hydrogen yield EoSs that vary by up to $\pm20\%$, significantly exceeding the accuracy required for constructing planetary models. Therefore, benchmarking against a higher level of theory is crucial to avoid qualitatively incorrect predictions.
We also find that the PBE functional, which forms the basis of currently utilized EoSs, does not perform excessively poorly.
{This means that these EoSs have an error (in pressure) of about $5\%$ at planetary conditions. The error also becomes smaller at higher densities.}
However, this result is likely fortuitous, as PBE significantly underestimates the metallization pressure by about $25\%$ and produces inaccurate molecular bond predictions.\cite{PhysRevB.89.184106}

The new EoS is derived from first principles and is thermodynamically consistent across a wide range of densities and temperatures.
Since knowledge of the hydrogen EoS is critical for determining Jupiter's internal structure, we also explored how our new EoS affects the predicted composition and internal structure of the planet.
Because the SCAN+vv10 hydrogen EoS is denser, it leads to a lower bulk metallicity for Jupiter.
The lower inferred envelope metallicity  in Jupiter further highlights the mismatch between the enrichment of Jupiter's atmosphere as measured by the Galileo probe and the one inferred from structure models. Although there were speculations that this disagreement could be resolved by improvements in the hydrogen EoS calculations \citep{2022Icar..37814937H}, our study shows that this is not the case. We therefore conclude that this inconsistency
cannot be resolved by improving the uncertainties in the hydrogen EoS.  Instead, it implies that the internal structure of Jupiter is more complex than typically assumed and that the atmospheric composition does not represent the bulk composition.
It is therefore possible that Jupiter's  outmost atmospheres is more metal-rich than its outer envelope \citep{DC19,howard2023b,2024ApJ...967....7M}.
This conclusion is of high importance for the interpretation of atmospheric measurements of giant planets' atmospheres in the solar system and around other stars.
Finally, our modified hydrogen EoS will also affect the inferred bulk compositions of Saturn and giant exoplanets and we hope to address this in future research.

\begin{acknowledgments}
%\vspace{2mm}
%\paragraph{{\textit{Acknowledgements}} $-$}
% FUGAKU
C.C., K.N., and G.M. are grateful for computational resources of the supercomputer Fugaku provided by RIKEN through the HPCI System Research Projects (Project IDs: hp230030).
% NIMS computer
K.N. is grateful for computational resources from the Numerical Materials Simulator at National Institute for Materials Science (NIMS).
% K.N. financial support
K.N. acknowledges financial supports from Grant-in-Aid for Early Career Scientists (Grant No.~JP21K17752), from MEXT Leading Initiative for Excellent Young Researchers (Grant No.~JPMXS0320220025), and from JST BOOST (Grant No.~JPMJBY24F3).
%CC & GM financial support
C.C, H.X., G.M. acknowledge financial support from the Swiss
National Science Foundation (grant PCEFP2\_203455).
% discussion ackn.
We acknowledge discussions with Michele Ceriotti, Marco Gibertini and Giacomo Tenti.
R.H. and S.H. acknowledge financial support from the Swiss
National Science Foundation (grant 200020\_215634).

% Code availability
The {\textit{ab initio}} QMC package used in this work, TurboRVB, is available from its GitHub repository [\url{https://github.com/sissaschool/turborvb}]. \\
The full SCAN+vv10 EoS is available from the GitHub repository [\url{https://github.com/cisar97/H_SCANvv10_EoS}].
\end{acknowledgments}

\appendix

\section{Details of DFT calculations}
\label{app:dft}
All the DFT simulations have been performed using QuantumESPRESSO \cite{Giannozzi2009_QE}, {(version 7.0, and code \textsf{pw.x})} with plane-wave basis set and $3\times3\times3$ Monkhorst-Pack grid for the k-points. For BLYP, B3LYP, PBE, PBEsol, vdW-DF, vdW-DF2, vdW-DF2-c09 xc functionals it has been used projected-augmented-wave (PAW) pseudopotentials \cite{BlochlPAW1004}: \texttt{H.pbe-kjpaw\_psl.1.0.0.UPF} from psl library  \cite{DalCorso2014_psl}. All the other xc functionals employ optimized norm-conserving Vanderbilt (ONCV) pseudopotentials \cite{Hamann1979_NormConserving}:  \texttt{H\_ONCV\_PBE-1.0.upf} from the SG15 library \cite{Hamann2013_SG15}. The SCAN+vv10 MD at densities $\ge$ 2.0 g/cc employ the Norm Conserving pseudopotential from PseudoDojo~\cite{pseudodojo2018} which has a smaller core radius than the SG15 one. \\
{All DFT and QMC systems are made of 128 atoms.}
{This system size, combined with adequate k-point sampling ($3\times3\times3$), has been shown to yield converged EoS values.\cite{lorenzen2010} While this system size may be small to quantitatively probe first-order phase transitions, where systems as large as 512 atoms have been used in direct DFT-MD simulations,\cite{bergermann2024nonmetal} such transitions occur at temperatures much lower than planetary conditions.}

For every xc functional the kinetic energy cutoffs for both the wavefunction and density have been studied through convergence of the internal energy and stresses; moreover the consistency between numerical and analytical pressure is checked as explained in Appendix \ref{app:qmc-validation}. All the xc functional using PAW pseudopotential have been employed with a wavefunction cutoff of 80 Ry and a density cutoff of 800 Ry. Calculations performed with HSE, PBE0, revPBE0, rVV10 and optB86B {{using the ONCV pseudo potential}} employ 100 and 400 Ry for the wavefunction and density cutoff respectively, moreover for HSE, PBE0, revPBE0 it has been employed a kinetic energy cutoff for the exact exchange operator of 100 Ry and a $1\times1\times1$ three-dimensional mesh for q sampling of the Fock operator. For B3LYP it has been used a kinetic energy cutoff for the exact exchange operator of 80 Ry and again  $1\times1\times1$ three-dimensional mesh for q sampling of the Fock operator. The meta-GGA xc functionals (i.e. SCAN, rSCAN, etc.) requires an higher wavefunction cutoff, due to their additional dependence on the kinetic energy density, of 140 Ry and 560 Ry for the density cutoff with {{the ONCV pseudo potential}}.

%It has been tried to increase just the FFT grid as suggested in Ref.\cite{yao2017_scanConvergence}, but it did not provide the converged results.
For all the SCANs functional a pseudopotential generated with SCAN xc has been tried \cite{yao2017_scanConvergence}, it has been tested on a smaller set of configurations (10 configuratiions for every $T-\rho$) providing results that are slightly better than the same functional with a PBE pseudopotential.
Nevertheless the SCAN pseusopotential requires a wavefunction and density cutoff of 280 and 1120 Ry to reach convergence, respectively, and the improvement in the results is not enough to justify a doubling in the computational time.

The Molecular Dynamics simulations with PBE xc functionals, used to sample the configurations for the benchmark against QMC have been performed with over damped Langevin dynamics with timestep ranging from 0.2 to 0.05 a.u. depending from $T-\rho$. The same set up has been used for the vdW-DF2 and PBEsol MD. The SCAN+vv10 MD has been performed using Stochastic Velocity Rescaling (SVR) thermostat \cite{bussi2007canonical}, with timestep ranging from 2 to 21 a.u. (0.1 to 1 fs) depending from $T-\rho$.
{We checked that SVR sampling is consistent with first-order Langevin dynamics. We further validated it against the commonly used Nose-Hoover thermostat.}

\begin{figure}[t]
    \centering
    \includegraphics[width=1.0\columnwidth]{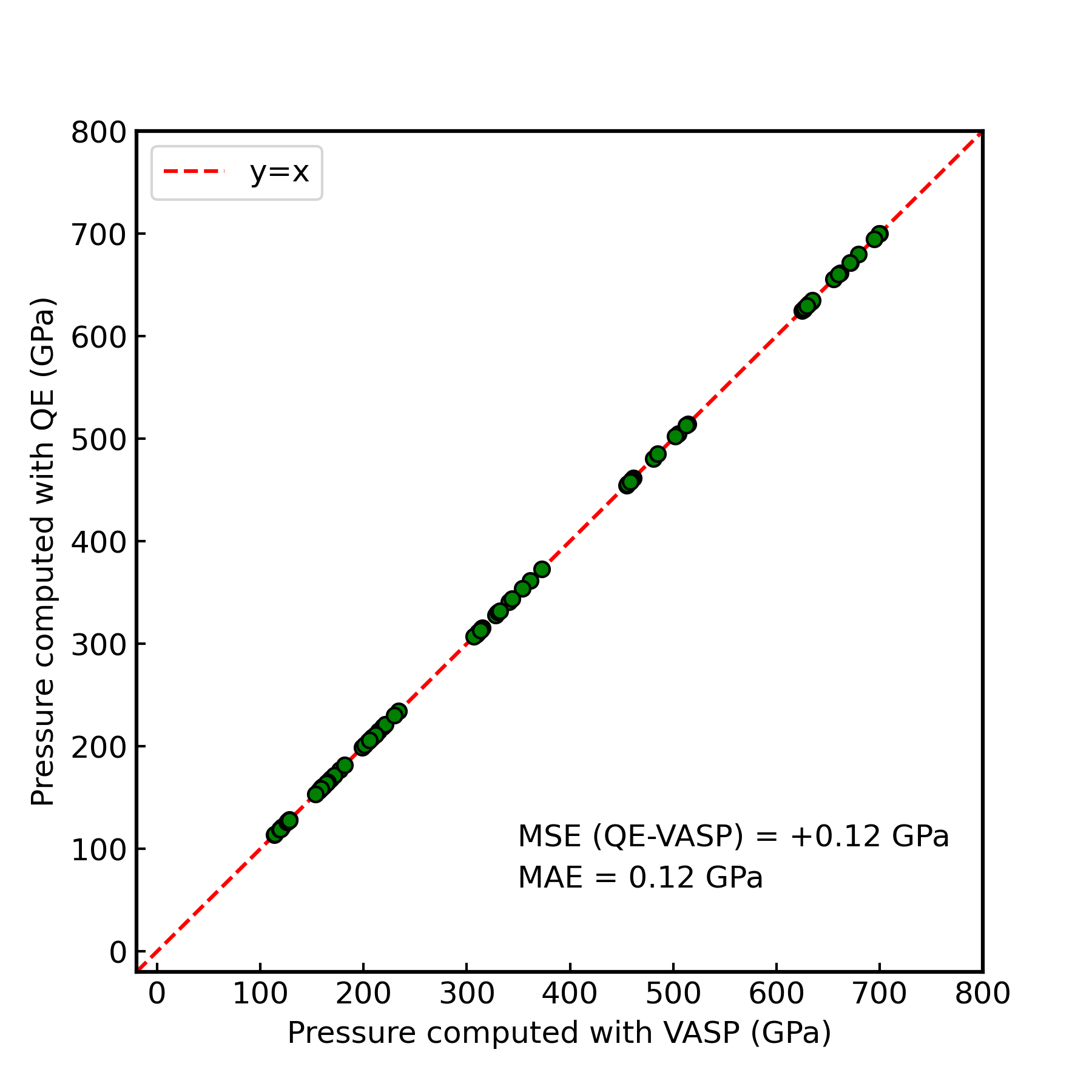}
    \caption{The comparison between pressures computed by QuantumESPRESSO with the PAW pseudo potential and those computed by all-electron calculations using VASP. The LDA-PZ functional was employed.
    MSE and MAE stand for mean signed error and mean absolute error, respectively.}
    \label{fig:qe-pz-paw-vs-vasp-pz-ae}
\end{figure}

The PAW and ONCV pseudopotentials used in this study were very carefully validated via the comparison with all-electron DFT calculations performed by Vienna Ab initio Simulation Package (VASP) code~{\cite{kresse1993vasp}}. In all-electron calculations by VASP, the kinetic energy cutoff was set to 10000 eV. The same smearing method, smearing parameter, and k-point were employed. Figure~{\ref{fig:qe-pz-paw-vs-vasp-pz-ae}} shows pressures computed by QuantumESPRESSO with the PAW pseudo potential and those computed by all-electron calculations using VASP, where LDA-PZ functional was employed. They give perfectly consistent pressures, indicating that the use of the PAW pseudo potential does not introduce bias.
Figure.~{\ref{fig:qe-pz-paw-vs-vasp-pz-ae}} shows pressures computed by QuantumESPRESSO with the PAW pseudo potential and those computed by QuantumESPRESSO with the ONCV pseudo potential, where GGA-PBE functional was employed. They give perfectly consistent pressures, indicating that the use of the ONCV pseudo potential does not introduce bias.

\begin{figure}[t]
    \centering
    \includegraphics[width=1.0\columnwidth]{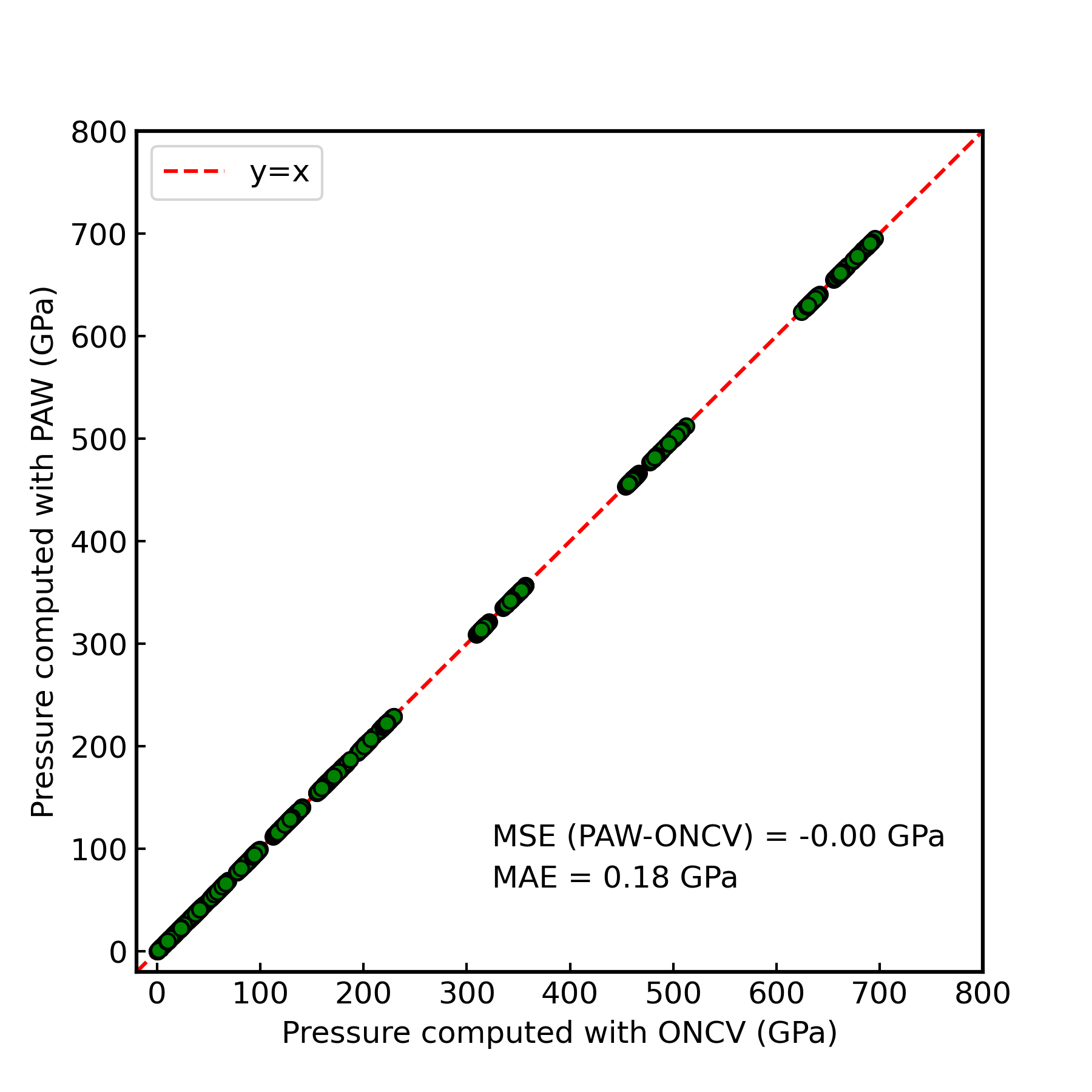}
    \caption{The comparison between pressures computed by QuantumESPRESSO with the PAW pseudo potential and those computed by QuantumESPRESSO with the ONCV pseudo potential. The GGA-PBE functional was employed.
    MSE and MAE stand for mean signed error and mean absolute error, respectively.}
    \label{fig:qe-pbe-paw-vs-qe-pbe-oncv}
\end{figure}

\begin{figure}[ht]
    \centering
    \includegraphics[width=1.0\columnwidth]{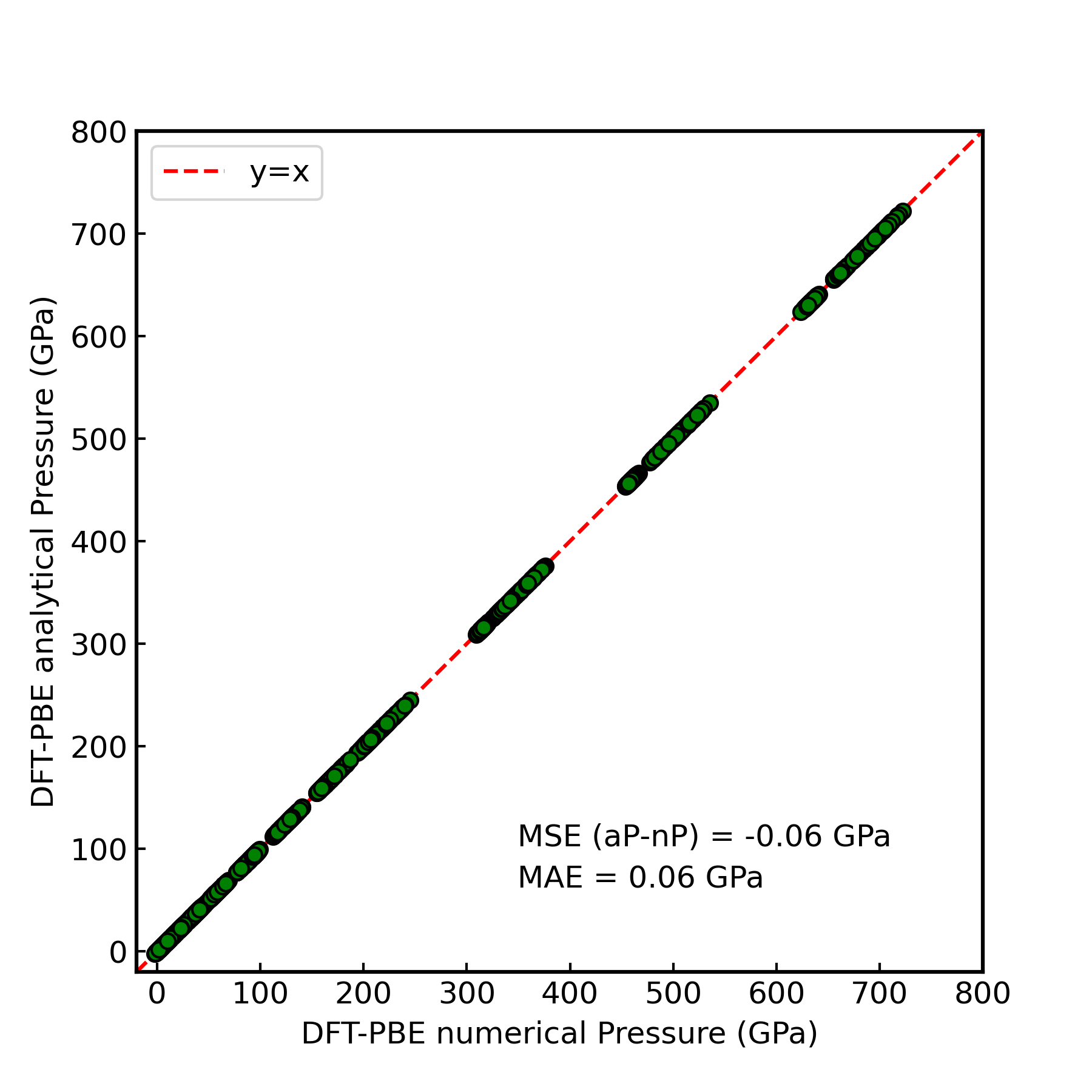}
    \caption{The comparison between pressures computed by QuantumESPRESSO via FDM (denoted as nP:numerical pressure) and those computed via the Hellmann–Feynman theorem (denoted as aP:analytical pressure). The GGA-PBE functional and the PAW pseudo potential were employed.
    MSE and MAE stand for mean signed error and mean absolute error, respectively.}
    \label{fig:qe_paw_nP_vs_qe_paw_aP}
\end{figure}

{We also test the influence of approximating the finite temperature electronic effects using Fermi-Dirac occupation of the electronic levels.}
 {We recalculate the pressure and energy using SCAN+vv10 setting the Fermi-Dirac smearing, for $T=5000,8000$ and $11000$.}
{We find that under conditions broadly relevant for Jupiter's interior, the difference in pressure is less than $1\%$. However, it remains to be shown that this difference does not appreciably affect the calculated density profile of the planet. We present this sensitivity analysis in the Appendix~\ref{app:robustness_model}.}

\section{Details of VMC and LRDMC calculations}
\label{app:vmc-and-lrdmc}
The all-electron VMC and LRDMC calculations for the liquid hydrogen were performed using TurboRVB~{\cite{nakano2020turborvb}} combined with the Python package TurboGenius~{\cite{nakano2023turbogenius}}. TurboRVB employs the Jastrow antisymmetrized geminal power (JAGP){~\cite{casula2003agpgeminal}} ansatz. The ansatz is composed of a Jastrow and an antisymmetric part ($\Psi = J \cdot {\Phi _{{\text{AGP}}}}$). The singlet antisymmetric part is denoted as the singlet antisymmetrized geminal power (AGPs), which reads:
\begin{eqnarray}
&{\Psi _{{\text{AGPs}}}}\left( {{{\mathbf{r}}_1}, \ldots ,{{\mathbf{r}}_N}} \right) = \nonumber \\
&{\hat A} \left[ {\Phi \left( {{\mathbf{r}}_1^ \uparrow ,{\mathbf{r}}_1^ \downarrow } \right)\Phi \left( {{\mathbf{r}}_2^ \uparrow ,{\mathbf{r}}_2^ \downarrow } \right) \cdots \Phi \left( {{\mathbf{r}}_{N/2}^ \uparrow ,{\mathbf{r}}_{N/2}^ \downarrow } \right)} \right],
\end{eqnarray}
where ${\hat A}$ is the antisymmetrization operator, and $\Phi \left( {{\mathbf{r}}_{}^ \uparrow ,{\mathbf{r}}_{}^ \downarrow } \right)$ is called the pairing function. The spatial part of the geminal function is expanded over the Gaussian-type atomic orbitals (GTOs):
%{\kn{(GTOs)}}:
%
\begin{equation}
{\Phi _{{\text{AGPs}}}}\left( {{{\mathbf{r}}_i},{{\mathbf{r}}_j}} \right) = \sum\limits_{l,m,a,b} {{f_{\left\{ {a,l} \right\},\left\{ {b,m} \right\}}}{\psi _{a,l}}\left( {{{\mathbf{r}}_i}} \right){\psi _{b,m}}\left( {{{\mathbf{r}}_j}} \right)},
\label{agp_expansion}
\end{equation}
where ${\psi _{a,l}}$ and ${\psi _{b,m}}$ are primitive GTOs, their indices $l$ and $m$ refer to different orbitals centered on atoms $a$ and $b$, and $i$ and $j$ are coordinates of spin-up and spin-down electrons, respectively. When the JAGPs is expanded over {$p$ = $N_{\rm el}/2$} molecular orbitals, the JAGPs coincides with the Jastrow--Slater determinant (JSD) ansatz. In this study, we restrict ourself to use the JSD anstaz. Indeed, the coefficients in the pairing functions (i.e., variational parameters in the antisymmetric part) are obtained by the built-in DFT code, named {\textsc{Prep}}, with the PZ-LDA functional and the coefficients are fixed during the VMC optimization step. This is the simplest choice, but it is reasonable in the majority of QMC applications. Only the coefficients in the Jastrow factor are optimized at the VMC level, which keeps the nodal surface of the trial wavefunction unchanged from that obtained at the DFT level. Indeed, the LRDMC calculations were done using the DFT nodal surfaces (i.e., the JSD ansatz with the DFT nodal surfaces).

The Jastrow term is composed of one-body, two-body, and three-body factors ($J = {J_1}{J_2}{J_3}$). The one-body and two-body factors are used to satisfy the electron--ion and electron--electron cusp conditions, respectively, and the three-/four-body factor is adopted to take the further electron--electron correlation into consideration. The one-body Jastrow factor reads:
\begin{equation}
{J_1}\left( {{{\mathbf{r}}_1}, \ldots , {{\mathbf{r}}_N}} \right) = \exp \left( {\sum\limits_{i,I,l} {g_{I,l}^{}\chi _{I,l}^{}\left( {{{\mathbf{r}}_i}} \right)} } \right) \cdot \prod\limits_i {{{\tilde J}_1}\left( {{{\mathbf{r}}_i}} \right)},
\label{onebody_jas}
\end{equation}
where
\begin{equation}
{\tilde J_1}\left( {\mathbf{r}} \right) = \exp \left( {\sum\limits_I { - {{\left( {2{Z_I}} \right)}^{3/4}}u\left( {2{Z_I}^{1/4}\left| {{\mathbf{r}} - {{\mathbf{R}}_I}} \right|} \right)} } \right),
\label{onebody_j_single}
\end{equation}
${{{\mathbf{r}}_i}}$ are the electron positions, ${{{\mathbf{R}}_I}}$ are the atomic positions with corresponding atomic number $Z_I$, $l$ runs over atomic orbitals ${\chi _{I,l}^J}$ centered on atom $I$, and ${u\left( r \right)}$ is a short-range function containing a variational parameter $b$:
\begin{equation}
u\left( r \right) = \frac{b}{2}\left( {1 - {e^{ - r/b}}} \right).
\label{onebody_u}
\end{equation}
The two-body Jastrow factor is defined by a long-range function as:
\begin{equation}
{J_2}\left( {{{\mathbf{r}}_1}, \ldots , {{\mathbf{r}}_N}} \right) = \exp \left( {\sum\limits_{i < j} {v\left( {\left| {{{\mathbf{r}}_i} - {{\mathbf{r}}_j}} \right|} \right)} } \right),
\label{twobody_jastrow}
\end{equation}
where $v\left( r \right)$ is:
\begin{equation}
v\left( r \right) = \frac{1}{2}r \cdot {\left( {1 + F \cdot r} \right)^{ - 1}}
\label{twobody_v}
\end{equation}
and $F$ is a variational parameter. The three-body Jastrow factors are:
\begin{equation}
{J_{3}}\left( {{{\mathbf{r}}_1}, \ldots , {{\mathbf{r}}_N}} \right) = \exp \left( {\sum\limits_{i < j} {{\Phi _{{\text{Jas}}}}\left( {{{\mathbf{r}}_i},{{\mathbf{r}}_j}} \right)} } \right),
\end{equation}
and
\begin{equation}
{\Phi _{{\text{Jas}}}}\left( {{{\mathbf{r}}_i},{{\mathbf{r}}_j}} \right) = \sum\limits_{l,m,a} {g_{l,m,a}^{}\chi _{a,l}^{{\text{Jas}}}\left( {{{\mathbf{r}}_i}} \right)\chi _{a,m}^{{\text{Jas}}}\left( {{{\mathbf{r}}_j}} \right)},
\label{threebody_jas}
\end{equation}
where the indices $l$ and $m$ again indicate different orbitals centered on corresponding atoms $a$.

In the antisymmetric part, the uncontracted cc-pVTZ basis set [4$s$2$p$1$d$] taken from the Basis-Set Exchange Library{~\cite{2019PRI_BSE}} modified with the one-body Jastrow factor~{\cite{mazzola2018} were employed. This is the same one used in previous high-pressure hydrogen studies~{\cite{tirelli2022high, Lorenzo.hydrogen.2023, tenti2024hugoniot}}. Moreover, Mazzola et al., reported that the modified cc-pVTZ basis was converged to the CBS limit (within 0.01~eV/atom)~{\cite{mazzola2018}}.
Hereafter, we describe the detail of the modification of the one-body orbitals. The original basis set was modified such that the $s$ orbitals whose exponents are larger than $8 \cdot Z^2$ (where $Z$ is the atomic number) are disregarded, and they are implicitly compensated by the homogeneous one-body Jastrow part [${\tilde J_1}\left( {\mathbf{r}} \right)$] to fulfill the electron--ion cusp condition explicitly~{\cite{mazzola2018, 2019NAK_J1_Na2}}.
Indeed, a single-particle orbital is modified as:
\begin{equation}
\tilde \phi _j^b\left( {{\mathbf{r}} - {{\mathbf{R}}_b}} \right) = \phi _j^b\left( {{\mathbf{r}} - {{\mathbf{R}}_b}} \right){{\tilde J}_1}\left( {\mathbf{r}} \right),
\label{onebody_j_single_DFT}
\end{equation}
where ${{\tilde J}_1}\left( {\mathbf{r}} \right)$ is the same as in Eq.~(\ref{onebody_j_single}). The parameter $b$ in Eq.~(\ref{onebody_u}) is set to 1.1 for all volume points, and optimized at the VMC level. In this way, each element of the modified basis set satisfies the electron--ion cusp conditions even at the DFT level~{\cite{mazzola2018, 2019NAK_J1_Na2}}.

In the inhomogeneous one-body and three-body Jastrow parts, [2$s$2$p$1$d$] basis set was employed, which is the same basis set as used in recent high-pressure hydrogen studies~{\cite{Lorenzo.hydrogen.2023, tenti2024hugoniot}}.
The variational parameters in the Jastrow factors were optimized the Stochastic Reconfiguration method described in Refs.~\cite{sorella1998sr, sorella2007sr} implemented in TurboRVB.

The LRDMC calculations for computing the PESs were performed by the single-grid scheme~\cite{casula2005lrdmc} with the lattice space $a$ = 0.30~Bohr. For several structures, we performed LRDMC calculations with the lattice space $a$ = 0.20~Bohr and confirmed that the obtained pressures are consistent with those obtained with the lattice space $a$ = 0.30~Bohr within their statistical uncertainties (within $2 \sigma$).

\begin{figure}[t]
    \centering
    \includegraphics[width=1.0\linewidth]{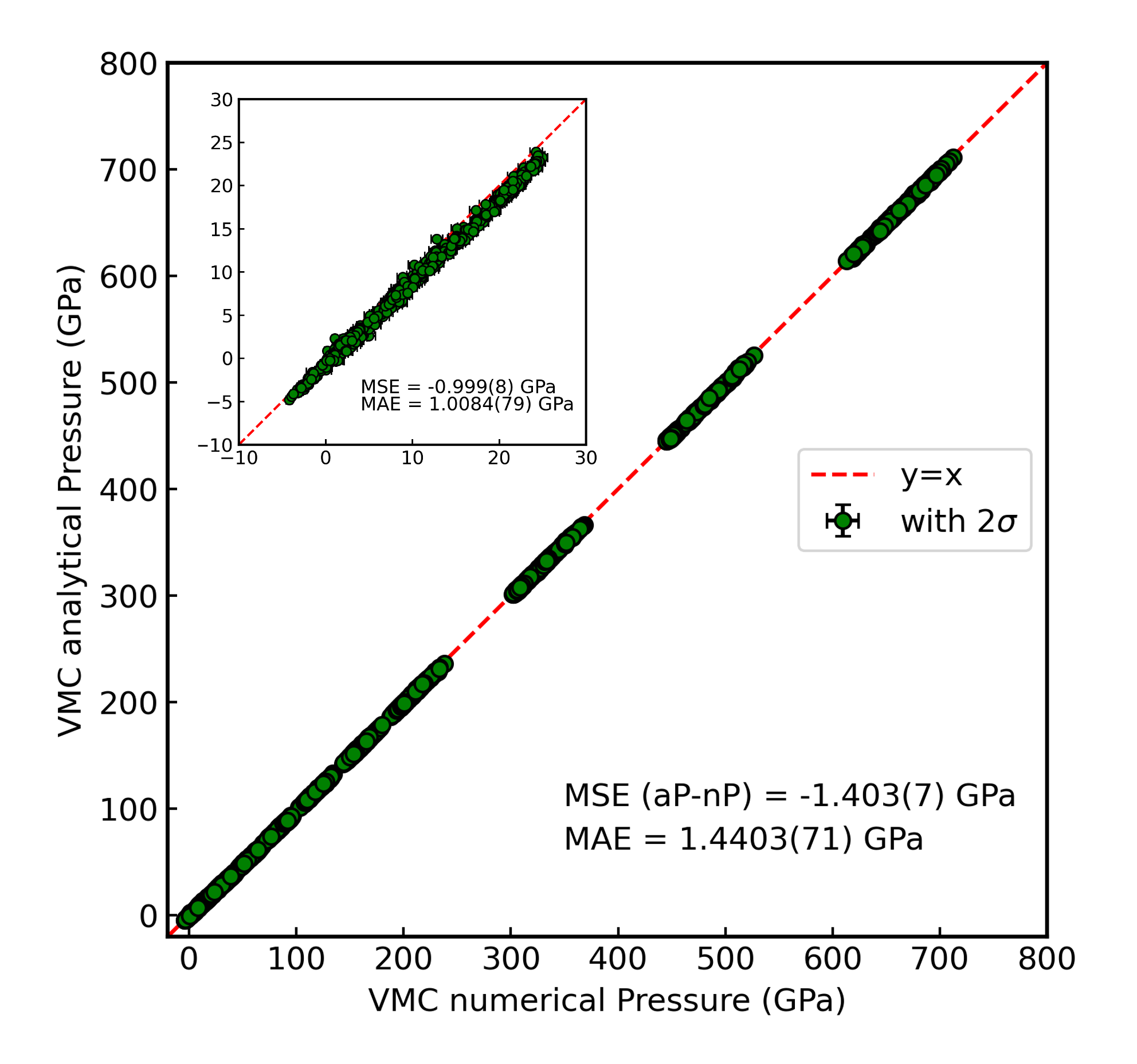}
    \caption{The comparison between Pressures computed via FDM (denoted as nP: numerical pressure) and those computed from the Hellmann–Feynman and Pulay terms (denoted as aP:analytical pressure) at the VMC level. Inset: Those in the range $\le$ 25 GPa.
    MSE and MAE stand for mean signed error and mean absolute error, respectively.
        }
    \label{fig:P_vmc_nP_vs_vmc_aP_all}
\end{figure}

\section{Difficulty in Computing Unbiased Atomic forces and Pressures by QMC}
\label{app:difficutly-in-qmc}
Although the computation of atomic forces and pressures are established and routinely used in the DFT framework, they are still under development in QMC calculations~{\cite{1986REY_qmcforce, 2000ASS_qmcforce, 2000FIL_qmcforce, 2005CHI_qmcforce, 2008BAD_qmcforce, 2011ASS_qmcforce, 2014MOR_qmcforce, 2021VAN_qmcforce, Tiihonen2021SCE, nakano2022SCE, nakano2024efficient}}.
There are three routes to compute {\it unbiased} forces and pressures. (i) fitting potential energy surfaces with respect to an atomic position or the volume (a.k.a. finite-difference method: FDM). (ii) computing the so-called Hellmann–Feynman and Pulay terms with a fully optimized WF. (iii) computing the Hellmann–Feynman and Pulay terms with a partially optimized WF and compensating the missing term appearing with the partially optimized WF.
(i) is the most straightforward computation and is applicable both for VMC and LRDMC calculations, and is the one adopted in this study. The significant drawback of this approach is that the computation of atomic forces is infeasible for large systems because 3$N_a$ times FDM fitting procedures are needed, where $N_a$ refers to the number of atoms in the system. Therefore, we computed only pressures and used them as reference data for bench-marking XC functionals.
(ii) has been successful for small molecules and solids~{\cite{nakano2022SCE,tenti2024hugoniot,slootman2024accurate}}, while it is impractical in this study because the number of variational parameters is too much to \textit{exactly} optimize all of them.
(iii) Nakano et al.~{\cite{nakano2024efficient}} recently proposed a method that guarantees unbiased forces and pressures even though not all variational parameters are optimized. Their proposed method have been successful for VMC force and pressure computations, while it is still under development for DMC forces and pressures.

\begin{figure}[t]
    \centering
    \includegraphics[width=1.0\linewidth]{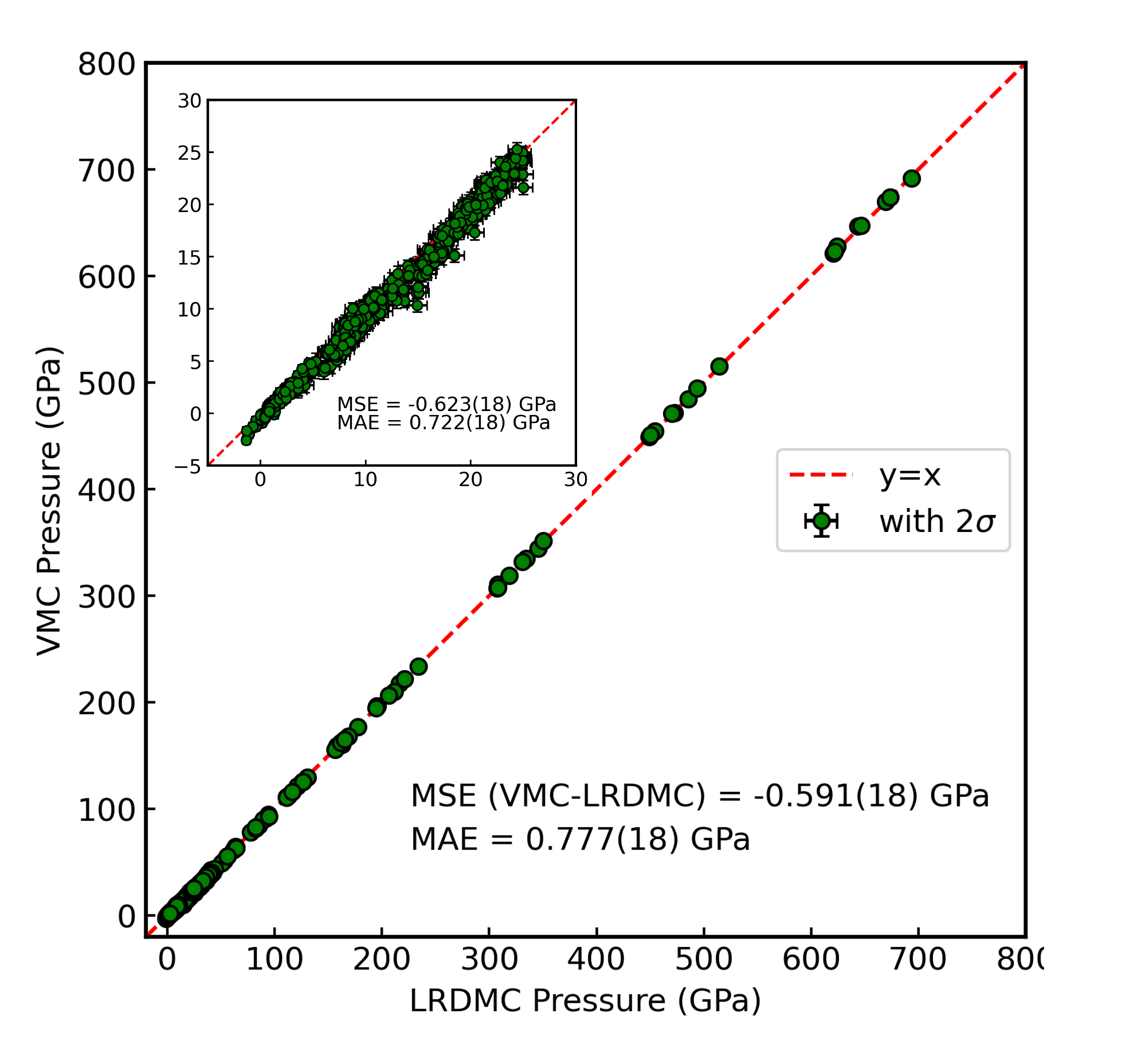}
    \caption{The comparison between Pressures computed by VMC and LRDMC. Inset: Those in the range $\le$ 25 GPa.
    MSE and MAE stand for mean signed error and mean absolute error, respectively.
        }
    \label{fig:P_vmc_nP_vs_lrdmc_nP_all}
\end{figure}

\section{{Details on the VMC and LRDMC pressures computations}}
\label{app:qmc-validation}
The VMC and LRDMC pressures were computed by fitting energy curves computed with 7 volumes ($r_s$) around the target one. The fittings were done using the third-order polynominal function.
First of all, using DFT, we confirmed that the chosen polynomial enables us to compute pressures correctly. More specifically, since the aforementioned bias problem does not exist in DFT, the pressures obtained via the PES fitting (denoted as numerical pressure) and those directly obtained via the Hellmann–Feynman theorem (denoted as analytical pressure) should be perfectly consistent.
Figure~{\ref{fig:qe_paw_nP_vs_qe_paw_aP}} shows the consistency checks for the randomly chosen configurations in the entire density region. The figure clearly shows that the fitting with the third-order polynomial gives pressures perfectly consistent with those directly obtained via the Hellmann–Feynman theorem. We also confirmed that all the PES obtained by VMC and LRDMC calculations are as smooth as those in the DFT calculations, thanks to accumulated statistics (In the VMC calculations, $\sim 1.2$~mHa/cell and $\sim 1.0$~mHa/cell for $r_s \leq 2.079$ and $r_s = 2.380$, respectively. In the LRDMC calculations, $\sim 2.0$~mHa/cell and $\sim 1.2$~mHa/cell for $r_s \leq 2.079$ and $r_s = 2.380$, respectively).

\begin{figure}[t]
    \centering
    \includegraphics[width=0.8\columnwidth]{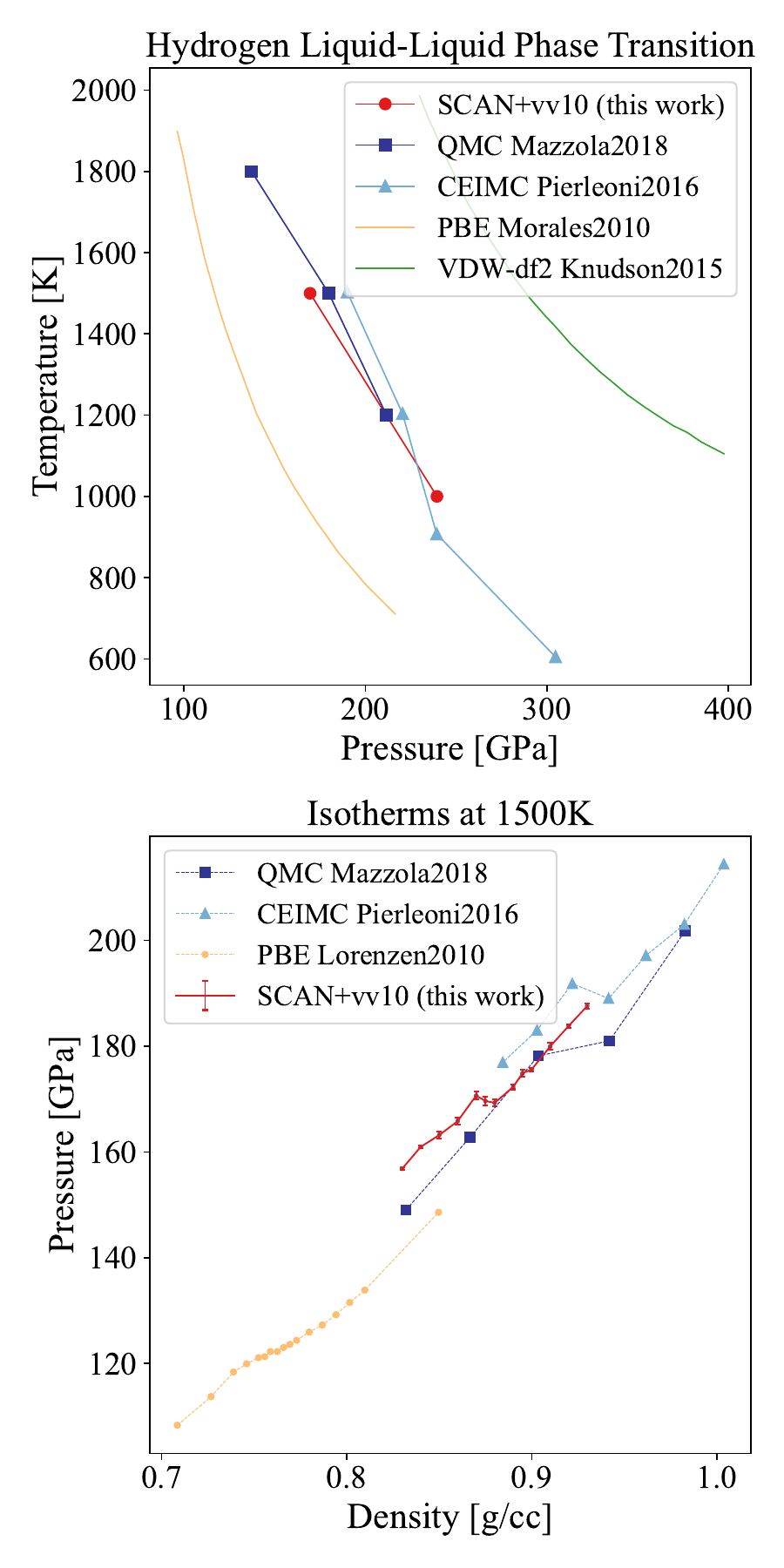}
    \caption{Liquid-liquid phase transition obtained with SCAN+vv10 ab-initio MD. \textit{Top:} Location of the phase boundary in the $P,T$ plane. We also plot the predictions made using PBE \cite{morales_evidence_2010},  vdW-DF2\cite{knudson2015direct} and QMC, using the CEIMC method\cite{pierleoni2016liquid} and a VMC-MD method which uses the same VMC code as the present work and but a slighlty different set-up\cite{mazzola2018}. \textit{Bottom:} the discontinuity of the EoS at 1500 K from the different theories.
    We chose to plot the 1500 K isotherm as we can find published data, from the above references, for this exact temperature.
        }
    \label{fig:llpt}
\end{figure}

Figure~{\ref{fig:P_vmc_nP_vs_vmc_aP_all}} plots the numerical and analytical VMC pressures collected from all 4512 configurations. The figure reveals that the biased arising from the parameters which are not variationally minimum amounts to MAE = $\sim$ 1.440(7)~GPa, which is constant among the studied $P$ and $r_s$, highlighting the importance in the smaller P and $r_s$ region.
Fig.~{\ref{fig:P_vmc_nP_vs_lrdmc_nP_all}} plots numerical (i.e. unbiased) VMC and LRDMC pressures collected from 832 configurations. The figure reveals that LRDMC and VMC pressures are consistent in the concerned pressure region (MAE = $\sim$ 0.78(2)~GPa and $\sim$ 0.72(2)~GPa for all $P$ and for $P \leq 25$ GPa, respectively), supporting the reliability our XC benchmark test referring to VMC Pressures as thier reference values.

\begin{figure}[t]
    \centering
    \includegraphics[width=0.9\columnwidth]{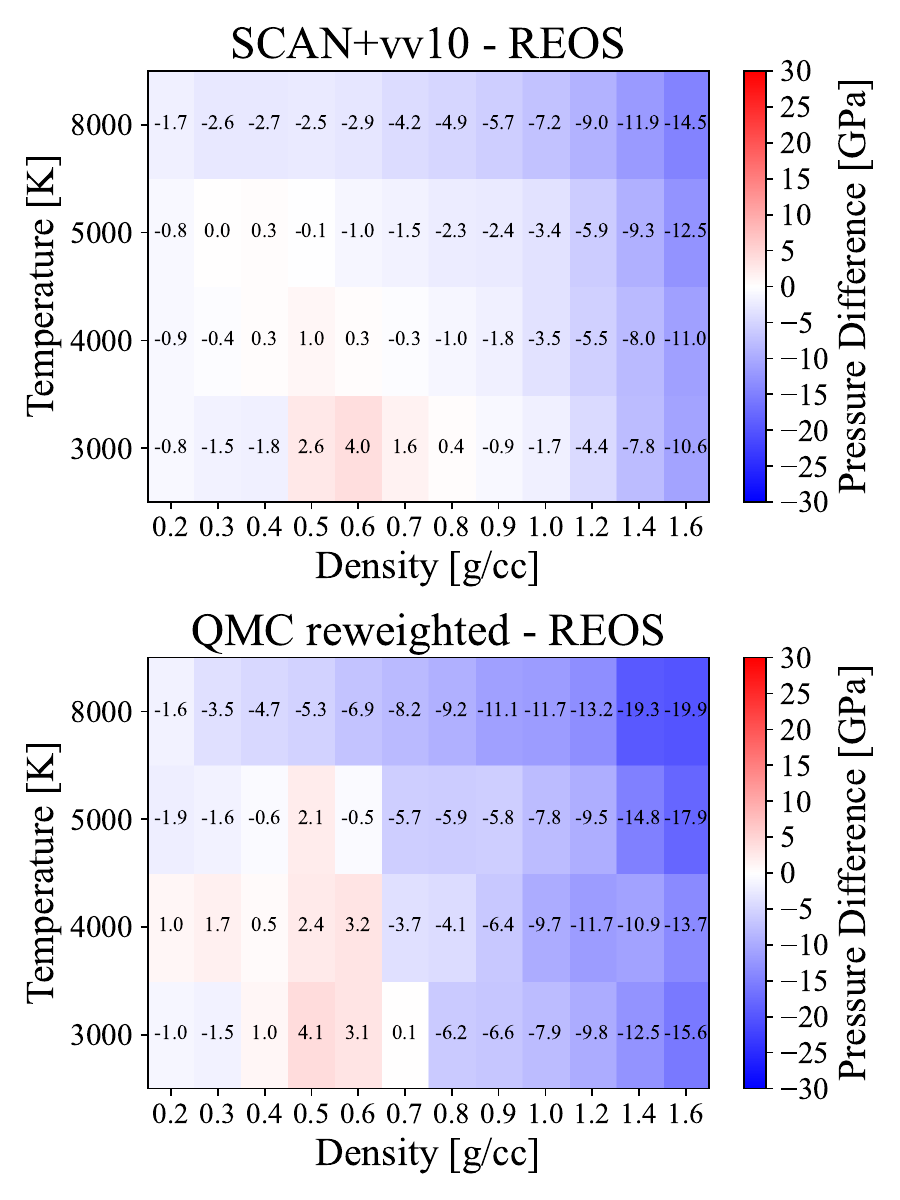}
    \caption{Comparison between the QMC EoS obtained by reweighting method (Eq.~\ref{eq:rewqmc}) and the SCAN EoS, for $T \ge 3000 K$.
 Shown is the difference with REOS.
    }
    \label{fig:QMCreweight}
\end{figure}

\section{Liquid-liquid phase transition}
\label{app:llpt}

In Fig.~\ref{fig:llpt} we show the agreement of SCAN+vv10 with previously published QMC data concerning the location of the liquid-liquid phase transition.
It is important to mention that all simulations reported adopt the classical-ions approximations as we do in the main text.
At about 1000 K nuclear quantum effects are expected to push the transition to lower pressure by about 30 GPa, therefore one needs to be careful to not mix data with and without inclusion of zero-point motion.

\section{Reweighting technique}
\label{app:reweighting}

In principle, it could be possible to compute average values for different level of theories, using only one configuration dataset $\bar{\bf x}_\alpha$, e.g. computing  $\langle O \rangle_B$ without running a full MD at theory $B$, but postprocessing the set $\bar{\bf x}_A$ using a procedure called \textit{reweighting}.
Formally, the expectation value of an operator $O$ can be computed using
\begin{equation}
\label{eq:rewqmc}
    \langle O \rangle_B = \frac{1}{M} \sum_{{\bf x}_i \in \bar{\bf x}_A} O_B({\bf x}_i) \omega_{A\rightarrow B}({\bf x}_i),
\end{equation}
where
%\begin{equation}
%\label{eq:boltzmann weights}
%   \omega_{A\rightarrow B}({\bf x}_i) = \frac{e^{-\beta (E_\textrm{QMC} ({\bf x}_i)-E_\textrm{PBE} ({\bf x}_i))}}{\sum_{{\bf x}_i \in \bar{\bf x}_A} e^{-\beta (E_\textrm{QMC} ({\bf x}_i)-E_\textrm{PBE} ({\bf x}_i))}} \quad .
%\end{equation}
\begin{equation}
\label{eq:boltzmann weights}
   \omega_{A\rightarrow B}({\bf x}_i) = \frac{e^{-\beta (E_B ({\bf x}_i)-E_A ({\bf x}_i))}}{\sum_{{\bf x}_i \in \bar{\bf x}_A} e^{-\beta (E_B ({\bf x}_i)-E_A ({\bf x}_i))}} \quad .
\end{equation}
With this procedure one enhance or decrease the significance for of a datapoint ${\bf x}_i$ sampled with theory $A$, for the theory $B$.
Unfortunately this procedure is statistically unstable, due to the large fluctuation of the weights, as the energy $E$ is an extensive quantity.
Clearly, the fluctuations are particularly severe at low temperature, due to the presence of the inverse temperature in the exponents.
The method becomes more accurate as the energy predicted by the two theories becomes equal.

We can try this procedure using $A=$ PBE and $B=$ VMC.
In Fig.~\ref{fig:QMCreweight} we plot the reweighted VMC pressure on the PBE configurations, and we compare with the SCAN+vv10 EoS. Note that we are reweighting the VMC pressures on PBE because the configurations used to build the QMC dataset come from PBE MDs.
For consistency we plot the differences with respect of REOS3.
We also plot only the region above 2000 K because it is less affected by statistical noise and is relevant for planetary science.

We notice that the two EoS are qualitative in agreement, both showing a region at intermediate densities where they predicts a lighter liquid compared to REOS, while it is denser at higher densities.
However, the statistical error of this method is of about 8 GPa. For this reason the VMC reweighted results cannot be used to build a precise EoS, but the signal it provides us is a further proof of the quality of the SCAN+vv10 results.
Notice that statistical noise in the EoS would prevent us from calculating entropy accurately.

\section{{Details on the  $e_0$ calculation for the Hugoniot}}
\label{app:e0_hugo}

One important factor to consider is the zero point energy (ZPE) of the hydrogen molecules at nearly zero temperature. To evaluate this, we use the very reasonable approximation at very low density that the total ZPE of the system is the sum of the ZPE of the isolated molecules.
\\
We focus now on this subproblem:
 we consider the SCAN+vv10 molecule, using a discrete variable representation (DVR) approach where we exactly diagonalize the associated Schr\"{o}dinger equation (in one dimension) in discretized space.\cite{harris1965calculation} We use the plane-wave basis to compute the kinetic operator, and a grid of 100 spacings in the range  0.5 to 2.5 Bohr}.
\\
We find the ZPE of the SCAN+vv10 molecule to be 0.007077 Ha. This translates into a value of about 1551 cm$^{-1}$, which is a reasonable value compared to experimental one of 1546.50(8) cm$^{-1}$.\cite{irikura2007experimental} The total energy of the zero-temperature D$_2$ molecule, obtained by adding this ZPE to the minimum classical ions energy at bond equilibrium, is -1.1626 Ha, i.e., -0.58131 Ha/atom (see Fig.~\ref{fig:vibrationD2}).
\\
\begin{figure}[t]
    \centering
    \includegraphics[width=1.0\columnwidth]{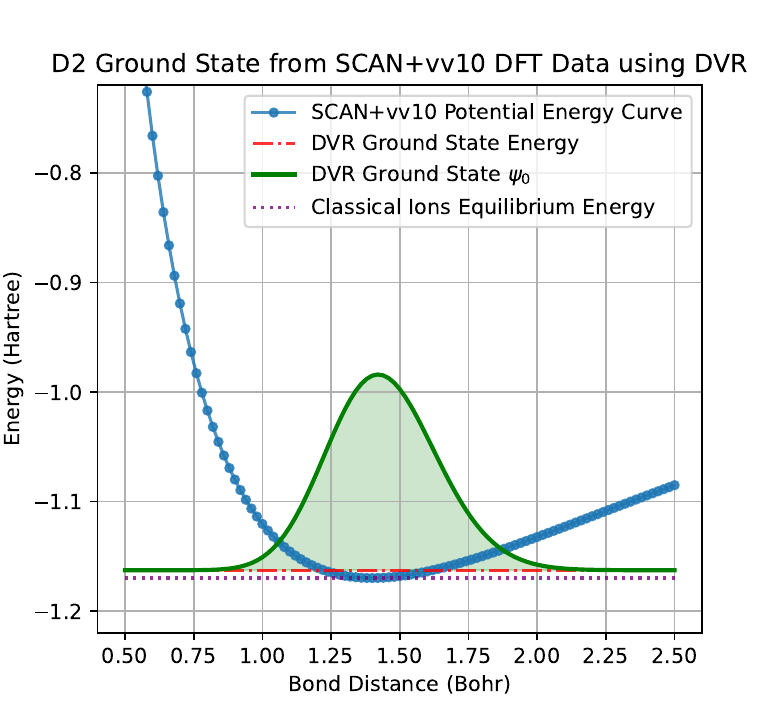}
    \caption{Vibrational ground state energy and wavefunction of D$_2$ calculated with a DVR method form the potential energy surface from SCAN+vv10 data.}
    \label{fig:vibrationD2}
\end{figure}

To further validate our methodology, we also fit our DFT data with a Morse potential and calculate the ZPE analytically\cite{dahl1988morse}; in this case, we find a vibrational energy of 0.00749 Ha. This comparison is useful to set the discretization mesh for the DVR method, i.e. comparing the DVR and the analytical results on same Morse potential function.
\\
The finite temperature corrections to the ZPE alone (i.e considering 22 K rather than absolute zero temperature) are negligible, since the first excited vibrational state energy is -1.14888 Ha.
\\
To find the $e_0$ parameter for the Hugoniot we add the ZPE, 0.003538 Ha/atom, to the total energy of the classical system at 22 K and at 0.167 g/cc. To compute this, we perform a MD simulation using 64 atoms using the classical nuclei approximation. The energy of this system is -0.58467(1) Ha/atom.
We find that this value is very close, up to 0.00007(1) Ha/atom, to the energy per atom of an isolated molecule. For instance, this difference is smaller compared to the approximation one would do in fitting the SCAN data with a Morse potential.
Adding the ZPE/atom found using the single-molecule approximation, we find the final estimate $e_0 =$-0.58114(1) Ha/atom.
\\
We perform a final consistency check and re-do the same workflow using PBE.
We find that the $e_0^\textrm{PBE}=$-0.58057(1) Ha/atom, which is in excellent agreement with the value of -0.58055(2) obtained from a full path-integral MD in Ref~\cite{tenti2024hugoniot}.
This validate the initial approximation of calculating the ZPE only using a single molecule.\\
Moreover, the difference in $e_0$ obtained with PBE vs SCAN gives us a meaningful energy spread that we can use to plot the systematic error bar for the Huguniot line in the main text.

\section{{Robustness of Jupiter's density profile}}
\label{app:robustness_model}

\begin{figure}[t]
    \centering
    \includegraphics[width=1.0\columnwidth]{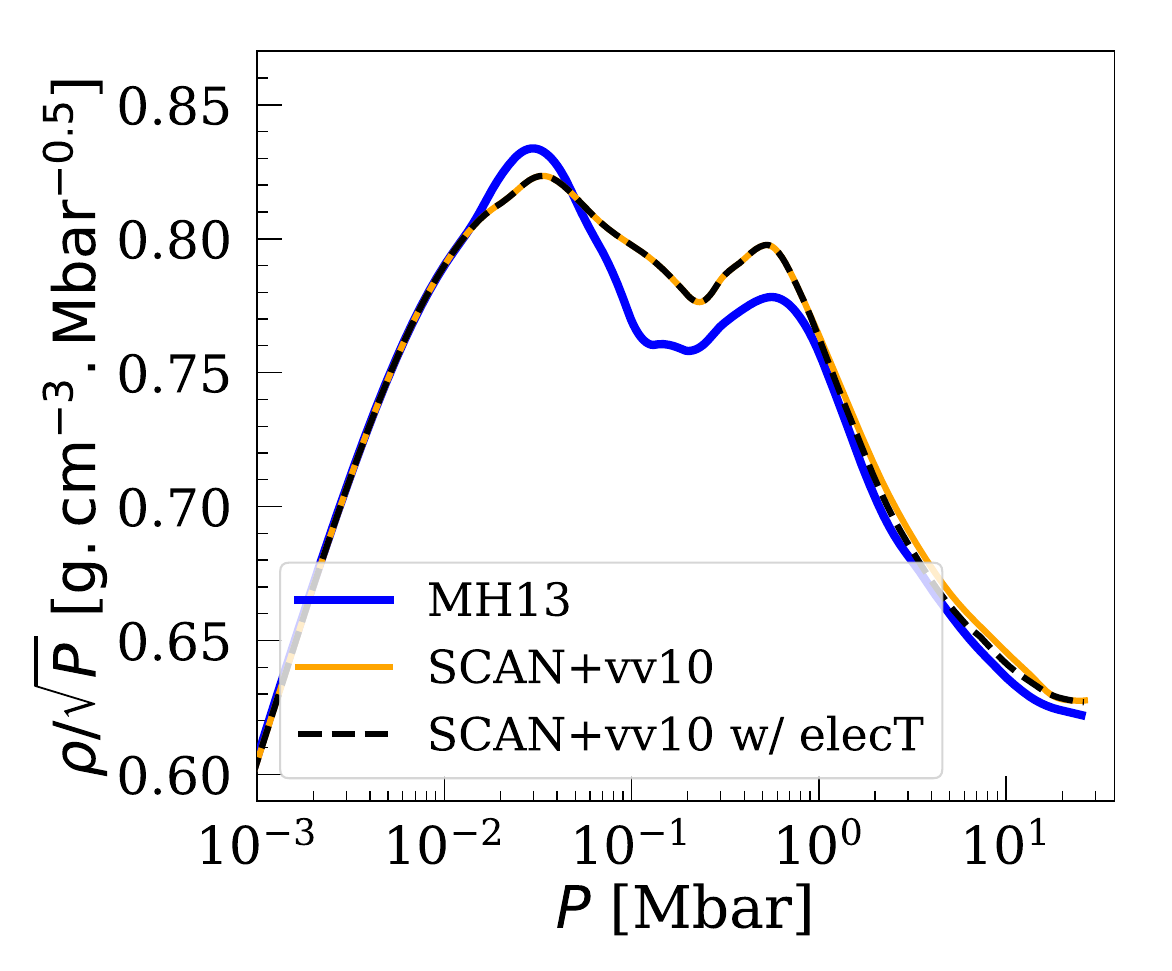}
    \caption{Same as Fig.~\ref{fig:saburo_XC_rhoP} but correcting for electronic temperature (dashed black line).}
    \label{fig:profile_elecT}
\end{figure}

\begin{figure}[tb]
    \centering
    \includegraphics[width=\linewidth]{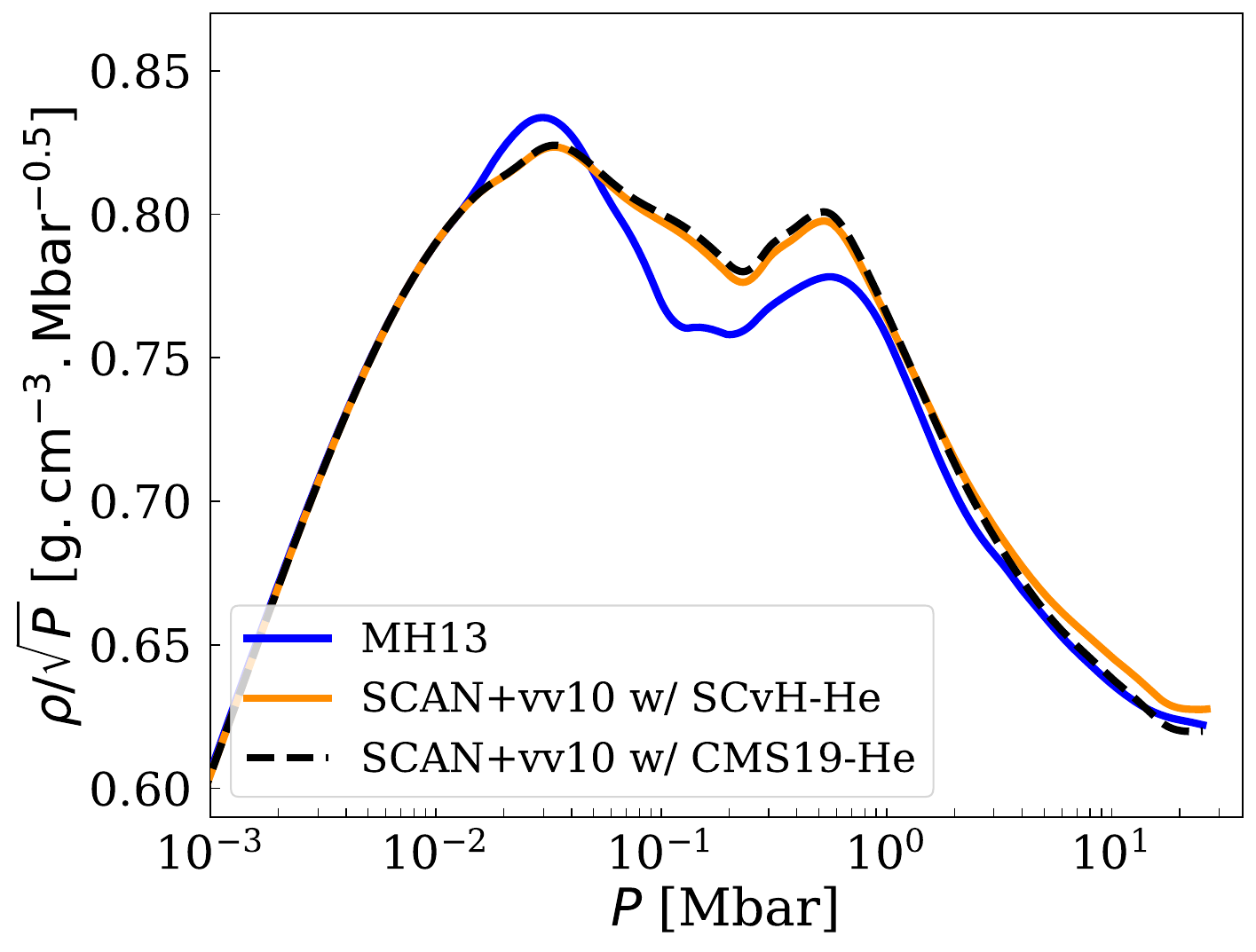}
    \caption[]{A comparison of adiabats obtained with different helium EoSs.}
    \label{fig:he-EOS}
\end{figure}

In this Appendix, we test the robustness of the calculated Jupiter density profile by varying certain details of the EoS.
First, we consider the electronic ground-state approximation. As mentioned in the main text, we adopt the Born-Oppenheimer approximation, where the ions move on the electronic ground-state potential surface.
A simple way to account for electronic thermal effects is to calculate observables, such as energy and pressure, using a Fermi-Dirac distribution for the molecular orbitals (see Appendix~\ref{app:dft}).
In Appendix~\ref{app:dft}, we report that the difference in pressure is less than $1\%$ under Jupiter-like conditions and can be parameterized as a smooth, increasing function of density and temperature.
We correct the values of $P$ and $E$ between $5000$ K and $11000$ K using the Fermi-Dirac smearing approach.
We then recalculate the entropy and, finally, the density profile of the planet (see Fig.~\ref{fig:saburo_XC_rhoP}).
While some differences are noticeable, they remain within $1\%$, and are evident only in the pressure range between 1 and 10 Mbar.
Notice that above 11000 K the Jupiter adiabat enters the interpolation region which connects the SCAN+vv10 data with CMS19.
Secondly, we examine the sensitivity of the results when using a different equation of state for He. The helium mass fraction in Jupiter's envelope varies between 0.23 and 0.28. To assess the sensitivity of our results to the helium EoS, we recalculated an adiabat (as in Sec.~\ref{s:models}) using our SCAN+vv10 hydrogen EoS in combination with the CMS19 helium EoS, instead of the SCvH helium EoS. The CMS19-He EoS differs from SCvH-He in the intermediate $T$--$\rho$ regime, where it is based on QMD simulations, and in the high density regime, where it is based on a fully ionized model \citep[see][and references therein]{2019ApJ...872...51C}. As shown in Fig.~\ref{fig:he-EOS}, the resulting difference in density between models using these two helium EoSs remains below 1\%. This confirms that the hydrogen EoS is the dominant factor controlling the density profile of Jupiter.

\section{Extended DFTxc benchmarks}
\label{app:more_benchmarks}

%\iffalse

In the following we show the results for the benchmark against VMC and LRDMC for energy and pressure for all the DFT xc functionals who has been tested.

\begin{figure*}[ht]
	\centering
    \subfigure[]{\includegraphics[width=0.49\linewidth]{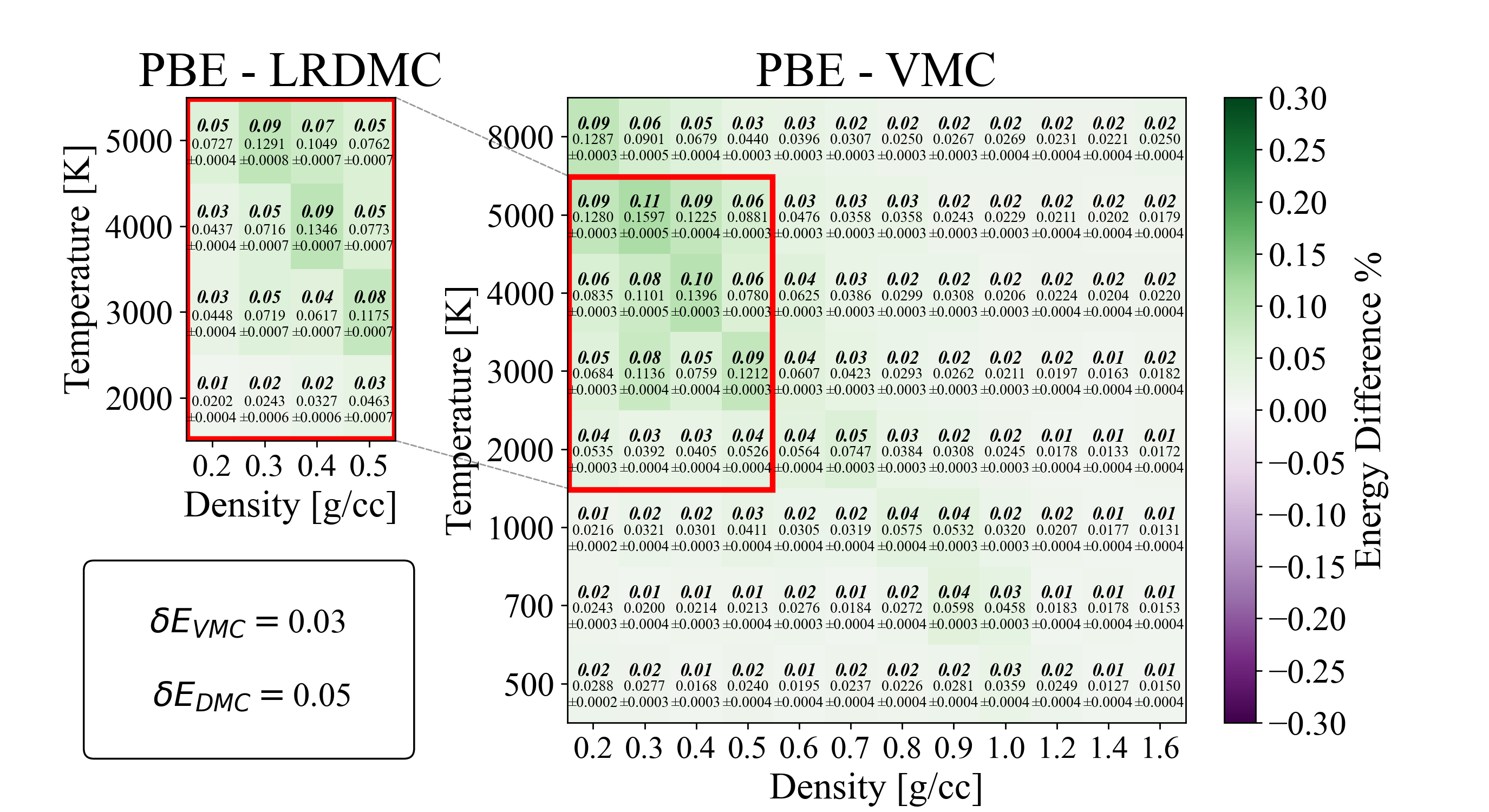}}
    \subfigure[]{\includegraphics[width=0.49\linewidth]{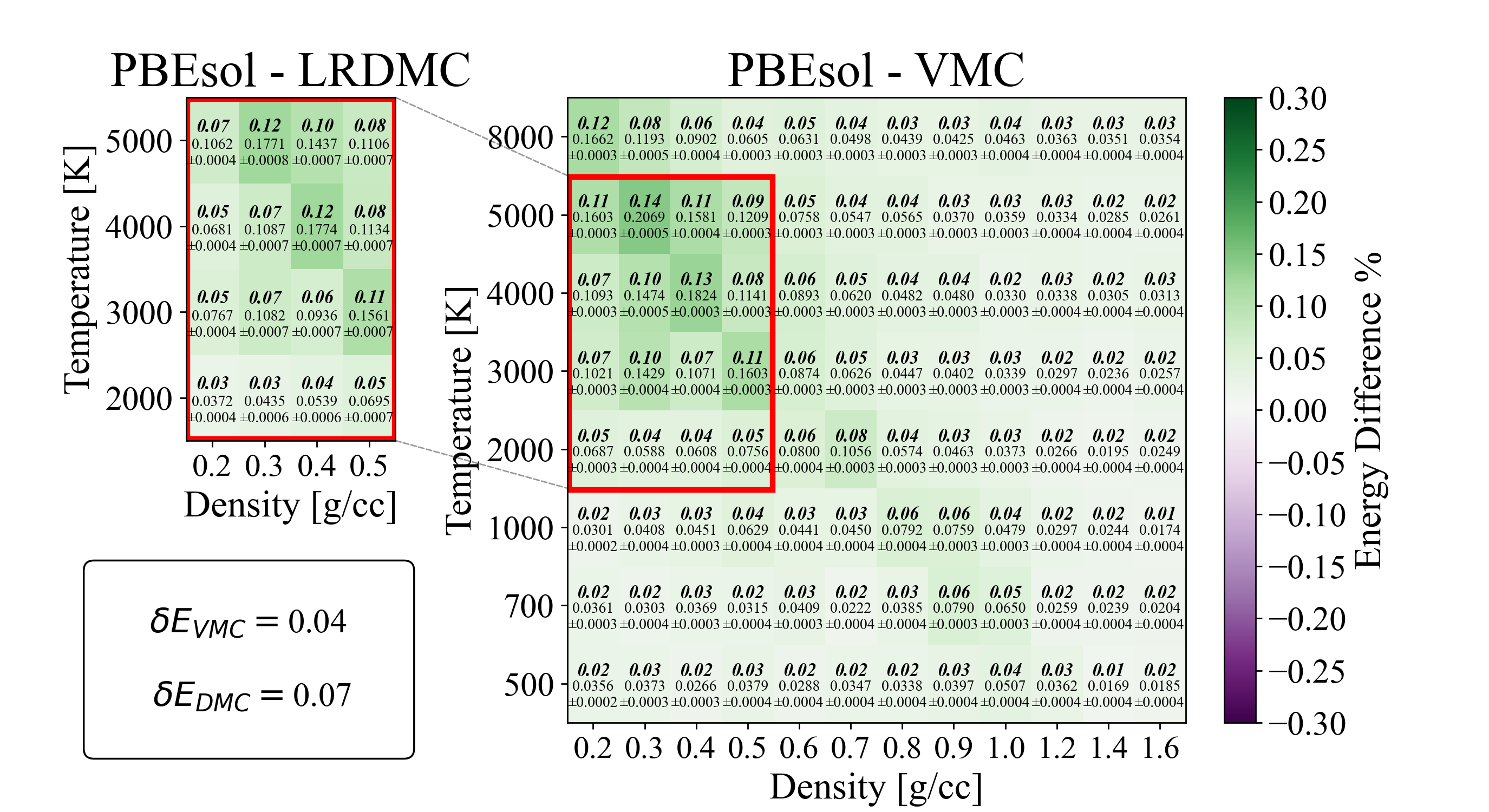}}
    \\
	\subfigure[]{\includegraphics[width=0.49\linewidth]{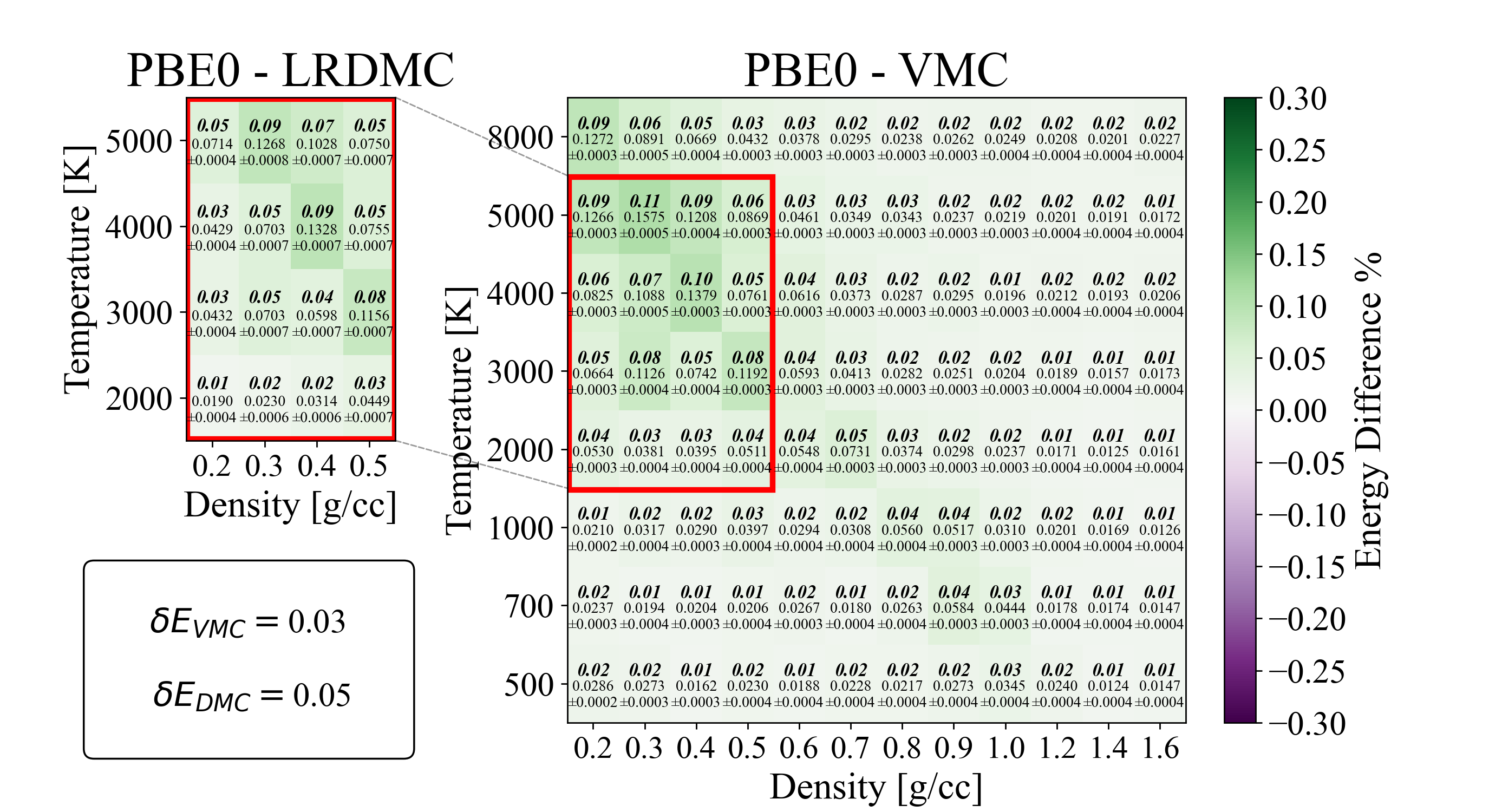}}
    \subfigure[]{\includegraphics[width=0.49\linewidth]{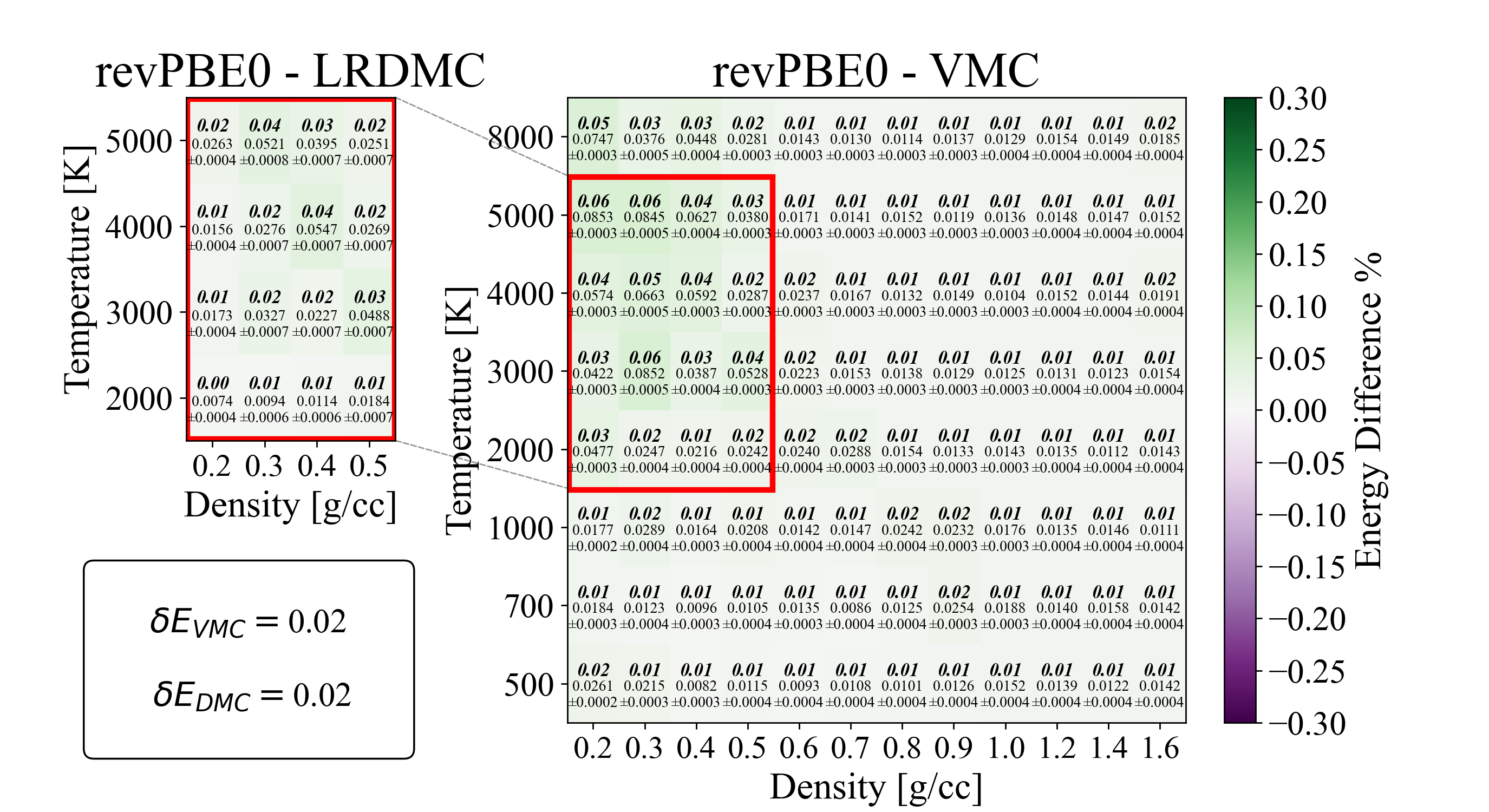}}
    \\
    \subfigure[]{\includegraphics[width=0.49\linewidth]{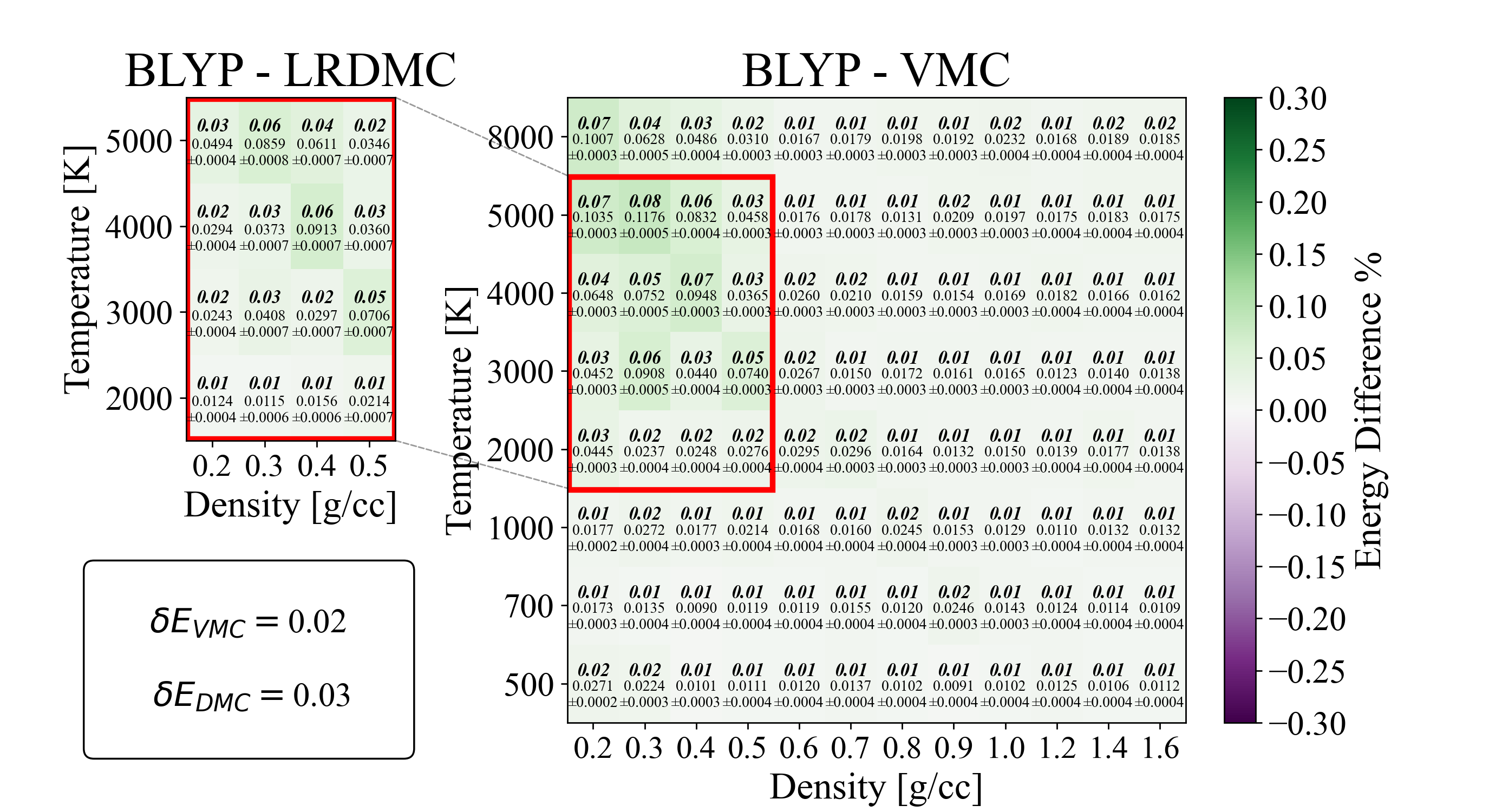}}
    \subfigure[]{\includegraphics[width=0.49\linewidth]{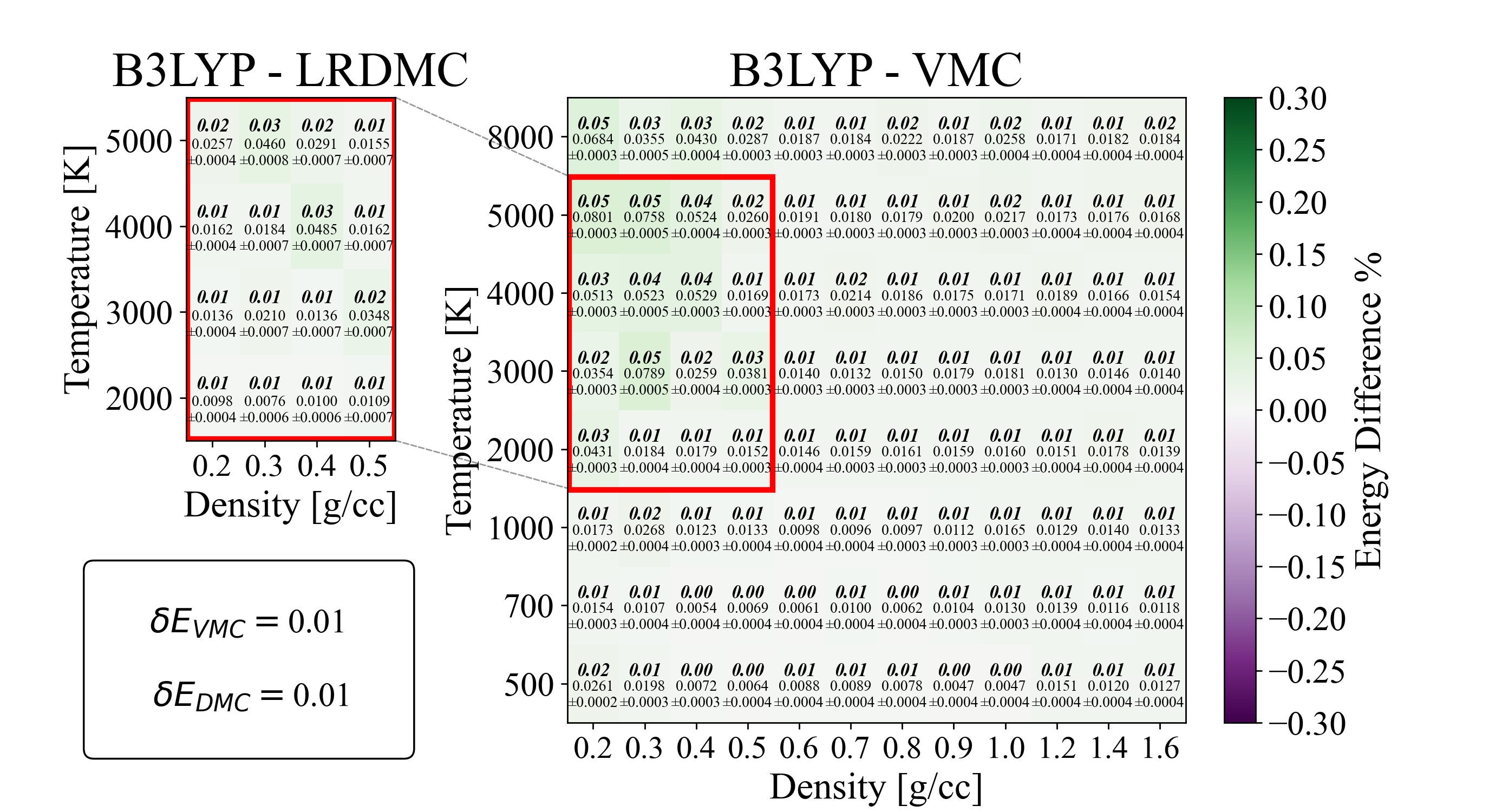}}
    \\
    \subfigure[]{\includegraphics[width=0.49\linewidth]{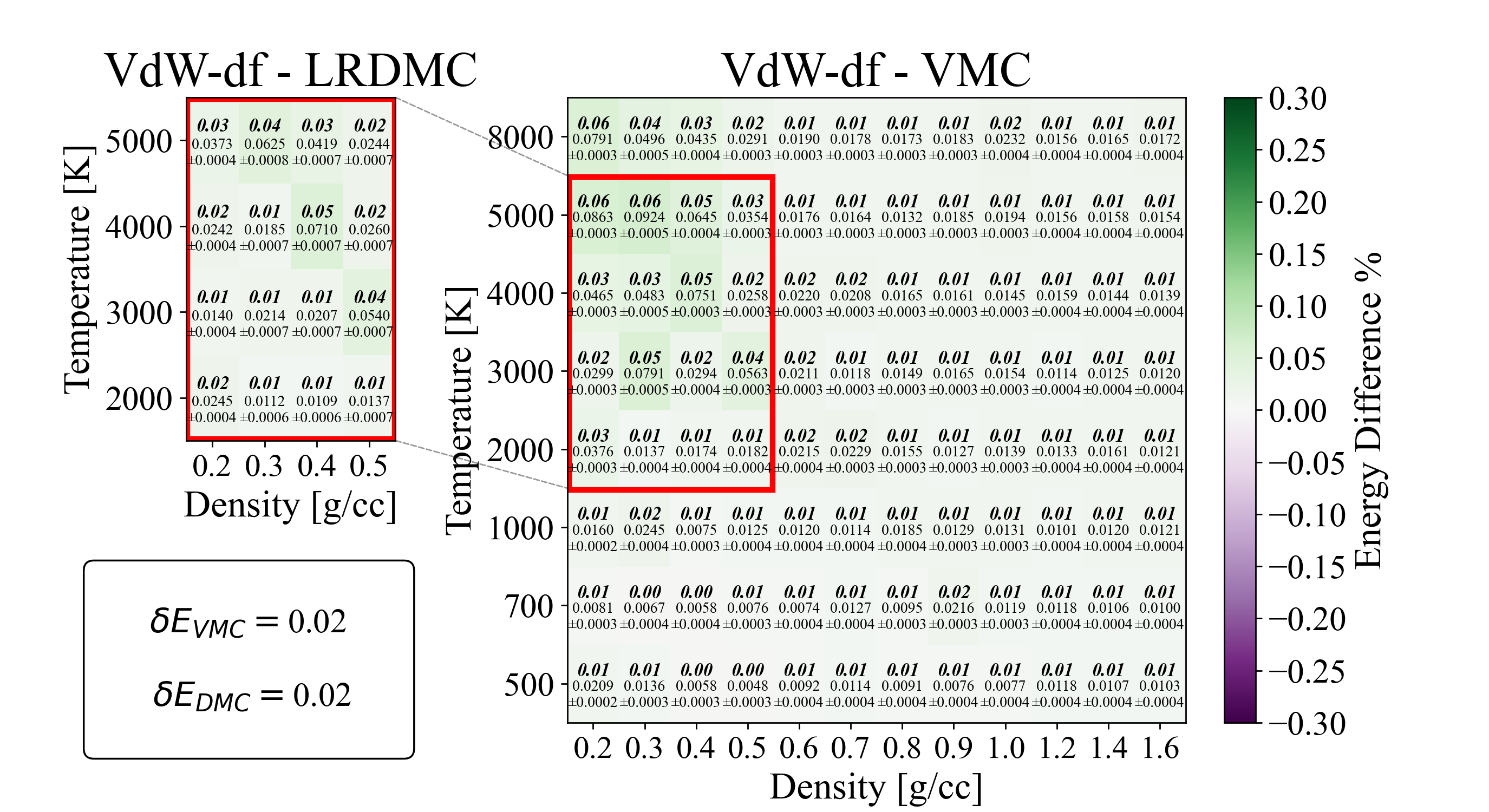}}
	\subfigure[]{\includegraphics[width=0.49\linewidth]{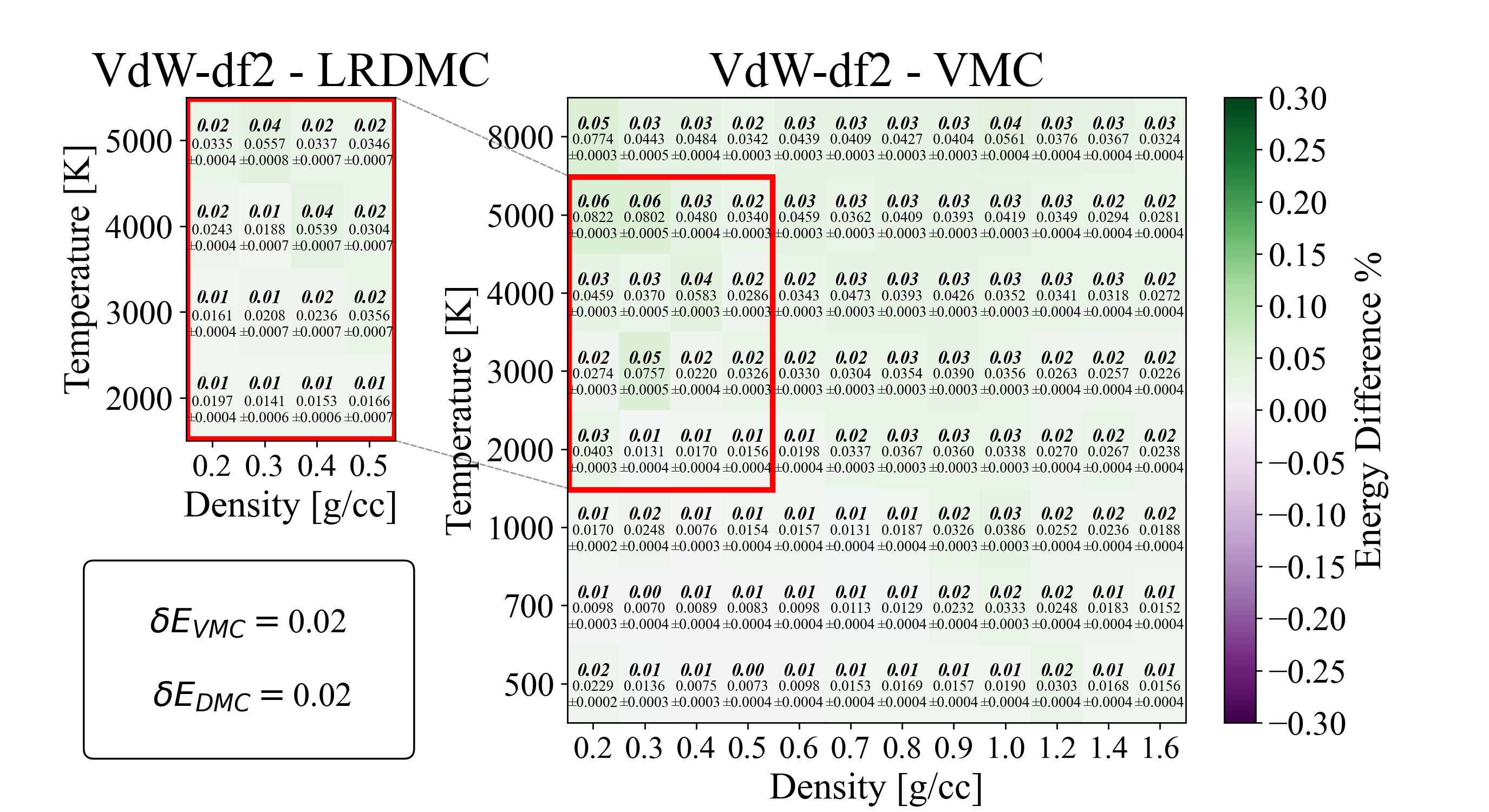}}

	\caption{\emph{Part 1.} Internal Energy benchmark of the DFT xc functional  against VMC and LRDMC for the $T-\rho$ of interest. In every cell the bold value is the relative difference, while the lower value is the absolute difference with error. These values are computed as in Eq.\ref{eq:energy_metric}. The values in the box of every image are computed as Eq.\ref{eq:scoreP}, exchanging $E$ with $P$ .}
    \label{fig:Energy_benchmark_1}
\end{figure*}

\begin{figure*}[ht]
	\centering
    \subfigure[]{\includegraphics[width=0.49\linewidth]{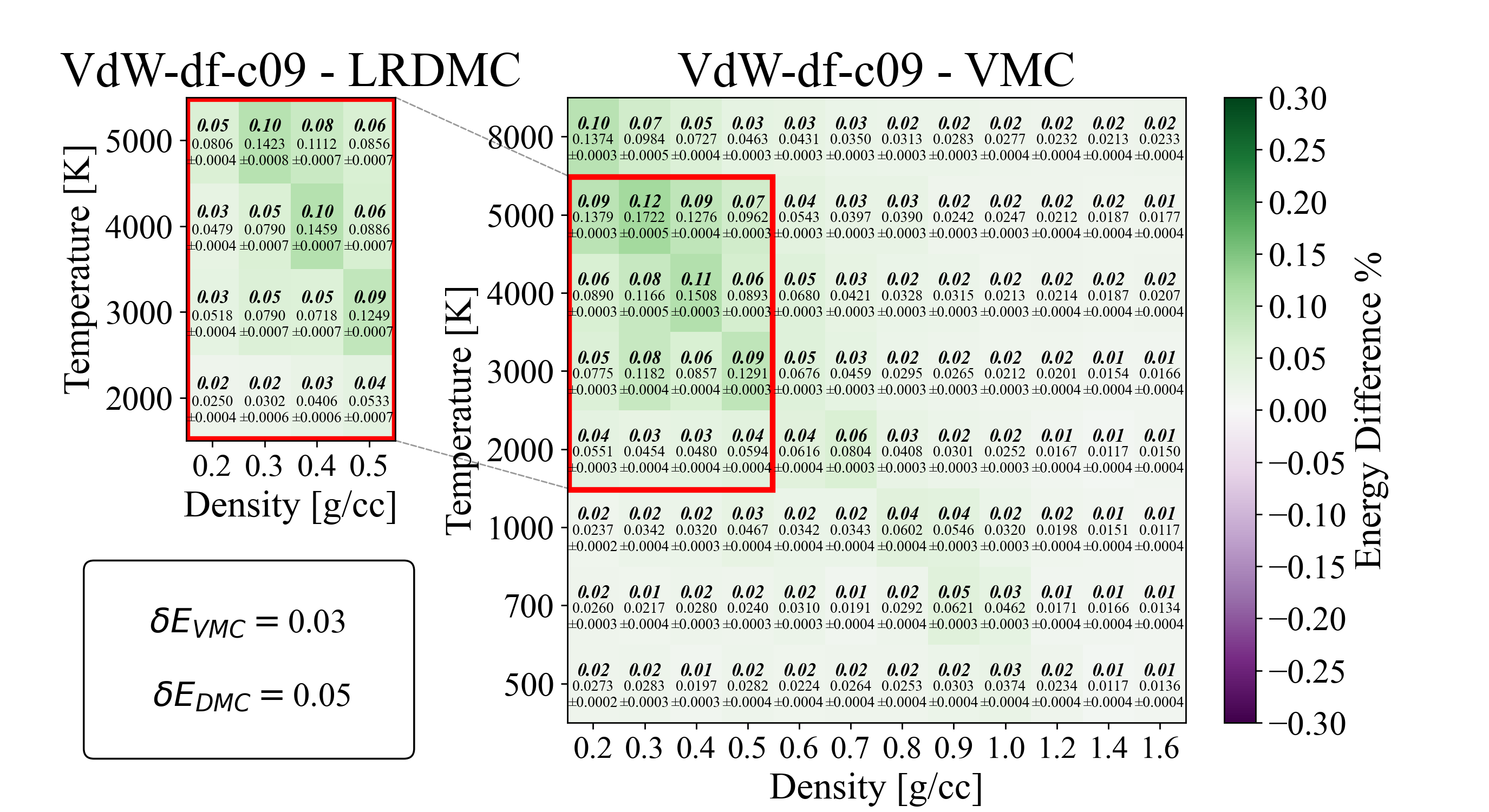}}
    \subfigure[]{\includegraphics[width=0.49\linewidth]{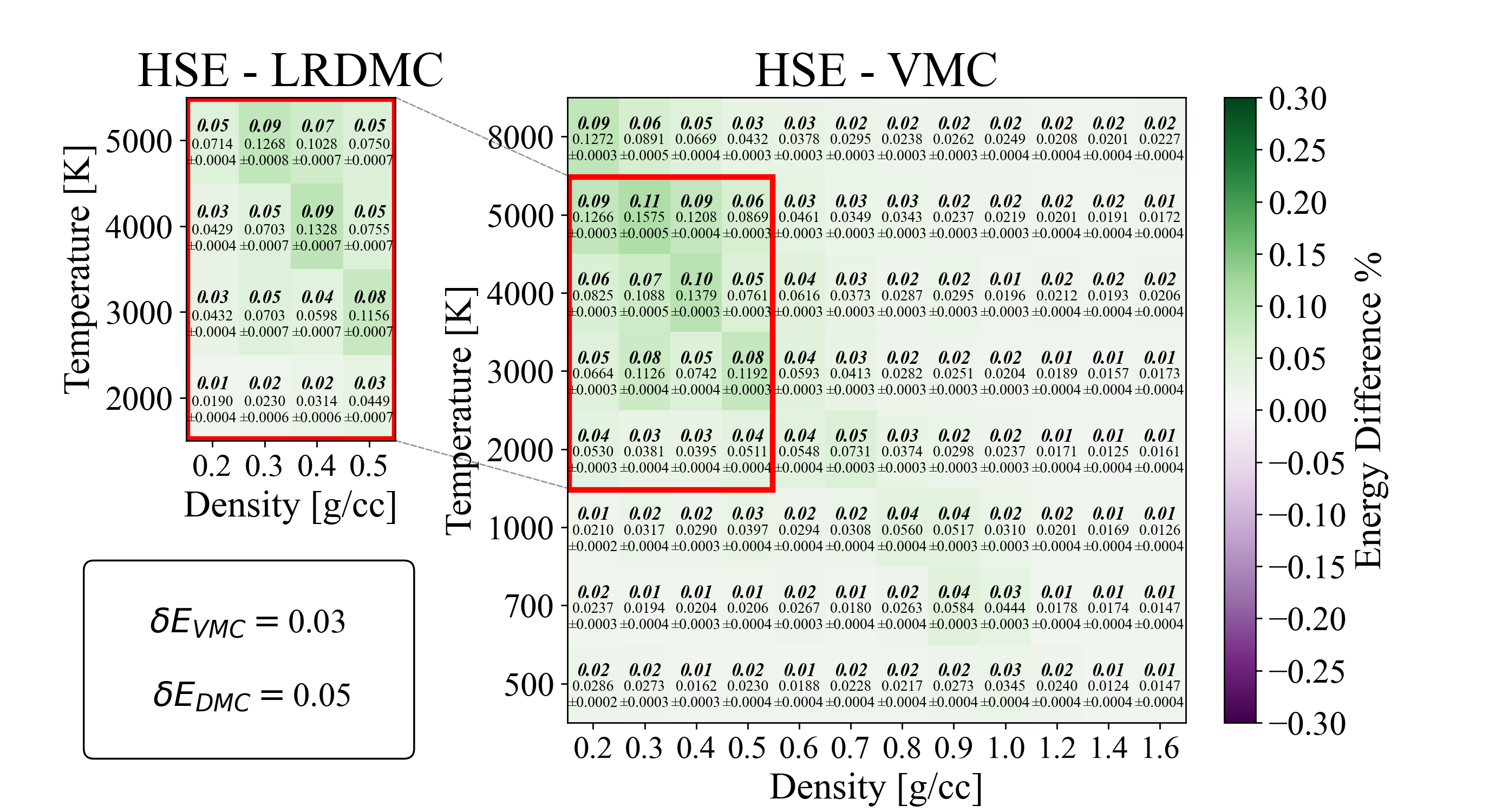}}
    \\
    \subfigure[]{\includegraphics[width=0.49\linewidth]{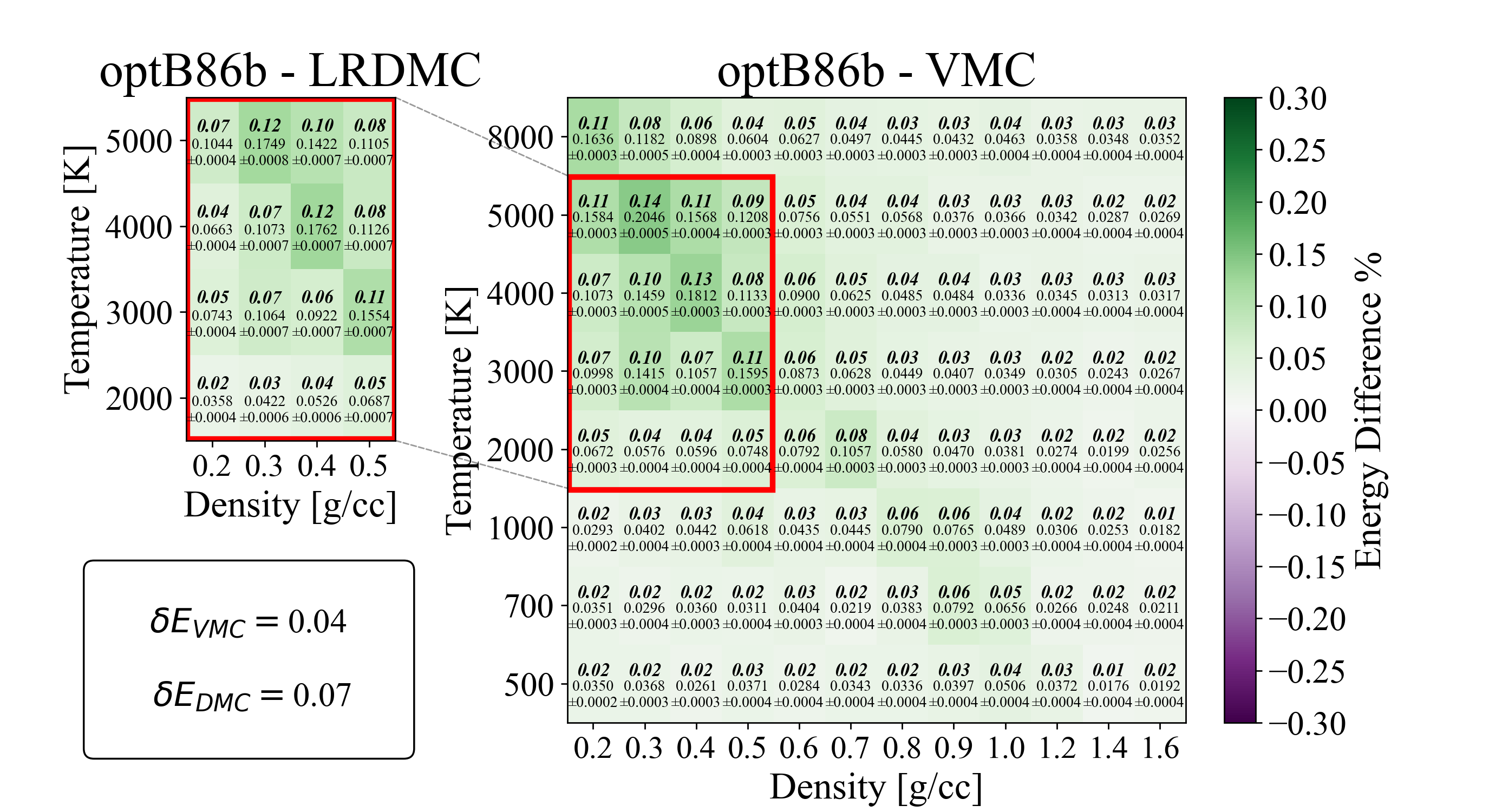}}
	\subfigure[]{\includegraphics[width=0.49\linewidth]{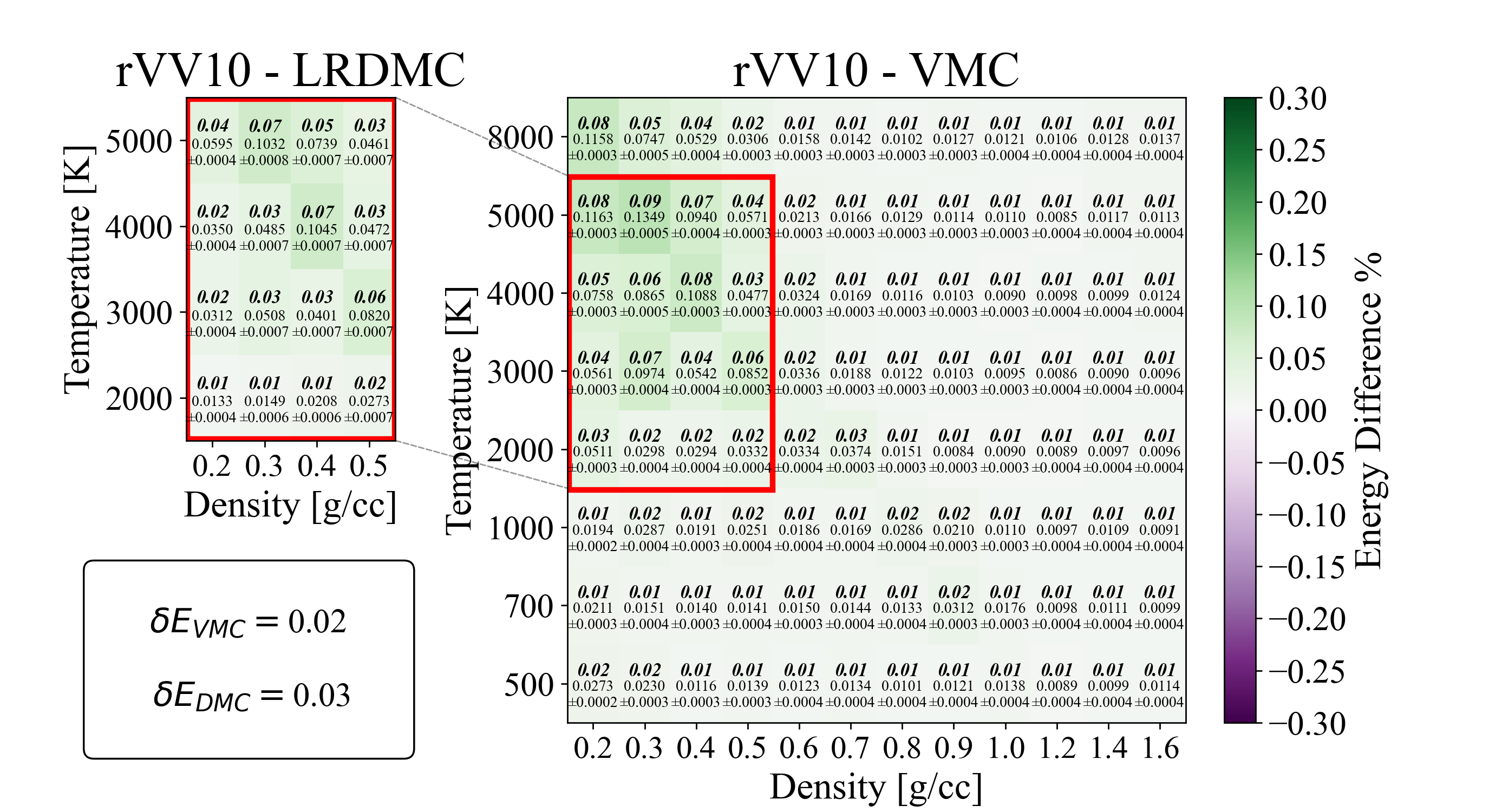}}
    \\
    \subfigure[]{\includegraphics[width=0.49\linewidth]{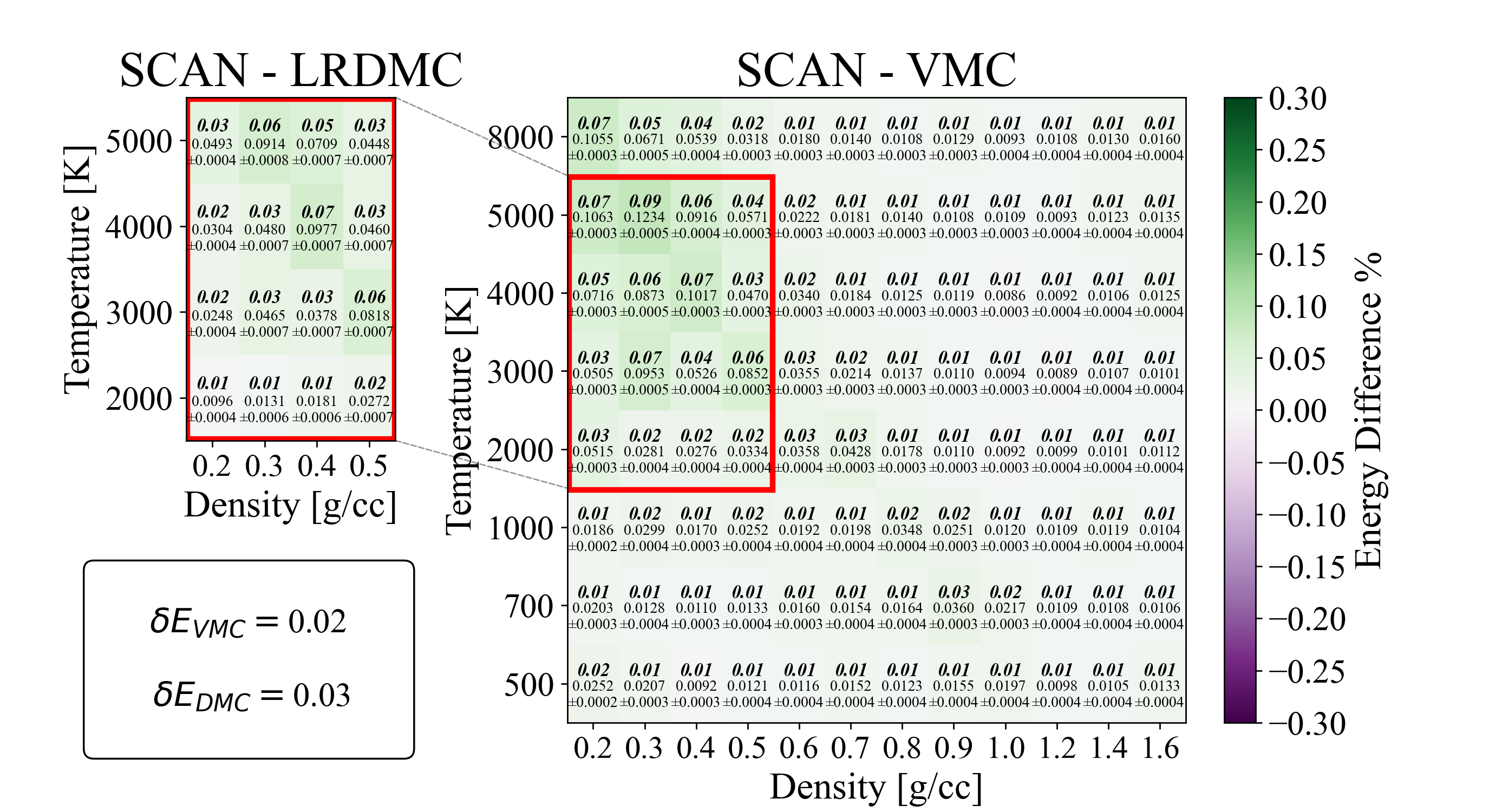}}
	\subfigure[]{\includegraphics[width=0.49\linewidth]{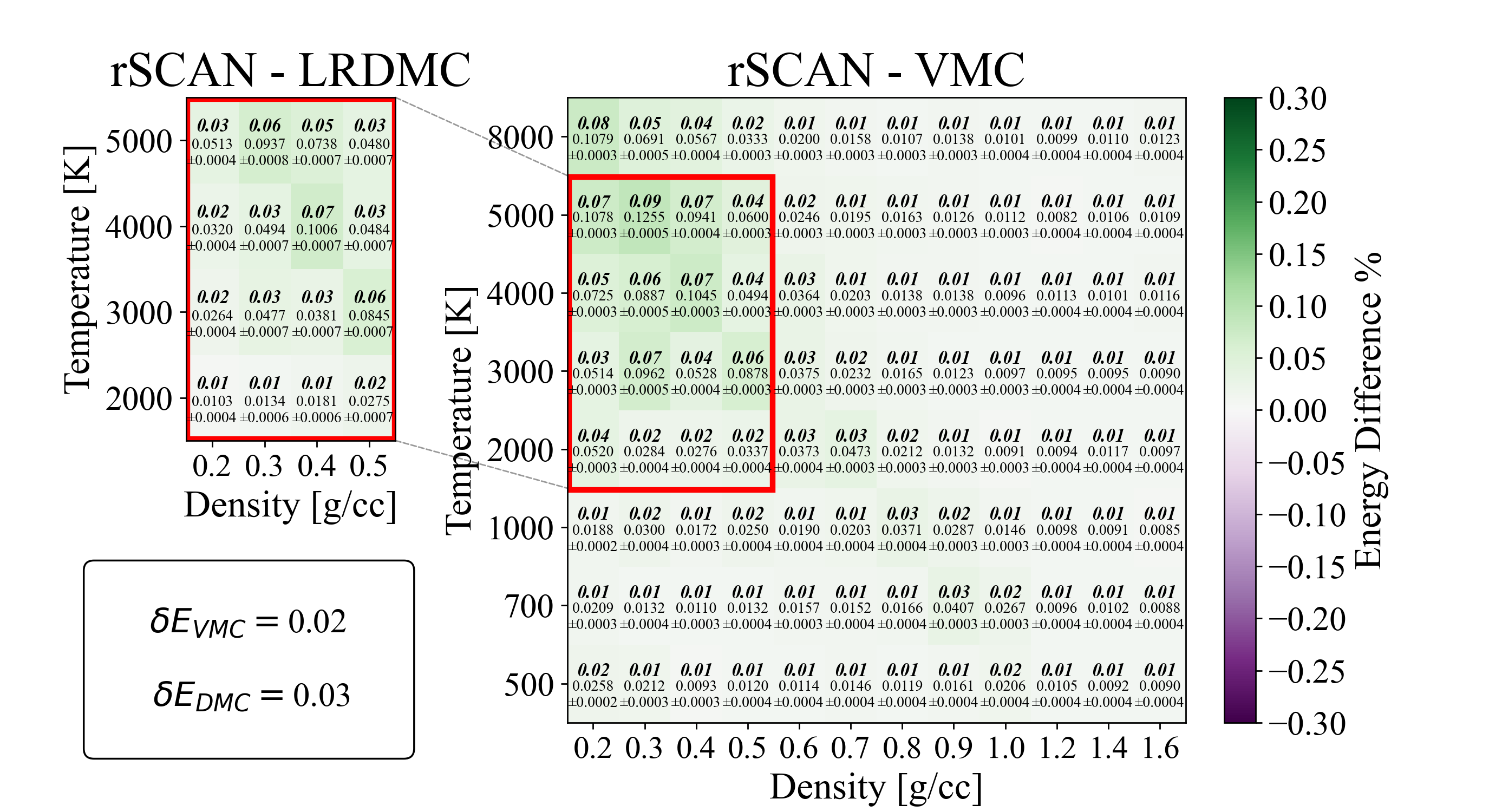}}
    \\
    \subfigure[]{\includegraphics[width=0.49\linewidth]{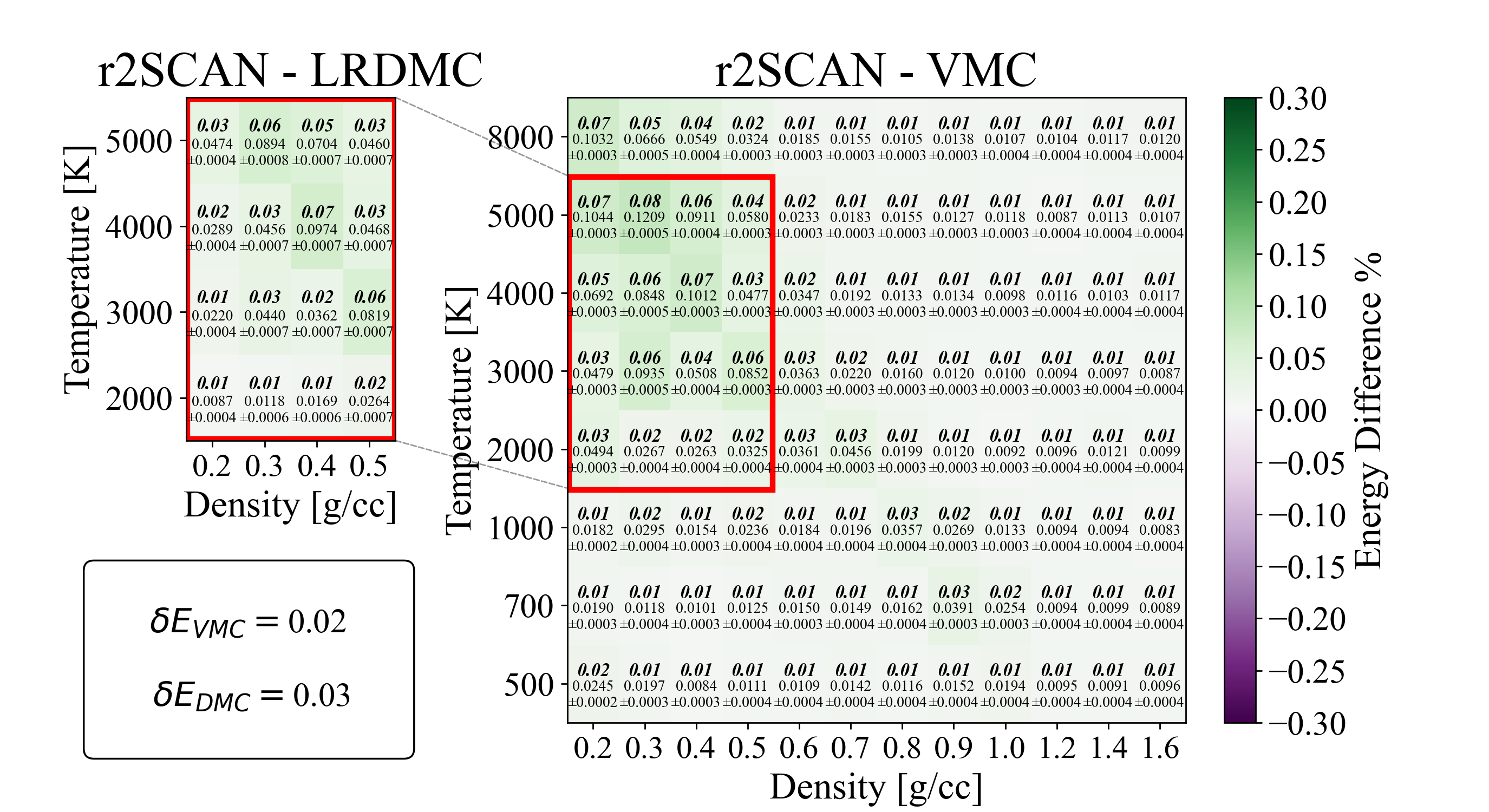}}
    \subfigure[]{\includegraphics[width=0.49\linewidth]{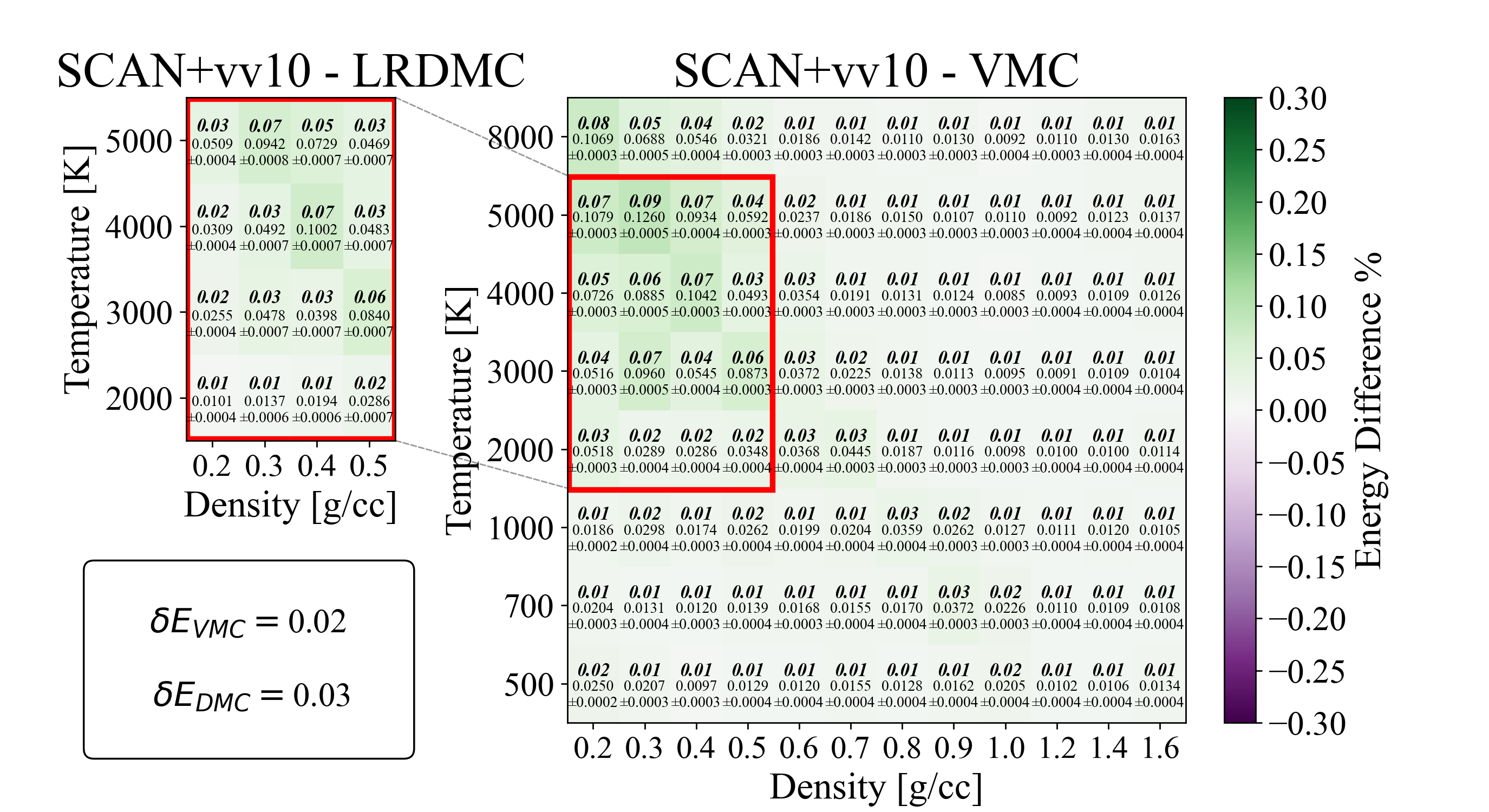}}

	\caption{\emph{Part 2.} Internal Energy benchmark of the DFT xc functional  against VMC and LRDMC for the $T-\rho$ of interest. In every cell the bold value is the relative difference, while the lower value is the absolute difference with error. These values are computed as in Eq.\ref{eq:energy_metric}. The values in the box of every image are computed as Eq.\ref{eq:scoreP}, exchanging $E$ with $P$.}
	\label{fig:Energy_benchmark_2}
\end{figure*}

\begin{figure*}[h]
	\centering
    \subfigure[]{\includegraphics[width=0.49\linewidth]{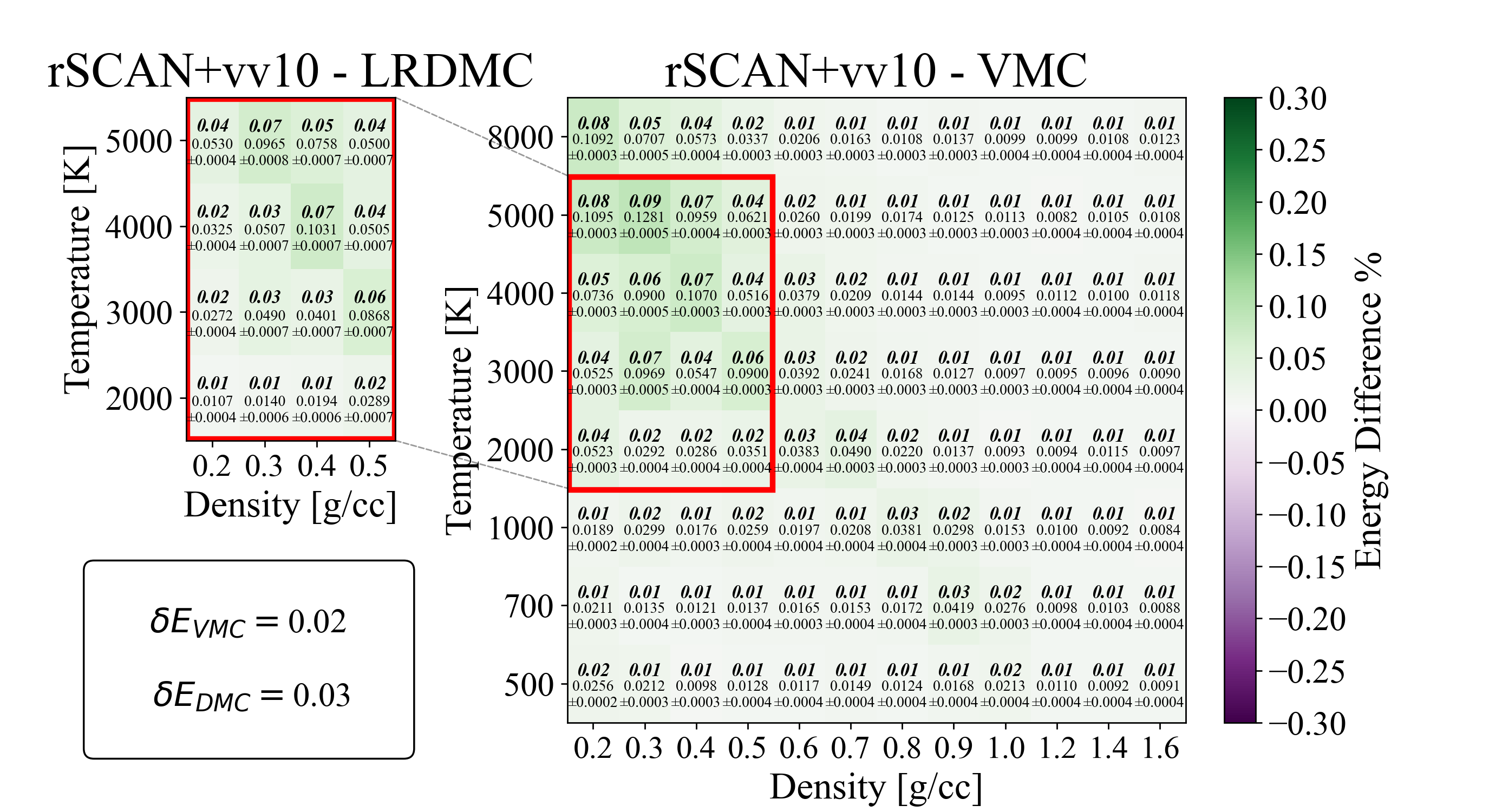}}
    \subfigure[]{\includegraphics[width=0.49\linewidth]{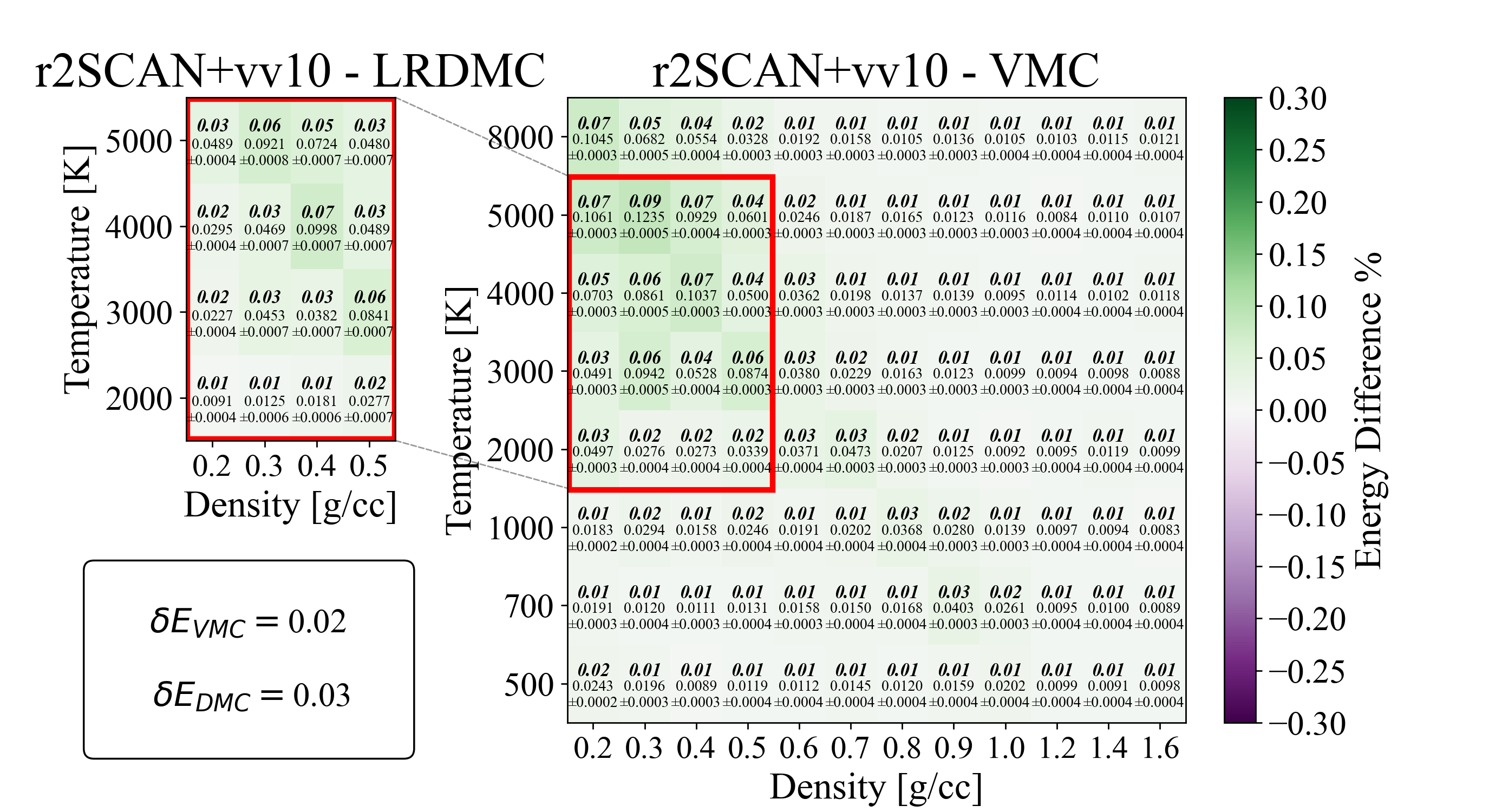}}

	\caption{\emph{Part 3.} Internal Energy benchmark of the DFT xc functional  against VMC and LRDMC for the $T-\rho$ of interest. In every cell the bold value is the relative difference, while the lower value is the absolute difference with error. These values are computed as in Eq.\ref{eq:energy_metric}. The values in the box of every image are computed as Eq.\ref{eq:scoreP}, exchanging $E$ with $P$.}
	\label{fig:Energy_benchmark_3}
\end{figure*}

\begin{figure*}[h]
	\centering
	\subfigure[]{\includegraphics[width=0.49\linewidth]{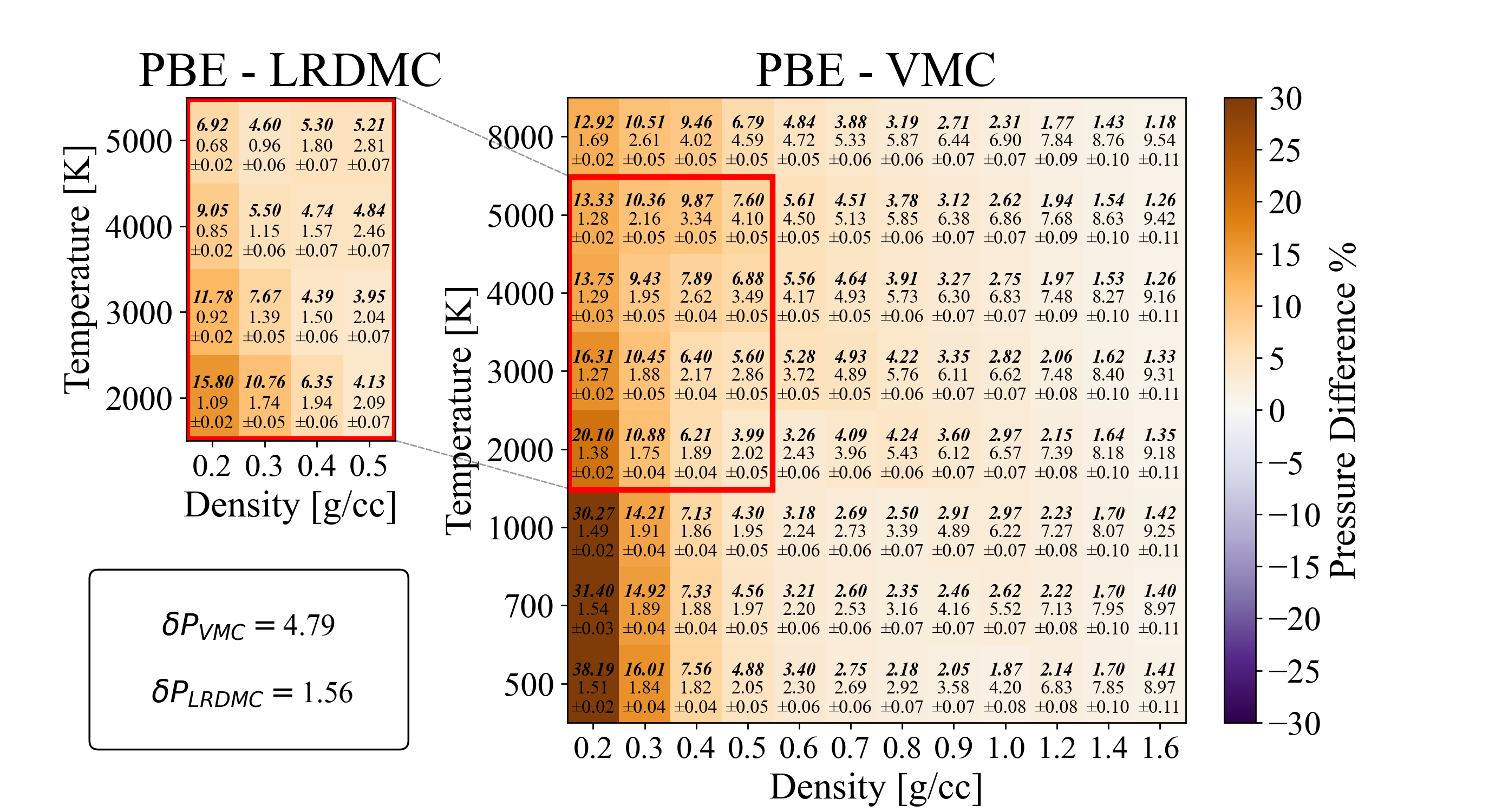}}
	\subfigure[]{\includegraphics[width=0.49\linewidth]{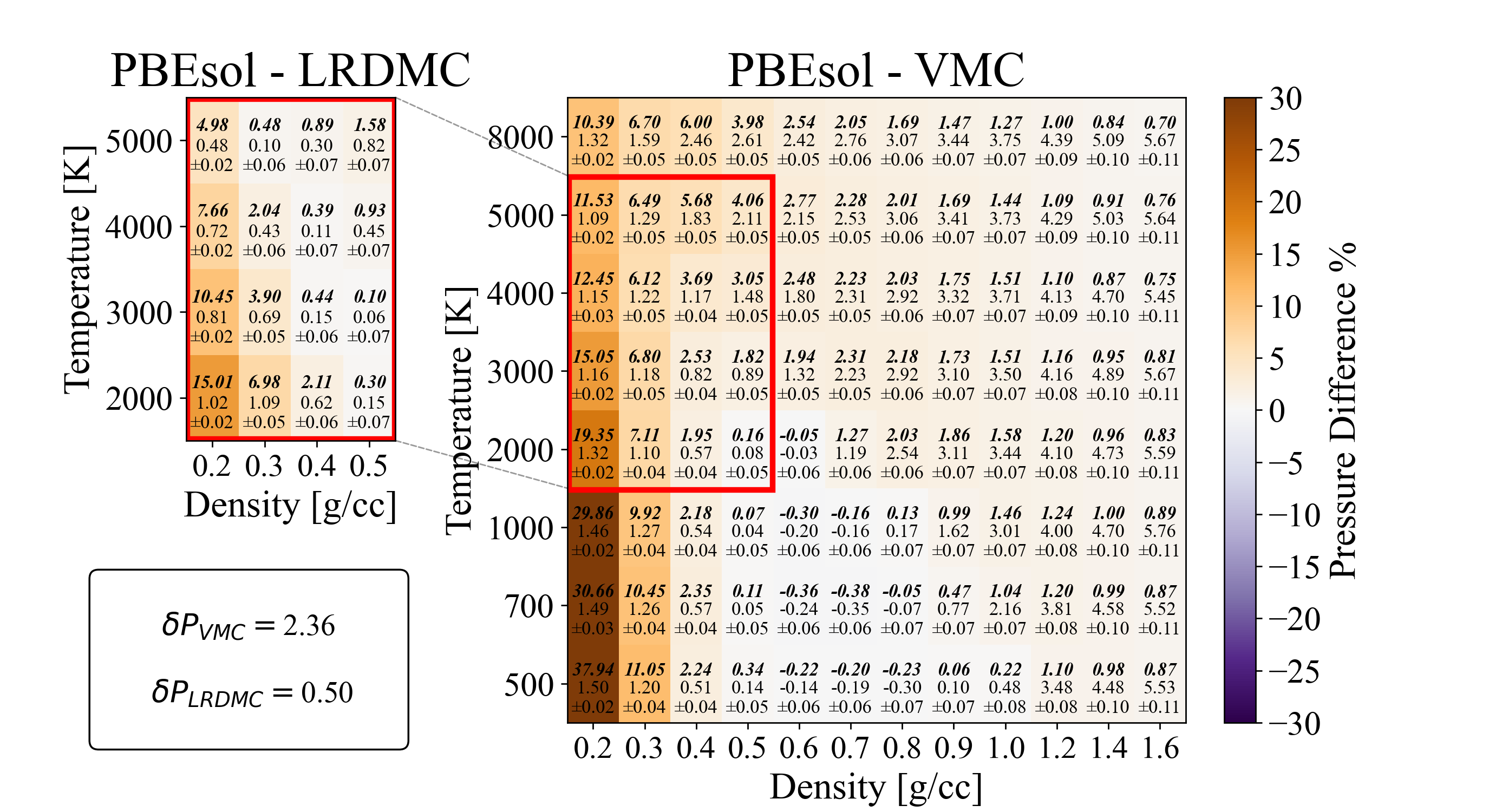}}
    \\
    \subfigure[]{\includegraphics[width=0.49\linewidth]{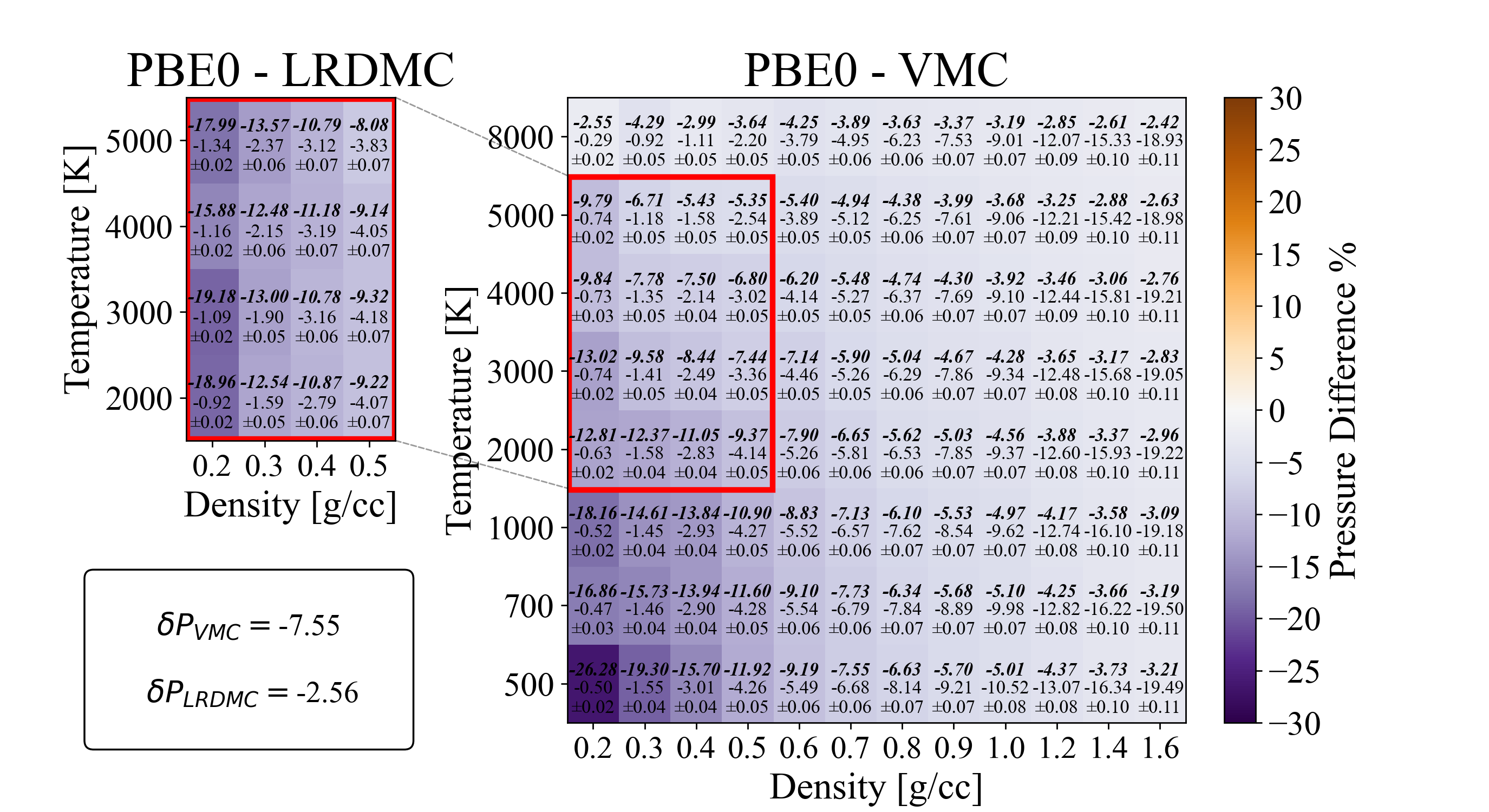}}
    \subfigure[]{\includegraphics[width=0.49\linewidth]{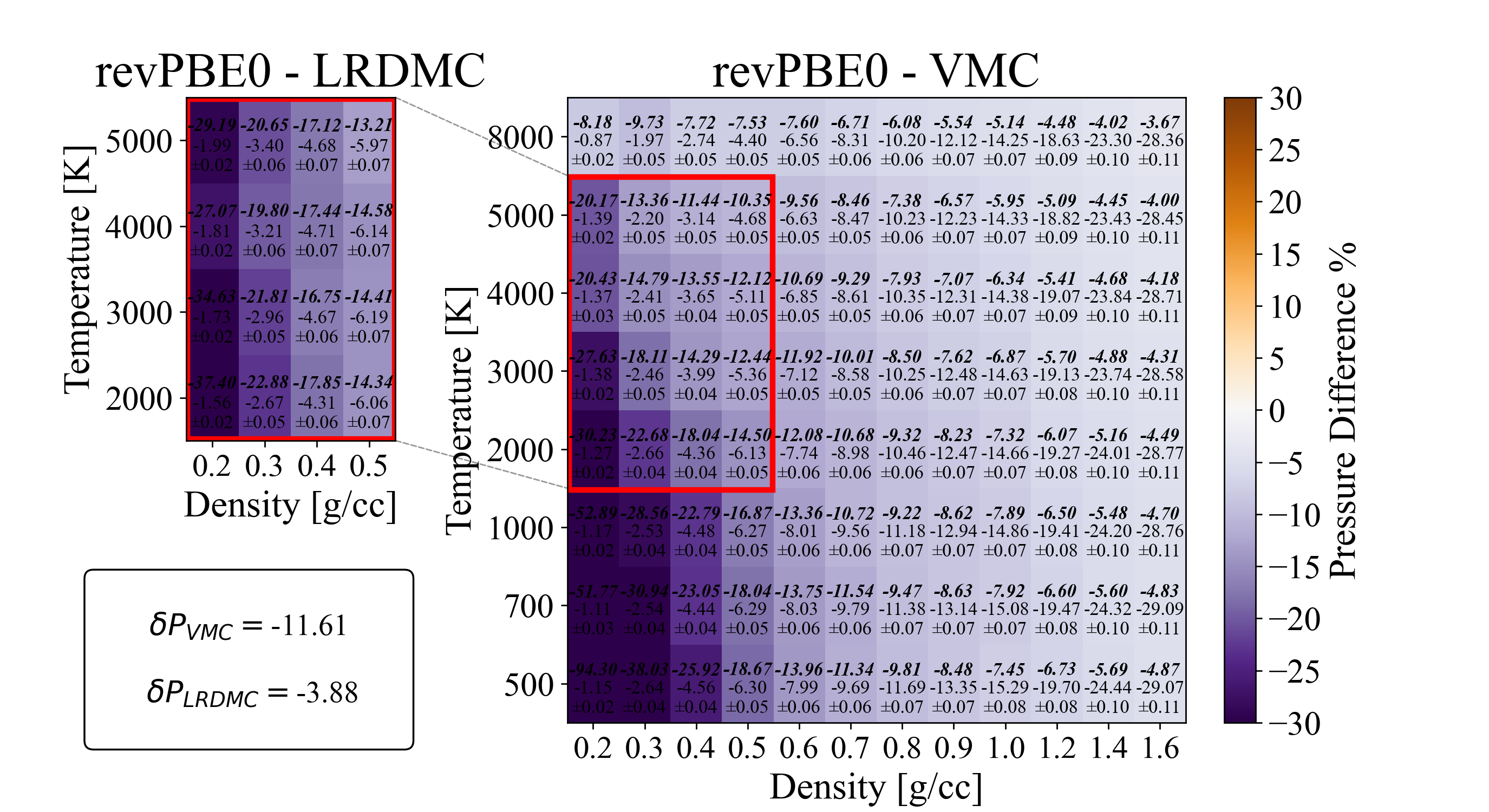}}
    \\
    \subfigure[]{\includegraphics[width=0.49\linewidth]{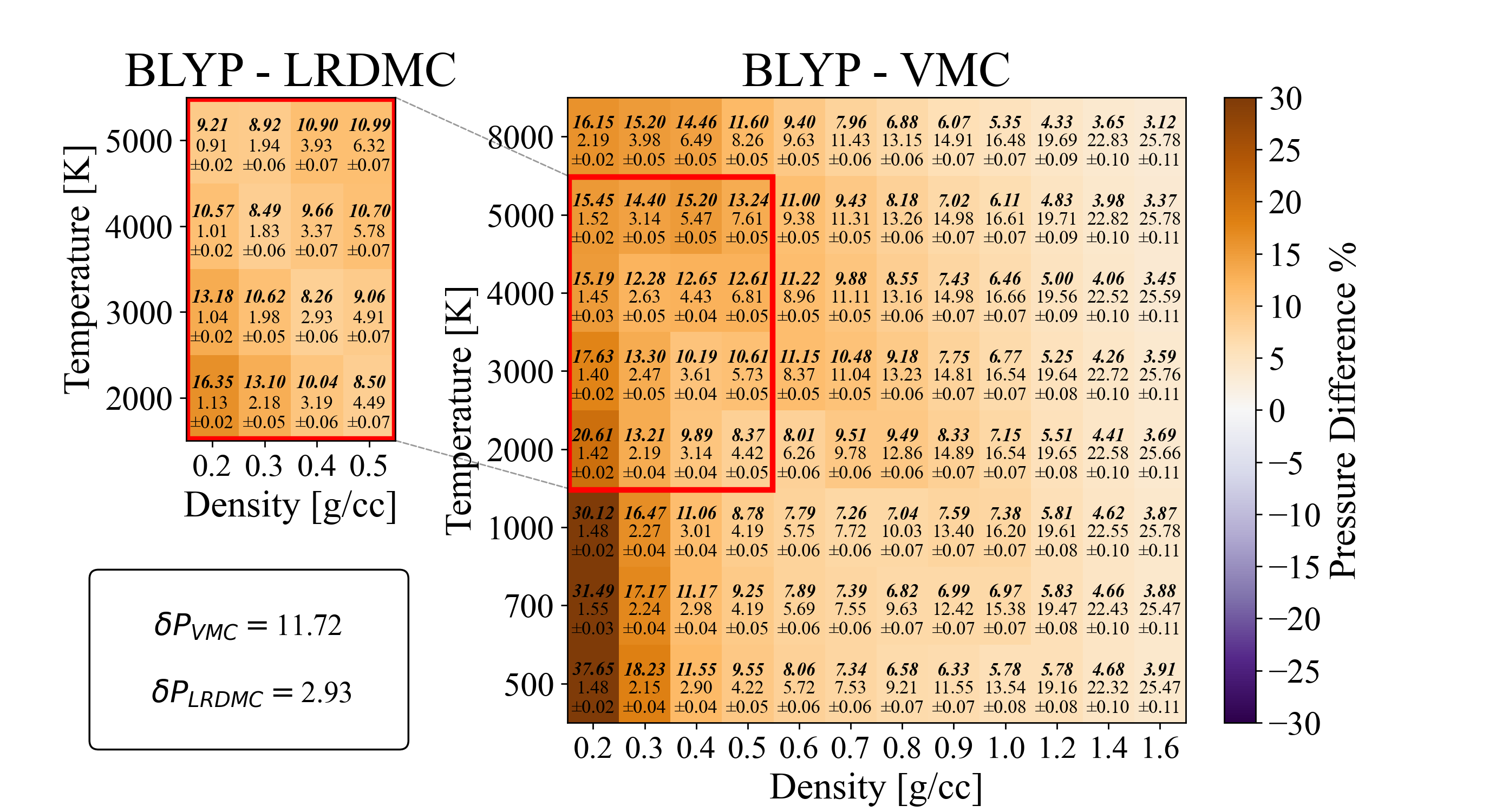}}
    \subfigure[]{\includegraphics[width=0.49\linewidth]{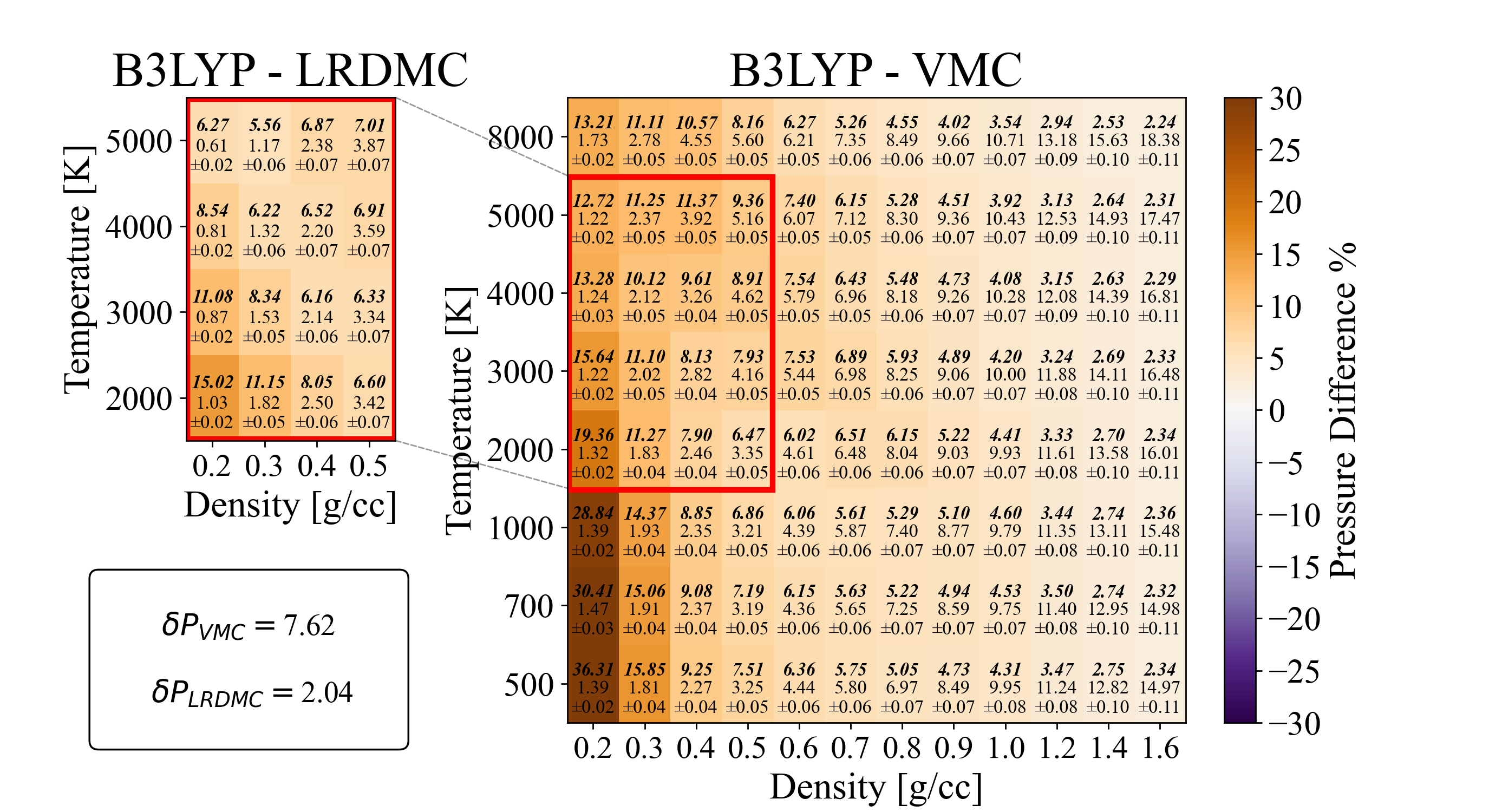}}
    \\
	\subfigure[]{\includegraphics[width=0.49\linewidth]{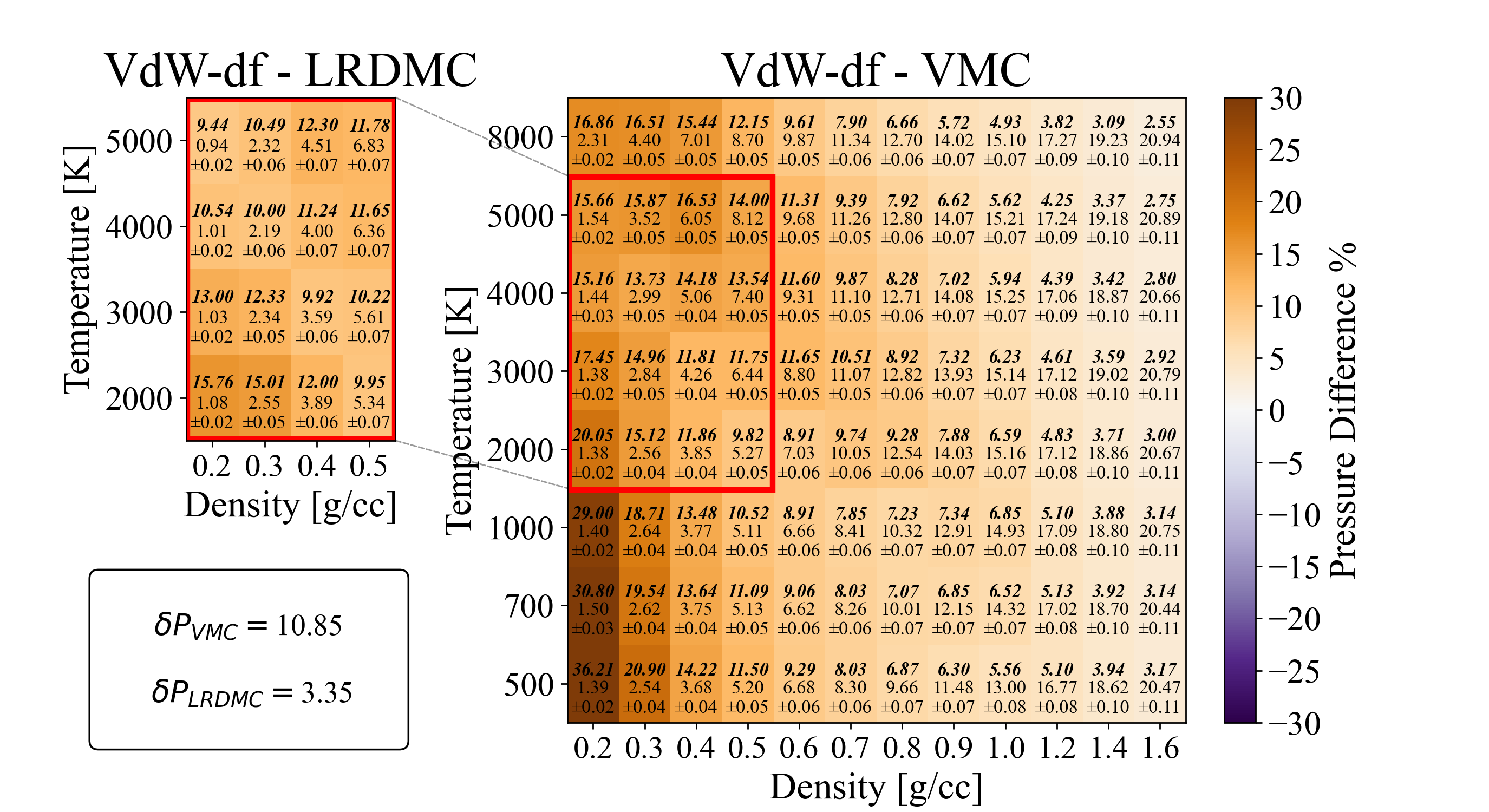}}
    \subfigure[]{\includegraphics[width=0.49\linewidth]{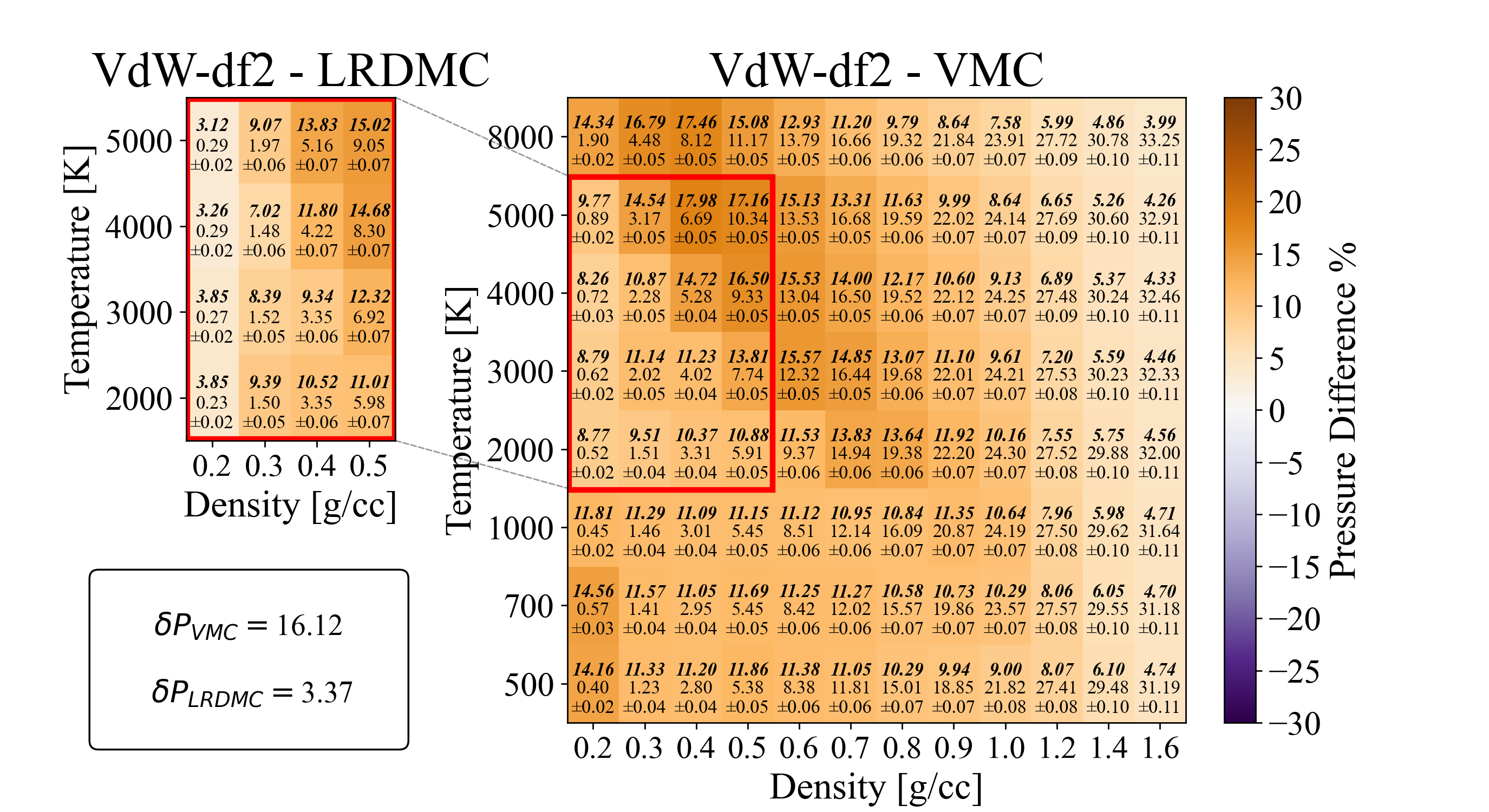}}
    \caption{\emph{Part 1.} Pressure benchmark of the DFT xc functional  against VMC and LRDMC for the $T-\rho$ of interest. In every cell the bold value is the relative difference, while the lower value is the absolute difference with error. These values are computed as in Eq.\ref{eq:pressure_metric}. The values in the box of every image are computed as Eq.\ref{eq:scoreP}.}
	\label{fig:Pressure_benchmark_1}
\end{figure*}

\begin{figure*}[h]
	\centering
    \subfigure[]{\includegraphics[width=0.49\linewidth]{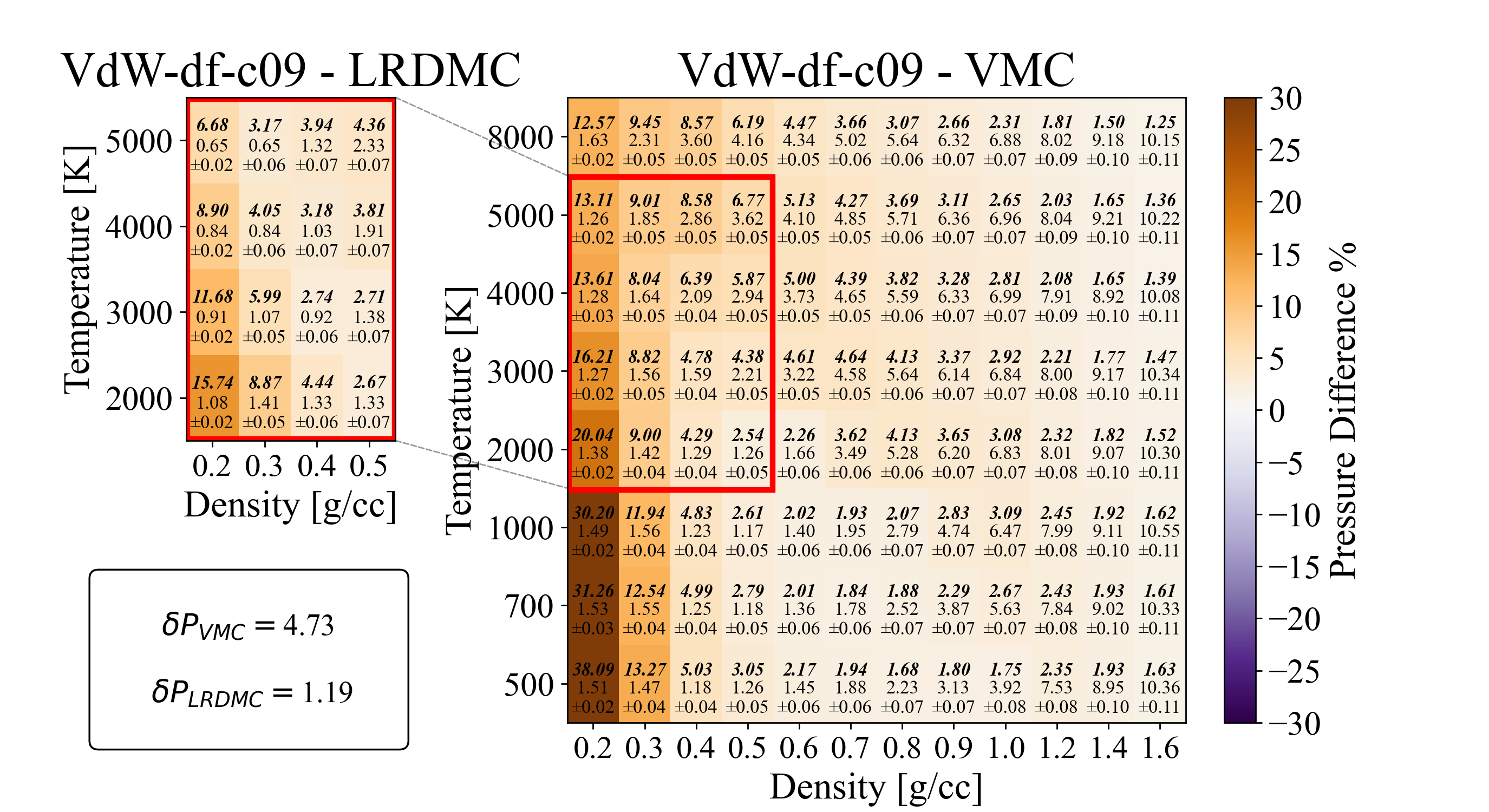}}
    \subfigure[]{\includegraphics[width=0.49\linewidth]{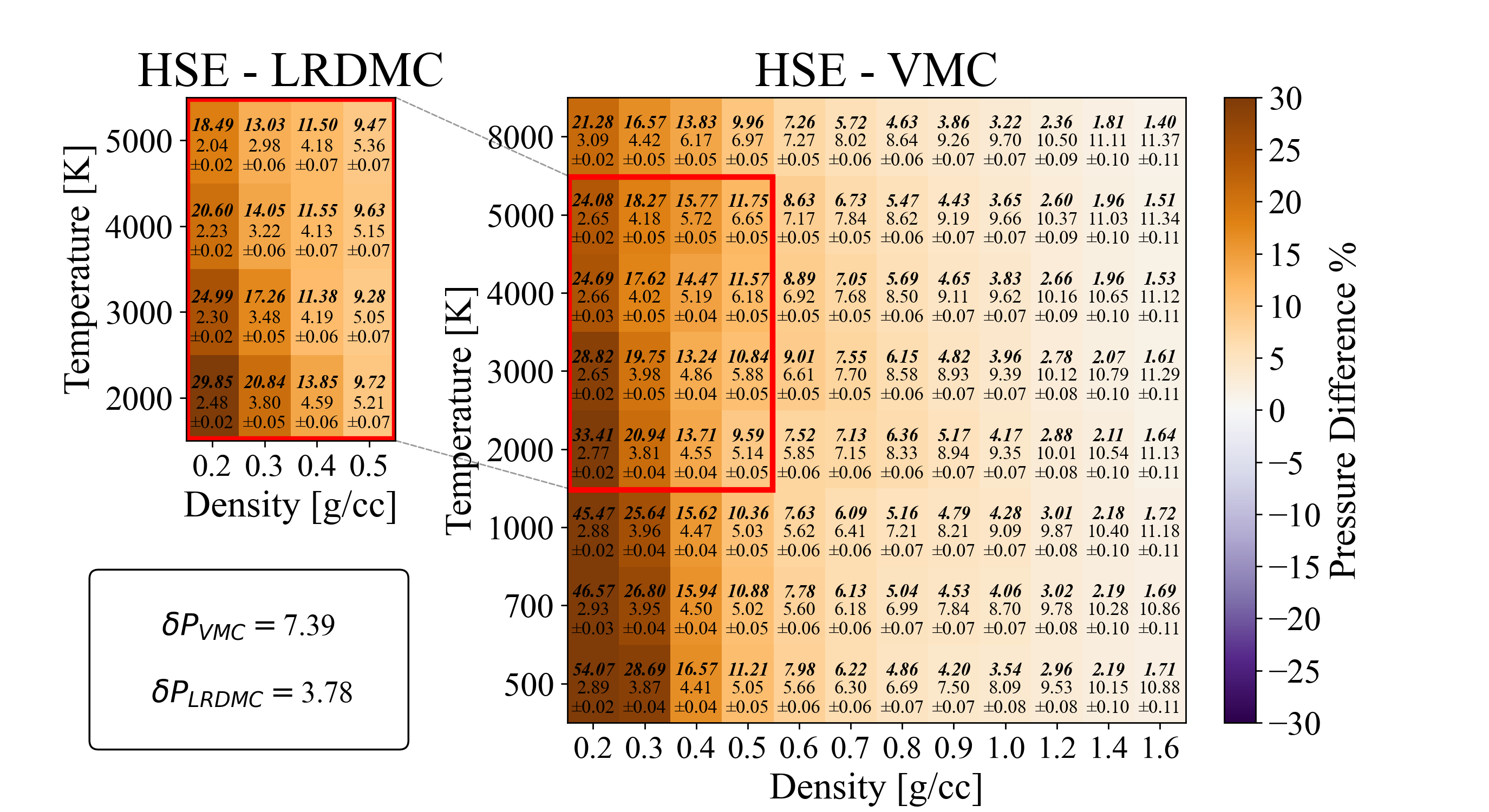}}
    \\
    \subfigure[]{\includegraphics[width=0.49\linewidth]{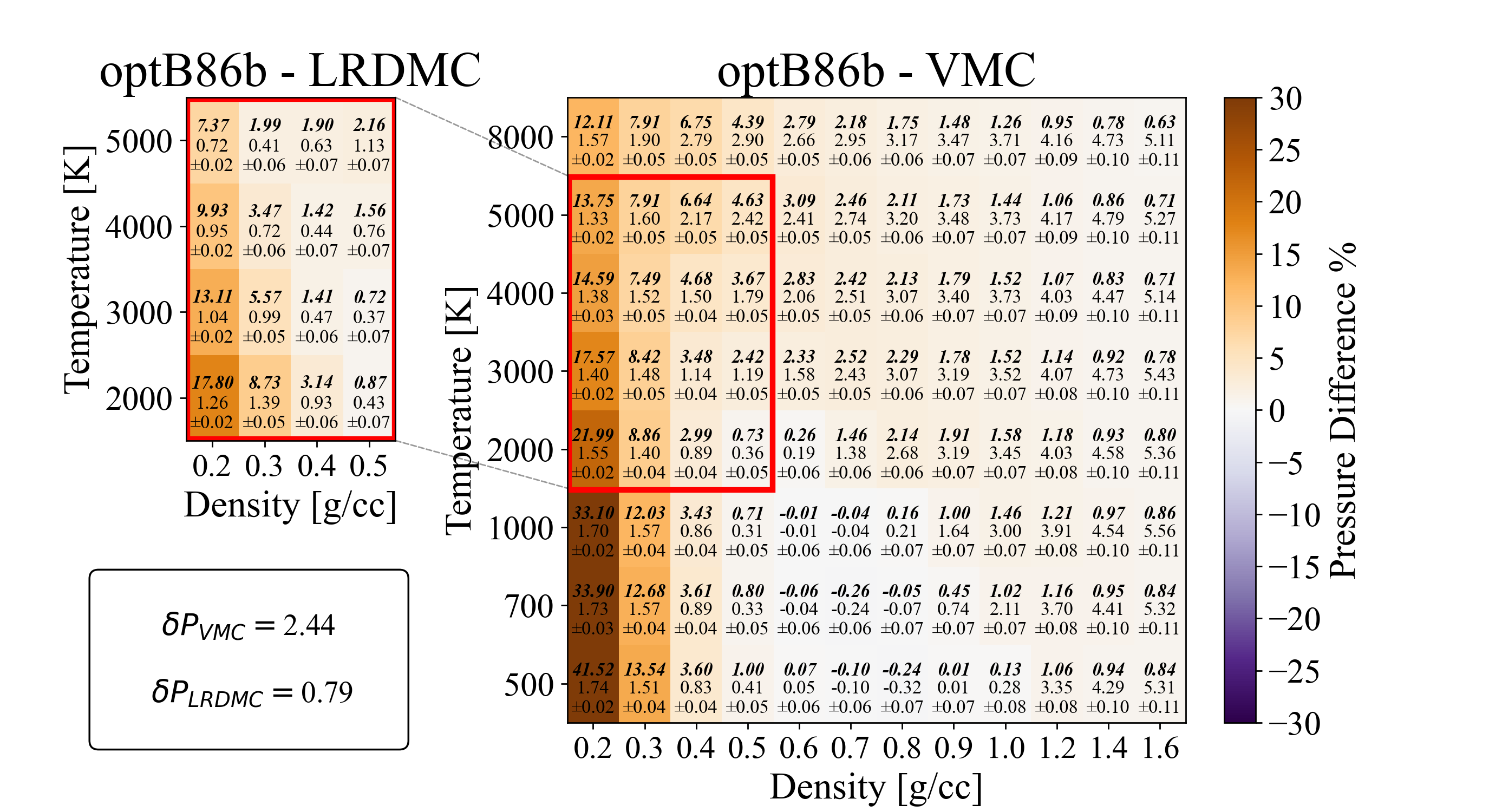}}
	\subfigure[]{\includegraphics[width=0.49\linewidth]{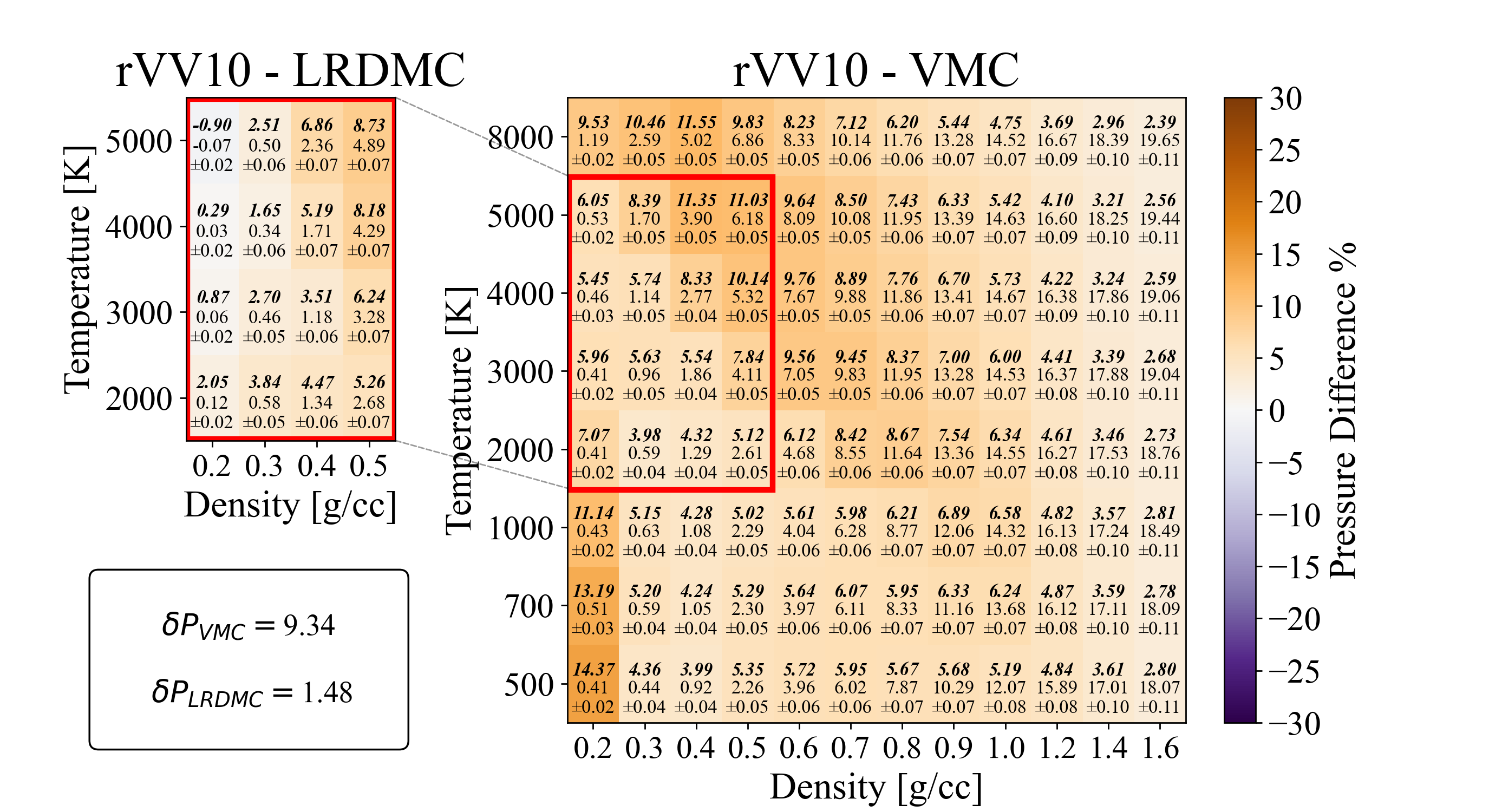}}
    \\
    \subfigure[]{\includegraphics[width=0.49\linewidth]{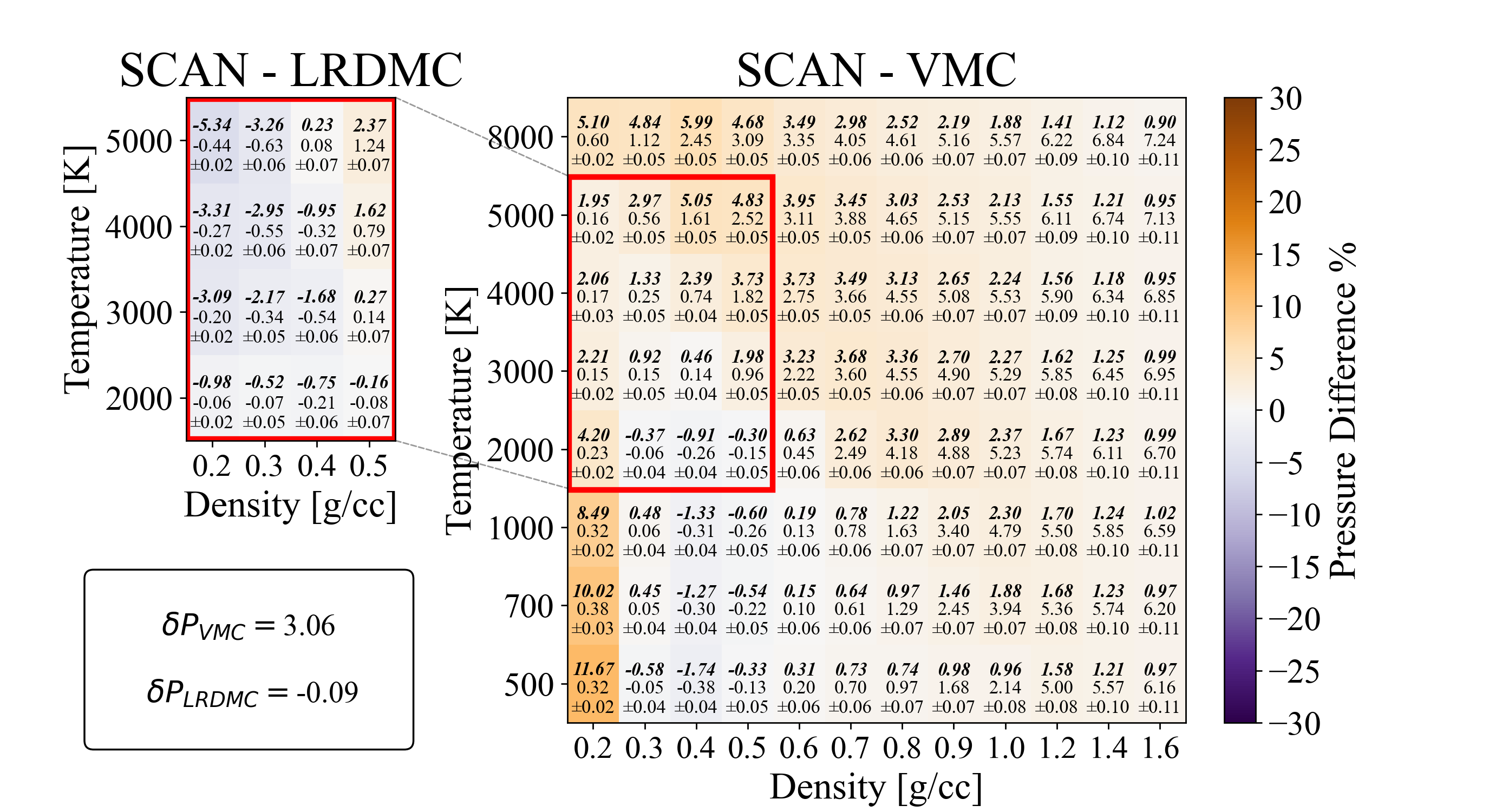}}
	\subfigure[]{\includegraphics[width=0.49\linewidth]{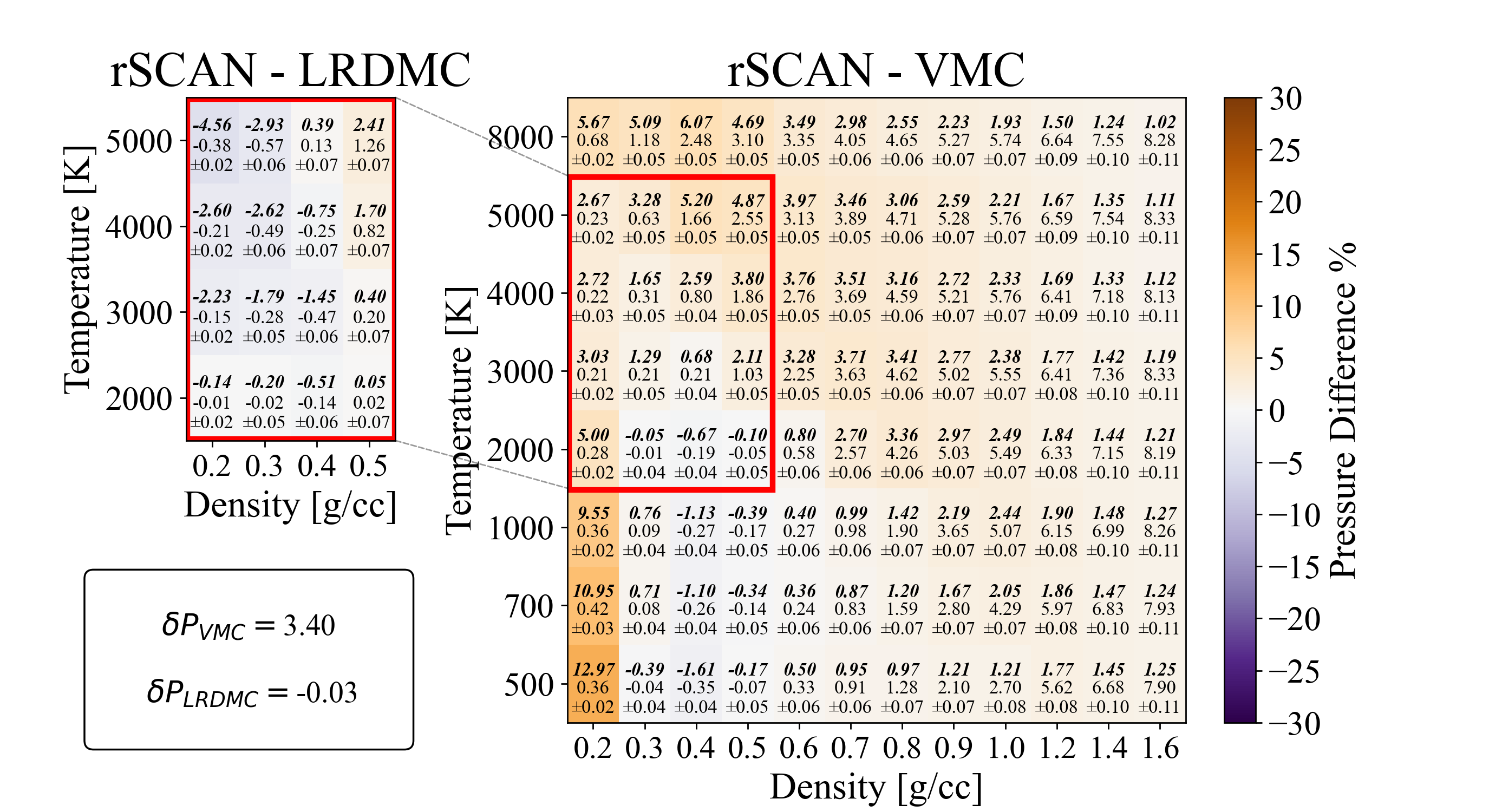}}
    \\
    \subfigure[]{\includegraphics[width=0.49\linewidth]{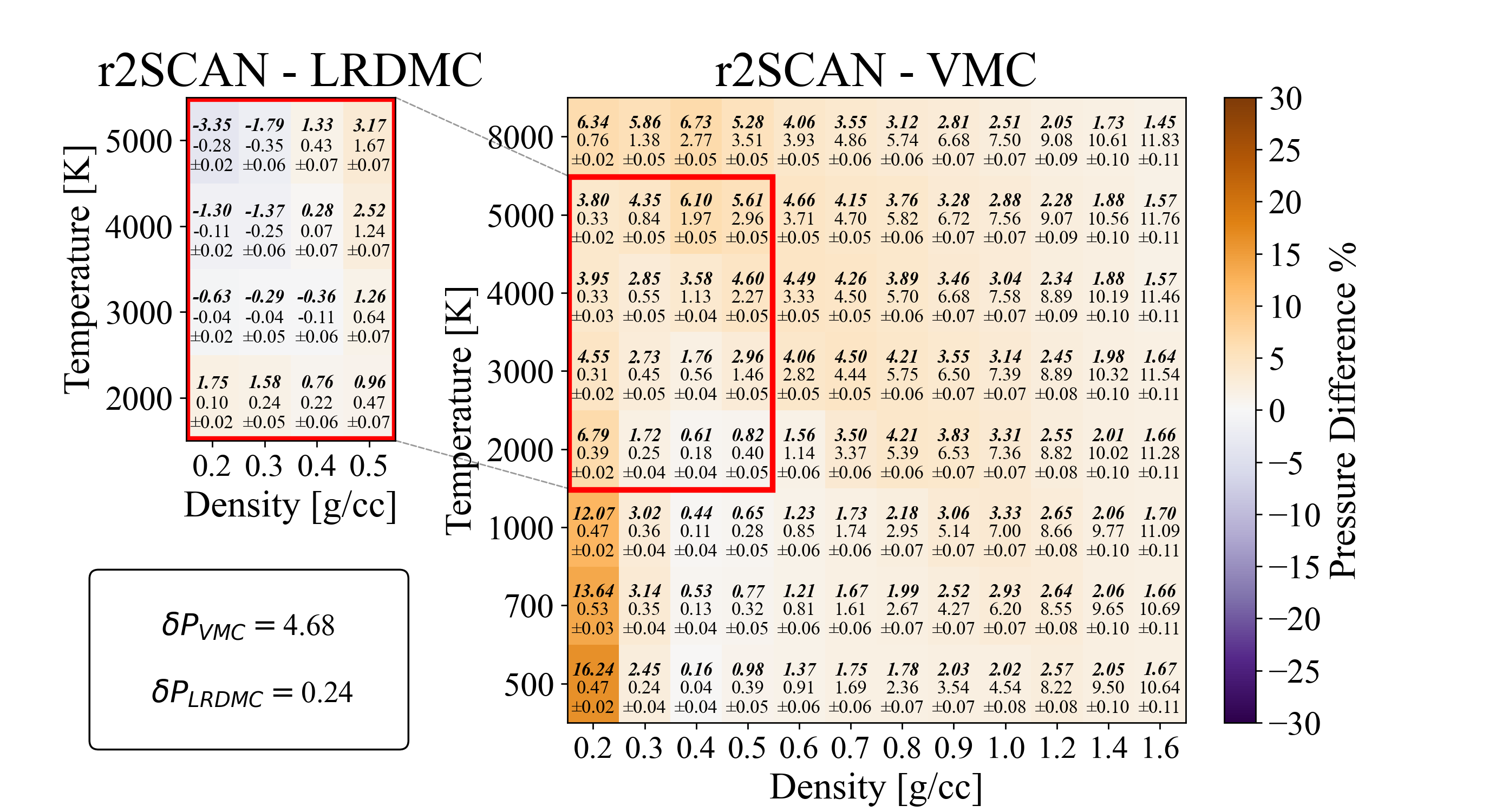}}
    \subfigure[]{\includegraphics[width=0.49\linewidth]{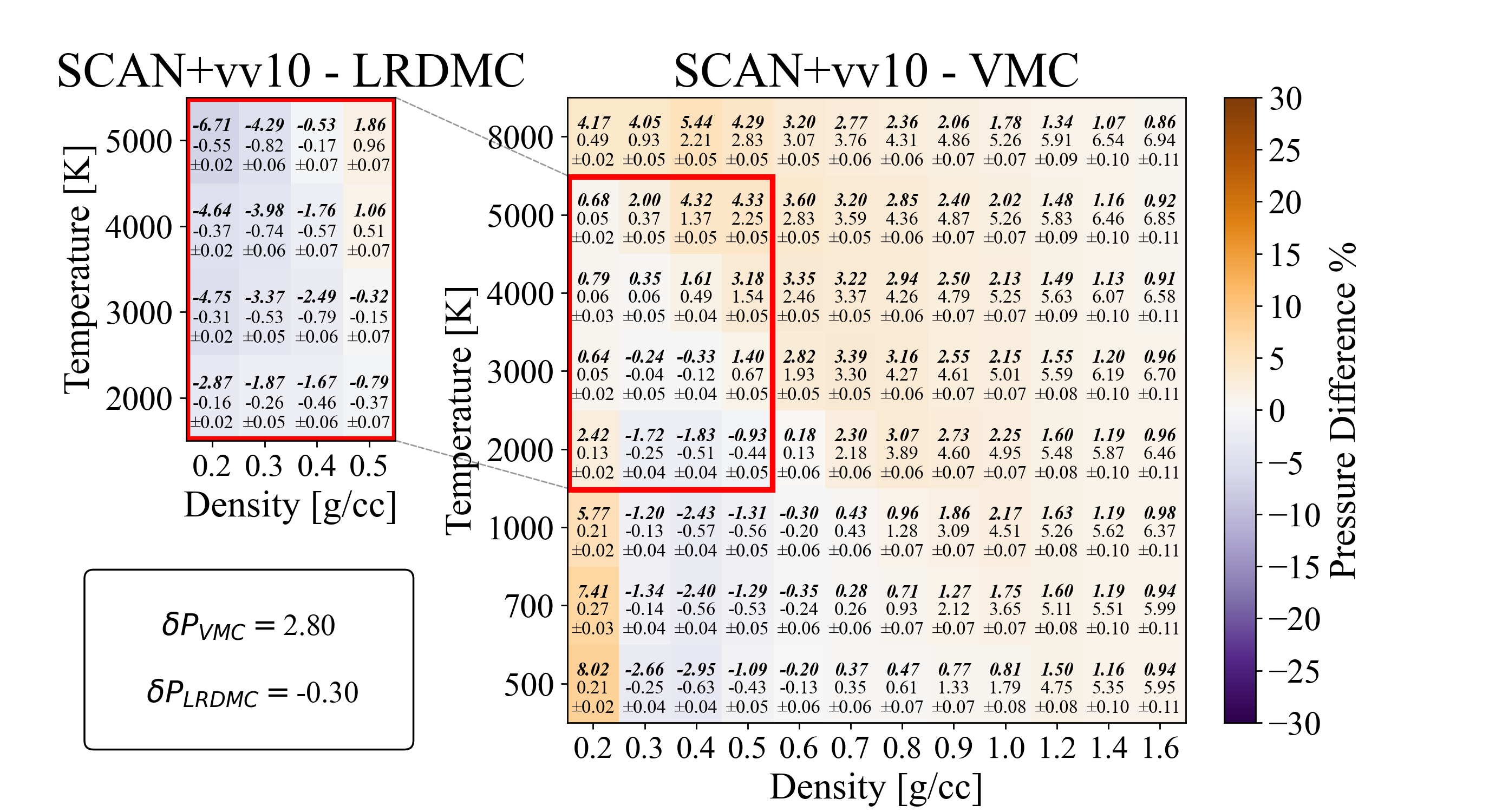}}

    \caption{\emph{Part 2.} Pressure benchmark of the DFT xc functional  against VMC and LRDMC for the $T-\rho$ of interest. In every cell the bold value is the relative difference, while the lower value is the absolute difference with error. These values are computed as in Eq.\ref{eq:pressure_metric}. The values in the box of every image are computed as Eq.\ref{eq:scoreP}.}
	\label{fig:Pressure_benchmark_2}
\end{figure*}

\begin{figure*}
        \centering
        \subfigure[]{\includegraphics[width=0.49\linewidth]{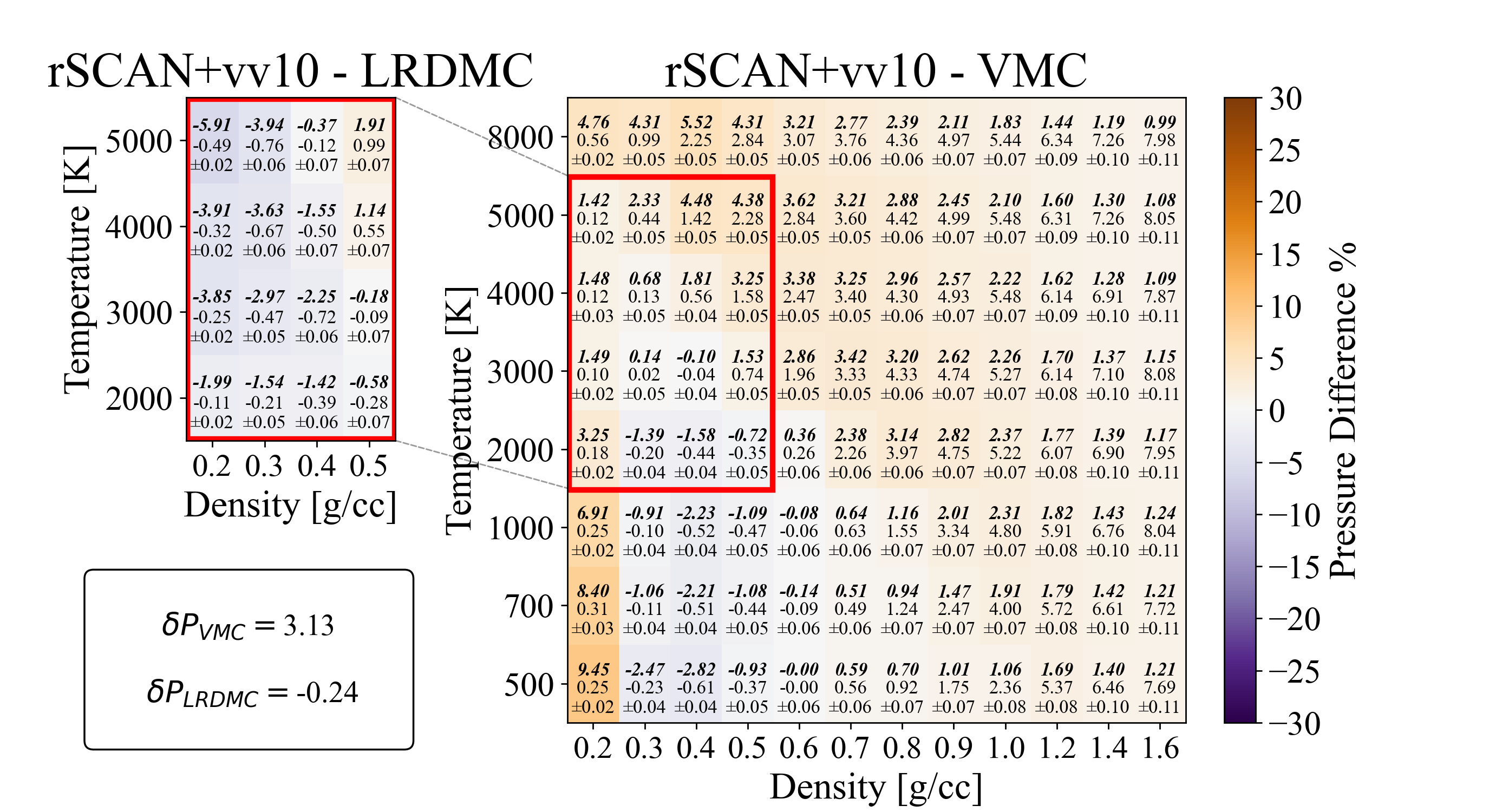}}
        \subfigure[]{\includegraphics[width=0.49\linewidth]{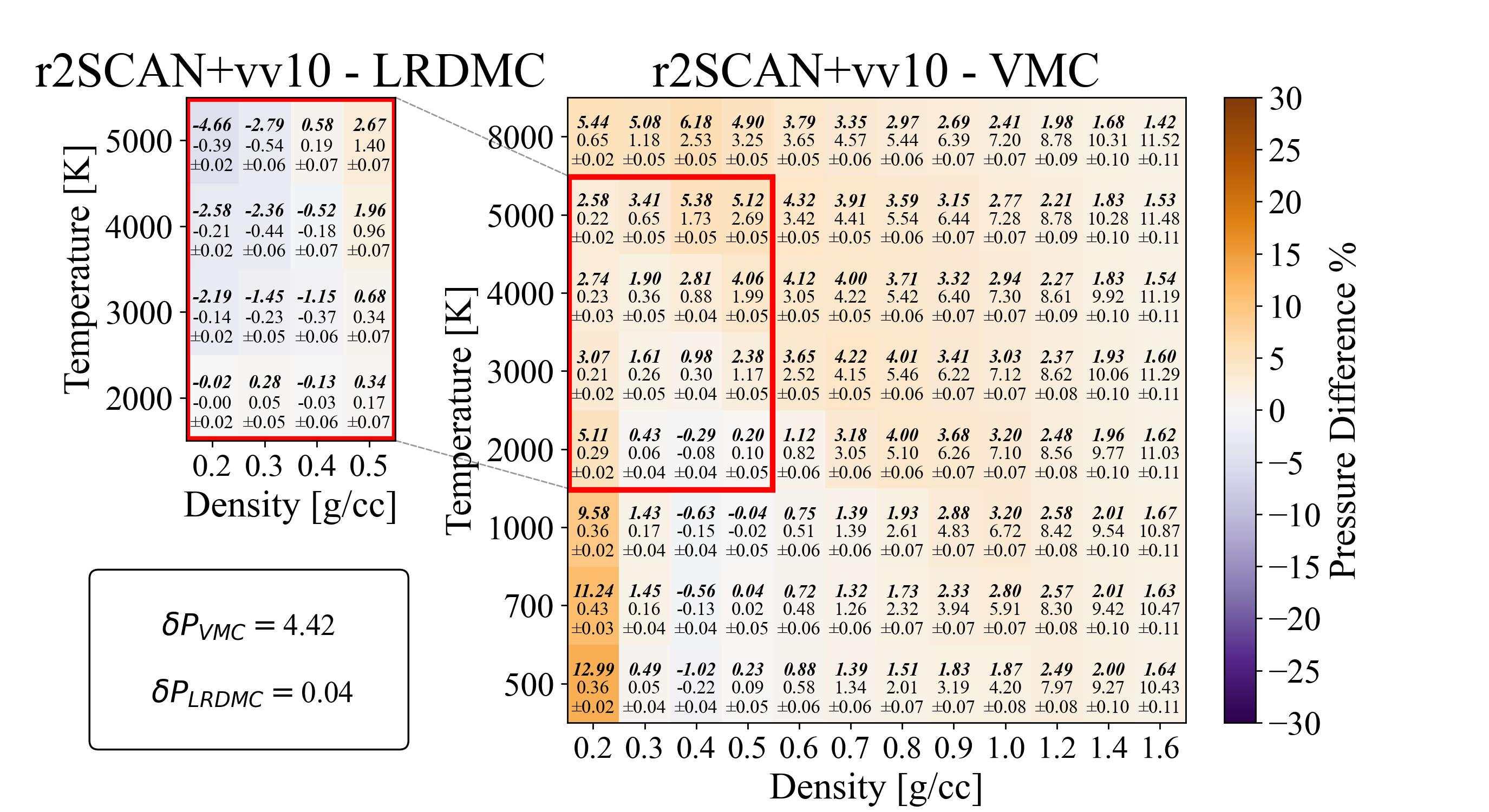}}
        \caption{\emph{Part 3.} Pressure benchmark of the DFT xc functional  against VMC and LRDMC for the $T-\rho$ of interest. In every cell the bold value is the relative difference, while the lower value is the absolute difference with error. These values are computed as in Eq.\ref{eq:pressure_metric}. The values in the box of every image are computed as Eq.\ref{eq:scoreP}.}
	\label{fig:Pressure_benchmark_3}
\end{figure*}

%\fi
\clearpage
\bibliography{main}

\begin{thebibliography}{165}%
\makeatletter
\providecommand \@ifxundefined [1]{%
 \@ifx{#1\undefined}
}%
\providecommand \@ifnum [1]{%
 \ifnum #1\expandafter \@firstoftwo
 \else \expandafter \@secondoftwo
 \fi
}%
\providecommand \@ifx [1]{%
 \ifx #1\expandafter \@firstoftwo
 \else \expandafter \@secondoftwo
 \fi
}%
\providecommand \natexlab [1]{#1}%
\providecommand \enquote  [1]{``#1''}%
\providecommand \bibnamefont  [1]{#1}%
\providecommand \bibfnamefont [1]{#1}%
\providecommand \citenamefont [1]{#1}%
\providecommand \href@noop [0]{\@secondoftwo}%
\providecommand \href [0]{\begingroup \@sanitize@url \@href}%
\providecommand \@href[1]{\@@startlink{#1}\@@href}%
\providecommand \@@href[1]{\endgroup#1\@@endlink}%
\providecommand \@sanitize@url [0]{\catcode `\\12\catcode `\$12\catcode `\&12\catcode `\#12\catcode `\^12\catcode `\_12\catcode `\%12\relax}%
\providecommand \@@startlink[1]{}%
\providecommand \@@endlink[0]{}%
\providecommand \url  [0]{\begingroup\@sanitize@url \@url }%
\providecommand \@url [1]{\endgroup\@href {#1}{\urlprefix }}%
\providecommand \urlprefix  [0]{URL }%
\providecommand \Eprint [0]{\href }%
\providecommand \doibase [0]{https://doi.org/}%
\providecommand \selectlanguage [0]{\@gobble}%
\providecommand \bibinfo  [0]{\@secondoftwo}%
\providecommand \bibfield  [0]{\@secondoftwo}%
\providecommand \translation [1]{[#1]}%
\providecommand \BibitemOpen [0]{}%
\providecommand \bibitemStop [0]{}%
\providecommand \bibitemNoStop [0]{.\EOS\space}%
\providecommand \EOS [0]{\spacefactor3000\relax}%
\providecommand \BibitemShut  [1]{\csname bibitem#1\endcsname}%
\let\auto@bib@innerbib\@empty
%</preamble>
\bibitem [{\citenamefont {{McMahon}}\ \emph {et~al.}(2012)\citenamefont {{McMahon}}, \citenamefont {Morales}, \citenamefont {Pierleoni},\ and\ \citenamefont {Ceperley}}]{mcmahon_properties_2012}%
  \BibitemOpen
  \bibfield  {author} {\bibinfo {author} {\bibfnamefont {J.~M.}\ \bibnamefont {{McMahon}}}, \bibinfo {author} {\bibfnamefont {M.~A.}\ \bibnamefont {Morales}}, \bibinfo {author} {\bibfnamefont {C.}~\bibnamefont {Pierleoni}},\ and\ \bibinfo {author} {\bibfnamefont {D.~M.}\ \bibnamefont {Ceperley}},\ }\bibfield  {title} {\bibinfo {title} {The properties of hydrogen and helium under extreme conditions},\ }\href {https://doi.org/10.1103/RevModPhys.84.1607} {\bibfield  {journal} {\bibinfo  {journal} {Reviews of Modern Physics}\ }\textbf {\bibinfo {volume} {84}},\ \bibinfo {pages} {1607} (\bibinfo {year} {2012})}\BibitemShut {NoStop}%
\bibitem [{\citenamefont {Helled}\ \emph {et~al.}(2020)\citenamefont {Helled}, \citenamefont {Mazzola},\ and\ \citenamefont {Redmer}}]{helled2020understanding}%
  \BibitemOpen
  \bibfield  {author} {\bibinfo {author} {\bibfnamefont {R.}~\bibnamefont {Helled}}, \bibinfo {author} {\bibfnamefont {G.}~\bibnamefont {Mazzola}},\ and\ \bibinfo {author} {\bibfnamefont {R.}~\bibnamefont {Redmer}},\ }\bibfield  {title} {\bibinfo {title} {Understanding dense hydrogen at planetary conditions},\ }\href@noop {} {\bibfield  {journal} {\bibinfo  {journal} {Nature Reviews Physics}\ }\textbf {\bibinfo {volume} {2}},\ \bibinfo {pages} {562} (\bibinfo {year} {2020})}\BibitemShut {NoStop}%
\bibitem [{\citenamefont {Wigner}\ and\ \citenamefont {Huntington}(1935)}]{wigner}%
  \BibitemOpen
  \bibfield  {author} {\bibinfo {author} {\bibfnamefont {E.}~\bibnamefont {Wigner}}\ and\ \bibinfo {author} {\bibfnamefont {H.~B.}\ \bibnamefont {Huntington}},\ }\bibfield  {title} {\bibinfo {title} {On the possibility of a metallic modification of hydrogen},\ }\href {https://doi.org/10.1063/1.1749590} {\bibfield  {journal} {\bibinfo  {journal} {The Journal of Chemical Physics}\ }\textbf {\bibinfo {volume} {3}},\ \bibinfo {pages} {764} (\bibinfo {year} {1935})}\BibitemShut {NoStop}%
\bibitem [{\citenamefont {Ginzburg}(2004)}]{RevModPhys.76.981}%
  \BibitemOpen
  \bibfield  {author} {\bibinfo {author} {\bibfnamefont {V.~L.}\ \bibnamefont {Ginzburg}},\ }\bibfield  {title} {\bibinfo {title} {Nobel lecture: On superconductivity and superfluidity (what i have and have not managed to do) as well as on the ``physical minimum'' at the beginning of the xxi century},\ }\href {https://doi.org/10.1103/RevModPhys.76.981} {\bibfield  {journal} {\bibinfo  {journal} {Rev. Mod. Phys.}\ }\textbf {\bibinfo {volume} {76}},\ \bibinfo {pages} {981} (\bibinfo {year} {2004})}\BibitemShut {NoStop}%
\bibitem [{\citenamefont {Dalladay-Simpson}\ \emph {et~al.}(2016)\citenamefont {Dalladay-Simpson}, \citenamefont {Howie},\ and\ \citenamefont {Gregoryanz}}]{dalladay2016evidence}%
  \BibitemOpen
  \bibfield  {author} {\bibinfo {author} {\bibfnamefont {P.}~\bibnamefont {Dalladay-Simpson}}, \bibinfo {author} {\bibfnamefont {R.~T.}\ \bibnamefont {Howie}},\ and\ \bibinfo {author} {\bibfnamefont {E.}~\bibnamefont {Gregoryanz}},\ }\bibfield  {title} {\bibinfo {title} {Evidence for a new phase of dense hydrogen above 325 gigapascals},\ }\href@noop {} {\bibfield  {journal} {\bibinfo  {journal} {Nature}\ }\textbf {\bibinfo {volume} {529}},\ \bibinfo {pages} {63} (\bibinfo {year} {2016})}\BibitemShut {NoStop}%
\bibitem [{\citenamefont {Dias}\ and\ \citenamefont {Silvera}(2017)}]{Diaseaal1579}%
  \BibitemOpen
  \bibfield  {author} {\bibinfo {author} {\bibfnamefont {R.~P.}\ \bibnamefont {Dias}}\ and\ \bibinfo {author} {\bibfnamefont {I.~F.}\ \bibnamefont {Silvera}},\ }\bibfield  {title} {\bibinfo {title} {Observation of the wigner-huntington transition to metallic hydrogen},\ }\bibfield  {journal} {\bibinfo  {journal} {Science}\ }\href {https://doi.org/10.1126/science.aal1579} {10.1126/science.aal1579} (\bibinfo {year} {2017})\BibitemShut {NoStop}%
\bibitem [{\citenamefont {Loubeyre}\ \emph {et~al.}(2020)\citenamefont {Loubeyre}, \citenamefont {Occelli},\ and\ \citenamefont {Dumas}}]{loubeyre2020synchrotron}%
  \BibitemOpen
  \bibfield  {author} {\bibinfo {author} {\bibfnamefont {P.}~\bibnamefont {Loubeyre}}, \bibinfo {author} {\bibfnamefont {F.}~\bibnamefont {Occelli}},\ and\ \bibinfo {author} {\bibfnamefont {P.}~\bibnamefont {Dumas}},\ }\bibfield  {title} {\bibinfo {title} {Synchrotron infrared spectroscopic evidence of the probable transition to metal hydrogen},\ }\href@noop {} {\bibfield  {journal} {\bibinfo  {journal} {Nature}\ }\textbf {\bibinfo {volume} {577}},\ \bibinfo {pages} {631} (\bibinfo {year} {2020})}\BibitemShut {NoStop}%
\bibitem [{\citenamefont {Weir}\ \emph {et~al.}(1996)\citenamefont {Weir}, \citenamefont {Mitchell},\ and\ \citenamefont {Nellis}}]{weir1996metallization}%
  \BibitemOpen
  \bibfield  {author} {\bibinfo {author} {\bibfnamefont {S.}~\bibnamefont {Weir}}, \bibinfo {author} {\bibfnamefont {A.}~\bibnamefont {Mitchell}},\ and\ \bibinfo {author} {\bibfnamefont {W.}~\bibnamefont {Nellis}},\ }\bibfield  {title} {\bibinfo {title} {Metallization of fluid molecular hydrogen at 140 gpa (1.4 mbar)},\ }\href@noop {} {\bibfield  {journal} {\bibinfo  {journal} {Physical Review Letters}\ }\textbf {\bibinfo {volume} {76}},\ \bibinfo {pages} {1860} (\bibinfo {year} {1996})}\BibitemShut {NoStop}%
\bibitem [{\citenamefont {Celliers}\ \emph {et~al.}(2018)\citenamefont {Celliers}, \citenamefont {Millot}, \citenamefont {Brygoo}, \citenamefont {McWilliams}, \citenamefont {Fratanduono}, \citenamefont {Rygg}, \citenamefont {Goncharov}, \citenamefont {Loubeyre}, \citenamefont {Eggert}, \citenamefont {Peterson}, \citenamefont {Meezan}, \citenamefont {Le~Pape}, \citenamefont {Collins}, \citenamefont {Jeanloz},\ and\ \citenamefont {Hemley}}]{Celliers677}%
  \BibitemOpen
  \bibfield  {author} {\bibinfo {author} {\bibfnamefont {P.~M.}\ \bibnamefont {Celliers}}, \bibinfo {author} {\bibfnamefont {M.}~\bibnamefont {Millot}}, \bibinfo {author} {\bibfnamefont {S.}~\bibnamefont {Brygoo}}, \bibinfo {author} {\bibfnamefont {R.~S.}\ \bibnamefont {McWilliams}}, \bibinfo {author} {\bibfnamefont {D.~E.}\ \bibnamefont {Fratanduono}}, \bibinfo {author} {\bibfnamefont {J.~R.}\ \bibnamefont {Rygg}}, \bibinfo {author} {\bibfnamefont {A.~F.}\ \bibnamefont {Goncharov}}, \bibinfo {author} {\bibfnamefont {P.}~\bibnamefont {Loubeyre}}, \bibinfo {author} {\bibfnamefont {J.~H.}\ \bibnamefont {Eggert}}, \bibinfo {author} {\bibfnamefont {J.~L.}\ \bibnamefont {Peterson}}, \bibinfo {author} {\bibfnamefont {N.~B.}\ \bibnamefont {Meezan}}, \bibinfo {author} {\bibfnamefont {S.}~\bibnamefont {Le~Pape}}, \bibinfo {author} {\bibfnamefont {G.~W.}\ \bibnamefont {Collins}}, \bibinfo {author} {\bibfnamefont {R.}~\bibnamefont {Jeanloz}},\ and\ \bibinfo {author} {\bibfnamefont {R.~J.}\ \bibnamefont {Hemley}},\
  }\bibfield  {title} {\bibinfo {title} {Insulator-metal transition in dense fluid deuterium},\ }\href {https://doi.org/10.1126/science.aat0970} {\bibfield  {journal} {\bibinfo  {journal} {Science}\ }\textbf {\bibinfo {volume} {361}},\ \bibinfo {pages} {677} (\bibinfo {year} {2018})}\BibitemShut {NoStop}%
\bibitem [{\citenamefont {Knudson}\ \emph {et~al.}(2015)\citenamefont {Knudson}, \citenamefont {Desjarlais}, \citenamefont {Becker}, \citenamefont {Lemke}, \citenamefont {Cochrane}, \citenamefont {Savage}, \citenamefont {Bliss}, \citenamefont {Mattsson},\ and\ \citenamefont {Redmer}}]{knudson2015direct}%
  \BibitemOpen
  \bibfield  {author} {\bibinfo {author} {\bibfnamefont {M.}~\bibnamefont {Knudson}}, \bibinfo {author} {\bibfnamefont {M.}~\bibnamefont {Desjarlais}}, \bibinfo {author} {\bibfnamefont {A.}~\bibnamefont {Becker}}, \bibinfo {author} {\bibfnamefont {R.}~\bibnamefont {Lemke}}, \bibinfo {author} {\bibfnamefont {K.}~\bibnamefont {Cochrane}}, \bibinfo {author} {\bibfnamefont {M.}~\bibnamefont {Savage}}, \bibinfo {author} {\bibfnamefont {D.}~\bibnamefont {Bliss}}, \bibinfo {author} {\bibfnamefont {T.}~\bibnamefont {Mattsson}},\ and\ \bibinfo {author} {\bibfnamefont {R.}~\bibnamefont {Redmer}},\ }\bibfield  {title} {\bibinfo {title} {Direct observation of an abrupt insulator-to-metal transition in dense liquid deuterium},\ }\href@noop {} {\bibfield  {journal} {\bibinfo  {journal} {Science}\ }\textbf {\bibinfo {volume} {348}},\ \bibinfo {pages} {1455} (\bibinfo {year} {2015})}\BibitemShut {NoStop}%
\bibitem [{\citenamefont {{L{\"u}hr}}\ \emph {et~al.}(2018)\citenamefont {{L{\"u}hr}}, \citenamefont {{Wicht}}, \citenamefont {{Gilder}},\ and\ \citenamefont {{Holschneider}}}]{2018ASSL..448.....L}%
  \BibitemOpen
  \bibfield  {author} {\bibinfo {author} {\bibfnamefont {H.}~\bibnamefont {{L{\"u}hr}}}, \bibinfo {author} {\bibfnamefont {J.}~\bibnamefont {{Wicht}}}, \bibinfo {author} {\bibfnamefont {S.~A.}\ \bibnamefont {{Gilder}}},\ and\ \bibinfo {author} {\bibfnamefont {M.}~\bibnamefont {{Holschneider}}},\ }\href {https://doi.org/10.1007/978-3-319-64292-5} {\emph {\bibinfo {title} {{Magnetic Fields in the Solar System}}}},\ Vol.\ \bibinfo {volume} {448}\ (\bibinfo  {publisher} {Springer},\ \bibinfo {year} {2018})\BibitemShut {NoStop}%
\bibitem [{\citenamefont {{Helled}}\ and\ \citenamefont {{Fortney}}(2020)}]{2020RSPTA.37890474H}%
  \BibitemOpen
  \bibfield  {author} {\bibinfo {author} {\bibfnamefont {R.}~\bibnamefont {{Helled}}}\ and\ \bibinfo {author} {\bibfnamefont {J.~J.}\ \bibnamefont {{Fortney}}},\ }\bibfield  {title} {\bibinfo {title} {{The interiors of Uranus and Neptune: current understanding and open questions}},\ }\href {https://doi.org/10.1098/rsta.2019.0474} {\bibfield  {journal} {\bibinfo  {journal} {Philosophical Transactions of the Royal Society of London Series A}\ }\textbf {\bibinfo {volume} {378}},\ \bibinfo {eid} {20190474} (\bibinfo {year} {2020})},\ \Eprint {https://arxiv.org/abs/2007.10783} {arXiv:2007.10783 [astro-ph.EP]} \BibitemShut {NoStop}%
\bibitem [{\citenamefont {{Howard}}\ \emph {et~al.}(2025)\citenamefont {{Howard}}, \citenamefont {{Helled}},\ and\ \citenamefont {{M{\"u}ller}}}]{2025A&A...693L...7H}%
  \BibitemOpen
  \bibfield  {author} {\bibinfo {author} {\bibfnamefont {S.}~\bibnamefont {{Howard}}}, \bibinfo {author} {\bibfnamefont {R.}~\bibnamefont {{Helled}}},\ and\ \bibinfo {author} {\bibfnamefont {S.}~\bibnamefont {{M{\"u}ller}}},\ }\bibfield  {title} {\bibinfo {title} {{Giant exoplanet composition: The impact of the hydrogen{\textendash}helium equation of state and interior structure}},\ }\href {https://doi.org/10.1051/0004-6361/202452783} {\bibfield  {journal} {\bibinfo  {journal} {Astronomy and Astrophysics}\ }\textbf {\bibinfo {volume} {693}},\ \bibinfo {eid} {L7} (\bibinfo {year} {2025})},\ \Eprint {https://arxiv.org/abs/2410.21382} {arXiv:2410.21382 [astro-ph.EP]} \BibitemShut {NoStop}%
\bibitem [{\citenamefont {{Durante}}\ \emph {et~al.}(2020)\citenamefont {{Durante}}, \citenamefont {{Parisi}}, \citenamefont {{Serra}}, \citenamefont {{Zannoni}}, \citenamefont {{Notaro}}, \citenamefont {{Racioppa}}, \citenamefont {{Buccino}}, \citenamefont {{Lari}}, \citenamefont {{Gomez Casajus}}, \citenamefont {{Iess}}, \citenamefont {{Folkner}}, \citenamefont {{Tommei}}, \citenamefont {{Tortora}},\ and\ \citenamefont {{Bolton}}}]{durante2020}%
  \BibitemOpen
  \bibfield  {author} {\bibinfo {author} {\bibfnamefont {D.}~\bibnamefont {{Durante}}}, \bibinfo {author} {\bibfnamefont {M.}~\bibnamefont {{Parisi}}}, \bibinfo {author} {\bibfnamefont {D.}~\bibnamefont {{Serra}}}, \bibinfo {author} {\bibfnamefont {M.}~\bibnamefont {{Zannoni}}}, \bibinfo {author} {\bibfnamefont {V.}~\bibnamefont {{Notaro}}}, \bibinfo {author} {\bibfnamefont {P.}~\bibnamefont {{Racioppa}}}, \bibinfo {author} {\bibfnamefont {D.~R.}\ \bibnamefont {{Buccino}}}, \bibinfo {author} {\bibfnamefont {G.}~\bibnamefont {{Lari}}}, \bibinfo {author} {\bibfnamefont {L.}~\bibnamefont {{Gomez Casajus}}}, \bibinfo {author} {\bibfnamefont {L.}~\bibnamefont {{Iess}}}, \bibinfo {author} {\bibfnamefont {W.~M.}\ \bibnamefont {{Folkner}}}, \bibinfo {author} {\bibfnamefont {G.}~\bibnamefont {{Tommei}}}, \bibinfo {author} {\bibfnamefont {P.}~\bibnamefont {{Tortora}}},\ and\ \bibinfo {author} {\bibfnamefont {S.~J.}\ \bibnamefont {{Bolton}}},\ }\bibfield  {title} {\bibinfo {title} {{Jupiter's Gravity Field Halfway
  Through the Juno Mission}},\ }\href {https://doi.org/10.1029/2019GL086572} {\bibfield  {journal} {\bibinfo  {journal} {Geophysical Research Letter}\ }\textbf {\bibinfo {volume} {47}},\ \bibinfo {eid} {e86572} (\bibinfo {year} {2020})}\BibitemShut {NoStop}%
\bibitem [{\citenamefont {{Iess}}\ \emph {et~al.}(2019)\citenamefont {{Iess}}, \citenamefont {{Militzer}}, \citenamefont {{Kaspi}}, \citenamefont {{Nicholson}}, \citenamefont {{Durante}}, \citenamefont {{Racioppa}}, \citenamefont {{Anabtawi}}, \citenamefont {{Galanti}}, \citenamefont {{Hubbard}}, \citenamefont {{Mariani}}, \citenamefont {{Tortora}}, \citenamefont {{Wahl}},\ and\ \citenamefont {{Zannoni}}}]{iess2019}%
  \BibitemOpen
  \bibfield  {author} {\bibinfo {author} {\bibfnamefont {L.}~\bibnamefont {{Iess}}}, \bibinfo {author} {\bibfnamefont {B.}~\bibnamefont {{Militzer}}}, \bibinfo {author} {\bibfnamefont {Y.}~\bibnamefont {{Kaspi}}}, \bibinfo {author} {\bibfnamefont {P.}~\bibnamefont {{Nicholson}}}, \bibinfo {author} {\bibfnamefont {D.}~\bibnamefont {{Durante}}}, \bibinfo {author} {\bibfnamefont {P.}~\bibnamefont {{Racioppa}}}, \bibinfo {author} {\bibfnamefont {A.}~\bibnamefont {{Anabtawi}}}, \bibinfo {author} {\bibfnamefont {E.}~\bibnamefont {{Galanti}}}, \bibinfo {author} {\bibfnamefont {W.}~\bibnamefont {{Hubbard}}}, \bibinfo {author} {\bibfnamefont {M.~J.}\ \bibnamefont {{Mariani}}}, \bibinfo {author} {\bibfnamefont {P.}~\bibnamefont {{Tortora}}}, \bibinfo {author} {\bibfnamefont {S.}~\bibnamefont {{Wahl}}},\ and\ \bibinfo {author} {\bibfnamefont {M.}~\bibnamefont {{Zannoni}}},\ }\bibfield  {title} {\bibinfo {title} {{Measurement and implications of Saturn's gravity field and ring mass}},\ }\href
  {https://doi.org/10.1126/science.aat2965} {\bibfield  {journal} {\bibinfo  {journal} {Science}\ }\textbf {\bibinfo {volume} {364}},\ \bibinfo {eid} {aat2965} (\bibinfo {year} {2019})}\BibitemShut {NoStop}%
\bibitem [{\citenamefont {{Helled}}\ and\ \citenamefont {{Howard}}(2024)}]{helledhoward2024}%
  \BibitemOpen
  \bibfield  {author} {\bibinfo {author} {\bibfnamefont {R.}~\bibnamefont {{Helled}}}\ and\ \bibinfo {author} {\bibfnamefont {S.}~\bibnamefont {{Howard}}},\ }\bibfield  {title} {\bibinfo {title} {{Giant planet interiors and atmospheres}},\ }\href {https://doi.org/10.48550/arXiv.2407.05853} {\bibfield  {journal} {\bibinfo  {journal} {arXiv e-prints}\ ,\ \bibinfo {eid} {arXiv:2407.05853}} (\bibinfo {year} {2024})},\ \Eprint {https://arxiv.org/abs/2407.05853} {arXiv:2407.05853 [astro-ph.EP]} \BibitemShut {NoStop}%
\bibitem [{\citenamefont {{Miguel}}\ \emph {et~al.}(2016)\citenamefont {{Miguel}}, \citenamefont {{Guillot}},\ and\ \citenamefont {{Fayon}}}]{Miguel2016}%
  \BibitemOpen
  \bibfield  {author} {\bibinfo {author} {\bibfnamefont {Y.}~\bibnamefont {{Miguel}}}, \bibinfo {author} {\bibfnamefont {T.}~\bibnamefont {{Guillot}}},\ and\ \bibinfo {author} {\bibfnamefont {L.}~\bibnamefont {{Fayon}}},\ }\bibfield  {title} {\bibinfo {title} {{Jupiter internal structure: the effect of different equations of state}},\ }\href {https://doi.org/10.1051/0004-6361/201629732} {\bibfield  {journal} {\bibinfo  {journal} {A\&A}\ }\textbf {\bibinfo {volume} {596}},\ \bibinfo {eid} {A114} (\bibinfo {year} {2016})}\BibitemShut {NoStop}%
\bibitem [{\citenamefont {{Howard}}\ \emph {et~al.}(2023{\natexlab{a}})\citenamefont {{Howard}}, \citenamefont {{Guillot}}, \citenamefont {{Bazot}}, \citenamefont {{Miguel}}, \citenamefont {{Stevenson}}, \citenamefont {{Galanti}}, \citenamefont {{Kaspi}}, \citenamefont {{Hubbard}}, \citenamefont {{Militzer}}, \citenamefont {{Helled}}, \citenamefont {{Nettelmann}}, \citenamefont {{Idini}},\ and\ \citenamefont {{Bolton}}}]{howard2023}%
  \BibitemOpen
  \bibfield  {author} {\bibinfo {author} {\bibfnamefont {S.}~\bibnamefont {{Howard}}}, \bibinfo {author} {\bibfnamefont {T.}~\bibnamefont {{Guillot}}}, \bibinfo {author} {\bibfnamefont {M.}~\bibnamefont {{Bazot}}}, \bibinfo {author} {\bibfnamefont {Y.}~\bibnamefont {{Miguel}}}, \bibinfo {author} {\bibfnamefont {D.~J.}\ \bibnamefont {{Stevenson}}}, \bibinfo {author} {\bibfnamefont {E.}~\bibnamefont {{Galanti}}}, \bibinfo {author} {\bibfnamefont {Y.}~\bibnamefont {{Kaspi}}}, \bibinfo {author} {\bibfnamefont {W.~B.}\ \bibnamefont {{Hubbard}}}, \bibinfo {author} {\bibfnamefont {B.}~\bibnamefont {{Militzer}}}, \bibinfo {author} {\bibfnamefont {R.}~\bibnamefont {{Helled}}}, \bibinfo {author} {\bibfnamefont {N.}~\bibnamefont {{Nettelmann}}}, \bibinfo {author} {\bibfnamefont {B.}~\bibnamefont {{Idini}}},\ and\ \bibinfo {author} {\bibfnamefont {S.}~\bibnamefont {{Bolton}}},\ }\bibfield  {title} {\bibinfo {title} {{Jupiter's interior from Juno: Equation-of-state uncertainties and dilute core extent}},\ }\href
  {https://doi.org/10.1051/0004-6361/202245625} {\bibfield  {journal} {\bibinfo  {journal} {Astronomy and Astrophysics}\ }\textbf {\bibinfo {volume} {672}},\ \bibinfo {eid} {A33} (\bibinfo {year} {2023}{\natexlab{a}})},\ \Eprint {https://arxiv.org/abs/2302.09082} {arXiv:2302.09082 [astro-ph.EP]} \BibitemShut {NoStop}%
\bibitem [{\citenamefont {{Miguel}}\ \emph {et~al.}(2022)\citenamefont {{Miguel}}, \citenamefont {{Bazot}}, \citenamefont {{Guillot}}, \citenamefont {{Howard}}, \citenamefont {{Galanti}}, \citenamefont {{Kaspi}}, \citenamefont {{Hubbard}}, \citenamefont {{Militzer}}, \citenamefont {{Helled}}, \citenamefont {{Atreya}}, \citenamefont {{Connerney}}, \citenamefont {{Durante}}, \citenamefont {{Kulowski}}, \citenamefont {{Lunine}}, \citenamefont {{Stevenson}},\ and\ \citenamefont {{Bolton}}}]{miguel2022}%
  \BibitemOpen
  \bibfield  {author} {\bibinfo {author} {\bibfnamefont {Y.}~\bibnamefont {{Miguel}}}, \bibinfo {author} {\bibfnamefont {M.}~\bibnamefont {{Bazot}}}, \bibinfo {author} {\bibfnamefont {T.}~\bibnamefont {{Guillot}}}, \bibinfo {author} {\bibfnamefont {S.}~\bibnamefont {{Howard}}}, \bibinfo {author} {\bibfnamefont {E.}~\bibnamefont {{Galanti}}}, \bibinfo {author} {\bibfnamefont {Y.}~\bibnamefont {{Kaspi}}}, \bibinfo {author} {\bibfnamefont {W.~B.}\ \bibnamefont {{Hubbard}}}, \bibinfo {author} {\bibfnamefont {B.}~\bibnamefont {{Militzer}}}, \bibinfo {author} {\bibfnamefont {R.}~\bibnamefont {{Helled}}}, \bibinfo {author} {\bibfnamefont {S.~K.}\ \bibnamefont {{Atreya}}}, \bibinfo {author} {\bibfnamefont {J.~E.~P.}\ \bibnamefont {{Connerney}}}, \bibinfo {author} {\bibfnamefont {D.}~\bibnamefont {{Durante}}}, \bibinfo {author} {\bibfnamefont {L.}~\bibnamefont {{Kulowski}}}, \bibinfo {author} {\bibfnamefont {J.~I.}\ \bibnamefont {{Lunine}}}, \bibinfo {author} {\bibfnamefont {D.}~\bibnamefont {{Stevenson}}},\ and\
  \bibinfo {author} {\bibfnamefont {S.}~\bibnamefont {{Bolton}}},\ }\bibfield  {title} {\bibinfo {title} {{Jupiter's inhomogeneous envelope}},\ }\href {https://doi.org/10.1051/0004-6361/202243207} {\bibfield  {journal} {\bibinfo  {journal} {Astronomy \& Astrophysics}\ }\textbf {\bibinfo {volume} {662}},\ \bibinfo {eid} {A18} (\bibinfo {year} {2022})},\ \Eprint {https://arxiv.org/abs/2203.01866} {arXiv:2203.01866 [astro-ph.EP]} \BibitemShut {NoStop}%
\bibitem [{\citenamefont {{Wahl}}(2017)}]{Wahl2017}%
  \BibitemOpen
  \bibfield  {author} {\bibinfo {author} {\bibfnamefont {S.~e.~a.}\ \bibnamefont {{Wahl}}},\ }\bibfield  {title} {\bibinfo {title} {{Comparing Jupiter interior structure models to Juno gravity measurements and the role of a dilute core}},\ }\href {https://doi.org/10.1002/2017GL073160} {\bibfield  {journal} {\bibinfo  {journal} {Geophys. Res. Lett.}\ }\textbf {\bibinfo {volume} {44}},\ \bibinfo {eid} {80} (\bibinfo {year} {2017})}\BibitemShut {NoStop}%
\bibitem [{\citenamefont {{Nettelmann}}\ \emph {et~al.}(2021)\citenamefont {{Nettelmann}}, \citenamefont {{Movshovitz}}, \citenamefont {{Ni}}, \citenamefont {{Fortney}}, \citenamefont {{Galanti}}, \citenamefont {{Kaspi}}, \citenamefont {{Helled}}, \citenamefont {{Mankovich}},\ and\ \citenamefont {{Bolton}}}]{nettelmann2021}%
  \BibitemOpen
  \bibfield  {author} {\bibinfo {author} {\bibfnamefont {N.}~\bibnamefont {{Nettelmann}}}, \bibinfo {author} {\bibfnamefont {N.}~\bibnamefont {{Movshovitz}}}, \bibinfo {author} {\bibfnamefont {D.}~\bibnamefont {{Ni}}}, \bibinfo {author} {\bibfnamefont {J.~J.}\ \bibnamefont {{Fortney}}}, \bibinfo {author} {\bibfnamefont {E.}~\bibnamefont {{Galanti}}}, \bibinfo {author} {\bibfnamefont {Y.}~\bibnamefont {{Kaspi}}}, \bibinfo {author} {\bibfnamefont {R.}~\bibnamefont {{Helled}}}, \bibinfo {author} {\bibfnamefont {C.~R.}\ \bibnamefont {{Mankovich}}},\ and\ \bibinfo {author} {\bibfnamefont {S.}~\bibnamefont {{Bolton}}},\ }\bibfield  {title} {\bibinfo {title} {{Theory of Figures to the Seventh Order and the Interiors of Jupiter and Saturn}},\ }\href {https://doi.org/10.3847/PSJ/ac390a} {\bibfield  {journal} {\bibinfo  {journal} {The Planetary Science Journal}\ }\textbf {\bibinfo {volume} {2}},\ \bibinfo {eid} {241} (\bibinfo {year} {2021})},\ \Eprint {https://arxiv.org/abs/2110.15452} {arXiv:2110.15452 [astro-ph.EP]}
  \BibitemShut {NoStop}%
\bibitem [{\citenamefont {{Militzer}}\ \emph {et~al.}(2022)\citenamefont {{Militzer}}, \citenamefont {{Hubbard}}, \citenamefont {{Wahl}}, \citenamefont {{Lunine}}, \citenamefont {{Galanti}}, \citenamefont {{Kaspi}}, \citenamefont {{Miguel}}, \citenamefont {{Guillot}}, \citenamefont {{Moore}}, \citenamefont {{Parisi}}, \citenamefont {{Connerney}}, \citenamefont {{Helled}}, \citenamefont {{Cao}}, \citenamefont {{Mankovich}}, \citenamefont {{Stevenson}}, \citenamefont {{Park}}, \citenamefont {{Wong}}, \citenamefont {{Atreya}}, \citenamefont {{Anderson}},\ and\ \citenamefont {{Bolton}}}]{militzer2022}%
  \BibitemOpen
  \bibfield  {author} {\bibinfo {author} {\bibfnamefont {B.}~\bibnamefont {{Militzer}}}, \bibinfo {author} {\bibfnamefont {W.~B.}\ \bibnamefont {{Hubbard}}}, \bibinfo {author} {\bibfnamefont {S.}~\bibnamefont {{Wahl}}}, \bibinfo {author} {\bibfnamefont {J.~I.}\ \bibnamefont {{Lunine}}}, \bibinfo {author} {\bibfnamefont {E.}~\bibnamefont {{Galanti}}}, \bibinfo {author} {\bibfnamefont {Y.}~\bibnamefont {{Kaspi}}}, \bibinfo {author} {\bibfnamefont {Y.}~\bibnamefont {{Miguel}}}, \bibinfo {author} {\bibfnamefont {T.}~\bibnamefont {{Guillot}}}, \bibinfo {author} {\bibfnamefont {K.~M.}\ \bibnamefont {{Moore}}}, \bibinfo {author} {\bibfnamefont {M.}~\bibnamefont {{Parisi}}}, \bibinfo {author} {\bibfnamefont {J.~E.~P.}\ \bibnamefont {{Connerney}}}, \bibinfo {author} {\bibfnamefont {R.}~\bibnamefont {{Helled}}}, \bibinfo {author} {\bibfnamefont {H.}~\bibnamefont {{Cao}}}, \bibinfo {author} {\bibfnamefont {C.}~\bibnamefont {{Mankovich}}}, \bibinfo {author} {\bibfnamefont {D.~J.}\ \bibnamefont {{Stevenson}}}, \bibinfo
  {author} {\bibfnamefont {R.~S.}\ \bibnamefont {{Park}}}, \bibinfo {author} {\bibfnamefont {M.}~\bibnamefont {{Wong}}}, \bibinfo {author} {\bibfnamefont {S.~K.}\ \bibnamefont {{Atreya}}}, \bibinfo {author} {\bibfnamefont {J.}~\bibnamefont {{Anderson}}},\ and\ \bibinfo {author} {\bibfnamefont {S.~J.}\ \bibnamefont {{Bolton}}},\ }\bibfield  {title} {\bibinfo {title} {{Juno Spacecraft Measurements of Jupiter's Gravity Imply a Dilute Core}},\ }\href {https://doi.org/10.3847/PSJ/ac7ec8} {\bibfield  {journal} {\bibinfo  {journal} {The Planetary Science Journal}\ }\textbf {\bibinfo {volume} {3}},\ \bibinfo {eid} {185} (\bibinfo {year} {2022})}\BibitemShut {NoStop}%
\bibitem [{\citenamefont {{Stevenson}}(2020)}]{stevenson2020}%
  \BibitemOpen
  \bibfield  {author} {\bibinfo {author} {\bibfnamefont {D.~J.}\ \bibnamefont {{Stevenson}}},\ }\bibfield  {title} {\bibinfo {title} {{Jupiter's Interior as Revealed by Juno}},\ }\href {https://doi.org/10.1146/annurev-earth-081619-052855} {\bibfield  {journal} {\bibinfo  {journal} {Annual Review of Earth and Planetary Sciences}\ }\textbf {\bibinfo {volume} {48}},\ \bibinfo {pages} {465} (\bibinfo {year} {2020})}\BibitemShut {NoStop}%
\bibitem [{\citenamefont {Knudson}\ \emph {et~al.}(2012)\citenamefont {Knudson}, \citenamefont {Desjarlais}, \citenamefont {Lemke}, \citenamefont {Mattsson}, \citenamefont {French}, \citenamefont {Nettelmann},\ and\ \citenamefont {Redmer}}]{PhysRevLett.108.091102}%
  \BibitemOpen
  \bibfield  {author} {\bibinfo {author} {\bibfnamefont {M.~D.}\ \bibnamefont {Knudson}}, \bibinfo {author} {\bibfnamefont {M.~P.}\ \bibnamefont {Desjarlais}}, \bibinfo {author} {\bibfnamefont {R.~W.}\ \bibnamefont {Lemke}}, \bibinfo {author} {\bibfnamefont {T.~R.}\ \bibnamefont {Mattsson}}, \bibinfo {author} {\bibfnamefont {M.}~\bibnamefont {French}}, \bibinfo {author} {\bibfnamefont {N.}~\bibnamefont {Nettelmann}},\ and\ \bibinfo {author} {\bibfnamefont {R.}~\bibnamefont {Redmer}},\ }\bibfield  {title} {\bibinfo {title} {Probing the interiors of the ice giants: Shock compression of water to 700 gpa and $3.8\text{ }\text{ }\mathbf{g}/{\mathrm{cm}}^{3}$},\ }\href {https://doi.org/10.1103/PhysRevLett.108.091102} {\bibfield  {journal} {\bibinfo  {journal} {Phys. Rev. Lett.}\ }\textbf {\bibinfo {volume} {108}},\ \bibinfo {pages} {091102} (\bibinfo {year} {2012})}\BibitemShut {NoStop}%
\bibitem [{\citenamefont {{Knudson}}\ \emph {et~al.}(2004)\citenamefont {{Knudson}}, \citenamefont {{Hanson}}, \citenamefont {{Bailey}}, \citenamefont {{Hall}}, \citenamefont {{Asay}},\ and\ \citenamefont {{Deeney}}}]{2004PhRvB..69n4209K}%
  \BibitemOpen
  \bibfield  {author} {\bibinfo {author} {\bibfnamefont {M.~D.}\ \bibnamefont {{Knudson}}}, \bibinfo {author} {\bibfnamefont {D.~L.}\ \bibnamefont {{Hanson}}}, \bibinfo {author} {\bibfnamefont {J.~E.}\ \bibnamefont {{Bailey}}}, \bibinfo {author} {\bibfnamefont {C.~A.}\ \bibnamefont {{Hall}}}, \bibinfo {author} {\bibfnamefont {J.~R.}\ \bibnamefont {{Asay}}},\ and\ \bibinfo {author} {\bibfnamefont {C.}~\bibnamefont {{Deeney}}},\ }\bibfield  {title} {\bibinfo {title} {{Principal Hugoniot, reverberating wave, and mechanical reshock measurements of liquid deuterium to 400 GPa using plate impact techniques}},\ }\href {https://doi.org/10.1103/PhysRevB.69.144209} {\bibfield  {journal} {\bibinfo  {journal} {\prb}\ }\textbf {\bibinfo {volume} {69}},\ \bibinfo {eid} {144209} (\bibinfo {year} {2004})}\BibitemShut {NoStop}%
\bibitem [{\citenamefont {{Hicks}}\ \emph {et~al.}(2009)\citenamefont {{Hicks}}, \citenamefont {{Boehly}}, \citenamefont {{Celliers}}, \citenamefont {{Eggert}}, \citenamefont {{Moon}}, \citenamefont {{Meyerhofer}},\ and\ \citenamefont {{Collins}}}]{2009PhRvB..79a4112H}%
  \BibitemOpen
  \bibfield  {author} {\bibinfo {author} {\bibfnamefont {D.~G.}\ \bibnamefont {{Hicks}}}, \bibinfo {author} {\bibfnamefont {T.~R.}\ \bibnamefont {{Boehly}}}, \bibinfo {author} {\bibfnamefont {P.~M.}\ \bibnamefont {{Celliers}}}, \bibinfo {author} {\bibfnamefont {J.~H.}\ \bibnamefont {{Eggert}}}, \bibinfo {author} {\bibfnamefont {S.~J.}\ \bibnamefont {{Moon}}}, \bibinfo {author} {\bibfnamefont {D.~D.}\ \bibnamefont {{Meyerhofer}}},\ and\ \bibinfo {author} {\bibfnamefont {G.~W.}\ \bibnamefont {{Collins}}},\ }\bibfield  {title} {\bibinfo {title} {{Laser-driven single shock compression of fluid deuterium from 45 to 220 GPa}},\ }\href {https://doi.org/10.1103/PhysRevB.79.014112} {\bibfield  {journal} {\bibinfo  {journal} {\prb}\ }\textbf {\bibinfo {volume} {79}},\ \bibinfo {eid} {014112} (\bibinfo {year} {2009})}\BibitemShut {NoStop}%
\bibitem [{\citenamefont {Saumon}\ \emph {et~al.}(1995)\citenamefont {Saumon}, \citenamefont {Chabrier},\ and\ \citenamefont {van Horn}}]{scvh1995}%
  \BibitemOpen
  \bibfield  {author} {\bibinfo {author} {\bibfnamefont {D.}~\bibnamefont {Saumon}}, \bibinfo {author} {\bibfnamefont {G.}~\bibnamefont {Chabrier}},\ and\ \bibinfo {author} {\bibfnamefont {H.~M.}\ \bibnamefont {van Horn}},\ }\bibfield  {title} {\bibinfo {title} {An equation of state for low-mass stars and giant planets},\ }\href@noop {} {\bibfield  {journal} {\bibinfo  {journal} {Astrophysical Journal Supplement v. 99, p. 713}\ }\textbf {\bibinfo {volume} {99}},\ \bibinfo {pages} {713} (\bibinfo {year} {1995})}\BibitemShut {NoStop}%
\bibitem [{\citenamefont {Nettelmann}\ \emph {et~al.}(2012)\citenamefont {Nettelmann}, \citenamefont {Becker}, \citenamefont {Holst},\ and\ \citenamefont {Redmer}}]{Nettelmann_2012}%
  \BibitemOpen
  \bibfield  {author} {\bibinfo {author} {\bibfnamefont {N.}~\bibnamefont {Nettelmann}}, \bibinfo {author} {\bibfnamefont {A.}~\bibnamefont {Becker}}, \bibinfo {author} {\bibfnamefont {B.}~\bibnamefont {Holst}},\ and\ \bibinfo {author} {\bibfnamefont {R.}~\bibnamefont {Redmer}},\ }\bibfield  {title} {\bibinfo {title} {Jupiter models with improved ab initio hydrogen equation of state (h-reos.2)},\ }\href {https://doi.org/10.1088/0004-637X/750/1/52} {\bibfield  {journal} {\bibinfo  {journal} {The Astrophysical Journal}\ }\textbf {\bibinfo {volume} {750}},\ \bibinfo {pages} {52} (\bibinfo {year} {2012})}\BibitemShut {NoStop}%
\bibitem [{\citenamefont {Militzer}\ and\ \citenamefont {Hubbard}(2013)}]{militzer2013ab}%
  \BibitemOpen
  \bibfield  {author} {\bibinfo {author} {\bibfnamefont {B.}~\bibnamefont {Militzer}}\ and\ \bibinfo {author} {\bibfnamefont {W.~B.}\ \bibnamefont {Hubbard}},\ }\bibfield  {title} {\bibinfo {title} {Ab initio equation of state for hydrogen-helium mixtures with recalibration of the giant-planet mass-radius relation},\ }\href@noop {} {\bibfield  {journal} {\bibinfo  {journal} {The Astrophysical Journal}\ }\textbf {\bibinfo {volume} {774}},\ \bibinfo {pages} {148} (\bibinfo {year} {2013})}\BibitemShut {NoStop}%
\bibitem [{\citenamefont {{Miguel, Y.}}\ \emph {et~al.}(2018)\citenamefont {{Miguel, Y.}}, \citenamefont {{Guillot, T.}},\ and\ \citenamefont {{Fayon, L.}}}]{Miguel2018erratum}%
  \BibitemOpen
  \bibfield  {author} {\bibinfo {author} {\bibnamefont {{Miguel, Y.}}}, \bibinfo {author} {\bibnamefont {{Guillot, T.}}},\ and\ \bibinfo {author} {\bibnamefont {{Fayon, L.}}},\ }\bibfield  {title} {\bibinfo {title} {Jupiter internal structure: the effect of different equations of state (corrigendum)},\ }\href {https://doi.org/10.1051/0004-6361/201629732e} {\bibfield  {journal} {\bibinfo  {journal} {A\&A}\ }\textbf {\bibinfo {volume} {618}},\ \bibinfo {pages} {C2} (\bibinfo {year} {2018})}\BibitemShut {NoStop}%
\bibitem [{\citenamefont {Xie}\ \emph {et~al.}(2025)\citenamefont {Xie}, \citenamefont {Howard},\ and\ \citenamefont {Mazzola}}]{xie2025}%
  \BibitemOpen
  \bibfield  {author} {\bibinfo {author} {\bibfnamefont {H.}~\bibnamefont {Xie}}, \bibinfo {author} {\bibfnamefont {S.}~\bibnamefont {Howard}},\ and\ \bibinfo {author} {\bibfnamefont {G.}~\bibnamefont {Mazzola}},\ }\href {https://arxiv.org/abs/2501.10594} {\bibinfo {title} {Accurate and thermodynamically consistent hydrogen equation of state for planetary modeling with flow matching}} (\bibinfo {year} {2025}),\ \Eprint {https://arxiv.org/abs/2501.10594} {arXiv:2501.10594 [astro-ph.EP]} \BibitemShut {NoStop}%
\bibitem [{\citenamefont {Chabrier}\ \emph {et~al.}(2019)\citenamefont {Chabrier}, \citenamefont {Mazevet},\ and\ \citenamefont {Soubiran}}]{CMS2019}%
  \BibitemOpen
  \bibfield  {author} {\bibinfo {author} {\bibfnamefont {G.}~\bibnamefont {Chabrier}}, \bibinfo {author} {\bibfnamefont {S.}~\bibnamefont {Mazevet}},\ and\ \bibinfo {author} {\bibfnamefont {F.}~\bibnamefont {Soubiran}},\ }\bibfield  {title} {\bibinfo {title} {A new equation of state for dense hydrogen--helium mixtures},\ }\href@noop {} {\bibfield  {journal} {\bibinfo  {journal} {The Astrophysical Journal}\ }\textbf {\bibinfo {volume} {872}},\ \bibinfo {pages} {51} (\bibinfo {year} {2019})}\BibitemShut {NoStop}%
\bibitem [{\citenamefont {Becker}\ \emph {et~al.}(2014)\citenamefont {Becker}, \citenamefont {Lorenzen}, \citenamefont {Fortney}, \citenamefont {Nettelmann}, \citenamefont {Sch{\"o}ttler},\ and\ \citenamefont {Redmer}}]{REOS_2014}%
  \BibitemOpen
  \bibfield  {author} {\bibinfo {author} {\bibfnamefont {A.}~\bibnamefont {Becker}}, \bibinfo {author} {\bibfnamefont {W.}~\bibnamefont {Lorenzen}}, \bibinfo {author} {\bibfnamefont {J.~J.}\ \bibnamefont {Fortney}}, \bibinfo {author} {\bibfnamefont {N.}~\bibnamefont {Nettelmann}}, \bibinfo {author} {\bibfnamefont {M.}~\bibnamefont {Sch{\"o}ttler}},\ and\ \bibinfo {author} {\bibfnamefont {R.}~\bibnamefont {Redmer}},\ }\bibfield  {title} {\bibinfo {title} {Ab initio equations of state for hydrogen (h-reos. 3) and helium (he-reos. 3) and their implications for the interior of brown dwarfs},\ }\href@noop {} {\bibfield  {journal} {\bibinfo  {journal} {The Astrophysical Journal Supplement Series}\ }\textbf {\bibinfo {volume} {215}},\ \bibinfo {pages} {21} (\bibinfo {year} {2014})}\BibitemShut {NoStop}%
\bibitem [{\citenamefont {Burke}(2012)}]{burke_perspective_2012}%
  \BibitemOpen
  \bibfield  {author} {\bibinfo {author} {\bibfnamefont {K.}~\bibnamefont {Burke}},\ }\bibfield  {title} {\bibinfo {title} {Perspective on density functional theory},\ }\href {https://doi.org/10.1063/1.4704546} {\bibfield  {journal} {\bibinfo  {journal} {The Journal of Chemical Physics}\ }\textbf {\bibinfo {volume} {136}},\ \bibinfo {pages} {150901} (\bibinfo {year} {2012})}\BibitemShut {NoStop}%
\bibitem [{\citenamefont {Azadi}\ and\ \citenamefont {Foulkes}(2013)}]{azadi_fate_2013}%
  \BibitemOpen
  \bibfield  {author} {\bibinfo {author} {\bibfnamefont {S.}~\bibnamefont {Azadi}}\ and\ \bibinfo {author} {\bibfnamefont {W.~M.~C.}\ \bibnamefont {Foulkes}},\ }\bibfield  {title} {\bibinfo {title} {Fate of density functional theory in the study of high-pressure solid hydrogen},\ }\href {https://doi.org/10.1103/PhysRevB.88.014115} {\bibfield  {journal} {\bibinfo  {journal} {Physical Review B}\ }\textbf {\bibinfo {volume} {88}},\ \bibinfo {pages} {014115} (\bibinfo {year} {2013})}\BibitemShut {NoStop}%
\bibitem [{\citenamefont {Mazzola}\ \emph {et~al.}(2018)\citenamefont {Mazzola}, \citenamefont {Helled},\ and\ \citenamefont {Sorella}}]{mazzola2018}%
  \BibitemOpen
  \bibfield  {author} {\bibinfo {author} {\bibfnamefont {G.}~\bibnamefont {Mazzola}}, \bibinfo {author} {\bibfnamefont {R.}~\bibnamefont {Helled}},\ and\ \bibinfo {author} {\bibfnamefont {S.}~\bibnamefont {Sorella}},\ }\bibfield  {title} {\bibinfo {title} {Phase diagram of hydrogen and a hydrogen-helium mixture at planetary conditions by quantum monte carlo simulations},\ }\href {https://doi.org/10.1103/PhysRevLett.120.025701} {\bibfield  {journal} {\bibinfo  {journal} {Phys. Rev. Lett.}\ }\textbf {\bibinfo {volume} {120}},\ \bibinfo {pages} {025701} (\bibinfo {year} {2018})}\BibitemShut {NoStop}%
\bibitem [{\citenamefont {Pierleoni}\ \emph {et~al.}(2016)\citenamefont {Pierleoni}, \citenamefont {Morales}, \citenamefont {Rillo}, \citenamefont {Holzmann},\ and\ \citenamefont {Ceperley}}]{pierleoni2016liquid}%
  \BibitemOpen
  \bibfield  {author} {\bibinfo {author} {\bibfnamefont {C.}~\bibnamefont {Pierleoni}}, \bibinfo {author} {\bibfnamefont {M.~A.}\ \bibnamefont {Morales}}, \bibinfo {author} {\bibfnamefont {G.}~\bibnamefont {Rillo}}, \bibinfo {author} {\bibfnamefont {M.}~\bibnamefont {Holzmann}},\ and\ \bibinfo {author} {\bibfnamefont {D.~M.}\ \bibnamefont {Ceperley}},\ }\bibfield  {title} {\bibinfo {title} {Liquid--liquid phase transition in hydrogen by coupled electron--ion monte carlo simulations},\ }\href@noop {} {\bibfield  {journal} {\bibinfo  {journal} {Proceedings of the National Academy of Sciences}\ }\textbf {\bibinfo {volume} {113}},\ \bibinfo {pages} {4953} (\bibinfo {year} {2016})}\BibitemShut {NoStop}%
\bibitem [{\citenamefont {Clay}\ \emph {et~al.}(2014)\citenamefont {Clay}, \citenamefont {Mcminis}, \citenamefont {McMahon}, \citenamefont {Pierleoni}, \citenamefont {Ceperley},\ and\ \citenamefont {Morales}}]{PhysRevB.89.184106}%
  \BibitemOpen
  \bibfield  {author} {\bibinfo {author} {\bibfnamefont {R.~C.}\ \bibnamefont {Clay}}, \bibinfo {author} {\bibfnamefont {J.}~\bibnamefont {Mcminis}}, \bibinfo {author} {\bibfnamefont {J.~M.}\ \bibnamefont {McMahon}}, \bibinfo {author} {\bibfnamefont {C.}~\bibnamefont {Pierleoni}}, \bibinfo {author} {\bibfnamefont {D.~M.}\ \bibnamefont {Ceperley}},\ and\ \bibinfo {author} {\bibfnamefont {M.~A.}\ \bibnamefont {Morales}},\ }\bibfield  {title} {\bibinfo {title} {Benchmarking exchange-correlation functionals for hydrogen at high pressures using quantum monte carlo},\ }\href {https://doi.org/10.1103/PhysRevB.89.184106} {\bibfield  {journal} {\bibinfo  {journal} {Phys. Rev. B}\ }\textbf {\bibinfo {volume} {89}},\ \bibinfo {pages} {184106} (\bibinfo {year} {2014})}\BibitemShut {NoStop}%
\bibitem [{\citenamefont {Knudson}\ \emph {et~al.}(2018)\citenamefont {Knudson}, \citenamefont {Desjarlais}, \citenamefont {Preising},\ and\ \citenamefont {Redmer}}]{PhysRevB.98.174110}%
  \BibitemOpen
  \bibfield  {author} {\bibinfo {author} {\bibfnamefont {M.~D.}\ \bibnamefont {Knudson}}, \bibinfo {author} {\bibfnamefont {M.~P.}\ \bibnamefont {Desjarlais}}, \bibinfo {author} {\bibfnamefont {M.}~\bibnamefont {Preising}},\ and\ \bibinfo {author} {\bibfnamefont {R.}~\bibnamefont {Redmer}},\ }\bibfield  {title} {\bibinfo {title} {Evaluation of exchange-correlation functionals with multiple-shock conductivity measurements in hydrogen and deuterium at the molecular-to-atomic transition},\ }\href {https://doi.org/10.1103/PhysRevB.98.174110} {\bibfield  {journal} {\bibinfo  {journal} {Phys. Rev. B}\ }\textbf {\bibinfo {volume} {98}},\ \bibinfo {pages} {174110} (\bibinfo {year} {2018})}\BibitemShut {NoStop}%
\bibitem [{\citenamefont {Bonitz}\ \emph {et~al.}(2024)\citenamefont {Bonitz}, \citenamefont {Vorberger}, \citenamefont {Bethkenhagen}, \citenamefont {Böhme}, \citenamefont {Ceperley}, \citenamefont {Filinov}, \citenamefont {Gawne}, \citenamefont {Graziani}, \citenamefont {Gregori}, \citenamefont {Hamann}, \citenamefont {Hansen}, \citenamefont {Holzmann}, \citenamefont {Hu}, \citenamefont {Kählert}, \citenamefont {Karasiev}, \citenamefont {Kleinschmidt}, \citenamefont {Kordts}, \citenamefont {Makait}, \citenamefont {Militzer}, \citenamefont {Moldabekov}, \citenamefont {Pierleoni}, \citenamefont {Preising}, \citenamefont {Ramakrishna}, \citenamefont {Redmer}, \citenamefont {Schwalbe}, \citenamefont {Svensson},\ and\ \citenamefont {Dornheim}}]{Bonitz_2024}%
  \BibitemOpen
  \bibfield  {author} {\bibinfo {author} {\bibfnamefont {M.}~\bibnamefont {Bonitz}}, \bibinfo {author} {\bibfnamefont {J.}~\bibnamefont {Vorberger}}, \bibinfo {author} {\bibfnamefont {M.}~\bibnamefont {Bethkenhagen}}, \bibinfo {author} {\bibfnamefont {M.~P.}\ \bibnamefont {Böhme}}, \bibinfo {author} {\bibfnamefont {D.~M.}\ \bibnamefont {Ceperley}}, \bibinfo {author} {\bibfnamefont {A.}~\bibnamefont {Filinov}}, \bibinfo {author} {\bibfnamefont {T.}~\bibnamefont {Gawne}}, \bibinfo {author} {\bibfnamefont {F.}~\bibnamefont {Graziani}}, \bibinfo {author} {\bibfnamefont {G.}~\bibnamefont {Gregori}}, \bibinfo {author} {\bibfnamefont {P.}~\bibnamefont {Hamann}}, \bibinfo {author} {\bibfnamefont {S.~B.}\ \bibnamefont {Hansen}}, \bibinfo {author} {\bibfnamefont {M.}~\bibnamefont {Holzmann}}, \bibinfo {author} {\bibfnamefont {S.~X.}\ \bibnamefont {Hu}}, \bibinfo {author} {\bibfnamefont {H.}~\bibnamefont {Kählert}}, \bibinfo {author} {\bibfnamefont {V.~V.}\ \bibnamefont {Karasiev}}, \bibinfo {author} {\bibfnamefont
  {U.}~\bibnamefont {Kleinschmidt}}, \bibinfo {author} {\bibfnamefont {L.}~\bibnamefont {Kordts}}, \bibinfo {author} {\bibfnamefont {C.}~\bibnamefont {Makait}}, \bibinfo {author} {\bibfnamefont {B.}~\bibnamefont {Militzer}}, \bibinfo {author} {\bibfnamefont {Z.~A.}\ \bibnamefont {Moldabekov}}, \bibinfo {author} {\bibfnamefont {C.}~\bibnamefont {Pierleoni}}, \bibinfo {author} {\bibfnamefont {M.}~\bibnamefont {Preising}}, \bibinfo {author} {\bibfnamefont {K.}~\bibnamefont {Ramakrishna}}, \bibinfo {author} {\bibfnamefont {R.}~\bibnamefont {Redmer}}, \bibinfo {author} {\bibfnamefont {S.}~\bibnamefont {Schwalbe}}, \bibinfo {author} {\bibfnamefont {P.}~\bibnamefont {Svensson}},\ and\ \bibinfo {author} {\bibfnamefont {T.}~\bibnamefont {Dornheim}},\ }\bibfield  {title} {\bibinfo {title} {Toward first principles-based simulations of dense hydrogen},\ }\bibfield  {journal} {\bibinfo  {journal} {Physics of Plasmas}\ }\textbf {\bibinfo {volume} {31}},\ \href {https://doi.org/10.1063/5.0219405} {10.1063/5.0219405}
  (\bibinfo {year} {2024})\BibitemShut {NoStop}%
\bibitem [{\citenamefont {Van De~Bund}\ \emph {et~al.}(2021)\citenamefont {Van De~Bund}, \citenamefont {Wiebe},\ and\ \citenamefont {Ackland}}]{van2021isotope}%
  \BibitemOpen
  \bibfield  {author} {\bibinfo {author} {\bibfnamefont {S.}~\bibnamefont {Van De~Bund}}, \bibinfo {author} {\bibfnamefont {H.}~\bibnamefont {Wiebe}},\ and\ \bibinfo {author} {\bibfnamefont {G.~J.}\ \bibnamefont {Ackland}},\ }\bibfield  {title} {\bibinfo {title} {Isotope quantum effects in the metallization transition in liquid hydrogen},\ }\href@noop {} {\bibfield  {journal} {\bibinfo  {journal} {Physical Review Letters}\ }\textbf {\bibinfo {volume} {126}},\ \bibinfo {pages} {225701} (\bibinfo {year} {2021})}\BibitemShut {NoStop}%
\bibitem [{\citenamefont {Zaghoo}\ \emph {et~al.}(2018)\citenamefont {Zaghoo}, \citenamefont {Husband},\ and\ \citenamefont {Silvera}}]{zaghoo2018striking}%
  \BibitemOpen
  \bibfield  {author} {\bibinfo {author} {\bibfnamefont {M.}~\bibnamefont {Zaghoo}}, \bibinfo {author} {\bibfnamefont {R.~J.}\ \bibnamefont {Husband}},\ and\ \bibinfo {author} {\bibfnamefont {I.~F.}\ \bibnamefont {Silvera}},\ }\bibfield  {title} {\bibinfo {title} {Striking isotope effect on the metallization phase lines of liquid hydrogen and deuterium},\ }\href@noop {} {\bibfield  {journal} {\bibinfo  {journal} {Physical Review B}\ }\textbf {\bibinfo {volume} {98}},\ \bibinfo {pages} {104102} (\bibinfo {year} {2018})}\BibitemShut {NoStop}%
\bibitem [{\citenamefont {Schuch}\ and\ \citenamefont {Verstraete}(2009)}]{schuch2009computational}%
  \BibitemOpen
  \bibfield  {author} {\bibinfo {author} {\bibfnamefont {N.}~\bibnamefont {Schuch}}\ and\ \bibinfo {author} {\bibfnamefont {F.}~\bibnamefont {Verstraete}},\ }\bibfield  {title} {\bibinfo {title} {Computational complexity of interacting electrons and fundamental limitations of density functional theory},\ }\href@noop {} {\bibfield  {journal} {\bibinfo  {journal} {Nature physics}\ }\textbf {\bibinfo {volume} {5}},\ \bibinfo {pages} {732} (\bibinfo {year} {2009})}\BibitemShut {NoStop}%
\bibitem [{\citenamefont {Jones}(2015)}]{dft2015RMP}%
  \BibitemOpen
  \bibfield  {author} {\bibinfo {author} {\bibfnamefont {R.~O.}\ \bibnamefont {Jones}},\ }\bibfield  {title} {\bibinfo {title} {Density functional theory: Its origins, rise to prominence, and future},\ }\href {https://doi.org/10.1103/RevModPhys.87.897} {\bibfield  {journal} {\bibinfo  {journal} {Rev. Mod. Phys.}\ }\textbf {\bibinfo {volume} {87}},\ \bibinfo {pages} {897} (\bibinfo {year} {2015})}\BibitemShut {NoStop}%
\bibitem [{\citenamefont {Lejaeghere}\ \emph {et~al.}(2016)\citenamefont {Lejaeghere}, \citenamefont {Bihlmayer}, \citenamefont {Bj{\"o}rkman}, \citenamefont {Blaha}, \citenamefont {Bl{\"u}gel}, \citenamefont {Blum}, \citenamefont {Caliste}, \citenamefont {Castelli}, \citenamefont {Clark}, \citenamefont {Dal~Corso} \emph {et~al.}}]{lejaeghere2016reproducibility}%
  \BibitemOpen
  \bibfield  {author} {\bibinfo {author} {\bibfnamefont {K.}~\bibnamefont {Lejaeghere}}, \bibinfo {author} {\bibfnamefont {G.}~\bibnamefont {Bihlmayer}}, \bibinfo {author} {\bibfnamefont {T.}~\bibnamefont {Bj{\"o}rkman}}, \bibinfo {author} {\bibfnamefont {P.}~\bibnamefont {Blaha}}, \bibinfo {author} {\bibfnamefont {S.}~\bibnamefont {Bl{\"u}gel}}, \bibinfo {author} {\bibfnamefont {V.}~\bibnamefont {Blum}}, \bibinfo {author} {\bibfnamefont {D.}~\bibnamefont {Caliste}}, \bibinfo {author} {\bibfnamefont {I.~E.}\ \bibnamefont {Castelli}}, \bibinfo {author} {\bibfnamefont {S.~J.}\ \bibnamefont {Clark}}, \bibinfo {author} {\bibfnamefont {A.}~\bibnamefont {Dal~Corso}}, \emph {et~al.},\ }\bibfield  {title} {\bibinfo {title} {Reproducibility in density functional theory calculations of solids},\ }\href@noop {} {\bibfield  {journal} {\bibinfo  {journal} {Science}\ }\textbf {\bibinfo {volume} {351}},\ \bibinfo {pages} {aad3000} (\bibinfo {year} {2016})}\BibitemShut {NoStop}%
\bibitem [{\citenamefont {Bosoni}\ \emph {et~al.}(2023)\citenamefont {Bosoni}, \citenamefont {Beal}, \citenamefont {Bercx}, \citenamefont {Blaha}, \citenamefont {Blügel}, \citenamefont {Bröder}, \citenamefont {Callsen}, \citenamefont {Cottenier}, \citenamefont {Degomme}, \citenamefont {Dikan}, \citenamefont {Eimre}, \citenamefont {Flage-Larsen}, \citenamefont {Fornari}, \citenamefont {Garcia}, \citenamefont {Genovese}, \citenamefont {Giantomassi}, \citenamefont {Huber}, \citenamefont {Janssen}, \citenamefont {Kastlunger}, \citenamefont {Krack}, \citenamefont {Kresse}, \citenamefont {Kühne}, \citenamefont {Lejaeghere}, \citenamefont {Madsen}, \citenamefont {Marsman}, \citenamefont {Marzari}, \citenamefont {Michalicek}, \citenamefont {Mirhosseini}, \citenamefont {Müller}, \citenamefont {Petretto}, \citenamefont {Pickard}, \citenamefont {Poncé}, \citenamefont {Rignanese}, \citenamefont {Rubel}, \citenamefont {Ruh}, \citenamefont {Sluydts}, \citenamefont {Vanpoucke}, \citenamefont {Vijay}, \citenamefont
  {Wolloch}, \citenamefont {Wortmann}, \citenamefont {Yakutovich}, \citenamefont {Yu}, \citenamefont {Zadoks}, \citenamefont {Zhu},\ and\ \citenamefont {Pizzi}}]{Bosoni_2023}%
  \BibitemOpen
  \bibfield  {author} {\bibinfo {author} {\bibfnamefont {E.}~\bibnamefont {Bosoni}}, \bibinfo {author} {\bibfnamefont {L.}~\bibnamefont {Beal}}, \bibinfo {author} {\bibfnamefont {M.}~\bibnamefont {Bercx}}, \bibinfo {author} {\bibfnamefont {P.}~\bibnamefont {Blaha}}, \bibinfo {author} {\bibfnamefont {S.}~\bibnamefont {Blügel}}, \bibinfo {author} {\bibfnamefont {J.}~\bibnamefont {Bröder}}, \bibinfo {author} {\bibfnamefont {M.}~\bibnamefont {Callsen}}, \bibinfo {author} {\bibfnamefont {S.}~\bibnamefont {Cottenier}}, \bibinfo {author} {\bibfnamefont {A.}~\bibnamefont {Degomme}}, \bibinfo {author} {\bibfnamefont {V.}~\bibnamefont {Dikan}}, \bibinfo {author} {\bibfnamefont {K.}~\bibnamefont {Eimre}}, \bibinfo {author} {\bibfnamefont {E.}~\bibnamefont {Flage-Larsen}}, \bibinfo {author} {\bibfnamefont {M.}~\bibnamefont {Fornari}}, \bibinfo {author} {\bibfnamefont {A.}~\bibnamefont {Garcia}}, \bibinfo {author} {\bibfnamefont {L.}~\bibnamefont {Genovese}}, \bibinfo {author} {\bibfnamefont {M.}~\bibnamefont
  {Giantomassi}}, \bibinfo {author} {\bibfnamefont {S.~P.}\ \bibnamefont {Huber}}, \bibinfo {author} {\bibfnamefont {H.}~\bibnamefont {Janssen}}, \bibinfo {author} {\bibfnamefont {G.}~\bibnamefont {Kastlunger}}, \bibinfo {author} {\bibfnamefont {M.}~\bibnamefont {Krack}}, \bibinfo {author} {\bibfnamefont {G.}~\bibnamefont {Kresse}}, \bibinfo {author} {\bibfnamefont {T.~D.}\ \bibnamefont {Kühne}}, \bibinfo {author} {\bibfnamefont {K.}~\bibnamefont {Lejaeghere}}, \bibinfo {author} {\bibfnamefont {G.~K.~H.}\ \bibnamefont {Madsen}}, \bibinfo {author} {\bibfnamefont {M.}~\bibnamefont {Marsman}}, \bibinfo {author} {\bibfnamefont {N.}~\bibnamefont {Marzari}}, \bibinfo {author} {\bibfnamefont {G.}~\bibnamefont {Michalicek}}, \bibinfo {author} {\bibfnamefont {H.}~\bibnamefont {Mirhosseini}}, \bibinfo {author} {\bibfnamefont {T.~M.~A.}\ \bibnamefont {Müller}}, \bibinfo {author} {\bibfnamefont {G.}~\bibnamefont {Petretto}}, \bibinfo {author} {\bibfnamefont {C.~J.}\ \bibnamefont {Pickard}}, \bibinfo {author}
  {\bibfnamefont {S.}~\bibnamefont {Poncé}}, \bibinfo {author} {\bibfnamefont {G.-M.}\ \bibnamefont {Rignanese}}, \bibinfo {author} {\bibfnamefont {O.}~\bibnamefont {Rubel}}, \bibinfo {author} {\bibfnamefont {T.}~\bibnamefont {Ruh}}, \bibinfo {author} {\bibfnamefont {M.}~\bibnamefont {Sluydts}}, \bibinfo {author} {\bibfnamefont {D.~E.~P.}\ \bibnamefont {Vanpoucke}}, \bibinfo {author} {\bibfnamefont {S.}~\bibnamefont {Vijay}}, \bibinfo {author} {\bibfnamefont {M.}~\bibnamefont {Wolloch}}, \bibinfo {author} {\bibfnamefont {D.}~\bibnamefont {Wortmann}}, \bibinfo {author} {\bibfnamefont {A.~V.}\ \bibnamefont {Yakutovich}}, \bibinfo {author} {\bibfnamefont {J.}~\bibnamefont {Yu}}, \bibinfo {author} {\bibfnamefont {A.}~\bibnamefont {Zadoks}}, \bibinfo {author} {\bibfnamefont {B.}~\bibnamefont {Zhu}},\ and\ \bibinfo {author} {\bibfnamefont {G.}~\bibnamefont {Pizzi}},\ }\bibfield  {title} {\bibinfo {title} {How to verify the precision of density-functional-theory implementations via reproducible and universal
  workflows},\ }\href {https://doi.org/10.1038/s42254-023-00655-3} {\bibfield  {journal} {\bibinfo  {journal} {Nature Reviews Physics}\ }\textbf {\bibinfo {volume} {6}},\ \bibinfo {pages} {45–58} (\bibinfo {year} {2023})}\BibitemShut {NoStop}%
\bibitem [{\citenamefont {Hohenberg}\ and\ \citenamefont {Kohn}(1964)}]{Hohenberg_Kohn1964}%
  \BibitemOpen
  \bibfield  {author} {\bibinfo {author} {\bibfnamefont {P.}~\bibnamefont {Hohenberg}}\ and\ \bibinfo {author} {\bibfnamefont {W.}~\bibnamefont {Kohn}},\ }\bibfield  {title} {\bibinfo {title} {Inhomogeneous electron gas},\ }\href {https://doi.org/10.1103/PhysRev.136.B864} {\bibfield  {journal} {\bibinfo  {journal} {Phys. Rev.}\ }\textbf {\bibinfo {volume} {136}},\ \bibinfo {pages} {B864} (\bibinfo {year} {1964})}\BibitemShut {NoStop}%
\bibitem [{\citenamefont {Kohn}\ and\ \citenamefont {Sham}(1965)}]{Kohn_Sham1965}%
  \BibitemOpen
  \bibfield  {author} {\bibinfo {author} {\bibfnamefont {W.}~\bibnamefont {Kohn}}\ and\ \bibinfo {author} {\bibfnamefont {L.~J.}\ \bibnamefont {Sham}},\ }\bibfield  {title} {\bibinfo {title} {Self-consistent equations including exchange and correlation effects},\ }\href {https://doi.org/10.1103/PhysRev.140.A1133} {\bibfield  {journal} {\bibinfo  {journal} {Phys. Rev.}\ }\textbf {\bibinfo {volume} {140}},\ \bibinfo {pages} {A1133} (\bibinfo {year} {1965})}\BibitemShut {NoStop}%
\bibitem [{\citenamefont {Martin}(2004)}]{martin_ElectronicStructure}%
  \BibitemOpen
  \bibfield  {author} {\bibinfo {author} {\bibfnamefont {R.~M.}\ \bibnamefont {Martin}},\ }\href {https://doi.org/10.1017/CBO9780511805769} {\emph {\bibinfo {title} {Electronic Structure: Basic Theory and Practical Methods}}}\ (\bibinfo  {publisher} {Cambridge University Press},\ \bibinfo {year} {2004})\BibitemShut {NoStop}%
\bibitem [{\citenamefont {Perdew}\ \emph {et~al.}(1996{\natexlab{a}})\citenamefont {Perdew}, \citenamefont {Burke},\ and\ \citenamefont {Ernzerhof}}]{PBE1996}%
  \BibitemOpen
  \bibfield  {author} {\bibinfo {author} {\bibfnamefont {J.~P.}\ \bibnamefont {Perdew}}, \bibinfo {author} {\bibfnamefont {K.}~\bibnamefont {Burke}},\ and\ \bibinfo {author} {\bibfnamefont {M.}~\bibnamefont {Ernzerhof}},\ }\bibfield  {title} {\bibinfo {title} {Generalized gradient approximation made simple},\ }\href {https://doi.org/10.1103/PhysRevLett.77.3865} {\bibfield  {journal} {\bibinfo  {journal} {Phys. Rev. Lett.}\ }\textbf {\bibinfo {volume} {77}},\ \bibinfo {pages} {3865} (\bibinfo {year} {1996}{\natexlab{a}})}\BibitemShut {NoStop}%
\bibitem [{\citenamefont {Chabrier}\ and\ \citenamefont {Debras}(2021)}]{CMS2021}%
  \BibitemOpen
  \bibfield  {author} {\bibinfo {author} {\bibfnamefont {G.}~\bibnamefont {Chabrier}}\ and\ \bibinfo {author} {\bibfnamefont {F.}~\bibnamefont {Debras}},\ }\bibfield  {title} {\bibinfo {title} {A new equation of state for dense hydrogen--helium mixtures. ii. taking into account hydrogen--helium interactions},\ }\href@noop {} {\bibfield  {journal} {\bibinfo  {journal} {The Astrophysical Journal}\ }\textbf {\bibinfo {volume} {917}},\ \bibinfo {pages} {4} (\bibinfo {year} {2021})}\BibitemShut {NoStop}%
\bibitem [{\citenamefont {Becke}(1988)}]{Becke1988_BLYP}%
  \BibitemOpen
  \bibfield  {author} {\bibinfo {author} {\bibfnamefont {A.~D.}\ \bibnamefont {Becke}},\ }\bibfield  {title} {\bibinfo {title} {Density-functional exchange-energy approximation with correct asymptotic behavior},\ }\href@noop {} {\bibfield  {journal} {\bibinfo  {journal} {Physical review A}\ }\textbf {\bibinfo {volume} {38}},\ \bibinfo {pages} {3098} (\bibinfo {year} {1988})}\BibitemShut {NoStop}%
\bibitem [{\citenamefont {Lee}\ \emph {et~al.}(1988)\citenamefont {Lee}, \citenamefont {Yang},\ and\ \citenamefont {Parr}}]{Lee1988_BLYP}%
  \BibitemOpen
  \bibfield  {author} {\bibinfo {author} {\bibfnamefont {C.}~\bibnamefont {Lee}}, \bibinfo {author} {\bibfnamefont {W.}~\bibnamefont {Yang}},\ and\ \bibinfo {author} {\bibfnamefont {R.~G.}\ \bibnamefont {Parr}},\ }\bibfield  {title} {\bibinfo {title} {Development of the colle-salvetti correlation-energy formula into a functional of the electron density},\ }\href@noop {} {\bibfield  {journal} {\bibinfo  {journal} {Physical review B}\ }\textbf {\bibinfo {volume} {37}},\ \bibinfo {pages} {785} (\bibinfo {year} {1988})}\BibitemShut {NoStop}%
\bibitem [{\citenamefont {Dion}\ \emph {et~al.}(2004)\citenamefont {Dion}, \citenamefont {Rydberg}, \citenamefont {Schr{\"o}der}, \citenamefont {Langreth},\ and\ \citenamefont {Lundqvist}}]{Dion2004_vdw-df}%
  \BibitemOpen
  \bibfield  {author} {\bibinfo {author} {\bibfnamefont {M.}~\bibnamefont {Dion}}, \bibinfo {author} {\bibfnamefont {H.}~\bibnamefont {Rydberg}}, \bibinfo {author} {\bibfnamefont {E.}~\bibnamefont {Schr{\"o}der}}, \bibinfo {author} {\bibfnamefont {D.~C.}\ \bibnamefont {Langreth}},\ and\ \bibinfo {author} {\bibfnamefont {B.~I.}\ \bibnamefont {Lundqvist}},\ }\bibfield  {title} {\bibinfo {title} {Van der waals density functional for general geometries},\ }\href@noop {} {\bibfield  {journal} {\bibinfo  {journal} {Physical review letters}\ }\textbf {\bibinfo {volume} {92}},\ \bibinfo {pages} {246401} (\bibinfo {year} {2004})}\BibitemShut {NoStop}%
\bibitem [{\citenamefont {Thonhauser}\ \emph {et~al.}(2015)\citenamefont {Thonhauser}, \citenamefont {Zuluaga}, \citenamefont {Arter}, \citenamefont {Berland}, \citenamefont {Schr{\"o}der},\ and\ \citenamefont {Hyldgaard}}]{Thonhauser2015_vdw-df}%
  \BibitemOpen
  \bibfield  {author} {\bibinfo {author} {\bibfnamefont {T.}~\bibnamefont {Thonhauser}}, \bibinfo {author} {\bibfnamefont {S.}~\bibnamefont {Zuluaga}}, \bibinfo {author} {\bibfnamefont {C.}~\bibnamefont {Arter}}, \bibinfo {author} {\bibfnamefont {K.}~\bibnamefont {Berland}}, \bibinfo {author} {\bibfnamefont {E.}~\bibnamefont {Schr{\"o}der}},\ and\ \bibinfo {author} {\bibfnamefont {P.}~\bibnamefont {Hyldgaard}},\ }\bibfield  {title} {\bibinfo {title} {Spin signature of nonlocal correlation binding in metal-organic frameworks},\ }\href@noop {} {\bibfield  {journal} {\bibinfo  {journal} {Physical review letters}\ }\textbf {\bibinfo {volume} {115}},\ \bibinfo {pages} {136402} (\bibinfo {year} {2015})}\BibitemShut {NoStop}%
\bibitem [{\citenamefont {Lee}\ \emph {et~al.}(2010)\citenamefont {Lee}, \citenamefont {Murray}, \citenamefont {Kong}, \citenamefont {Lundqvist},\ and\ \citenamefont {Langreth}}]{Lee2010_vdw-df2}%
  \BibitemOpen
  \bibfield  {author} {\bibinfo {author} {\bibfnamefont {K.}~\bibnamefont {Lee}}, \bibinfo {author} {\bibfnamefont {{\'E}.~D.}\ \bibnamefont {Murray}}, \bibinfo {author} {\bibfnamefont {L.}~\bibnamefont {Kong}}, \bibinfo {author} {\bibfnamefont {B.~I.}\ \bibnamefont {Lundqvist}},\ and\ \bibinfo {author} {\bibfnamefont {D.~C.}\ \bibnamefont {Langreth}},\ }\bibfield  {title} {\bibinfo {title} {Higher-accuracy van der waals density functional},\ }\href@noop {} {\bibfield  {journal} {\bibinfo  {journal} {Physical Review B—Condensed Matter and Materials Physics}\ }\textbf {\bibinfo {volume} {82}},\ \bibinfo {pages} {081101} (\bibinfo {year} {2010})}\BibitemShut {NoStop}%
\bibitem [{\citenamefont {Sun}\ \emph {et~al.}(2015)\citenamefont {Sun}, \citenamefont {Ruzsinszky},\ and\ \citenamefont {Perdew}}]{Sun2015_SCAN}%
  \BibitemOpen
  \bibfield  {author} {\bibinfo {author} {\bibfnamefont {J.}~\bibnamefont {Sun}}, \bibinfo {author} {\bibfnamefont {A.}~\bibnamefont {Ruzsinszky}},\ and\ \bibinfo {author} {\bibfnamefont {J.~P.}\ \bibnamefont {Perdew}},\ }\bibfield  {title} {\bibinfo {title} {Strongly constrained and appropriately normed semilocal density functional},\ }\href@noop {} {\bibfield  {journal} {\bibinfo  {journal} {Physical review letters}\ }\textbf {\bibinfo {volume} {115}},\ \bibinfo {pages} {036402} (\bibinfo {year} {2015})}\BibitemShut {NoStop}%
\bibitem [{\citenamefont {Yang}\ \emph {et~al.}(2020)\citenamefont {Yang}, \citenamefont {Liu}, \citenamefont {Lu},\ and\ \citenamefont {Lin}}]{PhysRevB.102.174109}%
  \BibitemOpen
  \bibfield  {author} {\bibinfo {author} {\bibfnamefont {H.-C.}\ \bibnamefont {Yang}}, \bibinfo {author} {\bibfnamefont {K.}~\bibnamefont {Liu}}, \bibinfo {author} {\bibfnamefont {Z.-Y.}\ \bibnamefont {Lu}},\ and\ \bibinfo {author} {\bibfnamefont {H.-Q.}\ \bibnamefont {Lin}},\ }\bibfield  {title} {\bibinfo {title} {First-principles study of solid hydrogen: Comparison among four exchange-correlation functionals},\ }\href {https://doi.org/10.1103/PhysRevB.102.174109} {\bibfield  {journal} {\bibinfo  {journal} {Phys. Rev. B}\ }\textbf {\bibinfo {volume} {102}},\ \bibinfo {pages} {174109} (\bibinfo {year} {2020})}\BibitemShut {NoStop}%
\bibitem [{\citenamefont {Sch\"ottler}\ and\ \citenamefont {Redmer}(2018)}]{PhysRevLett.120.115703}%
  \BibitemOpen
  \bibfield  {author} {\bibinfo {author} {\bibfnamefont {M.}~\bibnamefont {Sch\"ottler}}\ and\ \bibinfo {author} {\bibfnamefont {R.}~\bibnamefont {Redmer}},\ }\bibfield  {title} {\bibinfo {title} {Ab initio calculation of the miscibility diagram for hydrogen-helium mixtures},\ }\href {https://doi.org/10.1103/PhysRevLett.120.115703} {\bibfield  {journal} {\bibinfo  {journal} {Phys. Rev. Lett.}\ }\textbf {\bibinfo {volume} {120}},\ \bibinfo {pages} {115703} (\bibinfo {year} {2018})}\BibitemShut {NoStop}%
\bibitem [{\citenamefont {Azadi}\ and\ \citenamefont {Ackland}(2017)}]{azadi2017role}%
  \BibitemOpen
  \bibfield  {author} {\bibinfo {author} {\bibfnamefont {S.}~\bibnamefont {Azadi}}\ and\ \bibinfo {author} {\bibfnamefont {G.~J.}\ \bibnamefont {Ackland}},\ }\bibfield  {title} {\bibinfo {title} {The role of van der waals and exchange interactions in high-pressure solid hydrogen},\ }\href@noop {} {\bibfield  {journal} {\bibinfo  {journal} {Physical Chemistry Chemical Physics}\ }\textbf {\bibinfo {volume} {19}},\ \bibinfo {pages} {21829} (\bibinfo {year} {2017})}\BibitemShut {NoStop}%
\bibitem [{\citenamefont {Lu}\ \emph {et~al.}(2019)\citenamefont {Lu}, \citenamefont {Kang}, \citenamefont {Wang}, \citenamefont {Gao},\ and\ \citenamefont {Dai}}]{lu2019towards}%
  \BibitemOpen
  \bibfield  {author} {\bibinfo {author} {\bibfnamefont {B.}~\bibnamefont {Lu}}, \bibinfo {author} {\bibfnamefont {D.}~\bibnamefont {Kang}}, \bibinfo {author} {\bibfnamefont {D.}~\bibnamefont {Wang}}, \bibinfo {author} {\bibfnamefont {T.}~\bibnamefont {Gao}},\ and\ \bibinfo {author} {\bibfnamefont {J.}~\bibnamefont {Dai}},\ }\bibfield  {title} {\bibinfo {title} {Towards the same line of liquid--liquid phase transition of dense hydrogen from various theoretical predictions},\ }\href@noop {} {\bibfield  {journal} {\bibinfo  {journal} {Chinese Physics Letters}\ }\textbf {\bibinfo {volume} {36}},\ \bibinfo {pages} {103102} (\bibinfo {year} {2019})}\BibitemShut {NoStop}%
\bibitem [{\citenamefont {Hinz}\ \emph {et~al.}(2020)\citenamefont {Hinz}, \citenamefont {Karasiev}, \citenamefont {Hu}, \citenamefont {Zaghoo}, \citenamefont {Mej{\'\i}a-Rodr{\'\i}guez}, \citenamefont {Trickey},\ and\ \citenamefont {Calder{\'\i}n}}]{hinz2020fully}%
  \BibitemOpen
  \bibfield  {author} {\bibinfo {author} {\bibfnamefont {J.}~\bibnamefont {Hinz}}, \bibinfo {author} {\bibfnamefont {V.~V.}\ \bibnamefont {Karasiev}}, \bibinfo {author} {\bibfnamefont {S.}~\bibnamefont {Hu}}, \bibinfo {author} {\bibfnamefont {M.}~\bibnamefont {Zaghoo}}, \bibinfo {author} {\bibfnamefont {D.}~\bibnamefont {Mej{\'\i}a-Rodr{\'\i}guez}}, \bibinfo {author} {\bibfnamefont {S.}~\bibnamefont {Trickey}},\ and\ \bibinfo {author} {\bibfnamefont {L.}~\bibnamefont {Calder{\'\i}n}},\ }\bibfield  {title} {\bibinfo {title} {Fully consistent density functional theory determination of the insulator-metal transition boundary in warm dense hydrogen},\ }\href@noop {} {\bibfield  {journal} {\bibinfo  {journal} {Physical Review Research}\ }\textbf {\bibinfo {volume} {2}},\ \bibinfo {pages} {032065} (\bibinfo {year} {2020})}\BibitemShut {NoStop}%
\bibitem [{\citenamefont {Zaghoo}\ \emph {et~al.}(2016)\citenamefont {Zaghoo}, \citenamefont {Salamat},\ and\ \citenamefont {Silvera}}]{zaghoo2016evidence}%
  \BibitemOpen
  \bibfield  {author} {\bibinfo {author} {\bibfnamefont {M.}~\bibnamefont {Zaghoo}}, \bibinfo {author} {\bibfnamefont {A.}~\bibnamefont {Salamat}},\ and\ \bibinfo {author} {\bibfnamefont {I.~F.}\ \bibnamefont {Silvera}},\ }\bibfield  {title} {\bibinfo {title} {Evidence of a first-order phase transition to metallic hydrogen},\ }\href@noop {} {\bibfield  {journal} {\bibinfo  {journal} {Physical Review B}\ }\textbf {\bibinfo {volume} {93}},\ \bibinfo {pages} {155128} (\bibinfo {year} {2016})}\BibitemShut {NoStop}%
\bibitem [{\citenamefont {Zaghoo}\ and\ \citenamefont {Silvera}(2017)}]{zaghoo2017conductivity}%
  \BibitemOpen
  \bibfield  {author} {\bibinfo {author} {\bibfnamefont {M.}~\bibnamefont {Zaghoo}}\ and\ \bibinfo {author} {\bibfnamefont {I.~F.}\ \bibnamefont {Silvera}},\ }\bibfield  {title} {\bibinfo {title} {Conductivity and dissociation in liquid metallic hydrogen and implications for planetary interiors},\ }\href@noop {} {\bibfield  {journal} {\bibinfo  {journal} {Proceedings of the National Academy of Sciences}\ }\textbf {\bibinfo {volume} {114}},\ \bibinfo {pages} {11873} (\bibinfo {year} {2017})}\BibitemShut {NoStop}%
\bibitem [{\citenamefont {Ji}\ \emph {et~al.}(2019)\citenamefont {Ji}, \citenamefont {Li}, \citenamefont {Liu}, \citenamefont {Smith}, \citenamefont {Majumdar}, \citenamefont {Luo}, \citenamefont {Ahuja}, \citenamefont {Shu}, \citenamefont {Wang}, \citenamefont {Sinogeikin} \emph {et~al.}}]{ji2019ultrahigh}%
  \BibitemOpen
  \bibfield  {author} {\bibinfo {author} {\bibfnamefont {C.}~\bibnamefont {Ji}}, \bibinfo {author} {\bibfnamefont {B.}~\bibnamefont {Li}}, \bibinfo {author} {\bibfnamefont {W.}~\bibnamefont {Liu}}, \bibinfo {author} {\bibfnamefont {J.~S.}\ \bibnamefont {Smith}}, \bibinfo {author} {\bibfnamefont {A.}~\bibnamefont {Majumdar}}, \bibinfo {author} {\bibfnamefont {W.}~\bibnamefont {Luo}}, \bibinfo {author} {\bibfnamefont {R.}~\bibnamefont {Ahuja}}, \bibinfo {author} {\bibfnamefont {J.}~\bibnamefont {Shu}}, \bibinfo {author} {\bibfnamefont {J.}~\bibnamefont {Wang}}, \bibinfo {author} {\bibfnamefont {S.}~\bibnamefont {Sinogeikin}}, \emph {et~al.},\ }\bibfield  {title} {\bibinfo {title} {Ultrahigh-pressure isostructural electronic transitions in hydrogen},\ }\href@noop {} {\bibfield  {journal} {\bibinfo  {journal} {Nature}\ }\textbf {\bibinfo {volume} {573}},\ \bibinfo {pages} {558} (\bibinfo {year} {2019})}\BibitemShut {NoStop}%
\bibitem [{\citenamefont {Morales}\ \emph {et~al.}(2013)\citenamefont {Morales}, \citenamefont {{McMahon}}, \citenamefont {Pierleoni},\ and\ \citenamefont {Ceperley}}]{morales_nuclear_2013}%
  \BibitemOpen
  \bibfield  {author} {\bibinfo {author} {\bibfnamefont {M.~A.}\ \bibnamefont {Morales}}, \bibinfo {author} {\bibfnamefont {J.~M.}\ \bibnamefont {{McMahon}}}, \bibinfo {author} {\bibfnamefont {C.}~\bibnamefont {Pierleoni}},\ and\ \bibinfo {author} {\bibfnamefont {D.~M.}\ \bibnamefont {Ceperley}},\ }\bibfield  {title} {\bibinfo {title} {Nuclear quantum effects and nonlocal exchange-correlation functionals applied to liquid hydrogen at high pressure},\ }\href {https://doi.org/10.1103/PhysRevLett.110.065702} {\bibfield  {journal} {\bibinfo  {journal} {Physical Review Letters}\ }\textbf {\bibinfo {volume} {110}},\ \bibinfo {pages} {065702} (\bibinfo {year} {2013})}\BibitemShut {NoStop}%
\bibitem [{\citenamefont {Geng}\ \emph {et~al.}(2019)\citenamefont {Geng}, \citenamefont {Wu}, \citenamefont {Marqu\'es},\ and\ \citenamefont {Ackland}}]{Geng2019}%
  \BibitemOpen
  \bibfield  {author} {\bibinfo {author} {\bibfnamefont {H.~Y.}\ \bibnamefont {Geng}}, \bibinfo {author} {\bibfnamefont {Q.}~\bibnamefont {Wu}}, \bibinfo {author} {\bibfnamefont {M.}~\bibnamefont {Marqu\'es}},\ and\ \bibinfo {author} {\bibfnamefont {G.~J.}\ \bibnamefont {Ackland}},\ }\bibfield  {title} {\bibinfo {title} {Thermodynamic anomalies and three distinct liquid-liquid transitions in warm dense liquid hydrogen},\ }\href {https://doi.org/10.1103/PhysRevB.100.134109} {\bibfield  {journal} {\bibinfo  {journal} {Phys. Rev. B}\ }\textbf {\bibinfo {volume} {100}},\ \bibinfo {pages} {134109} (\bibinfo {year} {2019})}\BibitemShut {NoStop}%
\bibitem [{\citenamefont {Clay~III}\ \emph {et~al.}(2016)\citenamefont {Clay~III}, \citenamefont {Holzmann}, \citenamefont {Ceperley},\ and\ \citenamefont {Morales}}]{clay2016benchmarking}%
  \BibitemOpen
  \bibfield  {author} {\bibinfo {author} {\bibfnamefont {R.~C.}\ \bibnamefont {Clay~III}}, \bibinfo {author} {\bibfnamefont {M.}~\bibnamefont {Holzmann}}, \bibinfo {author} {\bibfnamefont {D.~M.}\ \bibnamefont {Ceperley}},\ and\ \bibinfo {author} {\bibfnamefont {M.~A.}\ \bibnamefont {Morales}},\ }\bibfield  {title} {\bibinfo {title} {Benchmarking density functionals for hydrogen-helium mixtures with quantum monte carlo: Energetics, pressures, and forces},\ }\href@noop {} {\bibfield  {journal} {\bibinfo  {journal} {Physical Review B}\ }\textbf {\bibinfo {volume} {93}},\ \bibinfo {pages} {035121} (\bibinfo {year} {2016})}\BibitemShut {NoStop}%
\bibitem [{\citenamefont {Cohen}\ \emph {et~al.}(2011)\citenamefont {Cohen}, \citenamefont {Mori-S{\'a}nchez},\ and\ \citenamefont {Yang}}]{cohen2011challenges}%
  \BibitemOpen
  \bibfield  {author} {\bibinfo {author} {\bibfnamefont {A.~J.}\ \bibnamefont {Cohen}}, \bibinfo {author} {\bibfnamefont {P.}~\bibnamefont {Mori-S{\'a}nchez}},\ and\ \bibinfo {author} {\bibfnamefont {W.}~\bibnamefont {Yang}},\ }\bibfield  {title} {\bibinfo {title} {Challenges for density functional theory},\ }\href@noop {} {\bibfield  {journal} {\bibinfo  {journal} {Chemical Reviews}\ }\textbf {\bibinfo {volume} {112}},\ \bibinfo {pages} {289} (\bibinfo {year} {2011})}\BibitemShut {NoStop}%
\bibitem [{\citenamefont {Wagner}\ \emph {et~al.}(2012)\citenamefont {Wagner}, \citenamefont {Stoudenmire}, \citenamefont {Burke},\ and\ \citenamefont {White}}]{wagner2012reference}%
  \BibitemOpen
  \bibfield  {author} {\bibinfo {author} {\bibfnamefont {L.~O.}\ \bibnamefont {Wagner}}, \bibinfo {author} {\bibfnamefont {E.}~\bibnamefont {Stoudenmire}}, \bibinfo {author} {\bibfnamefont {K.}~\bibnamefont {Burke}},\ and\ \bibinfo {author} {\bibfnamefont {S.~R.}\ \bibnamefont {White}},\ }\bibfield  {title} {\bibinfo {title} {Reference electronic structure calculations in one dimension},\ }\href@noop {} {\bibfield  {journal} {\bibinfo  {journal} {Physical Chemistry Chemical Physics}\ }\textbf {\bibinfo {volume} {14}},\ \bibinfo {pages} {8581} (\bibinfo {year} {2012})}\BibitemShut {NoStop}%
\bibitem [{\citenamefont {Motta}\ \emph {et~al.}(2017)\citenamefont {Motta}, \citenamefont {Ceperley}, \citenamefont {Chan}, \citenamefont {Gomez}, \citenamefont {Gull}, \citenamefont {Guo}, \citenamefont {Jim\'enez-Hoyos}, \citenamefont {Lan}, \citenamefont {Li}, \citenamefont {Ma}, \citenamefont {Millis}, \citenamefont {Prokof'ev}, \citenamefont {Ray}, \citenamefont {Scuseria}, \citenamefont {Sorella}, \citenamefont {Stoudenmire}, \citenamefont {Sun}, \citenamefont {Tupitsyn}, \citenamefont {White}, \citenamefont {Zgid},\ and\ \citenamefont {Zhang}}]{Motta_benchmark}%
  \BibitemOpen
  \bibfield  {author} {\bibinfo {author} {\bibfnamefont {M.}~\bibnamefont {Motta}}, \bibinfo {author} {\bibfnamefont {D.~M.}\ \bibnamefont {Ceperley}}, \bibinfo {author} {\bibfnamefont {G.~K.-L.}\ \bibnamefont {Chan}}, \bibinfo {author} {\bibfnamefont {J.~A.}\ \bibnamefont {Gomez}}, \bibinfo {author} {\bibfnamefont {E.}~\bibnamefont {Gull}}, \bibinfo {author} {\bibfnamefont {S.}~\bibnamefont {Guo}}, \bibinfo {author} {\bibfnamefont {C.~A.}\ \bibnamefont {Jim\'enez-Hoyos}}, \bibinfo {author} {\bibfnamefont {T.~N.}\ \bibnamefont {Lan}}, \bibinfo {author} {\bibfnamefont {J.}~\bibnamefont {Li}}, \bibinfo {author} {\bibfnamefont {F.}~\bibnamefont {Ma}}, \bibinfo {author} {\bibfnamefont {A.~J.}\ \bibnamefont {Millis}}, \bibinfo {author} {\bibfnamefont {N.~V.}\ \bibnamefont {Prokof'ev}}, \bibinfo {author} {\bibfnamefont {U.}~\bibnamefont {Ray}}, \bibinfo {author} {\bibfnamefont {G.~E.}\ \bibnamefont {Scuseria}}, \bibinfo {author} {\bibfnamefont {S.}~\bibnamefont {Sorella}}, \bibinfo {author} {\bibfnamefont {E.~M.}\
  \bibnamefont {Stoudenmire}}, \bibinfo {author} {\bibfnamefont {Q.}~\bibnamefont {Sun}}, \bibinfo {author} {\bibfnamefont {I.~S.}\ \bibnamefont {Tupitsyn}}, \bibinfo {author} {\bibfnamefont {S.~R.}\ \bibnamefont {White}}, \bibinfo {author} {\bibfnamefont {D.}~\bibnamefont {Zgid}},\ and\ \bibinfo {author} {\bibfnamefont {S.}~\bibnamefont {Zhang}} (\bibinfo {collaboration} {Simons Collaboration on the Many-Electron Problem}),\ }\bibfield  {title} {\bibinfo {title} {Towards the solution of the many-electron problem in real materials: Equation of state of the hydrogen chain with state-of-the-art many-body methods},\ }\href {https://doi.org/10.1103/PhysRevX.7.031059} {\bibfield  {journal} {\bibinfo  {journal} {Phys. Rev. X}\ }\textbf {\bibinfo {volume} {7}},\ \bibinfo {pages} {031059} (\bibinfo {year} {2017})}\BibitemShut {NoStop}%
\bibitem [{\citenamefont {Dubeckyy}\ \emph {et~al.}(2016)\citenamefont {Dubeckyy}, \citenamefont {Mitas},\ and\ \citenamefont {Jurecka}}]{dubecky2016noncovalent}%
  \BibitemOpen
  \bibfield  {author} {\bibinfo {author} {\bibfnamefont {M.}~\bibnamefont {Dubeckyy}}, \bibinfo {author} {\bibfnamefont {L.}~\bibnamefont {Mitas}},\ and\ \bibinfo {author} {\bibfnamefont {P.}~\bibnamefont {Jurecka}},\ }\bibfield  {title} {\bibinfo {title} {Noncovalent interactions by quantum monte carlo},\ }\href@noop {} {\bibfield  {journal} {\bibinfo  {journal} {Chemical reviews}\ }\textbf {\bibinfo {volume} {116}},\ \bibinfo {pages} {5188} (\bibinfo {year} {2016})}\BibitemShut {NoStop}%
\bibitem [{\citenamefont {Gillan}\ \emph {et~al.}(2016)\citenamefont {Gillan}, \citenamefont {Alf{\`e}},\ and\ \citenamefont {Michaelides}}]{doi:10.1063/1.4944633}%
  \BibitemOpen
  \bibfield  {author} {\bibinfo {author} {\bibfnamefont {M.~J.}\ \bibnamefont {Gillan}}, \bibinfo {author} {\bibfnamefont {D.}~\bibnamefont {Alf{\`e}}},\ and\ \bibinfo {author} {\bibfnamefont {A.}~\bibnamefont {Michaelides}},\ }\bibfield  {title} {\bibinfo {title} {Perspective: How good is dft for water?},\ }\href {https://doi.org/10.1063/1.4944633} {\bibfield  {journal} {\bibinfo  {journal} {The Journal of Chemical Physics}\ }\textbf {\bibinfo {volume} {144}},\ \bibinfo {pages} {130901} (\bibinfo {year} {2016})}\BibitemShut {NoStop}%
\bibitem [{\citenamefont {Foulkes}\ \emph {et~al.}(2001)\citenamefont {Foulkes}, \citenamefont {Mitas}, \citenamefont {Needs},\ and\ \citenamefont {Rajagopal}}]{foulkes_quantum_2001}%
  \BibitemOpen
  \bibfield  {author} {\bibinfo {author} {\bibfnamefont {W.~M.~C.}\ \bibnamefont {Foulkes}}, \bibinfo {author} {\bibfnamefont {L.}~\bibnamefont {Mitas}}, \bibinfo {author} {\bibfnamefont {R.~J.}\ \bibnamefont {Needs}},\ and\ \bibinfo {author} {\bibfnamefont {G.}~\bibnamefont {Rajagopal}},\ }\bibfield  {title} {\bibinfo {title} {Quantum monte carlo simulations of solids},\ }\href {https://doi.org/10.1103/RevModPhys.73.33} {\bibfield  {journal} {\bibinfo  {journal} {Reviews of Modern Physics}\ }\textbf {\bibinfo {volume} {73}},\ \bibinfo {pages} {33} (\bibinfo {year} {2001})}\BibitemShut {NoStop}%
\bibitem [{\citenamefont {Nakano}\ \emph {et~al.}(2020)\citenamefont {Nakano}, \citenamefont {Attaccalite}, \citenamefont {Barborini}, \citenamefont {Capriotti}, \citenamefont {Casula}, \citenamefont {Coccia}, \citenamefont {Dagrada}, \citenamefont {Genovese}, \citenamefont {Luo}, \citenamefont {Mazzola}, \citenamefont {Zen},\ and\ \citenamefont {Sorella}}]{nakano2020turborvb}%
  \BibitemOpen
  \bibfield  {author} {\bibinfo {author} {\bibfnamefont {K.}~\bibnamefont {Nakano}}, \bibinfo {author} {\bibfnamefont {C.}~\bibnamefont {Attaccalite}}, \bibinfo {author} {\bibfnamefont {M.}~\bibnamefont {Barborini}}, \bibinfo {author} {\bibfnamefont {L.}~\bibnamefont {Capriotti}}, \bibinfo {author} {\bibfnamefont {M.}~\bibnamefont {Casula}}, \bibinfo {author} {\bibfnamefont {E.}~\bibnamefont {Coccia}}, \bibinfo {author} {\bibfnamefont {M.}~\bibnamefont {Dagrada}}, \bibinfo {author} {\bibfnamefont {C.}~\bibnamefont {Genovese}}, \bibinfo {author} {\bibfnamefont {Y.}~\bibnamefont {Luo}}, \bibinfo {author} {\bibfnamefont {G.}~\bibnamefont {Mazzola}}, \bibinfo {author} {\bibfnamefont {A.}~\bibnamefont {Zen}},\ and\ \bibinfo {author} {\bibfnamefont {S.}~\bibnamefont {Sorella}},\ }\bibfield  {title} {\bibinfo {title} {Turborvb: A many-body toolkit for ab initio electronic simulations by quantum monte carlo},\ }\href {https://doi.org/10.1063/5.0005037} {\bibfield  {journal} {\bibinfo  {journal} {J. Chem. Phys.}\
  }\textbf {\bibinfo {volume} {152}},\ \bibinfo {pages} {204121} (\bibinfo {year} {2020})}\BibitemShut {NoStop}%
\bibitem [{\citenamefont {Szabo}\ and\ \citenamefont {Ostlund}(2012)}]{szabo2012modern}%
  \BibitemOpen
  \bibfield  {author} {\bibinfo {author} {\bibfnamefont {A.}~\bibnamefont {Szabo}}\ and\ \bibinfo {author} {\bibfnamefont {N.~S.}\ \bibnamefont {Ostlund}},\ }\href@noop {} {\emph {\bibinfo {title} {Modern quantum chemistry: introduction to advanced electronic structure theory}}}\ (\bibinfo  {publisher} {Courier Corporation},\ \bibinfo {year} {2012})\BibitemShut {NoStop}%
\bibitem [{\citenamefont {Riley}\ \emph {et~al.}(2010)\citenamefont {Riley}, \citenamefont {Pitonak}, \citenamefont {Jurecka},\ and\ \citenamefont {Hobza}}]{doi:10.1021/cr1000173}%
  \BibitemOpen
  \bibfield  {author} {\bibinfo {author} {\bibfnamefont {K.~E.}\ \bibnamefont {Riley}}, \bibinfo {author} {\bibfnamefont {M.}~\bibnamefont {Pitonak}}, \bibinfo {author} {\bibfnamefont {P.}~\bibnamefont {Jurecka}},\ and\ \bibinfo {author} {\bibfnamefont {P.}~\bibnamefont {Hobza}},\ }\bibfield  {title} {\bibinfo {title} {Stabilization and structure calculations for noncovalent interactions in extended molecular systems based on wave function and density functional theories},\ }\href {https://doi.org/10.1021/cr1000173} {\bibfield  {journal} {\bibinfo  {journal} {Chemical Reviews}\ }\textbf {\bibinfo {volume} {110}},\ \bibinfo {pages} {5023} (\bibinfo {year} {2010})},\ \bibinfo {note} {pMID: 20486691},\ \Eprint {https://arxiv.org/abs/http://dx.doi.org/10.1021/cr1000173} {http://dx.doi.org/10.1021/cr1000173} \BibitemShut {NoStop}%
\bibitem [{\citenamefont {Casula}\ \emph {et~al.}(2005)\citenamefont {Casula}, \citenamefont {Filippi},\ and\ \citenamefont {Sorella}}]{casula2005lrdmc}%
  \BibitemOpen
  \bibfield  {author} {\bibinfo {author} {\bibfnamefont {M.}~\bibnamefont {Casula}}, \bibinfo {author} {\bibfnamefont {C.}~\bibnamefont {Filippi}},\ and\ \bibinfo {author} {\bibfnamefont {S.}~\bibnamefont {Sorella}},\ }\bibfield  {title} {\bibinfo {title} {{Diffusion Monte Carlo method with lattice regularization}},\ }\href@noop {} {\bibfield  {journal} {\bibinfo  {journal} {Phys. Rev. Lett.}\ }\textbf {\bibinfo {volume} {95}},\ \bibinfo {pages} {1} (\bibinfo {year} {2005})}\BibitemShut {NoStop}%
\bibitem [{\citenamefont {Nakano}\ \emph {et~al.}(2023)\citenamefont {Nakano}, \citenamefont {Kohulák}, \citenamefont {Raghav}, \citenamefont {Casula},\ and\ \citenamefont {Sorella}}]{nakano2023turbogenius}%
  \BibitemOpen
  \bibfield  {author} {\bibinfo {author} {\bibfnamefont {K.}~\bibnamefont {Nakano}}, \bibinfo {author} {\bibfnamefont {O.}~\bibnamefont {Kohulák}}, \bibinfo {author} {\bibfnamefont {A.}~\bibnamefont {Raghav}}, \bibinfo {author} {\bibfnamefont {M.}~\bibnamefont {Casula}},\ and\ \bibinfo {author} {\bibfnamefont {S.}~\bibnamefont {Sorella}},\ }\bibfield  {title} {\bibinfo {title} {{TurboGenius: Python suite for high-throughput calculations of ab initio quantum Monte Carlo methods}},\ }\href {https://doi.org/10.1063/5.0179003} {\bibfield  {journal} {\bibinfo  {journal} {J. Chem. Phys.}\ }\textbf {\bibinfo {volume} {159}},\ \bibinfo {pages} {224801} (\bibinfo {year} {2023})}\BibitemShut {NoStop}%
\bibitem [{\citenamefont {Mazzola}\ \emph {et~al.}(2012)\citenamefont {Mazzola}, \citenamefont {Zen},\ and\ \citenamefont {Sorella}}]{mazzola_finite-temperature_2012}%
  \BibitemOpen
  \bibfield  {author} {\bibinfo {author} {\bibfnamefont {G.}~\bibnamefont {Mazzola}}, \bibinfo {author} {\bibfnamefont {A.}~\bibnamefont {Zen}},\ and\ \bibinfo {author} {\bibfnamefont {S.}~\bibnamefont {Sorella}},\ }\bibfield  {title} {\bibinfo {title} {Finite-temperature electronic simulations without the born-oppenheimer constraint},\ }\href {https://doi.org/10.1063/1.4755992} {\bibfield  {journal} {\bibinfo  {journal} {The Journal of Chemical Physics}\ }\textbf {\bibinfo {volume} {137}},\ \bibinfo {pages} {134112} (\bibinfo {year} {2012})}\BibitemShut {NoStop}%
\bibitem [{\citenamefont {Hermann}\ \emph {et~al.}(2020)\citenamefont {Hermann}, \citenamefont {Schätzle},\ and\ \citenamefont {Noé}}]{Hermann_2020}%
  \BibitemOpen
  \bibfield  {author} {\bibinfo {author} {\bibfnamefont {J.}~\bibnamefont {Hermann}}, \bibinfo {author} {\bibfnamefont {Z.}~\bibnamefont {Schätzle}},\ and\ \bibinfo {author} {\bibfnamefont {F.}~\bibnamefont {Noé}},\ }\bibfield  {title} {\bibinfo {title} {Deep-neural-network solution of the electronic schrödinger equation},\ }\href {https://doi.org/10.1038/s41557-020-0544-y} {\bibfield  {journal} {\bibinfo  {journal} {Nature Chemistry}\ }\textbf {\bibinfo {volume} {12}},\ \bibinfo {pages} {891–897} (\bibinfo {year} {2020})}\BibitemShut {NoStop}%
\bibitem [{\citenamefont {Pescia}\ \emph {et~al.}(2024)\citenamefont {Pescia}, \citenamefont {Nys}, \citenamefont {Kim}, \citenamefont {Lovato},\ and\ \citenamefont {Carleo}}]{PhysRevB.110.035108}%
  \BibitemOpen
  \bibfield  {author} {\bibinfo {author} {\bibfnamefont {G.}~\bibnamefont {Pescia}}, \bibinfo {author} {\bibfnamefont {J.}~\bibnamefont {Nys}}, \bibinfo {author} {\bibfnamefont {J.}~\bibnamefont {Kim}}, \bibinfo {author} {\bibfnamefont {A.}~\bibnamefont {Lovato}},\ and\ \bibinfo {author} {\bibfnamefont {G.}~\bibnamefont {Carleo}},\ }\bibfield  {title} {\bibinfo {title} {Message-passing neural quantum states for the homogeneous electron gas},\ }\href {https://doi.org/10.1103/PhysRevB.110.035108} {\bibfield  {journal} {\bibinfo  {journal} {Phys. Rev. B}\ }\textbf {\bibinfo {volume} {110}},\ \bibinfo {pages} {035108} (\bibinfo {year} {2024})}\BibitemShut {NoStop}%
\bibitem [{\citenamefont {Pfau}\ \emph {et~al.}(2020)\citenamefont {Pfau}, \citenamefont {Spencer}, \citenamefont {Matthews},\ and\ \citenamefont {Foulkes}}]{PhysRevResearch.2.033429}%
  \BibitemOpen
  \bibfield  {author} {\bibinfo {author} {\bibfnamefont {D.}~\bibnamefont {Pfau}}, \bibinfo {author} {\bibfnamefont {J.~S.}\ \bibnamefont {Spencer}}, \bibinfo {author} {\bibfnamefont {A.~G. D.~G.}\ \bibnamefont {Matthews}},\ and\ \bibinfo {author} {\bibfnamefont {W.~M.~C.}\ \bibnamefont {Foulkes}},\ }\bibfield  {title} {\bibinfo {title} {Ab initio solution of the many-electron schr\"odinger equation with deep neural networks},\ }\href {https://doi.org/10.1103/PhysRevResearch.2.033429} {\bibfield  {journal} {\bibinfo  {journal} {Phys. Rev. Res.}\ }\textbf {\bibinfo {volume} {2}},\ \bibinfo {pages} {033429} (\bibinfo {year} {2020})}\BibitemShut {NoStop}%
\bibitem [{\citenamefont {Sorella}(1998)}]{sorella1998sr}%
  \BibitemOpen
  \bibfield  {author} {\bibinfo {author} {\bibfnamefont {S.}~\bibnamefont {Sorella}},\ }\bibfield  {title} {\bibinfo {title} {Green function monte carlo with stochastic reconfiguration},\ }\href@noop {} {\bibfield  {journal} {\bibinfo  {journal} {Phys. Rev. Lett.}\ }\textbf {\bibinfo {volume} {80}},\ \bibinfo {pages} {4558} (\bibinfo {year} {1998})}\BibitemShut {NoStop}%
\bibitem [{\citenamefont {Sorella}\ \emph {et~al.}(2007)\citenamefont {Sorella}, \citenamefont {Casula},\ and\ \citenamefont {Rocca}}]{sorella2007sr}%
  \BibitemOpen
  \bibfield  {author} {\bibinfo {author} {\bibfnamefont {S.}~\bibnamefont {Sorella}}, \bibinfo {author} {\bibfnamefont {M.}~\bibnamefont {Casula}},\ and\ \bibinfo {author} {\bibfnamefont {D.}~\bibnamefont {Rocca}},\ }\bibfield  {title} {\bibinfo {title} {Weak binding between two aromatic rings: Feeling the van der waals attraction by quantum monte carlo methods},\ }\href@noop {} {\bibfield  {journal} {\bibinfo  {journal} {J. Chem. Phys.}\ }\textbf {\bibinfo {volume} {127}},\ \bibinfo {pages} {014105} (\bibinfo {year} {2007})}\BibitemShut {NoStop}%
\bibitem [{\citenamefont {Monacelli}\ \emph {et~al.}(2023)\citenamefont {Monacelli}, \citenamefont {Casula}, \citenamefont {Nakano}, \citenamefont {Sorella},\ and\ \citenamefont {Mauri}}]{Lorenzo.hydrogen.2023}%
  \BibitemOpen
  \bibfield  {author} {\bibinfo {author} {\bibfnamefont {L.}~\bibnamefont {Monacelli}}, \bibinfo {author} {\bibfnamefont {M.}~\bibnamefont {Casula}}, \bibinfo {author} {\bibfnamefont {K.}~\bibnamefont {Nakano}}, \bibinfo {author} {\bibfnamefont {S.}~\bibnamefont {Sorella}},\ and\ \bibinfo {author} {\bibfnamefont {F.}~\bibnamefont {Mauri}},\ }\bibfield  {title} {\bibinfo {title} {Quantum phase diagram of high-pressure hydrogen},\ }\href {https://doi.org/10.1038/s41567-023-01960-5} {\bibfield  {journal} {\bibinfo  {journal} {Nat. Phys.}\ }\textbf {\bibinfo {volume} {19}},\ \bibinfo {pages} {845} (\bibinfo {year} {2023})}\BibitemShut {NoStop}%
\bibitem [{\citenamefont {Raghav}\ \emph {et~al.}(2023)\citenamefont {Raghav}, \citenamefont {Maezono}, \citenamefont {Hongo}, \citenamefont {Sorella},\ and\ \citenamefont {Nakano}}]{Raghav2023.g2set.benchmark}%
  \BibitemOpen
  \bibfield  {author} {\bibinfo {author} {\bibfnamefont {A.}~\bibnamefont {Raghav}}, \bibinfo {author} {\bibfnamefont {R.}~\bibnamefont {Maezono}}, \bibinfo {author} {\bibfnamefont {K.}~\bibnamefont {Hongo}}, \bibinfo {author} {\bibfnamefont {S.}~\bibnamefont {Sorella}},\ and\ \bibinfo {author} {\bibfnamefont {K.}~\bibnamefont {Nakano}},\ }\bibfield  {title} {\bibinfo {title} {Toward chemical accuracy using the jastrow correlated antisymmetrized geminal power ansatz},\ }\href {https://doi.org/10.1021/acs.jctc.2c01141} {\bibfield  {journal} {\bibinfo  {journal} {J. Chem. Theory Comput.}\ }\textbf {\bibinfo {volume} {19}},\ \bibinfo {pages} {2222} (\bibinfo {year} {2023})}\BibitemShut {NoStop}%
\bibitem [{\citenamefont {Becca}\ and\ \citenamefont {Sorella}(2017)}]{Becca2017}%
  \BibitemOpen
  \bibfield  {author} {\bibinfo {author} {\bibfnamefont {F.}~\bibnamefont {Becca}}\ and\ \bibinfo {author} {\bibfnamefont {S.}~\bibnamefont {Sorella}},\ }\href {https://doi.org/10.1017/9781316417041} {\emph {\bibinfo {title} {Quantum Monte Carlo Approaches for Correlated Systems}}},\ \bibinfo {edition} {1st}\ ed.\ (\bibinfo  {publisher} {Cambridge University Press},\ \bibinfo {year} {2017})\BibitemShut {NoStop}%
\bibitem [{\citenamefont {Mazzola}\ \emph {et~al.}(2014)\citenamefont {Mazzola}, \citenamefont {Yunoki},\ and\ \citenamefont {Sorella}}]{mazzola_unexpectedly_2014}%
  \BibitemOpen
  \bibfield  {author} {\bibinfo {author} {\bibfnamefont {G.}~\bibnamefont {Mazzola}}, \bibinfo {author} {\bibfnamefont {S.}~\bibnamefont {Yunoki}},\ and\ \bibinfo {author} {\bibfnamefont {S.}~\bibnamefont {Sorella}},\ }\bibfield  {title} {\bibinfo {title} {Unexpectedly high pressure for molecular dissociation in liquid hydrogen by electronic simulation},\ }\href {https://doi.org/10.1038/ncomms4487} {\bibfield  {journal} {\bibinfo  {journal} {Nature Communications}\ }\textbf {\bibinfo {volume} {5}},\ \bibinfo {pages} {3487} (\bibinfo {year} {2014})}\BibitemShut {NoStop}%
\bibitem [{\citenamefont {Zen}\ \emph {et~al.}(2015)\citenamefont {Zen}, \citenamefont {Luo}, \citenamefont {Mazzola}, \citenamefont {Guidoni},\ and\ \citenamefont {Sorella}}]{zen2015ab}%
  \BibitemOpen
  \bibfield  {author} {\bibinfo {author} {\bibfnamefont {A.}~\bibnamefont {Zen}}, \bibinfo {author} {\bibfnamefont {Y.}~\bibnamefont {Luo}}, \bibinfo {author} {\bibfnamefont {G.}~\bibnamefont {Mazzola}}, \bibinfo {author} {\bibfnamefont {L.}~\bibnamefont {Guidoni}},\ and\ \bibinfo {author} {\bibfnamefont {S.}~\bibnamefont {Sorella}},\ }\bibfield  {title} {\bibinfo {title} {Ab initio molecular dynamics simulation of liquid water by quantum monte carlo},\ }\href@noop {} {\bibfield  {journal} {\bibinfo  {journal} {The Journal of chemical physics}\ }\textbf {\bibinfo {volume} {142}},\ \bibinfo {pages} {144111} (\bibinfo {year} {2015})}\BibitemShut {NoStop}%
\bibitem [{\citenamefont {Cheng}\ \emph {et~al.}(2020)\citenamefont {Cheng}, \citenamefont {Mazzola}, \citenamefont {Pickard},\ and\ \citenamefont {Ceriotti}}]{cheng2020evidence}%
  \BibitemOpen
  \bibfield  {author} {\bibinfo {author} {\bibfnamefont {B.}~\bibnamefont {Cheng}}, \bibinfo {author} {\bibfnamefont {G.}~\bibnamefont {Mazzola}}, \bibinfo {author} {\bibfnamefont {C.~J.}\ \bibnamefont {Pickard}},\ and\ \bibinfo {author} {\bibfnamefont {M.}~\bibnamefont {Ceriotti}},\ }\bibfield  {title} {\bibinfo {title} {Evidence for supercritical behaviour of high-pressure liquid hydrogen},\ }\href@noop {} {\bibfield  {journal} {\bibinfo  {journal} {Nature}\ }\textbf {\bibinfo {volume} {585}},\ \bibinfo {pages} {217} (\bibinfo {year} {2020})}\BibitemShut {NoStop}%
\bibitem [{\citenamefont {Tirelli}\ \emph {et~al.}(2022)\citenamefont {Tirelli}, \citenamefont {Tenti}, \citenamefont {Nakano},\ and\ \citenamefont {Sorella}}]{tirelli2022high}%
  \BibitemOpen
  \bibfield  {author} {\bibinfo {author} {\bibfnamefont {A.}~\bibnamefont {Tirelli}}, \bibinfo {author} {\bibfnamefont {G.}~\bibnamefont {Tenti}}, \bibinfo {author} {\bibfnamefont {K.}~\bibnamefont {Nakano}},\ and\ \bibinfo {author} {\bibfnamefont {S.}~\bibnamefont {Sorella}},\ }\bibfield  {title} {\bibinfo {title} {High-pressure hydrogen by machine learning and quantum monte carlo},\ }\href@noop {} {\bibfield  {journal} {\bibinfo  {journal} {Physical Review B}\ }\textbf {\bibinfo {volume} {106}},\ \bibinfo {pages} {L041105} (\bibinfo {year} {2022})}\BibitemShut {NoStop}%
\bibitem [{\citenamefont {Niu}\ \emph {et~al.}(2023)\citenamefont {Niu}, \citenamefont {Yang}, \citenamefont {Jensen}, \citenamefont {Holzmann}, \citenamefont {Pierleoni},\ and\ \citenamefont {Ceperley}}]{PhysRevLett.130.076102}%
  \BibitemOpen
  \bibfield  {author} {\bibinfo {author} {\bibfnamefont {H.}~\bibnamefont {Niu}}, \bibinfo {author} {\bibfnamefont {Y.}~\bibnamefont {Yang}}, \bibinfo {author} {\bibfnamefont {S.}~\bibnamefont {Jensen}}, \bibinfo {author} {\bibfnamefont {M.}~\bibnamefont {Holzmann}}, \bibinfo {author} {\bibfnamefont {C.}~\bibnamefont {Pierleoni}},\ and\ \bibinfo {author} {\bibfnamefont {D.~M.}\ \bibnamefont {Ceperley}},\ }\bibfield  {title} {\bibinfo {title} {Stable solid molecular hydrogen above 900 k from a machine-learned potential trained with diffusion quantum monte carlo},\ }\href {https://doi.org/10.1103/PhysRevLett.130.076102} {\bibfield  {journal} {\bibinfo  {journal} {Phys. Rev. Lett.}\ }\textbf {\bibinfo {volume} {130}},\ \bibinfo {pages} {076102} (\bibinfo {year} {2023})}\BibitemShut {NoStop}%
\bibitem [{\citenamefont {Goswami}\ \emph {et~al.}(2024)\citenamefont {Goswami}, \citenamefont {Jensen}, \citenamefont {Yang}, \citenamefont {Holzmann}, \citenamefont {Pierleoni},\ and\ \citenamefont {Ceperley}}]{goswami2024hightemperaturemeltingdense}%
  \BibitemOpen
  \bibfield  {author} {\bibinfo {author} {\bibfnamefont {S.}~\bibnamefont {Goswami}}, \bibinfo {author} {\bibfnamefont {S.}~\bibnamefont {Jensen}}, \bibinfo {author} {\bibfnamefont {Y.}~\bibnamefont {Yang}}, \bibinfo {author} {\bibfnamefont {M.}~\bibnamefont {Holzmann}}, \bibinfo {author} {\bibfnamefont {C.}~\bibnamefont {Pierleoni}},\ and\ \bibinfo {author} {\bibfnamefont {D.~M.}\ \bibnamefont {Ceperley}},\ }\href {https://arxiv.org/abs/2411.15665} {\bibinfo {title} {High temperature melting of dense molecular hydrogen from machine-learning interatomic potentials trained on quantum monte carlo}} (\bibinfo {year} {2024}),\ \Eprint {https://arxiv.org/abs/2411.15665} {arXiv:2411.15665 [physics.chem-ph]} \BibitemShut {NoStop}%
\bibitem [{\citenamefont {Nakano}\ \emph {et~al.}(2024)\citenamefont {Nakano}, \citenamefont {Casula},\ and\ \citenamefont {Tenti}}]{nakano2024efficient}%
  \BibitemOpen
  \bibfield  {author} {\bibinfo {author} {\bibfnamefont {K.}~\bibnamefont {Nakano}}, \bibinfo {author} {\bibfnamefont {M.}~\bibnamefont {Casula}},\ and\ \bibinfo {author} {\bibfnamefont {G.}~\bibnamefont {Tenti}},\ }\bibfield  {title} {\bibinfo {title} {Efficient calculation of unbiased atomic forces in ab initio variational monte carlo},\ }\href@noop {} {\bibfield  {journal} {\bibinfo  {journal} {Physical Review B}\ }\textbf {\bibinfo {volume} {109}},\ \bibinfo {pages} {205151} (\bibinfo {year} {2024})}\BibitemShut {NoStop}%
\bibitem [{\citenamefont {Pierleoni}\ \emph {et~al.}(2004)\citenamefont {Pierleoni}, \citenamefont {Ceperley},\ and\ \citenamefont {Holzmann}}]{pierleoni_coupled_2004}%
  \BibitemOpen
  \bibfield  {author} {\bibinfo {author} {\bibfnamefont {C.}~\bibnamefont {Pierleoni}}, \bibinfo {author} {\bibfnamefont {D.~M.}\ \bibnamefont {Ceperley}},\ and\ \bibinfo {author} {\bibfnamefont {M.}~\bibnamefont {Holzmann}},\ }\bibfield  {title} {\bibinfo {title} {Coupled electron-ion monte carlo calculations of dense metallic hydrogen},\ }\href {https://doi.org/10.1103/PhysRevLett.93.146402} {\bibfield  {journal} {\bibinfo  {journal} {Physical Review Letters}\ }\textbf {\bibinfo {volume} {93}},\ \bibinfo {pages} {146402} (\bibinfo {year} {2004})}\BibitemShut {NoStop}%
\bibitem [{\citenamefont {Xie}\ \emph {et~al.}(2023)\citenamefont {Xie}, \citenamefont {Li}, \citenamefont {Wang}, \citenamefont {Zhang},\ and\ \citenamefont {Wang}}]{PhysRevLett.131.126501}%
  \BibitemOpen
  \bibfield  {author} {\bibinfo {author} {\bibfnamefont {H.}~\bibnamefont {Xie}}, \bibinfo {author} {\bibfnamefont {Z.-H.}\ \bibnamefont {Li}}, \bibinfo {author} {\bibfnamefont {H.}~\bibnamefont {Wang}}, \bibinfo {author} {\bibfnamefont {L.}~\bibnamefont {Zhang}},\ and\ \bibinfo {author} {\bibfnamefont {L.}~\bibnamefont {Wang}},\ }\bibfield  {title} {\bibinfo {title} {Deep variational free energy approach to dense hydrogen},\ }\href {https://doi.org/10.1103/PhysRevLett.131.126501} {\bibfield  {journal} {\bibinfo  {journal} {Phys. Rev. Lett.}\ }\textbf {\bibinfo {volume} {131}},\ \bibinfo {pages} {126501} (\bibinfo {year} {2023})}\BibitemShut {NoStop}%
\bibitem [{\citenamefont {Linteau}\ \emph {et~al.}(2025)\citenamefont {Linteau}, \citenamefont {Moroni}, \citenamefont {Carleo},\ and\ \citenamefont {Holzmann}}]{linteau2025universal}%
  \BibitemOpen
  \bibfield  {author} {\bibinfo {author} {\bibfnamefont {D.}~\bibnamefont {Linteau}}, \bibinfo {author} {\bibfnamefont {S.}~\bibnamefont {Moroni}}, \bibinfo {author} {\bibfnamefont {G.}~\bibnamefont {Carleo}},\ and\ \bibinfo {author} {\bibfnamefont {M.}~\bibnamefont {Holzmann}},\ }\bibfield  {title} {\bibinfo {title} {Universal neural wave functions for high-pressure hydrogen},\ }\href@noop {} {\bibfield  {journal} {\bibinfo  {journal} {arXiv preprint arXiv:2504.07062}\ } (\bibinfo {year} {2025})}\BibitemShut {NoStop}%
\bibitem [{\citenamefont {Marx}\ and\ \citenamefont {Hutter}(2009)}]{marx2009ab}%
  \BibitemOpen
  \bibfield  {author} {\bibinfo {author} {\bibfnamefont {D.}~\bibnamefont {Marx}}\ and\ \bibinfo {author} {\bibfnamefont {J.}~\bibnamefont {Hutter}},\ }\href@noop {} {\emph {\bibinfo {title} {Ab initio molecular dynamics: basic theory and advanced methods}}}\ (\bibinfo  {publisher} {Cambridge University Press},\ \bibinfo {year} {2009})\BibitemShut {NoStop}%
\bibitem [{\citenamefont {Allen}\ and\ \citenamefont {Tildesley}(2017)}]{allen2017computer}%
  \BibitemOpen
  \bibfield  {author} {\bibinfo {author} {\bibfnamefont {M.~P.}\ \bibnamefont {Allen}}\ and\ \bibinfo {author} {\bibfnamefont {D.~J.}\ \bibnamefont {Tildesley}},\ }\href@noop {} {\emph {\bibinfo {title} {Computer simulation of liquids}}}\ (\bibinfo  {publisher} {Oxford university press},\ \bibinfo {year} {2017})\BibitemShut {NoStop}%
\bibitem [{\citenamefont {Perdew}\ \emph {et~al.}(2008)\citenamefont {Perdew}, \citenamefont {Ruzsinszky}, \citenamefont {Csonka}, \citenamefont {Vydrov}, \citenamefont {Scuseria}, \citenamefont {Constantin}, \citenamefont {Zhou},\ and\ \citenamefont {Burke}}]{PBEsol}%
  \BibitemOpen
  \bibfield  {author} {\bibinfo {author} {\bibfnamefont {J.~P.}\ \bibnamefont {Perdew}}, \bibinfo {author} {\bibfnamefont {A.}~\bibnamefont {Ruzsinszky}}, \bibinfo {author} {\bibfnamefont {G.~I.}\ \bibnamefont {Csonka}}, \bibinfo {author} {\bibfnamefont {O.~A.}\ \bibnamefont {Vydrov}}, \bibinfo {author} {\bibfnamefont {G.~E.}\ \bibnamefont {Scuseria}}, \bibinfo {author} {\bibfnamefont {L.~A.}\ \bibnamefont {Constantin}}, \bibinfo {author} {\bibfnamefont {X.}~\bibnamefont {Zhou}},\ and\ \bibinfo {author} {\bibfnamefont {K.}~\bibnamefont {Burke}},\ }\bibfield  {title} {\bibinfo {title} {Restoring the density-gradient expansion for exchange in solids and surfaces},\ }\href@noop {} {\bibfield  {journal} {\bibinfo  {journal} {Physical review letters}\ }\textbf {\bibinfo {volume} {100}},\ \bibinfo {pages} {136406} (\bibinfo {year} {2008})}\BibitemShut {NoStop}%
\bibitem [{\citenamefont {Peng}\ \emph {et~al.}(2016)\citenamefont {Peng}, \citenamefont {Yang}, \citenamefont {Perdew},\ and\ \citenamefont {Sun}}]{Peng2016_scan+vv10}%
  \BibitemOpen
  \bibfield  {author} {\bibinfo {author} {\bibfnamefont {H.}~\bibnamefont {Peng}}, \bibinfo {author} {\bibfnamefont {Z.-H.}\ \bibnamefont {Yang}}, \bibinfo {author} {\bibfnamefont {J.~P.}\ \bibnamefont {Perdew}},\ and\ \bibinfo {author} {\bibfnamefont {J.}~\bibnamefont {Sun}},\ }\bibfield  {title} {\bibinfo {title} {Versatile van der waals density functional based on a meta-generalized gradient approximation},\ }\href@noop {} {\bibfield  {journal} {\bibinfo  {journal} {Physical Review X}\ }\textbf {\bibinfo {volume} {6}},\ \bibinfo {pages} {041005} (\bibinfo {year} {2016})}\BibitemShut {NoStop}%
\bibitem [{\citenamefont {Csonka}\ \emph {et~al.}(2009)\citenamefont {Csonka}, \citenamefont {Perdew}, \citenamefont {Ruzsinszky}, \citenamefont {Philipsen}, \citenamefont {Leb\`egue}, \citenamefont {Paier}, \citenamefont {Vydrov},\ and\ \citenamefont {\'Angy\'an}}]{PhysRevB.79.155107}%
  \BibitemOpen
  \bibfield  {author} {\bibinfo {author} {\bibfnamefont {G.~I.}\ \bibnamefont {Csonka}}, \bibinfo {author} {\bibfnamefont {J.~P.}\ \bibnamefont {Perdew}}, \bibinfo {author} {\bibfnamefont {A.}~\bibnamefont {Ruzsinszky}}, \bibinfo {author} {\bibfnamefont {P.~H.~T.}\ \bibnamefont {Philipsen}}, \bibinfo {author} {\bibfnamefont {S.}~\bibnamefont {Leb\`egue}}, \bibinfo {author} {\bibfnamefont {J.}~\bibnamefont {Paier}}, \bibinfo {author} {\bibfnamefont {O.~A.}\ \bibnamefont {Vydrov}},\ and\ \bibinfo {author} {\bibfnamefont {J.~G.}\ \bibnamefont {\'Angy\'an}},\ }\bibfield  {title} {\bibinfo {title} {Assessing the performance of recent density functionals for bulk solids},\ }\href {https://doi.org/10.1103/PhysRevB.79.155107} {\bibfield  {journal} {\bibinfo  {journal} {Phys. Rev. B}\ }\textbf {\bibinfo {volume} {79}},\ \bibinfo {pages} {155107} (\bibinfo {year} {2009})}\BibitemShut {NoStop}%
\bibitem [{\citenamefont {Bonev}\ \emph {et~al.}(2004)\citenamefont {Bonev}, \citenamefont {Schwegler}, \citenamefont {Galli},\ and\ \citenamefont {Ogitsu}}]{bonev2004quantum}%
  \BibitemOpen
  \bibfield  {author} {\bibinfo {author} {\bibfnamefont {S.}~\bibnamefont {Bonev}}, \bibinfo {author} {\bibfnamefont {E.}~\bibnamefont {Schwegler}}, \bibinfo {author} {\bibfnamefont {G.}~\bibnamefont {Galli}},\ and\ \bibinfo {author} {\bibfnamefont {T.}~\bibnamefont {Ogitsu}},\ }\bibfield  {title} {\bibinfo {title} {A quantum fluid of metallic hydrogen suggested by first-principles calculations},\ }\href@noop {} {\bibfield  {journal} {\bibinfo  {journal} {Nature}\ }\textbf {\bibinfo {volume} {431}},\ \bibinfo {pages} {669} (\bibinfo {year} {2004})}\BibitemShut {NoStop}%
\bibitem [{\citenamefont {Morales}\ \emph {et~al.}(2010)\citenamefont {Morales}, \citenamefont {Pierleoni}, \citenamefont {Schwegler},\ and\ \citenamefont {Ceperley}}]{morales_evidence_2010}%
  \BibitemOpen
  \bibfield  {author} {\bibinfo {author} {\bibfnamefont {M.~A.}\ \bibnamefont {Morales}}, \bibinfo {author} {\bibfnamefont {C.}~\bibnamefont {Pierleoni}}, \bibinfo {author} {\bibfnamefont {E.}~\bibnamefont {Schwegler}},\ and\ \bibinfo {author} {\bibfnamefont {D.~M.}\ \bibnamefont {Ceperley}},\ }\bibfield  {title} {\bibinfo {title} {Evidence for a first-order liquid-liquid transition in high-pressure hydrogen from ab initio simulations},\ }\href {https://doi.org/10.1073/pnas.1007309107} {\bibfield  {journal} {\bibinfo  {journal} {Proceedings of the National Academy of Sciences}\ }\textbf {\bibinfo {volume} {107}},\ \bibinfo {pages} {12799} (\bibinfo {year} {2010})}\BibitemShut {NoStop}%
\bibitem [{\citenamefont {Scandolo}(2003)}]{scandolo2003liquid}%
  \BibitemOpen
  \bibfield  {author} {\bibinfo {author} {\bibfnamefont {S.}~\bibnamefont {Scandolo}},\ }\bibfield  {title} {\bibinfo {title} {Liquid--liquid phase transition in compressed hydrogen from first-principles simulations},\ }\href@noop {} {\bibfield  {journal} {\bibinfo  {journal} {Proceedings of the National Academy of Sciences}\ }\textbf {\bibinfo {volume} {100}},\ \bibinfo {pages} {3051} (\bibinfo {year} {2003})}\BibitemShut {NoStop}%
\bibitem [{\citenamefont {Lorenzen}\ \emph {et~al.}(2010)\citenamefont {Lorenzen}, \citenamefont {Holst},\ and\ \citenamefont {Redmer}}]{lorenzen2010}%
  \BibitemOpen
  \bibfield  {author} {\bibinfo {author} {\bibfnamefont {W.}~\bibnamefont {Lorenzen}}, \bibinfo {author} {\bibfnamefont {B.}~\bibnamefont {Holst}},\ and\ \bibinfo {author} {\bibfnamefont {R.}~\bibnamefont {Redmer}},\ }\bibfield  {title} {\bibinfo {title} {First-order liquid-liquid phase transition in dense hydrogen},\ }\href {https://doi.org/10.1103/PhysRevB.82.195107} {\bibfield  {journal} {\bibinfo  {journal} {Phys. Rev. B}\ }\textbf {\bibinfo {volume} {82}},\ \bibinfo {pages} {195107} (\bibinfo {year} {2010})}\BibitemShut {NoStop}%
\bibitem [{\citenamefont {Tamblyn}\ and\ \citenamefont {Bonev}(2010)}]{tamblyn2010structure}%
  \BibitemOpen
  \bibfield  {author} {\bibinfo {author} {\bibfnamefont {I.}~\bibnamefont {Tamblyn}}\ and\ \bibinfo {author} {\bibfnamefont {S.~A.}\ \bibnamefont {Bonev}},\ }\bibfield  {title} {\bibinfo {title} {Structure and phase boundaries of compressed liquid hydrogen},\ }\href@noop {} {\bibfield  {journal} {\bibinfo  {journal} {Physical review letters}\ }\textbf {\bibinfo {volume} {104}},\ \bibinfo {pages} {065702} (\bibinfo {year} {2010})}\BibitemShut {NoStop}%
\bibitem [{\citenamefont {Bergermann}\ \emph {et~al.}(2024)\citenamefont {Bergermann}, \citenamefont {Kleindienst},\ and\ \citenamefont {Redmer}}]{bergermann2024nonmetal}%
  \BibitemOpen
  \bibfield  {author} {\bibinfo {author} {\bibfnamefont {A.}~\bibnamefont {Bergermann}}, \bibinfo {author} {\bibfnamefont {L.}~\bibnamefont {Kleindienst}},\ and\ \bibinfo {author} {\bibfnamefont {R.}~\bibnamefont {Redmer}},\ }\bibfield  {title} {\bibinfo {title} {Nonmetal-to-metal transition in liquid hydrogen using density functional theory and the heyd--scuseria--ernzerhof exchange-correlation functional},\ }\href@noop {} {\bibfield  {journal} {\bibinfo  {journal} {The Journal of Chemical Physics}\ }\textbf {\bibinfo {volume} {161}} (\bibinfo {year} {2024})}\BibitemShut {NoStop}%
\bibitem [{\citenamefont {Azadi}\ \emph {et~al.}(2017)\citenamefont {Azadi}, \citenamefont {Drummond},\ and\ \citenamefont {Foulkes}}]{azadi2017nature}%
  \BibitemOpen
  \bibfield  {author} {\bibinfo {author} {\bibfnamefont {S.}~\bibnamefont {Azadi}}, \bibinfo {author} {\bibfnamefont {N.~D.}\ \bibnamefont {Drummond}},\ and\ \bibinfo {author} {\bibfnamefont {W.~M.~C.}\ \bibnamefont {Foulkes}},\ }\bibfield  {title} {\bibinfo {title} {Nature of the metallization transition in solid hydrogen},\ }\href@noop {} {\bibfield  {journal} {\bibinfo  {journal} {Physical Review B}\ }\textbf {\bibinfo {volume} {95}},\ \bibinfo {pages} {035142} (\bibinfo {year} {2017})}\BibitemShut {NoStop}%
\bibitem [{\citenamefont {Tiihonen}\ \emph {et~al.}(2021)\citenamefont {Tiihonen}, \citenamefont {Clay~III},\ and\ \citenamefont {Krogel}}]{Tiihonen2021SCE}%
  \BibitemOpen
  \bibfield  {author} {\bibinfo {author} {\bibfnamefont {J.}~\bibnamefont {Tiihonen}}, \bibinfo {author} {\bibfnamefont {R.~C.}\ \bibnamefont {Clay~III}},\ and\ \bibinfo {author} {\bibfnamefont {J.~T.}\ \bibnamefont {Krogel}},\ }\bibfield  {title} {\bibinfo {title} {Toward quantum monte carlo forces on heavier ions: Scaling properties},\ }\href {https://doi.org/10.1063/5.0052266} {\bibfield  {journal} {\bibinfo  {journal} {J. Chem. Phys.}\ }\textbf {\bibinfo {volume} {154}},\ \bibinfo {pages} {204111} (\bibinfo {year} {2021})}\BibitemShut {NoStop}%
\bibitem [{\citenamefont {Nakano}\ \emph {et~al.}(2022)\citenamefont {Nakano}, \citenamefont {Raghav},\ and\ \citenamefont {Sorella}}]{nakano2022SCE}%
  \BibitemOpen
  \bibfield  {author} {\bibinfo {author} {\bibfnamefont {K.}~\bibnamefont {Nakano}}, \bibinfo {author} {\bibfnamefont {A.}~\bibnamefont {Raghav}},\ and\ \bibinfo {author} {\bibfnamefont {S.}~\bibnamefont {Sorella}},\ }\bibfield  {title} {\bibinfo {title} {Space-warp coordinate transformation for efficient ionic force calculations in quantum monte carlo},\ }\href {https://doi.org/10.1063/5.0076302} {\bibfield  {journal} {\bibinfo  {journal} {J. Chem. Phys.}\ }\textbf {\bibinfo {volume} {156}},\ \bibinfo {pages} {034101} (\bibinfo {year} {2022})}\BibitemShut {NoStop}%
\bibitem [{\citenamefont {Perdew}\ \emph {et~al.}(1996{\natexlab{b}})\citenamefont {Perdew}, \citenamefont {Ernzerhof},\ and\ \citenamefont {Burke}}]{Perdew1996_pbe0}%
  \BibitemOpen
  \bibfield  {author} {\bibinfo {author} {\bibfnamefont {J.~P.}\ \bibnamefont {Perdew}}, \bibinfo {author} {\bibfnamefont {M.}~\bibnamefont {Ernzerhof}},\ and\ \bibinfo {author} {\bibfnamefont {K.}~\bibnamefont {Burke}},\ }\bibfield  {title} {\bibinfo {title} {Rationale for mixing exact exchange with density functional approximations},\ }\href@noop {} {\bibfield  {journal} {\bibinfo  {journal} {The Journal of chemical physics}\ }\textbf {\bibinfo {volume} {105}},\ \bibinfo {pages} {9982} (\bibinfo {year} {1996}{\natexlab{b}})}\BibitemShut {NoStop}%
\bibitem [{\citenamefont {Guido}\ \emph {et~al.}(2013)\citenamefont {Guido}, \citenamefont {Brémond}, \citenamefont {Adamo},\ and\ \citenamefont {Cortona}}]{revPBE0}%
  \BibitemOpen
  \bibfield  {author} {\bibinfo {author} {\bibfnamefont {C.~A.}\ \bibnamefont {Guido}}, \bibinfo {author} {\bibfnamefont {E.}~\bibnamefont {Brémond}}, \bibinfo {author} {\bibfnamefont {C.}~\bibnamefont {Adamo}},\ and\ \bibinfo {author} {\bibfnamefont {P.}~\bibnamefont {Cortona}},\ }\bibfield  {title} {\bibinfo {title} {Communication: One third: A new recipe for the pbe0 paradigm},\ }\href {https://doi.org/10.1063/1.4775591} {\bibfield  {journal} {\bibinfo  {journal} {The Journal of Chemical Physics}\ }\textbf {\bibinfo {volume} {138}},\ \bibinfo {pages} {021104} (\bibinfo {year} {2013})},\ \Eprint {https://arxiv.org/abs/https://doi.org/10.1063/1.4775591} {https://doi.org/10.1063/1.4775591} \BibitemShut {NoStop}%
\bibitem [{\citenamefont {Stephens}\ \emph {et~al.}(1994)\citenamefont {Stephens}, \citenamefont {Devlin}, \citenamefont {Chabalowski},\ and\ \citenamefont {Frisch}}]{B3LYP}%
  \BibitemOpen
  \bibfield  {author} {\bibinfo {author} {\bibfnamefont {P.~J.}\ \bibnamefont {Stephens}}, \bibinfo {author} {\bibfnamefont {F.~J.}\ \bibnamefont {Devlin}}, \bibinfo {author} {\bibfnamefont {C.~F.}\ \bibnamefont {Chabalowski}},\ and\ \bibinfo {author} {\bibfnamefont {M.~J.}\ \bibnamefont {Frisch}},\ }\bibfield  {title} {\bibinfo {title} {Ab initio calculation of vibrational absorption and circular dichroism spectra using density functional force fields},\ }\href {https://doi.org/10.1021/j100096a001} {\bibfield  {journal} {\bibinfo  {journal} {The Journal of Physical Chemistry}\ }\textbf {\bibinfo {volume} {98}},\ \bibinfo {pages} {11623} (\bibinfo {year} {1994})},\ \Eprint {https://arxiv.org/abs/https://doi.org/10.1021/j100096a001} {https://doi.org/10.1021/j100096a001} \BibitemShut {NoStop}%
\bibitem [{\citenamefont {Berland}\ \emph {et~al.}(2017)\citenamefont {Berland}, \citenamefont {Jiao}, \citenamefont {Lee}, \citenamefont {Rangel}, \citenamefont {Neaton},\ and\ \citenamefont {Hyldgaard}}]{Berland2017_vdW-c09}%
  \BibitemOpen
  \bibfield  {author} {\bibinfo {author} {\bibfnamefont {K.}~\bibnamefont {Berland}}, \bibinfo {author} {\bibfnamefont {Y.}~\bibnamefont {Jiao}}, \bibinfo {author} {\bibfnamefont {J.-H.}\ \bibnamefont {Lee}}, \bibinfo {author} {\bibfnamefont {T.}~\bibnamefont {Rangel}}, \bibinfo {author} {\bibfnamefont {J.~B.}\ \bibnamefont {Neaton}},\ and\ \bibinfo {author} {\bibfnamefont {P.}~\bibnamefont {Hyldgaard}},\ }\bibfield  {title} {\bibinfo {title} {Assessment of two hybrid van der waals density functionals for covalent and non-covalent binding of molecules},\ }\href@noop {} {\bibfield  {journal} {\bibinfo  {journal} {The Journal of Chemical Physics}\ }\textbf {\bibinfo {volume} {146}} (\bibinfo {year} {2017})}\BibitemShut {NoStop}%
\bibitem [{\citenamefont {Klimeš}\ \emph {et~al.}(2009)\citenamefont {Klimeš}, \citenamefont {Bowler},\ and\ \citenamefont {Michaelides}}]{optb86b}%
  \BibitemOpen
  \bibfield  {author} {\bibinfo {author} {\bibfnamefont {J.}~\bibnamefont {Klimeš}}, \bibinfo {author} {\bibfnamefont {D.~R.}\ \bibnamefont {Bowler}},\ and\ \bibinfo {author} {\bibfnamefont {A.}~\bibnamefont {Michaelides}},\ }\bibfield  {title} {\bibinfo {title} {Chemical accuracy for the van der waals density functional},\ }\href {https://doi.org/10.1088/0953-8984/22/2/022201} {\bibfield  {journal} {\bibinfo  {journal} {Journal of Physics: Condensed Matter}\ }\textbf {\bibinfo {volume} {22}},\ \bibinfo {pages} {022201} (\bibinfo {year} {2009})}\BibitemShut {NoStop}%
\bibitem [{\citenamefont {Vydrov}\ and\ \citenamefont {Van~Voorhis}(2010)}]{Vydrov2010_vv10}%
  \BibitemOpen
  \bibfield  {author} {\bibinfo {author} {\bibfnamefont {O.~A.}\ \bibnamefont {Vydrov}}\ and\ \bibinfo {author} {\bibfnamefont {T.}~\bibnamefont {Van~Voorhis}},\ }\bibfield  {title} {\bibinfo {title} {Nonlocal van der waals density functional: The simpler the better},\ }\href@noop {} {\bibfield  {journal} {\bibinfo  {journal} {The Journal of chemical physics}\ }\textbf {\bibinfo {volume} {133}} (\bibinfo {year} {2010})}\BibitemShut {NoStop}%
\bibitem [{\citenamefont {Sabatini}\ \emph {et~al.}(2013)\citenamefont {Sabatini}, \citenamefont {Gorni},\ and\ \citenamefont {De~Gironcoli}}]{Sabatini2013_rvv10}%
  \BibitemOpen
  \bibfield  {author} {\bibinfo {author} {\bibfnamefont {R.}~\bibnamefont {Sabatini}}, \bibinfo {author} {\bibfnamefont {T.}~\bibnamefont {Gorni}},\ and\ \bibinfo {author} {\bibfnamefont {S.}~\bibnamefont {De~Gironcoli}},\ }\bibfield  {title} {\bibinfo {title} {Nonlocal van der waals density functional made simple and efficient},\ }\href@noop {} {\bibfield  {journal} {\bibinfo  {journal} {Physical Review B—Condensed Matter and Materials Physics}\ }\textbf {\bibinfo {volume} {87}},\ \bibinfo {pages} {041108} (\bibinfo {year} {2013})}\BibitemShut {NoStop}%
\bibitem [{\citenamefont {Heyd}\ \emph {et~al.}(2003)\citenamefont {Heyd}, \citenamefont {Scuseria},\ and\ \citenamefont {Ernzerhof}}]{Heyd2003_HSE}%
  \BibitemOpen
  \bibfield  {author} {\bibinfo {author} {\bibfnamefont {J.}~\bibnamefont {Heyd}}, \bibinfo {author} {\bibfnamefont {G.~E.}\ \bibnamefont {Scuseria}},\ and\ \bibinfo {author} {\bibfnamefont {M.}~\bibnamefont {Ernzerhof}},\ }\bibfield  {title} {\bibinfo {title} {Hybrid functionals based on a screened coulomb potential},\ }\href@noop {} {\bibfield  {journal} {\bibinfo  {journal} {The Journal of chemical physics}\ }\textbf {\bibinfo {volume} {118}},\ \bibinfo {pages} {8207} (\bibinfo {year} {2003})}\BibitemShut {NoStop}%
\bibitem [{\citenamefont {Bart{\'o}k}\ and\ \citenamefont {Yates}(2019)}]{Bartok2019_rscan}%
  \BibitemOpen
  \bibfield  {author} {\bibinfo {author} {\bibfnamefont {A.~P.}\ \bibnamefont {Bart{\'o}k}}\ and\ \bibinfo {author} {\bibfnamefont {J.~R.}\ \bibnamefont {Yates}},\ }\bibfield  {title} {\bibinfo {title} {Regularized scan functional},\ }\href@noop {} {\bibfield  {journal} {\bibinfo  {journal} {The Journal of chemical physics}\ }\textbf {\bibinfo {volume} {150}} (\bibinfo {year} {2019})}\BibitemShut {NoStop}%
\bibitem [{\citenamefont {Furness}\ \emph {et~al.}(2020)\citenamefont {Furness}, \citenamefont {Kaplan}, \citenamefont {Ning}, \citenamefont {Perdew},\ and\ \citenamefont {Sun}}]{Furness2020_r2scan}%
  \BibitemOpen
  \bibfield  {author} {\bibinfo {author} {\bibfnamefont {J.~W.}\ \bibnamefont {Furness}}, \bibinfo {author} {\bibfnamefont {A.~D.}\ \bibnamefont {Kaplan}}, \bibinfo {author} {\bibfnamefont {J.}~\bibnamefont {Ning}}, \bibinfo {author} {\bibfnamefont {J.~P.}\ \bibnamefont {Perdew}},\ and\ \bibinfo {author} {\bibfnamefont {J.}~\bibnamefont {Sun}},\ }\bibfield  {title} {\bibinfo {title} {Accurate and numerically efficient r2scan meta-generalized gradient approximation},\ }\href@noop {} {\bibfield  {journal} {\bibinfo  {journal} {The journal of physical chemistry letters}\ }\textbf {\bibinfo {volume} {11}},\ \bibinfo {pages} {8208} (\bibinfo {year} {2020})}\BibitemShut {NoStop}%
\bibitem [{\citenamefont {Knudson}\ and\ \citenamefont {Desjarlais}(2017)}]{PhysRevLett.118.035501}%
  \BibitemOpen
  \bibfield  {author} {\bibinfo {author} {\bibfnamefont {M.~D.}\ \bibnamefont {Knudson}}\ and\ \bibinfo {author} {\bibfnamefont {M.~P.}\ \bibnamefont {Desjarlais}},\ }\bibfield  {title} {\bibinfo {title} {High-precision shock wave measurements of deuterium: Evaluation of exchange-correlation functionals at the molecular-to-atomic transition},\ }\href {https://doi.org/10.1103/PhysRevLett.118.035501} {\bibfield  {journal} {\bibinfo  {journal} {Phys. Rev. Lett.}\ }\textbf {\bibinfo {volume} {118}},\ \bibinfo {pages} {035501} (\bibinfo {year} {2017})}\BibitemShut {NoStop}%
\bibitem [{\citenamefont {{Brygoo}}\ \emph {et~al.}(2015)\citenamefont {{Brygoo}}, \citenamefont {{Millot}}, \citenamefont {{Loubeyre}}, \citenamefont {{Lazicki}}, \citenamefont {{Hamel}}, \citenamefont {{Qi}}, \citenamefont {{Celliers}}, \citenamefont {{Coppari}}, \citenamefont {{Eggert}}, \citenamefont {{Fratanduono}}, \citenamefont {{Hicks}}, \citenamefont {{Rygg}}, \citenamefont {{Smith}}, \citenamefont {{Swift}}, \citenamefont {{Collins}},\ and\ \citenamefont {{Jeanloz}}}]{2015JAP...118s5901B}%
  \BibitemOpen
  \bibfield  {author} {\bibinfo {author} {\bibfnamefont {S.}~\bibnamefont {{Brygoo}}}, \bibinfo {author} {\bibfnamefont {M.}~\bibnamefont {{Millot}}}, \bibinfo {author} {\bibfnamefont {P.}~\bibnamefont {{Loubeyre}}}, \bibinfo {author} {\bibfnamefont {A.~E.}\ \bibnamefont {{Lazicki}}}, \bibinfo {author} {\bibfnamefont {S.}~\bibnamefont {{Hamel}}}, \bibinfo {author} {\bibfnamefont {T.}~\bibnamefont {{Qi}}}, \bibinfo {author} {\bibfnamefont {P.~M.}\ \bibnamefont {{Celliers}}}, \bibinfo {author} {\bibfnamefont {F.}~\bibnamefont {{Coppari}}}, \bibinfo {author} {\bibfnamefont {J.~H.}\ \bibnamefont {{Eggert}}}, \bibinfo {author} {\bibfnamefont {D.~E.}\ \bibnamefont {{Fratanduono}}}, \bibinfo {author} {\bibfnamefont {D.~G.}\ \bibnamefont {{Hicks}}}, \bibinfo {author} {\bibfnamefont {J.~R.}\ \bibnamefont {{Rygg}}}, \bibinfo {author} {\bibfnamefont {R.~F.}\ \bibnamefont {{Smith}}}, \bibinfo {author} {\bibfnamefont {D.~C.}\ \bibnamefont {{Swift}}}, \bibinfo {author} {\bibfnamefont {G.~W.}\ \bibnamefont {{Collins}}},\
  and\ \bibinfo {author} {\bibfnamefont {R.}~\bibnamefont {{Jeanloz}}},\ }\bibfield  {title} {\bibinfo {title} {{Analysis of laser shock experiments on precompressed samples using a quartz reference and application to warm dense hydrogen and helium}},\ }\href {https://doi.org/10.1063/1.4935295} {\bibfield  {journal} {\bibinfo  {journal} {Journal of Applied Physics}\ }\textbf {\bibinfo {volume} {118}},\ \bibinfo {eid} {195901} (\bibinfo {year} {2015})}\BibitemShut {NoStop}%
\bibitem [{\citenamefont {Clay}\ \emph {et~al.}(2019)\citenamefont {Clay}, \citenamefont {Desjarlais},\ and\ \citenamefont {Shulenburger}}]{PhysRevB.100.075103}%
  \BibitemOpen
  \bibfield  {author} {\bibinfo {author} {\bibfnamefont {R.~C.}\ \bibnamefont {Clay}}, \bibinfo {author} {\bibfnamefont {M.~P.}\ \bibnamefont {Desjarlais}},\ and\ \bibinfo {author} {\bibfnamefont {L.}~\bibnamefont {Shulenburger}},\ }\bibfield  {title} {\bibinfo {title} {Deuterium hugoniot: Pitfalls of thermodynamic sampling beyond density functional theory},\ }\href {https://doi.org/10.1103/PhysRevB.100.075103} {\bibfield  {journal} {\bibinfo  {journal} {Phys. Rev. B}\ }\textbf {\bibinfo {volume} {100}},\ \bibinfo {pages} {075103} (\bibinfo {year} {2019})}\BibitemShut {NoStop}%
\bibitem [{\citenamefont {Ruggeri}\ \emph {et~al.}(2020)\citenamefont {Ruggeri}, \citenamefont {Holzmann}, \citenamefont {Ceperley},\ and\ \citenamefont {Pierleoni}}]{ruggeri2020quantum}%
  \BibitemOpen
  \bibfield  {author} {\bibinfo {author} {\bibfnamefont {M.}~\bibnamefont {Ruggeri}}, \bibinfo {author} {\bibfnamefont {M.}~\bibnamefont {Holzmann}}, \bibinfo {author} {\bibfnamefont {D.~M.}\ \bibnamefont {Ceperley}},\ and\ \bibinfo {author} {\bibfnamefont {C.}~\bibnamefont {Pierleoni}},\ }\bibfield  {title} {\bibinfo {title} {Quantum monte carlo determination of the principal hugoniot of deuterium},\ }\href@noop {} {\bibfield  {journal} {\bibinfo  {journal} {Physical Review B}\ }\textbf {\bibinfo {volume} {102}},\ \bibinfo {pages} {144108} (\bibinfo {year} {2020})}\BibitemShut {NoStop}%
\bibitem [{\citenamefont {Tenti}\ \emph {et~al.}(2024)\citenamefont {Tenti}, \citenamefont {Nakano}, \citenamefont {Tirelli}, \citenamefont {Sorella},\ and\ \citenamefont {Casula}}]{tenti2024hugoniot}%
  \BibitemOpen
  \bibfield  {author} {\bibinfo {author} {\bibfnamefont {G.}~\bibnamefont {Tenti}}, \bibinfo {author} {\bibfnamefont {K.}~\bibnamefont {Nakano}}, \bibinfo {author} {\bibfnamefont {A.}~\bibnamefont {Tirelli}}, \bibinfo {author} {\bibfnamefont {S.}~\bibnamefont {Sorella}},\ and\ \bibinfo {author} {\bibfnamefont {M.}~\bibnamefont {Casula}},\ }\bibfield  {title} {\bibinfo {title} {Principal deuterium hugoniot via quantum monte carlo and $\mathrm{\ensuremath{\Delta}}$-learning},\ }\href {https://doi.org/10.1103/PhysRevB.110.L041107} {\bibfield  {journal} {\bibinfo  {journal} {Phys. Rev. B}\ }\textbf {\bibinfo {volume} {110}},\ \bibinfo {pages} {L041107} (\bibinfo {year} {2024})}\BibitemShut {NoStop}%
\bibitem [{\citenamefont {Knudson}\ and\ \citenamefont {Desjarlais}(2021)}]{Knudson_2021}%
  \BibitemOpen
  \bibfield  {author} {\bibinfo {author} {\bibfnamefont {M.~D.}\ \bibnamefont {Knudson}}\ and\ \bibinfo {author} {\bibfnamefont {M.~P.}\ \bibnamefont {Desjarlais}},\ }\bibfield  {title} {\bibinfo {title} {Interplay of high-precision shock wave experiments with first-principles theory to explore molecular systems at extreme conditions: A perspective},\ }\href {https://doi.org/10.1063/5.0050878} {\bibfield  {journal} {\bibinfo  {journal} {Journal of Applied Physics}\ }\textbf {\bibinfo {volume} {129}},\ \bibinfo {pages} {210904} (\bibinfo {year} {2021})}\BibitemShut {NoStop}%
\bibitem [{\citenamefont {{Nellis}}\ \emph {et~al.}(1983)\citenamefont {{Nellis}}, \citenamefont {{Mitchell}}, \citenamefont {{van Thiel}}, \citenamefont {{Devine}}, \citenamefont {{Trainor}},\ and\ \citenamefont {{Brown}}}]{1983JChPh..79.1480N}%
  \BibitemOpen
  \bibfield  {author} {\bibinfo {author} {\bibfnamefont {W.~J.}\ \bibnamefont {{Nellis}}}, \bibinfo {author} {\bibfnamefont {A.~C.}\ \bibnamefont {{Mitchell}}}, \bibinfo {author} {\bibfnamefont {M.}~\bibnamefont {{van Thiel}}}, \bibinfo {author} {\bibfnamefont {G.~J.}\ \bibnamefont {{Devine}}}, \bibinfo {author} {\bibfnamefont {R.~J.}\ \bibnamefont {{Trainor}}},\ and\ \bibinfo {author} {\bibfnamefont {N.}~\bibnamefont {{Brown}}},\ }\bibfield  {title} {\bibinfo {title} {{Equation of state data for molecular hydrogen and deuterium at shock pressures in the range 2-76 Gpa (20-760 kbar)}},\ }\href {https://doi.org/10.1063/1.445938} {\bibfield  {journal} {\bibinfo  {journal} {\jcp}\ }\textbf {\bibinfo {volume} {79}},\ \bibinfo {pages} {1480} (\bibinfo {year} {1983})}\BibitemShut {NoStop}%
\bibitem [{\citenamefont {Cox}\ \emph {et~al.}(1989)\citenamefont {Cox}, \citenamefont {Wagman},\ and\ \citenamefont {Medvedev}}]{cox1989codata}%
  \BibitemOpen
  \bibfield  {author} {\bibinfo {author} {\bibfnamefont {J.~D.}\ \bibnamefont {Cox}}, \bibinfo {author} {\bibfnamefont {D.~D.}\ \bibnamefont {Wagman}},\ and\ \bibinfo {author} {\bibfnamefont {V.~A.}\ \bibnamefont {Medvedev}},\ }\bibfield  {title} {\bibinfo {title} {Codata key values for thermodynamics},\ }in\ \href {https://doi.org/10.1002/bbpc.19900940121} {\emph {\bibinfo {booktitle} {Series on Thermodynamic Properties}}}\ (\bibinfo {year} {1989})\BibitemShut {NoStop}%
\bibitem [{\citenamefont {Lemmon}\ \emph {et~al.}(2023)\citenamefont {Lemmon}, \citenamefont {Bell}, \citenamefont {Huber},\ and\ \citenamefont {McLinden}}]{Lemmon_Thermophysical_Properties}%
  \BibitemOpen
  \bibfield  {author} {\bibinfo {author} {\bibfnamefont {E.~W.}\ \bibnamefont {Lemmon}}, \bibinfo {author} {\bibfnamefont {I.~H.}\ \bibnamefont {Bell}}, \bibinfo {author} {\bibfnamefont {M.~L.}\ \bibnamefont {Huber}},\ and\ \bibinfo {author} {\bibfnamefont {M.~O.}\ \bibnamefont {McLinden}},\ }\bibfield  {title} {\bibinfo {title} {Thermophysical properties of fluid systems},\ }in\ \href {https://doi.org/10.18434/T4D303} {\emph {\bibinfo {booktitle} {NIST Chemistry WebBook, NIST Standard Reference Database Number 69}}},\ \bibinfo {editor} {edited by\ \bibinfo {editor} {\bibfnamefont {P.}~\bibnamefont {Linstrom}}\ and\ \bibinfo {editor} {\bibfnamefont {W.}~\bibnamefont {Mallard}}}\ (\bibinfo  {publisher} {National Institute of Standards and Technology, Gaithersburg MD, 20899},\ \bibinfo {year} {2023})\BibitemShut {NoStop}%
\bibitem [{\citenamefont {Saumon}\ and\ \citenamefont {Chabrier}()}]{private_comm}%
  \BibitemOpen
  \bibfield  {author} {\bibinfo {author} {\bibfnamefont {D.}~\bibnamefont {Saumon}}\ and\ \bibinfo {author} {\bibfnamefont {G.}~\bibnamefont {Chabrier}},\ }\href@noop {} {}\bibinfo {note} {{private communication}}\BibitemShut {NoStop}%
\bibitem [{\citenamefont {{Mankovich}}\ and\ \citenamefont {{Fuller}}(2021)}]{mankovich2021}%
  \BibitemOpen
  \bibfield  {author} {\bibinfo {author} {\bibfnamefont {C.~R.}\ \bibnamefont {{Mankovich}}}\ and\ \bibinfo {author} {\bibfnamefont {J.}~\bibnamefont {{Fuller}}},\ }\bibfield  {title} {\bibinfo {title} {{A diffuse core in Saturn revealed by ring seismology}},\ }\href {https://doi.org/10.1038/s41550-021-01448-3} {\bibfield  {journal} {\bibinfo  {journal} {Nature Astronomy}\ }\textbf {\bibinfo {volume} {5}},\ \bibinfo {pages} {1103} (\bibinfo {year} {2021})},\ \Eprint {https://arxiv.org/abs/2104.13385} {arXiv:2104.13385 [astro-ph.EP]} \BibitemShut {NoStop}%
\bibitem [{\citenamefont {Van~Setten}\ \emph {et~al.}(2018)\citenamefont {Van~Setten}, \citenamefont {Giantomassi}, \citenamefont {Bousquet}, \citenamefont {Verstraete}, \citenamefont {Hamann}, \citenamefont {Gonze},\ and\ \citenamefont {Rignanese}}]{pseudodojo2018}%
  \BibitemOpen
  \bibfield  {author} {\bibinfo {author} {\bibfnamefont {M.~J.}\ \bibnamefont {Van~Setten}}, \bibinfo {author} {\bibfnamefont {M.}~\bibnamefont {Giantomassi}}, \bibinfo {author} {\bibfnamefont {E.}~\bibnamefont {Bousquet}}, \bibinfo {author} {\bibfnamefont {M.~J.}\ \bibnamefont {Verstraete}}, \bibinfo {author} {\bibfnamefont {D.~R.}\ \bibnamefont {Hamann}}, \bibinfo {author} {\bibfnamefont {X.}~\bibnamefont {Gonze}},\ and\ \bibinfo {author} {\bibfnamefont {G.-M.}\ \bibnamefont {Rignanese}},\ }\bibfield  {title} {\bibinfo {title} {The pseudodojo: Training and grading a 85 element optimized norm-conserving pseudopotential table},\ }\href@noop {} {\bibfield  {journal} {\bibinfo  {journal} {Computer Physics Communications}\ }\textbf {\bibinfo {volume} {226}},\ \bibinfo {pages} {39} (\bibinfo {year} {2018})}\BibitemShut {NoStop}%
\bibitem [{\citenamefont {Cozza}\ \emph {et~al.}(2025)\citenamefont {Cozza}, \citenamefont {Nakano}, \citenamefont {Howard}, \citenamefont {Xie}, \citenamefont {Helled},\ and\ \citenamefont {Mazzola}}]{H_SCANvv10_EoS}%
  \BibitemOpen
  \bibfield  {author} {\bibinfo {author} {\bibfnamefont {C.}~\bibnamefont {Cozza}}, \bibinfo {author} {\bibfnamefont {K.}~\bibnamefont {Nakano}}, \bibinfo {author} {\bibfnamefont {S.}~\bibnamefont {Howard}}, \bibinfo {author} {\bibfnamefont {H.}~\bibnamefont {Xie}}, \bibinfo {author} {\bibfnamefont {R.}~\bibnamefont {Helled}},\ and\ \bibinfo {author} {\bibfnamefont {G.}~\bibnamefont {Mazzola}},\ }\href@noop {} {\bibinfo {title} {H\_scanvv10\_eos: Hydrogen equation of state data}},\ \bibinfo {howpublished} {\url{https://github.com/cisar97/H_SCANvv10_EoS}} (\bibinfo {year} {2025})\BibitemShut {NoStop}%
\bibitem [{\citenamefont {{Hubbard}}(1975)}]{hubbard1975}%
  \BibitemOpen
  \bibfield  {author} {\bibinfo {author} {\bibfnamefont {W.~B.}\ \bibnamefont {{Hubbard}}},\ }\bibfield  {title} {\bibinfo {title} {{Gravitational field of a rotating planet with a polytropic index of unity}},\ }\href@noop {} {\bibfield  {journal} {\bibinfo  {journal} {Astronomicheskii Zhurnal}\ }\textbf {\bibinfo {volume} {18}},\ \bibinfo {pages} {621} (\bibinfo {year} {1975})}\BibitemShut {NoStop}%
\bibitem [{\citenamefont {{Guillot}}\ and\ \citenamefont {{Morel}}(1995)}]{guillot1995}%
  \BibitemOpen
  \bibfield  {author} {\bibinfo {author} {\bibfnamefont {T.}~\bibnamefont {{Guillot}}}\ and\ \bibinfo {author} {\bibfnamefont {P.}~\bibnamefont {{Morel}}},\ }\bibfield  {title} {\bibinfo {title} {{CEPAM: a code for modeling the interiors of giant planets.}},\ }\href@noop {} {\bibfield  {journal} {\bibinfo  {journal} {Astronomy and Astrophysics Suppl.}\ }\textbf {\bibinfo {volume} {109}},\ \bibinfo {pages} {109} (\bibinfo {year} {1995})}\BibitemShut {NoStop}%
\bibitem [{\citenamefont {{Helled}}\ \emph {et~al.}(2022)\citenamefont {{Helled}}, \citenamefont {{Stevenson}}, \citenamefont {{Lunine}}, \citenamefont {{Bolton}}, \citenamefont {{Nettelmann}}, \citenamefont {{Atreya}}, \citenamefont {{Guillot}}, \citenamefont {{Militzer}}, \citenamefont {{Miguel}},\ and\ \citenamefont {{Hubbard}}}]{2022Icar..37814937H}%
  \BibitemOpen
  \bibfield  {author} {\bibinfo {author} {\bibfnamefont {R.}~\bibnamefont {{Helled}}}, \bibinfo {author} {\bibfnamefont {D.~J.}\ \bibnamefont {{Stevenson}}}, \bibinfo {author} {\bibfnamefont {J.~I.}\ \bibnamefont {{Lunine}}}, \bibinfo {author} {\bibfnamefont {S.~J.}\ \bibnamefont {{Bolton}}}, \bibinfo {author} {\bibfnamefont {N.}~\bibnamefont {{Nettelmann}}}, \bibinfo {author} {\bibfnamefont {S.}~\bibnamefont {{Atreya}}}, \bibinfo {author} {\bibfnamefont {T.}~\bibnamefont {{Guillot}}}, \bibinfo {author} {\bibfnamefont {B.}~\bibnamefont {{Militzer}}}, \bibinfo {author} {\bibfnamefont {Y.}~\bibnamefont {{Miguel}}},\ and\ \bibinfo {author} {\bibfnamefont {W.~B.}\ \bibnamefont {{Hubbard}}},\ }\bibfield  {title} {\bibinfo {title} {{Revelations on Jupiter's formation, evolution and interior: Challenges from Juno results}},\ }\href {https://doi.org/10.1016/j.icarus.2022.114937} {\bibfield  {journal} {\bibinfo  {journal} {Icarus}\ }\textbf {\bibinfo {volume} {378}},\ \bibinfo {eid} {114937} (\bibinfo {year}
  {2022})},\ \Eprint {https://arxiv.org/abs/2202.10041} {arXiv:2202.10041 [astro-ph.EP]} \BibitemShut {NoStop}%
\bibitem [{\citenamefont {{Debras}}\ and\ \citenamefont {{Chabrier}}(2019)}]{DC19}%
  \BibitemOpen
  \bibfield  {author} {\bibinfo {author} {\bibfnamefont {F.}~\bibnamefont {{Debras}}}\ and\ \bibinfo {author} {\bibfnamefont {G.}~\bibnamefont {{Chabrier}}},\ }\bibfield  {title} {\bibinfo {title} {{New Models of Jupiter in the Context of Juno and Galileo}},\ }\href {https://doi.org/10.3847/1538-4357/aaff65} {\bibfield  {journal} {\bibinfo  {journal} {ApJ}\ }\textbf {\bibinfo {volume} {872}},\ \bibinfo {eid} {100} (\bibinfo {year} {2019})}\BibitemShut {NoStop}%
\bibitem [{\citenamefont {{Howard}}\ \emph {et~al.}(2023{\natexlab{b}})\citenamefont {{Howard}}, \citenamefont {{Guillot}}, \citenamefont {{Markham}}, \citenamefont {{Helled}}, \citenamefont {{M{\"u}ller}}, \citenamefont {{Stevenson}}, \citenamefont {{Lunine}}, \citenamefont {{Miguel}},\ and\ \citenamefont {{Nettelmann}}}]{howard2023b}%
  \BibitemOpen
  \bibfield  {author} {\bibinfo {author} {\bibfnamefont {S.}~\bibnamefont {{Howard}}}, \bibinfo {author} {\bibfnamefont {T.}~\bibnamefont {{Guillot}}}, \bibinfo {author} {\bibfnamefont {S.}~\bibnamefont {{Markham}}}, \bibinfo {author} {\bibfnamefont {R.}~\bibnamefont {{Helled}}}, \bibinfo {author} {\bibfnamefont {S.}~\bibnamefont {{M{\"u}ller}}}, \bibinfo {author} {\bibfnamefont {D.~J.}\ \bibnamefont {{Stevenson}}}, \bibinfo {author} {\bibfnamefont {J.~I.}\ \bibnamefont {{Lunine}}}, \bibinfo {author} {\bibfnamefont {Y.}~\bibnamefont {{Miguel}}},\ and\ \bibinfo {author} {\bibfnamefont {N.}~\bibnamefont {{Nettelmann}}},\ }\bibfield  {title} {\bibinfo {title} {{Exploring the hypothesis of an inverted Z gradient inside Jupiter}},\ }\href {https://doi.org/10.1051/0004-6361/202348129} {\bibfield  {journal} {\bibinfo  {journal} {Astronomy and Astrophysics}\ }\textbf {\bibinfo {volume} {680}},\ \bibinfo {eid} {L2} (\bibinfo {year} {2023}{\natexlab{b}})},\ \Eprint {https://arxiv.org/abs/2311.07646} {arXiv:2311.07646
  [astro-ph.EP]} \BibitemShut {NoStop}%
\bibitem [{\citenamefont {{M{\"u}ller}}\ and\ \citenamefont {{Helled}}(2024)}]{2024ApJ...967....7M}%
  \BibitemOpen
  \bibfield  {author} {\bibinfo {author} {\bibfnamefont {S.}~\bibnamefont {{M{\"u}ller}}}\ and\ \bibinfo {author} {\bibfnamefont {R.}~\bibnamefont {{Helled}}},\ }\bibfield  {title} {\bibinfo {title} {{Can Jupiter's Atmospheric Metallicity Be Different from the Deep Interior?}},\ }\href {https://doi.org/10.3847/1538-4357/ad3738} {\bibfield  {journal} {\bibinfo  {journal} {\apj}\ }\textbf {\bibinfo {volume} {967}},\ \bibinfo {eid} {7} (\bibinfo {year} {2024})},\ \Eprint {https://arxiv.org/abs/2403.16273} {arXiv:2403.16273 [astro-ph.EP]} \BibitemShut {NoStop}%
\bibitem [{\citenamefont {Giannozzi}\ \emph {et~al.}(2009)\citenamefont {Giannozzi}, \citenamefont {Baroni}, \citenamefont {Bonini}, \citenamefont {Calandra}, \citenamefont {Car}, \citenamefont {Cavazzoni}, \citenamefont {Ceresoli}, \citenamefont {Chiarotti}, \citenamefont {Cococcioni}, \citenamefont {Dabo} \emph {et~al.}}]{Giannozzi2009_QE}%
  \BibitemOpen
  \bibfield  {author} {\bibinfo {author} {\bibfnamefont {P.}~\bibnamefont {Giannozzi}}, \bibinfo {author} {\bibfnamefont {S.}~\bibnamefont {Baroni}}, \bibinfo {author} {\bibfnamefont {N.}~\bibnamefont {Bonini}}, \bibinfo {author} {\bibfnamefont {M.}~\bibnamefont {Calandra}}, \bibinfo {author} {\bibfnamefont {R.}~\bibnamefont {Car}}, \bibinfo {author} {\bibfnamefont {C.}~\bibnamefont {Cavazzoni}}, \bibinfo {author} {\bibfnamefont {D.}~\bibnamefont {Ceresoli}}, \bibinfo {author} {\bibfnamefont {G.~L.}\ \bibnamefont {Chiarotti}}, \bibinfo {author} {\bibfnamefont {M.}~\bibnamefont {Cococcioni}}, \bibinfo {author} {\bibfnamefont {I.}~\bibnamefont {Dabo}}, \emph {et~al.},\ }\bibfield  {title} {\bibinfo {title} {Quantum espresso: a modular and open-source software project for quantum simulations of materials},\ }\href@noop {} {\bibfield  {journal} {\bibinfo  {journal} {Journal of physics: Condensed matter}\ }\textbf {\bibinfo {volume} {21}},\ \bibinfo {pages} {395502} (\bibinfo {year} {2009})}\BibitemShut {NoStop}%
\bibitem [{\citenamefont {Bl\"ochl}(1994)}]{BlochlPAW1004}%
  \BibitemOpen
  \bibfield  {author} {\bibinfo {author} {\bibfnamefont {P.~E.}\ \bibnamefont {Bl\"ochl}},\ }\bibfield  {title} {\bibinfo {title} {Projector augmented-wave method},\ }\href {https://doi.org/10.1103/PhysRevB.50.17953} {\bibfield  {journal} {\bibinfo  {journal} {Phys. Rev. B}\ }\textbf {\bibinfo {volume} {50}},\ \bibinfo {pages} {17953} (\bibinfo {year} {1994})}\BibitemShut {NoStop}%
\bibitem [{\citenamefont {{Dal Corso}}(2014)}]{DalCorso2014_psl}%
  \BibitemOpen
  \bibfield  {author} {\bibinfo {author} {\bibfnamefont {A.}~\bibnamefont {{Dal Corso}}},\ }\bibfield  {title} {\bibinfo {title} {Pseudopotentials periodic table: From h to pu},\ }\href {https://doi.org/https://doi.org/10.1016/j.commatsci.2014.07.043} {\bibfield  {journal} {\bibinfo  {journal} {Computational Materials Science}\ }\textbf {\bibinfo {volume} {95}},\ \bibinfo {pages} {337} (\bibinfo {year} {2014})}\BibitemShut {NoStop}%
\bibitem [{\citenamefont {Hamann}\ \emph {et~al.}(1979)\citenamefont {Hamann}, \citenamefont {Schl\"uter},\ and\ \citenamefont {Chiang}}]{Hamann1979_NormConserving}%
  \BibitemOpen
  \bibfield  {author} {\bibinfo {author} {\bibfnamefont {D.~R.}\ \bibnamefont {Hamann}}, \bibinfo {author} {\bibfnamefont {M.}~\bibnamefont {Schl\"uter}},\ and\ \bibinfo {author} {\bibfnamefont {C.}~\bibnamefont {Chiang}},\ }\bibfield  {title} {\bibinfo {title} {Norm-conserving pseudopotentials},\ }\href {https://doi.org/10.1103/PhysRevLett.43.1494} {\bibfield  {journal} {\bibinfo  {journal} {Phys. Rev. Lett.}\ }\textbf {\bibinfo {volume} {43}},\ \bibinfo {pages} {1494} (\bibinfo {year} {1979})}\BibitemShut {NoStop}%
\bibitem [{\citenamefont {Hamann}(2013)}]{Hamann2013_SG15}%
  \BibitemOpen
  \bibfield  {author} {\bibinfo {author} {\bibfnamefont {D.~R.}\ \bibnamefont {Hamann}},\ }\bibfield  {title} {\bibinfo {title} {Optimized norm-conserving vanderbilt pseudopotentials},\ }\href {https://doi.org/10.1103/PhysRevB.88.085117} {\bibfield  {journal} {\bibinfo  {journal} {Phys. Rev. B}\ }\textbf {\bibinfo {volume} {88}},\ \bibinfo {pages} {085117} (\bibinfo {year} {2013})}\BibitemShut {NoStop}%
\bibitem [{\citenamefont {Yao}\ and\ \citenamefont {Kanai}(2017)}]{yao2017_scanConvergence}%
  \BibitemOpen
  \bibfield  {author} {\bibinfo {author} {\bibfnamefont {Y.}~\bibnamefont {Yao}}\ and\ \bibinfo {author} {\bibfnamefont {Y.}~\bibnamefont {Kanai}},\ }\bibfield  {title} {\bibinfo {title} {Plane-wave pseudopotential implementation and performance of scan meta-gga exchange-correlation functional for extended systems},\ }\href@noop {} {\bibfield  {journal} {\bibinfo  {journal} {The Journal of chemical physics}\ }\textbf {\bibinfo {volume} {146}},\ \bibinfo {pages} {224105} (\bibinfo {year} {2017})}\BibitemShut {NoStop}%
\bibitem [{\citenamefont {Bussi}\ \emph {et~al.}(2007)\citenamefont {Bussi}, \citenamefont {Donadio},\ and\ \citenamefont {Parrinello}}]{bussi2007canonical}%
  \BibitemOpen
  \bibfield  {author} {\bibinfo {author} {\bibfnamefont {G.}~\bibnamefont {Bussi}}, \bibinfo {author} {\bibfnamefont {D.}~\bibnamefont {Donadio}},\ and\ \bibinfo {author} {\bibfnamefont {M.}~\bibnamefont {Parrinello}},\ }\bibfield  {title} {\bibinfo {title} {Canonical sampling through velocity rescaling},\ }\href@noop {} {\bibfield  {journal} {\bibinfo  {journal} {The Journal of chemical physics}\ }\textbf {\bibinfo {volume} {126}},\ \bibinfo {pages} {014101} (\bibinfo {year} {2007})}\BibitemShut {NoStop}%
\bibitem [{\citenamefont {Kresse}\ and\ \citenamefont {Hafner}(1993)}]{kresse1993vasp}%
  \BibitemOpen
  \bibfield  {author} {\bibinfo {author} {\bibfnamefont {G.}~\bibnamefont {Kresse}}\ and\ \bibinfo {author} {\bibfnamefont {J.}~\bibnamefont {Hafner}},\ }\bibfield  {title} {\bibinfo {title} {Ab initio molecular dynamics for liquid metals},\ }\href@noop {} {\bibfield  {journal} {\bibinfo  {journal} {Physical review B}\ }\textbf {\bibinfo {volume} {47}},\ \bibinfo {pages} {558} (\bibinfo {year} {1993})}\BibitemShut {NoStop}%
\bibitem [{\citenamefont {Casula}\ and\ \citenamefont {Sorella}(2003)}]{casula2003agpgeminal}%
  \BibitemOpen
  \bibfield  {author} {\bibinfo {author} {\bibfnamefont {M.}~\bibnamefont {Casula}}\ and\ \bibinfo {author} {\bibfnamefont {S.}~\bibnamefont {Sorella}},\ }\bibfield  {title} {\bibinfo {title} {Geminal wave functions with jastrow correlation: A first application to atoms},\ }\href@noop {} {\bibfield  {journal} {\bibinfo  {journal} {J. Chem. Phys.}\ }\textbf {\bibinfo {volume} {119}},\ \bibinfo {pages} {6500} (\bibinfo {year} {2003})}\BibitemShut {NoStop}%
\bibitem [{\citenamefont {Pritchard}\ \emph {et~al.}(2019)\citenamefont {Pritchard}, \citenamefont {Altarawy}, \citenamefont {Didier}, \citenamefont {Gibson},\ and\ \citenamefont {Windus}}]{2019PRI_BSE}%
  \BibitemOpen
  \bibfield  {author} {\bibinfo {author} {\bibfnamefont {B.~P.}\ \bibnamefont {Pritchard}}, \bibinfo {author} {\bibfnamefont {D.}~\bibnamefont {Altarawy}}, \bibinfo {author} {\bibfnamefont {B.~T.}\ \bibnamefont {Didier}}, \bibinfo {author} {\bibfnamefont {T.~D.}\ \bibnamefont {Gibson}},\ and\ \bibinfo {author} {\bibfnamefont {T.~L.}\ \bibnamefont {Windus}},\ }\bibfield  {title} {\bibinfo {title} {A new basis set exchange: An open, up-to-date resource for the molecular sciences community},\ }\href {https://doi.org/10.1021/acs.jcim.9b00725} {\bibfield  {journal} {\bibinfo  {journal} {J. Chem. Inf. Model.}\ }\textbf {\bibinfo {volume} {59}},\ \bibinfo {pages} {4814} (\bibinfo {year} {2019})}\BibitemShut {NoStop}%
\bibitem [{\citenamefont {Nakano}\ \emph {et~al.}(2019)\citenamefont {Nakano}, \citenamefont {Maezono},\ and\ \citenamefont {Sorella}}]{2019NAK_J1_Na2}%
  \BibitemOpen
  \bibfield  {author} {\bibinfo {author} {\bibfnamefont {K.}~\bibnamefont {Nakano}}, \bibinfo {author} {\bibfnamefont {R.}~\bibnamefont {Maezono}},\ and\ \bibinfo {author} {\bibfnamefont {S.}~\bibnamefont {Sorella}},\ }\bibfield  {title} {\bibinfo {title} {All-electron quantum monte carlo with jastrow single determinant ansatz: Application to the sodium dimer},\ }\href@noop {} {\bibfield  {journal} {\bibinfo  {journal} {J. Chem. Theory Comput.}\ }\textbf {\bibinfo {volume} {15}},\ \bibinfo {pages} {4044} (\bibinfo {year} {2019})}\BibitemShut {NoStop}%
\bibitem [{\citenamefont {Reynolds}\ \emph {et~al.}(1986)\citenamefont {Reynolds}, \citenamefont {Barnett}, \citenamefont {Hammond}, \citenamefont {Grimes},\ and\ \citenamefont {Lester~Jr}}]{1986REY_qmcforce}%
  \BibitemOpen
  \bibfield  {author} {\bibinfo {author} {\bibfnamefont {P.}~\bibnamefont {Reynolds}}, \bibinfo {author} {\bibfnamefont {R.}~\bibnamefont {Barnett}}, \bibinfo {author} {\bibfnamefont {B.}~\bibnamefont {Hammond}}, \bibinfo {author} {\bibfnamefont {R.}~\bibnamefont {Grimes}},\ and\ \bibinfo {author} {\bibfnamefont {W.}~\bibnamefont {Lester~Jr}},\ }\bibfield  {title} {\bibinfo {title} {Quantum chemistry by quantum monte carlo: Beyond ground-state energy calculations},\ }\href@noop {} {\bibfield  {journal} {\bibinfo  {journal} {Int. J. Quantum Chem.}\ }\textbf {\bibinfo {volume} {29}},\ \bibinfo {pages} {589} (\bibinfo {year} {1986})}\BibitemShut {NoStop}%
\bibitem [{\citenamefont {Assaraf}\ and\ \citenamefont {Caffarel}(2000)}]{2000ASS_qmcforce}%
  \BibitemOpen
  \bibfield  {author} {\bibinfo {author} {\bibfnamefont {R.}~\bibnamefont {Assaraf}}\ and\ \bibinfo {author} {\bibfnamefont {M.}~\bibnamefont {Caffarel}},\ }\bibfield  {title} {\bibinfo {title} {Computing forces with quantum monte carlo},\ }\href@noop {} {\bibfield  {journal} {\bibinfo  {journal} {J. Chem. Phys.}\ }\textbf {\bibinfo {volume} {113}},\ \bibinfo {pages} {4028} (\bibinfo {year} {2000})}\BibitemShut {NoStop}%
\bibitem [{\citenamefont {Filippi}\ and\ \citenamefont {Umrigar}(2000)}]{2000FIL_qmcforce}%
  \BibitemOpen
  \bibfield  {author} {\bibinfo {author} {\bibfnamefont {C.}~\bibnamefont {Filippi}}\ and\ \bibinfo {author} {\bibfnamefont {C.~J.}\ \bibnamefont {Umrigar}},\ }\bibfield  {title} {\bibinfo {title} {Correlated sampling in quantum monte carlo: A route to forces},\ }\href {https://doi.org/10.1103/PhysRevB.61.R16291} {\bibfield  {journal} {\bibinfo  {journal} {Phys. Rev. B}\ }\textbf {\bibinfo {volume} {61}},\ \bibinfo {pages} {R16291} (\bibinfo {year} {2000})}\BibitemShut {NoStop}%
\bibitem [{\citenamefont {Chiesa}\ \emph {et~al.}(2005)\citenamefont {Chiesa}, \citenamefont {Ceperley},\ and\ \citenamefont {Zhang}}]{2005CHI_qmcforce}%
  \BibitemOpen
  \bibfield  {author} {\bibinfo {author} {\bibfnamefont {S.}~\bibnamefont {Chiesa}}, \bibinfo {author} {\bibfnamefont {D.~M.}\ \bibnamefont {Ceperley}},\ and\ \bibinfo {author} {\bibfnamefont {S.}~\bibnamefont {Zhang}},\ }\bibfield  {title} {\bibinfo {title} {Accurate, efficient, and simple forces computed with quantum monte carlo methods},\ }\href {https://doi.org/10.1103/PhysRevLett.94.036404} {\bibfield  {journal} {\bibinfo  {journal} {Phys. Rev. Lett.}\ }\textbf {\bibinfo {volume} {94}},\ \bibinfo {pages} {036404} (\bibinfo {year} {2005})}\BibitemShut {NoStop}%
\bibitem [{\citenamefont {Badinski}\ and\ \citenamefont {Needs}(2008)}]{2008BAD_qmcforce}%
  \BibitemOpen
  \bibfield  {author} {\bibinfo {author} {\bibfnamefont {A.}~\bibnamefont {Badinski}}\ and\ \bibinfo {author} {\bibfnamefont {R.~J.}\ \bibnamefont {Needs}},\ }\bibfield  {title} {\bibinfo {title} {Total forces in the diffusion monte carlo method with nonlocal pseudopotentials},\ }\href {https://doi.org/10.1103/PhysRevB.78.035134} {\bibfield  {journal} {\bibinfo  {journal} {Phys. Rev. B}\ }\textbf {\bibinfo {volume} {78}},\ \bibinfo {pages} {035134} (\bibinfo {year} {2008})}\BibitemShut {NoStop}%
\bibitem [{\citenamefont {Assaraf}\ \emph {et~al.}(2011)\citenamefont {Assaraf}, \citenamefont {Caffarel},\ and\ \citenamefont {Kollias}}]{2011ASS_qmcforce}%
  \BibitemOpen
  \bibfield  {author} {\bibinfo {author} {\bibfnamefont {R.}~\bibnamefont {Assaraf}}, \bibinfo {author} {\bibfnamefont {M.}~\bibnamefont {Caffarel}},\ and\ \bibinfo {author} {\bibfnamefont {A.~C.}\ \bibnamefont {Kollias}},\ }\bibfield  {title} {\bibinfo {title} {Chaotic versus nonchaotic stochastic dynamics in monte carlo simulations: A route for accurate energy differences in $n$-body systems},\ }\href {https://doi.org/10.1103/PhysRevLett.106.150601} {\bibfield  {journal} {\bibinfo  {journal} {Phys. Rev. Lett.}\ }\textbf {\bibinfo {volume} {106}},\ \bibinfo {pages} {150601} (\bibinfo {year} {2011})}\BibitemShut {NoStop}%
\bibitem [{\citenamefont {Moroni}\ \emph {et~al.}(2014)\citenamefont {Moroni}, \citenamefont {Saccani},\ and\ \citenamefont {Filippi}}]{2014MOR_qmcforce}%
  \BibitemOpen
  \bibfield  {author} {\bibinfo {author} {\bibfnamefont {S.}~\bibnamefont {Moroni}}, \bibinfo {author} {\bibfnamefont {S.}~\bibnamefont {Saccani}},\ and\ \bibinfo {author} {\bibfnamefont {C.}~\bibnamefont {Filippi}},\ }\bibfield  {title} {\bibinfo {title} {Practical schemes for accurate forces in quantum monte carlo},\ }\href {https://doi.org/10.1021/ct500780r} {\bibfield  {journal} {\bibinfo  {journal} {J. Chem. Theory Comput.}\ }\textbf {\bibinfo {volume} {10}},\ \bibinfo {pages} {4823} (\bibinfo {year} {2014})}\BibitemShut {NoStop}%
\bibitem [{\citenamefont {Van~Rhijn}\ \emph {et~al.}(2021)\citenamefont {Van~Rhijn}, \citenamefont {Filippi}, \citenamefont {De~Palo},\ and\ \citenamefont {Moroni}}]{2021VAN_qmcforce}%
  \BibitemOpen
  \bibfield  {author} {\bibinfo {author} {\bibfnamefont {J.}~\bibnamefont {Van~Rhijn}}, \bibinfo {author} {\bibfnamefont {C.}~\bibnamefont {Filippi}}, \bibinfo {author} {\bibfnamefont {S.}~\bibnamefont {De~Palo}},\ and\ \bibinfo {author} {\bibfnamefont {S.}~\bibnamefont {Moroni}},\ }\bibfield  {title} {\bibinfo {title} {Energy derivatives in real-space diffusion monte carlo},\ }\href {https://doi.org/10.1021/acs.jctc.1c00496} {\bibfield  {journal} {\bibinfo  {journal} {J. Chem. Theory Comput.}\ }\textbf {\bibinfo {volume} {18}},\ \bibinfo {pages} {118} (\bibinfo {year} {2021})}\BibitemShut {NoStop}%
\bibitem [{\citenamefont {Slootman}\ \emph {et~al.}(2024)\citenamefont {Slootman}, \citenamefont {Poltavsky}, \citenamefont {Shinde}, \citenamefont {Cocomello}, \citenamefont {Moroni}, \citenamefont {Tkatchenko},\ and\ \citenamefont {Filippi}}]{slootman2024accurate}%
  \BibitemOpen
  \bibfield  {author} {\bibinfo {author} {\bibfnamefont {E.}~\bibnamefont {Slootman}}, \bibinfo {author} {\bibfnamefont {I.}~\bibnamefont {Poltavsky}}, \bibinfo {author} {\bibfnamefont {R.}~\bibnamefont {Shinde}}, \bibinfo {author} {\bibfnamefont {J.}~\bibnamefont {Cocomello}}, \bibinfo {author} {\bibfnamefont {S.}~\bibnamefont {Moroni}}, \bibinfo {author} {\bibfnamefont {A.}~\bibnamefont {Tkatchenko}},\ and\ \bibinfo {author} {\bibfnamefont {C.}~\bibnamefont {Filippi}},\ }\bibfield  {title} {\bibinfo {title} {Accurate quantum monte carlo forces for machine-learned force fields: Ethanol as a benchmark},\ }\href@noop {} {\bibfield  {journal} {\bibinfo  {journal} {Journal of chemical theory and computation}\ }\textbf {\bibinfo {volume} {20}},\ \bibinfo {pages} {6020} (\bibinfo {year} {2024})}\BibitemShut {NoStop}%
\bibitem [{\citenamefont {Harris}\ \emph {et~al.}(1965)\citenamefont {Harris}, \citenamefont {Engerholm},\ and\ \citenamefont {Gwinn}}]{harris1965calculation}%
  \BibitemOpen
  \bibfield  {author} {\bibinfo {author} {\bibfnamefont {D.~O.}\ \bibnamefont {Harris}}, \bibinfo {author} {\bibfnamefont {G.~G.}\ \bibnamefont {Engerholm}},\ and\ \bibinfo {author} {\bibfnamefont {W.~D.}\ \bibnamefont {Gwinn}},\ }\bibfield  {title} {\bibinfo {title} {Calculation of matrix elements for one-dimensional quantum-mechanical problems and the application to anharmonic oscillators},\ }\href@noop {} {\bibfield  {journal} {\bibinfo  {journal} {Journal of Chemical Physics}\ }\textbf {\bibinfo {volume} {43}},\ \bibinfo {pages} {1515} (\bibinfo {year} {1965})}\BibitemShut {NoStop}%
\bibitem [{\citenamefont {Irikura}(2007)}]{irikura2007experimental}%
  \BibitemOpen
  \bibfield  {author} {\bibinfo {author} {\bibfnamefont {K.~K.}\ \bibnamefont {Irikura}},\ }\bibfield  {title} {\bibinfo {title} {Experimental vibrational zero-point energies: Diatomic molecules},\ }\href@noop {} {\bibfield  {journal} {\bibinfo  {journal} {Journal of physical and chemical reference data}\ }\textbf {\bibinfo {volume} {36}},\ \bibinfo {pages} {389} (\bibinfo {year} {2007})}\BibitemShut {NoStop}%
\bibitem [{\citenamefont {Dahl}\ and\ \citenamefont {Springborg}(1988)}]{dahl1988morse}%
  \BibitemOpen
  \bibfield  {author} {\bibinfo {author} {\bibfnamefont {J.~P.}\ \bibnamefont {Dahl}}\ and\ \bibinfo {author} {\bibfnamefont {M.}~\bibnamefont {Springborg}},\ }\bibfield  {title} {\bibinfo {title} {The morse oscillator in position space, momentum space, and phase space},\ }\href@noop {} {\bibfield  {journal} {\bibinfo  {journal} {The Journal of chemical physics}\ }\textbf {\bibinfo {volume} {88}},\ \bibinfo {pages} {4535} (\bibinfo {year} {1988})}\BibitemShut {NoStop}%
\bibitem [{\citenamefont {{Chabrier}}\ \emph {et~al.}(2019)\citenamefont {{Chabrier}}, \citenamefont {{Mazevet}},\ and\ \citenamefont {{Soubiran}}}]{2019ApJ...872...51C}%
  \BibitemOpen
  \bibfield  {author} {\bibinfo {author} {\bibfnamefont {G.}~\bibnamefont {{Chabrier}}}, \bibinfo {author} {\bibfnamefont {S.}~\bibnamefont {{Mazevet}}},\ and\ \bibinfo {author} {\bibfnamefont {F.}~\bibnamefont {{Soubiran}}},\ }\bibfield  {title} {\bibinfo {title} {{A New Equation of State for Dense Hydrogen-Helium Mixtures}},\ }\href {https://doi.org/10.3847/1538-4357/aaf99f} {\bibfield  {journal} {\bibinfo  {journal} {ApJ}\ }\textbf {\bibinfo {volume} {872}},\ \bibinfo {eid} {51} (\bibinfo {year} {2019})}\BibitemShut {NoStop}%
\end{thebibliography}%

\end{document}